\documentclass{cernyrep}
\usepackage{graphicx}
\usepackage{colordvi}
\usepackage{color}
\usepackage{latexsym}
\usepackage{epsfig}
\usepackage{amsmath}
\newcommand{\nc}{\newcommand}
\nc{\beq}{\begin{equation}} \nc{\eeq}{\end{equation}}
\nc{\beqa}{\begin{eqnarray}} \nc{\eeqa}{\end{eqnarray}}
\nc{\tb}{\tan\beta}
\nc{\bsg}{b\to s\gamma}
\nc{\btaunu}{b\to\tau\nu}
\nc{\bsmm}{B_s\to\mu\mu}
\nc{\gl}{\tilde g}
\nc{\sq}{\tilde q}
\nc{\mzero}{m_0}
\nc{\mhalf}{m_{1/2}}

\begin{document} \thispagestyle{empty}
\begin{center}
{\Large \bf  IS (Low Energy) SUSY STILL ALIVE?}
\renewcommand{\thefootnote}{\fnsymbol{footnote}}\footnote[2]{Lectures
given at the European School of High-Energy Physics, June 2012, Anjou, France}
\renewcommand{\thefootnote}{\arabic{footnote}}
\vspace{10mm}

{\large \bf A.V.~Gladyshev and D.I.~Kazakov } \\[5mm]

{\it Bogoliubov Laboratory of Theoretical Physics, JINR,
Dubna, Russia \\[0.1cm]
and\\[0.1cm]
Institute for Theoretical and Experimental Physics, Moscow, Russia}

\vspace{12mm}

\begin{center}
\includegraphics[width=0.407\textwidth]{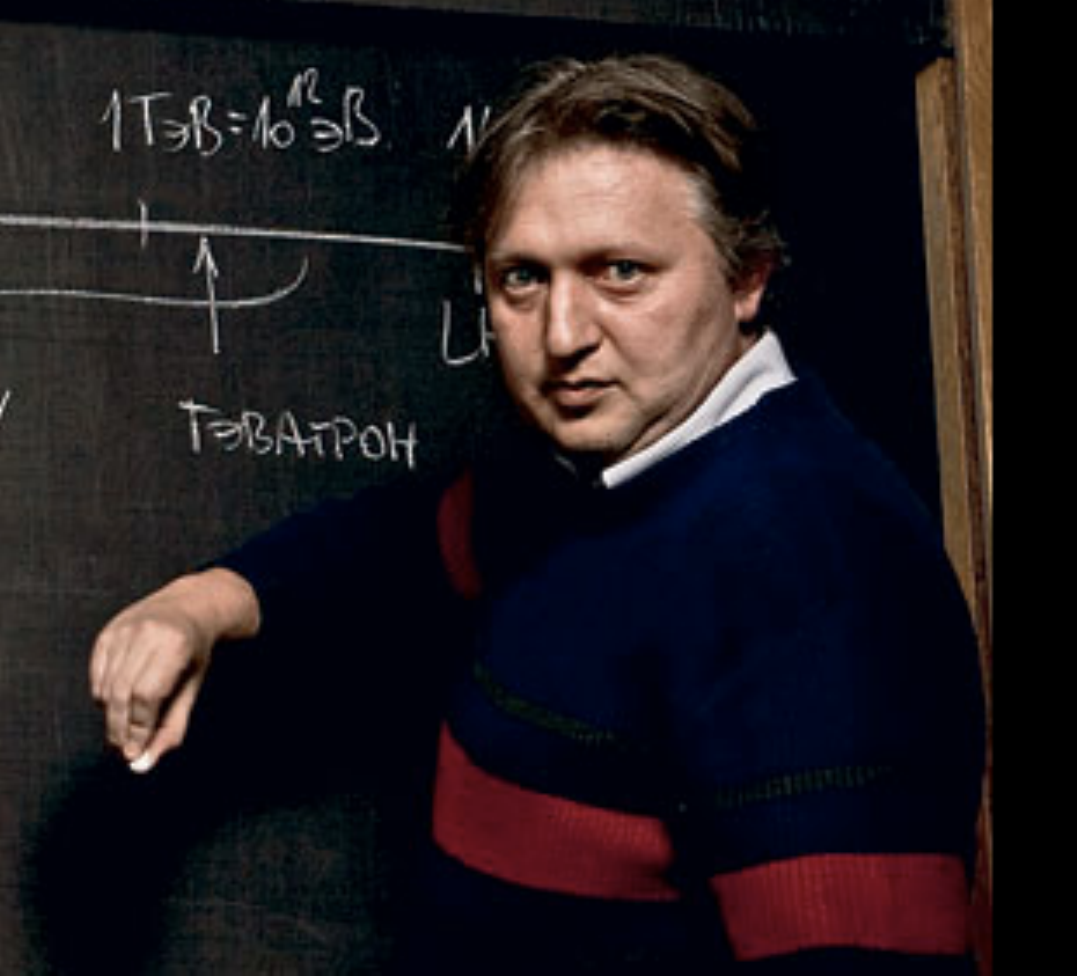}
\includegraphics[width=0.40\textwidth]{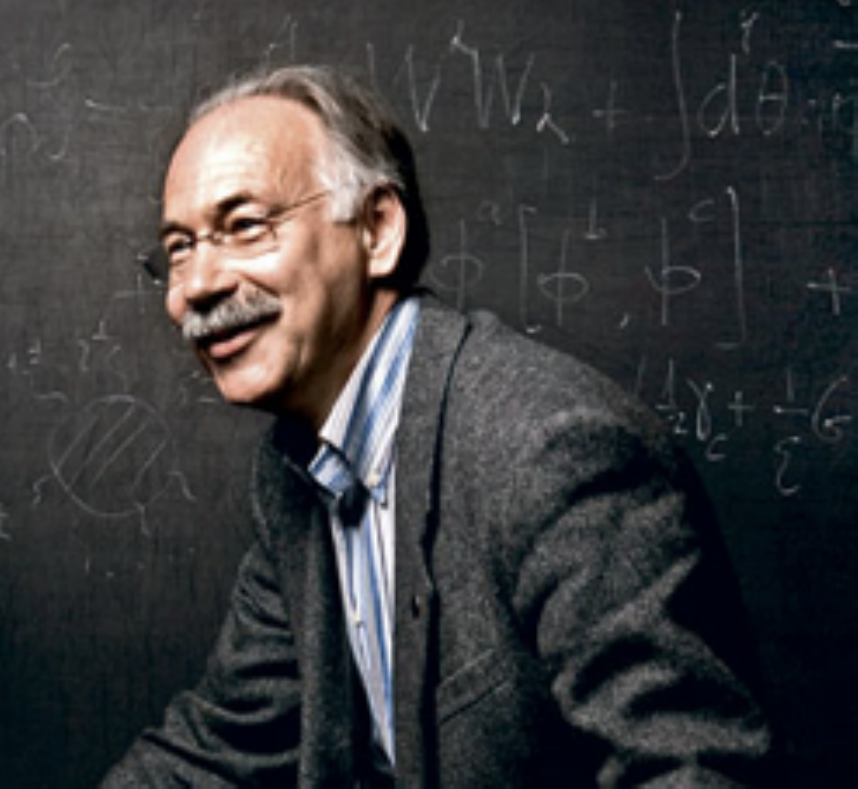}
\end{center}
(courtesy of A.Maishev, "Russian Reporter")
\vspace{8mm}

\begin{abstract}
Supersymmetry, a new symmetry that relates bosons and fermions
in particle physics, still escapes observation. Search for
supersymmetry is one of the main aims of the  Large Hadron
Collider. The other possible manifestation of supersymmetry is
the Dark Matter in the Universe. The present lectures contain a
brief introduction to supersymmetry in particle physics. The
main notions of supersymmetry  are introduced. The
supersymmetric extension of the Standard Model -- the
Minimal Supersymmetric Standard Model --  is considered in more
detail. Phenomenological features of the Minimal Supersymmetric
Standard Model as well as possible experimental signatures of
supersymmetry at the Large Hadron Collider are described. The
present limits on supersymmetric particles are presented and
the allowed region of parameter space of the MSSM is shown.
%
%\vspace{1pc}
\end{abstract}
\end{center}

\thispagestyle{empty}

%\vspace{8mm}

\tableofcontents 
%\vglue 0.34cm {\bf References} \hfill {\bf 49}
\vspace{8mm}
%\newpage

%--------------------------------------------
\section{Introduction: What is supersymmetry}

{\it Supersymmetry} is a {\it boson-fermion} symmetry that is
aimed to unify all forces in Nature including gravity within a
singe framework~\cite{super,Rev,WessB,sspace,Books}. Modern views on
supersymmetry in particle physics are based on a string
paradigm, though low energy manifestations of supersymmetry
(SUSY) can be possibly found at modern colliders and in
non-accelerator experiments.

Supersymmetry emerged from attempts to generalize the Poincar\'e
algebra to mix representations with different spin~\cite{super}.
It happened to be a problematic task due to ``no-go'' theorems
preventing such generalizations~\cite{theorem}. The way out was
found by introducing so-called graded Lie algebras, i.~e. adding
anti-commutators to usual commutators of the Lorentz algebra.
Such a generalization, described below, appeared to be the only
possible one within the relativistic field theory.

If $Q$ is a generator of the SUSY algebra, then acting on a
boson state it produces a fermion one and vice versa
$$
\bar Q \, | \text{boson} \rangle = | \text{fermion} \rangle , \ \
Q \, | \text{fermion} \rangle = | \text{boson} \rangle.
$$

Since the bosons commute with each other and the fermions
anticommute, one immediately finds that the SUSY generators
should also anticommute, they must be {\em fermionic}, i.~e.
they must change the spin by a half-odd amount and change the
statistics. The key element of the SUSY algebra is
\begin{equation}
\label{ant}
\{Q_\alpha, \bar{Q}_{\dot \alpha}\}=2\,\sigma_{\alpha \dot
\alpha}^\mu P_\mu
\end{equation}
where $Q$ and $\bar Q$ are the generators of the supersymmetry
transformation and $P_\mu$ is the generator of translation,
the four-momentum.

In what follows we describe the SUSY algebra in more detail and
construct its representations which are needed to build the
SUSY generalization of the Standard Model (SM) of fundamental
interactions. Such a generalization is based on a softly broken
SUSY quantum filed theory and contains the SM as the low energy
theory.

Supersymmetry promises to solve some problems of the Standard
Model and of Grand Unified Theories. In what follows we describe
supersymmetry as the nearest option for the new physics on the
TeV scale.

%------------------------------------------------
\section{Motivation for SUSY in particle physics}
%\setcounter{equation} 0

%------------------------------------
\subsection{Unification with gravity}

The {\em general idea} is a unification of all forces of Nature
including quantum gravity. However, the graviton has the spin 2,
while other gauge bosons (the photon, gluons, $W$ and $Z$ weak
bosons) have the spin 1. Therefore, they correspond to different
representations of the Poincar\'e algebra. To mix them one can
use supersymmetry transformations. Starting with the graviton
state of the spin 2 and acting by the SUSY generators we get the
following chain of states:
$$
\text{spin} \ 2 \ \rightarrow \
\text{spin} \ \frac 32 \ \rightarrow \
\text{spin} \ 1 \ \rightarrow \
\text{spin} \ \frac 12 \ \rightarrow \
\text{spin} \ 0.
$$
Thus, the partial unification of matter (the fermions) with
forces (the bosons) naturally arises from an attempt to unify
gravity with the other interactions.

Taking infinitesimal transformations
$\delta_\epsilon = \epsilon^\alpha Q_\alpha, \
\bar{\delta}_{\bar \epsilon} =
\bar{Q}_{\dot \alpha}{\bar \epsilon}^{\dot \alpha},$
and using Eqn.~(\ref{ant}) one gets
\begin{equation}
\{\delta_\epsilon,\bar{\delta}_{\bar \epsilon} \} =
2 \, (\epsilon \sigma^\mu \bar \epsilon )P_\mu ,
\label{com}
\end{equation}
where $\epsilon, \bar\epsilon$ are transformation parameters.
Choosing $\epsilon$ to be local, i.~e. the function of the
space-time point $\epsilon = \epsilon(x)$, one finds from
Eqn.~(\ref{com}) that the anticommutator of two SUSY
transformations is a local coordinate translation, and the
theory which is invariant under the local coordinate
transformation is the General Relativity. Thus, making SUSY
local, one naturally obtains the General Relativity, or the
theory of gravity, or supergravity~\cite{Rev}.

%--------------------------------------------
\subsection{Unification of gauge couplings}

According to the  Grand Unification  {\em hypothesis}, the gauge
symmetry increases with the energy~\cite{GUT}. All known
interactions are different branches of the unique interaction
associated with a simple gauge group. The unification (or
splitting) occurs at the high energy. To reach this goal one has
to consider how the couplings change with the energy. It is
described by renormalization group equations. In the SM the
strong and weak couplings associated with the non-Abelian gauge
groups decrease with the energy, while the electromagnetic one
associated with the Abelian group on the contrary increases.
Thus, it is possible that at some energy scale they are
equal.

After the precise measurement of the $SU(3)\times SU(2) \times
U(1)$ coupling constants, it has become possible to check the
unification numerically. The three coupling constants to be
compared are
\begin{equation}
\begin{split}
\alpha_1 &=(5/3)g^{\prime2}/(4\pi)=5\alpha/(3\cos^2\theta_W),\\
\alpha_2 &= g^2/(4\pi)=\alpha/\sin^2\theta_W, \\
\alpha_3 &= g_s^2/(4\pi)
\end{split}
\end{equation}
where $g',~g$ and $g_s$ are the usual $U(1)$, $SU(2)$ and $SU(3)$
couplings and $\alpha$ is the fine structure constant.
The factor of 5/3 in $\alpha_1$ has been included for proper
normalization of the generators.

In the modified minimal subtraction ($\overline{MS}$) scheme, the
world averaged values of the couplings at the $Z^0$ energy are
obtained from the fit to the LEP and Tevatron data~\cite{SM}:
\begin{align}
\begin{split}
\alpha^{-1}(M_Z)             &= 128.978\pm 0.027 \\
\sin^2\theta_{\overline{MS}} &= 0.23146\pm 0.00017 \\
\alpha_s                     &= 0.1184\pm 0.0031,
\label{worave}
\end{split}
\intertext{that gives}
\begin{split}
\alpha_1(M_Z) &= 0.017, \\
\alpha_2(M_Z) &= 0.034, \\
\alpha_3(M_Z) &= 0.118\pm 0.003.
\end{split}
\end{align}
Assuming that the SM is valid up to the unification
scale, one can then use the known RG equations for the three
couplings. In the leading order they are:
\begin{equation}
\frac{d\tilde{\alpha}_i}{dt} =  b_i\tilde{\alpha}_i^2, \ \ \ \ \
\tilde{\alpha}_i = \frac{\alpha_i}{4\pi}, \ \ \ \ \
t = \log \left( \frac{Q^2}{\mu^2} \right),
\label{alpha}
\end{equation}
where the coefficients for the SM are $b_i=(41/10, -19/6, -7)$.

\begin{figure}
\begin{center}
\leavevmode
\includegraphics[width=0.84\textwidth]{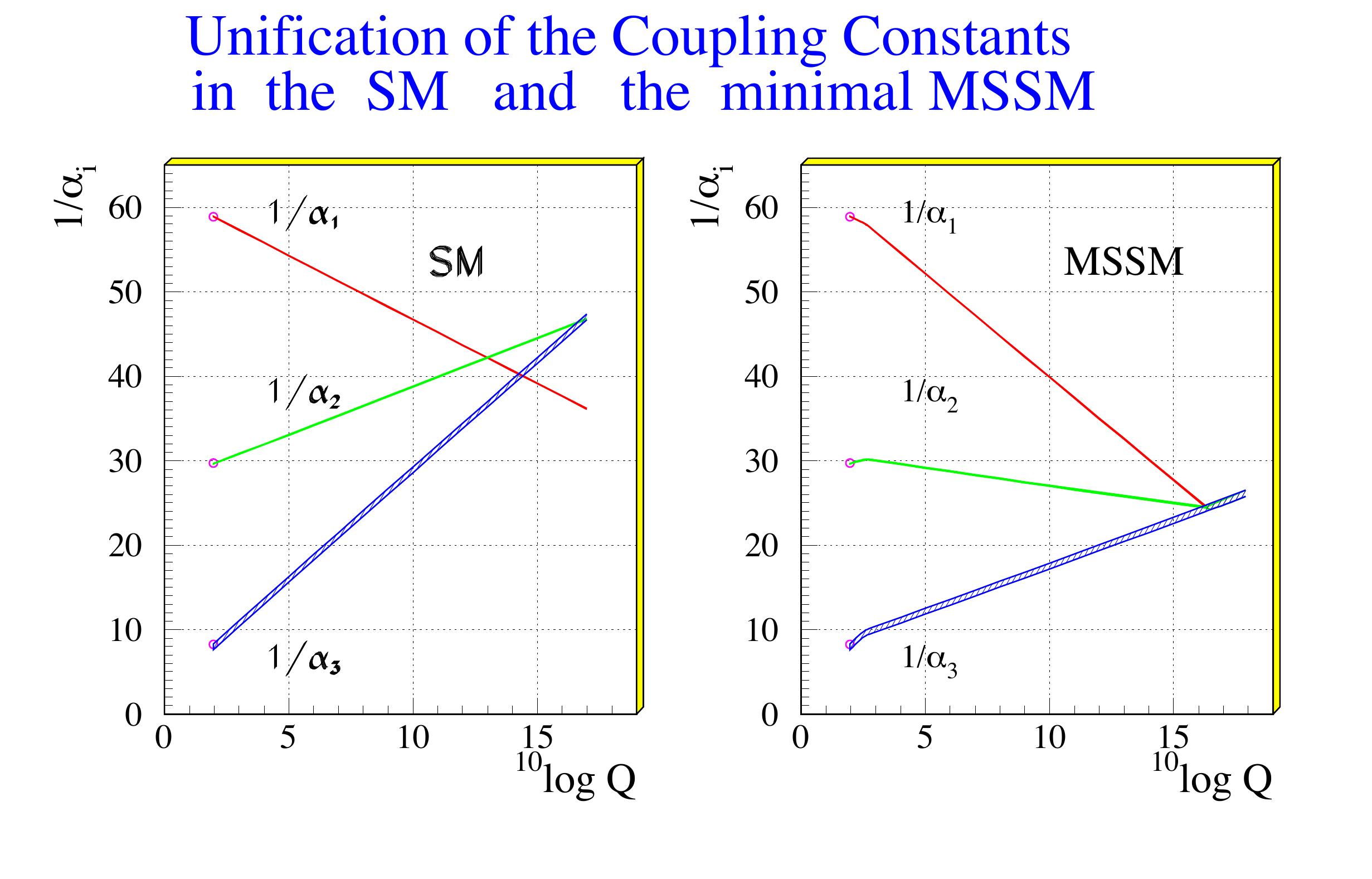}
\vspace*{-10mm}
\end{center}
\caption{The evolution of the inverse of the three coupling
constants in the Standard Model (left) and in the supersymmetric
extension of the SM (MSSM) (right).}
\label{unif}
\end{figure}

The solution to Eqn.~(\ref{alpha}) is very simple
\begin{equation}
\frac{1}{\tilde{\alpha}_i(Q^2)} =
\frac{1}{\tilde{\alpha}_i(\mu^2)} -
b_i \log \left( \frac{Q^2}{\mu^2} \right).
\label{alphasol}
\end{equation}
The result is demonstrated in Fig.~\ref{unif} showing the
evolution of the inverse of the couplings as a function of the
logarithm of energy. In this presentation, the evolution becomes
a straight line in the first order. The second order corrections
are small and do not cause any visible deviation from the
straight line. Fig.~\ref{unif} clearly demonstrates that within
the SM the coupling constant unification at a single point is
impossible. It is excluded by more than 8 standard deviations.
This result means that the unification can only be obtained if
the new physics enters between the electroweak and the Planck
scales.

%----------------- Unification of couplings
%

In the SUSY case, the slopes of the RG evolution curves are
modified. The coefficients $b_i$ in Eqn.~(\ref{alpha}) now are
$b_i = (33/5, 1, -3)$. The SUSY particles are assumed to
contribute effectively to the running of the coupling constants
only for the energies above the typical SUSY mass scale. It
turns out that within the SUSY model the perfect unification can
be obtained as it is shown in Fig.~\ref{unif}. From the fit
requiring the unification one finds for the break point
$M_{SUSY}$ and the unification point $M_{GUT}$~\cite{ABF}
\begin{equation}
\begin{split}
M_{SUSY} &= 10^{3.4\pm 0.9\pm 0.4}~\text{GeV}, \\
M_{GUT} &= 10^{15.8\pm 0.3\pm 0.1}~\text{GeV}, \\
\alpha^{-1}_{GUT} &= 26.3 \pm 1.9 \pm 1.0.
\end{split}
\label{MSUSY}
\end{equation}
The first error originates from the uncertainty in the coupling
constant, while the second one is due to the uncertainty in the
mass splitting between the SUSY particles.

This observation was considered as the first ``evidence'' for
supersymmetry, especially since $M_{SUSY}$ was found in the
range preferred by the fine-tuning arguments.

%---------------------------------------------
\subsection{Solution to the hierarchy problem}

The appearance of two different scales $V \gg v$ in the GUT
theory, namely, $M_{GUT}$ and $M_W$, leads to a very  serious
problem which is called the {\em hierarchy problem}. There are
two aspects of this problem.

The first one is the very existence of the hierarchy. To get
the desired spontaneous symmetry breaking pattern, one needs
\begin{equation}
\begin{split}
m_H        &\sim v \sim 10^2~\text{GeV} \\[1mm]
m_{\Sigma} &\sim V \sim 10^{16}~\text{GeV}
\end{split} \ \ \ \ \
%\text{\raisebox{2pt}{$\frac{m_H}{m_\Sigma} \sim 10^{-14} \ll 1$}},
\ \ \ \frac{m_H}{m_\Sigma} \sim 10^{-14} \ll 1,
\label{hier}
\end{equation}
where $H$ and $\Sigma$ are the Higgs fields responsible for the
spontaneous breaking of the $SU(2)$ and GUT group,
respectively. The question arises of how to get so small number
in a natural way.

\begin{figure}[htb]
\begin{center}
\leavevmode
\includegraphics[width=0.44\textwidth]{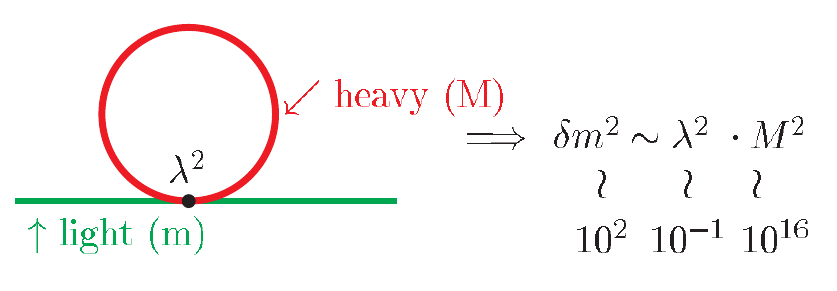}
\end{center}
\vspace{-5mm}
\caption{Radiative correction to the light Higgs boson mass}
\label{fig:hierar}
\end{figure}

The second aspect of the hierarchy problem is connected with
the preservation of the given hierarchy. Even if we choose the
hierarchy like in Eqn.~(\ref{hier}) the radiative corrections
will destroy~it! To see this, let us consider the radiative
correction to the light Higgs mass given by the Feynman
diagram shown in Fig.~\ref{fig:hierar}.
\begin{figure}[htb]
\begin{center}
\leavevmode
\includegraphics[width=0.49\textwidth]{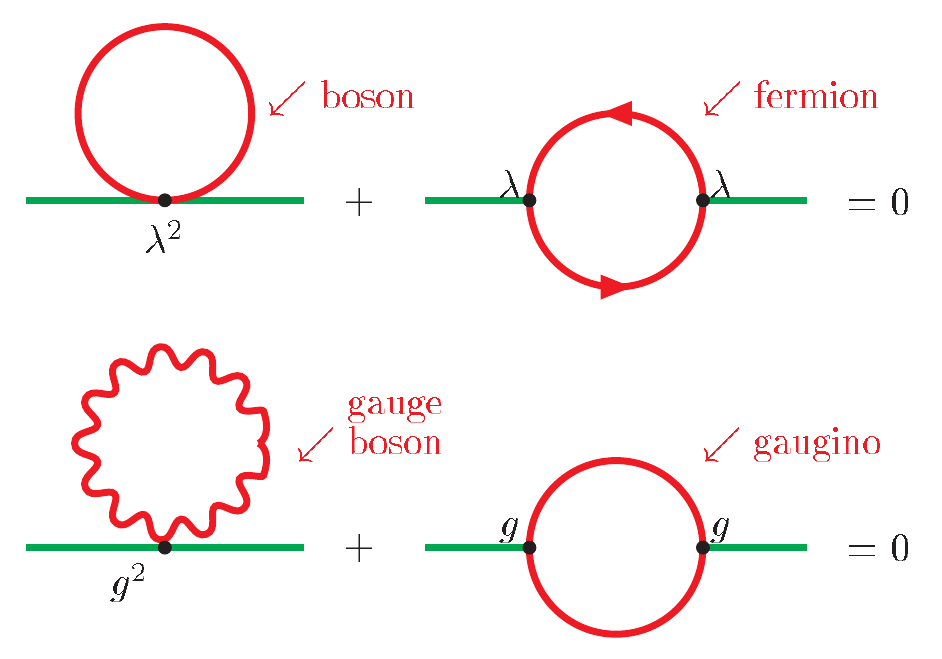}
\end{center}
\vspace{-6mm}
\caption{Cancellation of the quadratic terms (divergencies).}
\label{fig:cancel}
\end{figure}
This correction which is proportional to the mass squared of the heavy
particle, obviously, spoils the hierarchy if it is not
cancelled. This very accurate cancelation with a precision
$\sim 10^{-14}$ needs a fine-tuning of the coupling constants.

The only known way of achieving this kind of cancelation of
quadratic terms (also known as the cancelation of the
quadratic divergencies) is supersymmetry. Moreover, SUSY
automatically cancels the quadratic corrections in all orders
of the perturbation theory. This is due to the contributions
of superpartners of ordinary particles. The contribution from
boson loops cancels those from the fermion ones because of an
additional factor ($-1$) coming from the Fermi statistics,
as shown in Fig.~\ref{fig:cancel}.

One can see here two types of contribution. The first line is
the contribution of the heavy Higgs boson and its superpartner
(higgsino). The strength of the interaction is given by the
Yukawa coupling constant $\lambda$. The second line represents
the gauge interaction proportional to the gauge coupling
constant $g$ with the contribution from the heavy gauge boson
and its heavy superpartner (gaugino).

In both cases the cancelation of the quadratic terms takes
place. This cancelation is true up to the SUSY breaking scale,
$M_{SUSY}$, which should not be very large ($\leq$ 1 TeV) to
make the fine-tuning natural. Indeed, let us take the Higgs
boson mass. Requiring for consistency of the perturbation
theory that the radiative corrections to the Higgs boson mass
do not exceed the mass itself gives
\begin{equation}
\delta M_h^2 \sim g^2 M^2_{SUSY} \sim M_h^2.
\label{del}
\end{equation}
So, if $M_h \sim 10^2$ GeV and $g \sim 10^{-1}$, one needs
$M_{SUSY} \sim 10^3$ GeV in order that the relation~(\ref{del})
is valid. Thus, we again get the same rough estimate of
$M_{SUSY} \sim $ 1 TeV as from the gauge coupling unification
above.

That is why it is usually said that supersymmetry solves the
hierarchy problem. We show below how SUSY can also explain the
origin of the hierarchy.

%--------------------------------------
\subsection{Astrophysics and Cosmology}

\begin{figure}[htb]
\begin{center}
\leavevmode
\includegraphics[width=0.51\textwidth]{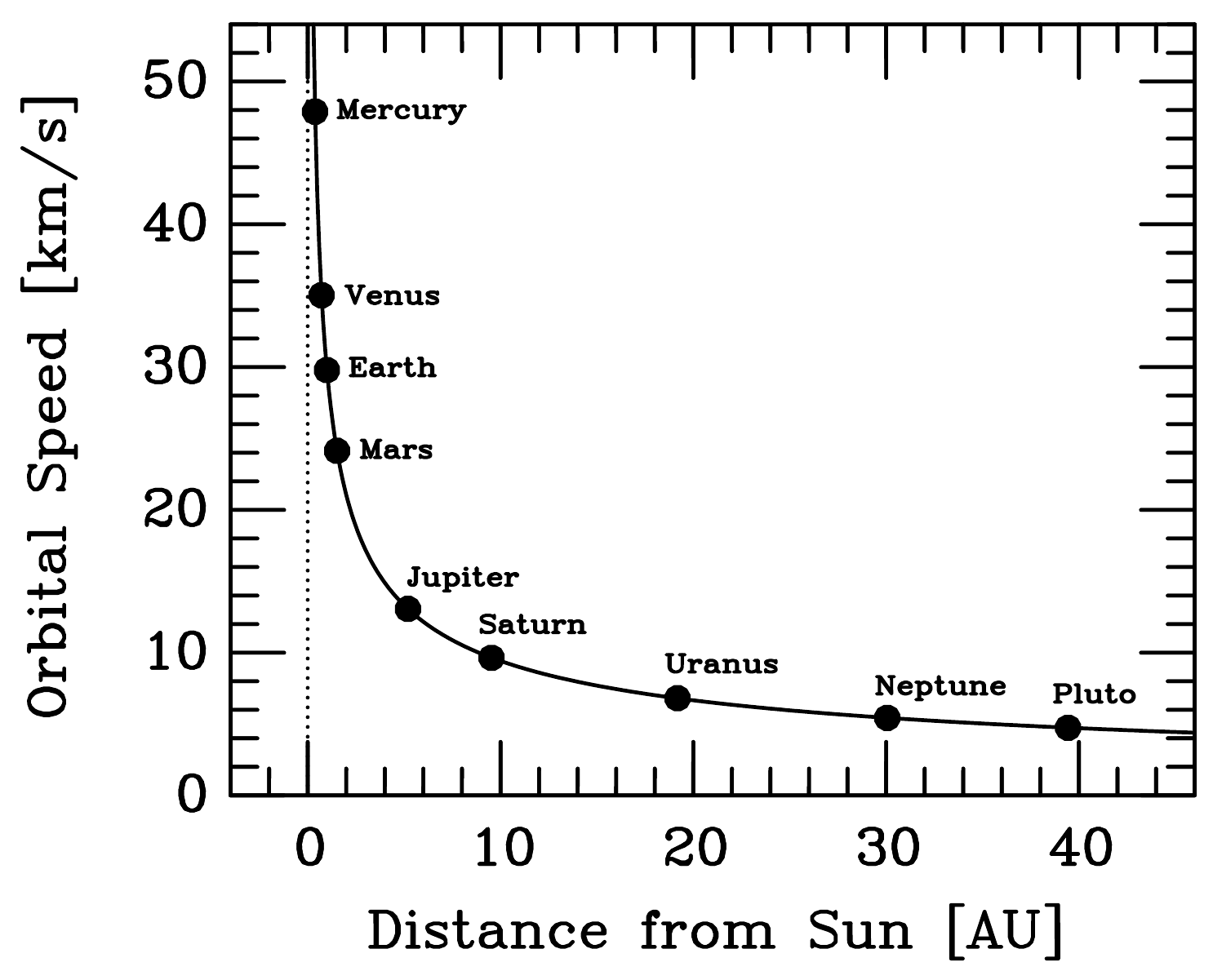}
\hspace*{0.05\textwidth}
\raisebox{8.1pt}{\includegraphics[width=0.41\textwidth]{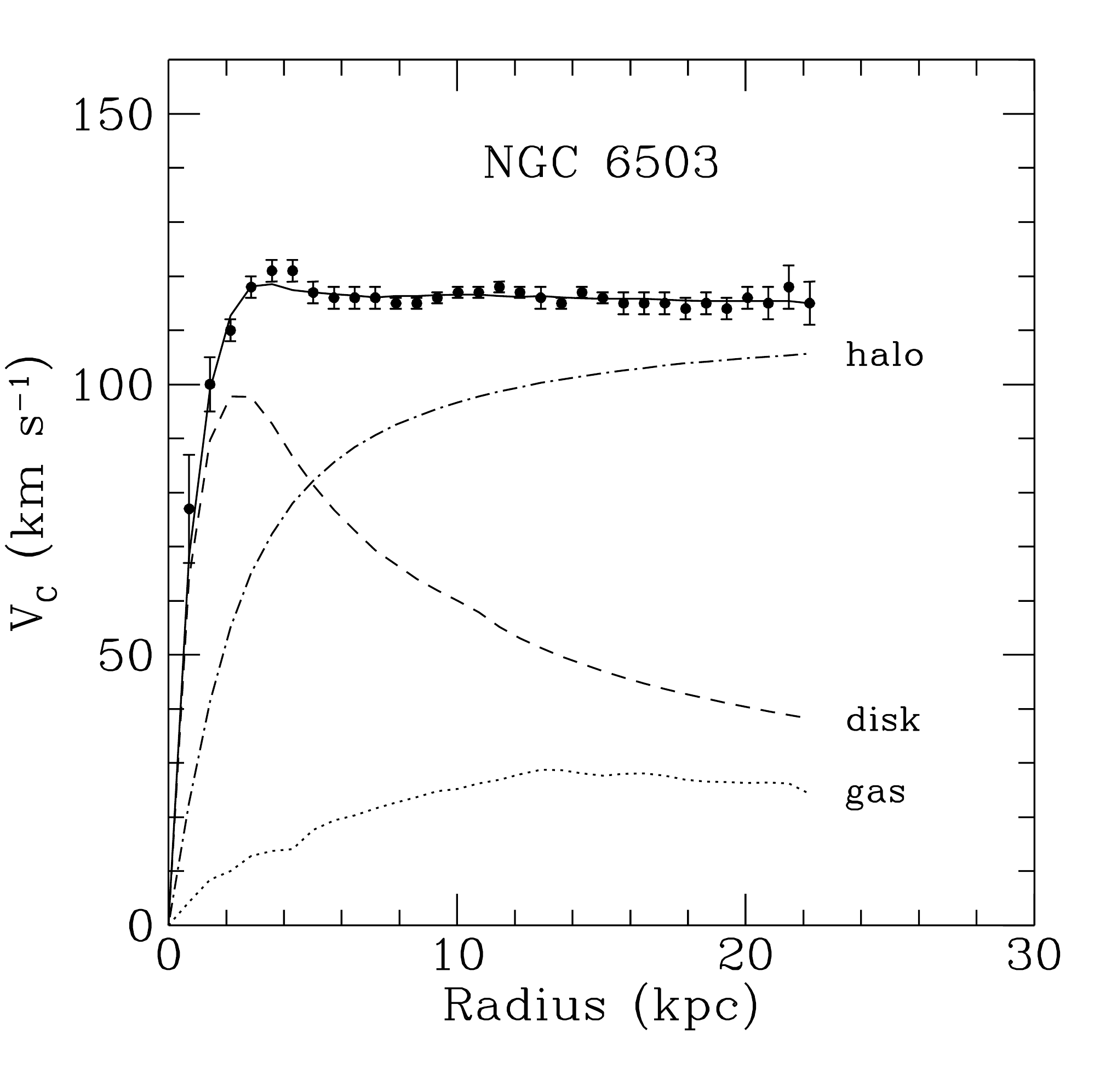}}
\vspace{-4mm}
\end{center}
\caption{Rotation curves for the solar system and the spiral
galaxy.}
\label{gal}
\end{figure}

The shining matter is not the only one in the Universe.
Considerable amount of the energy budget consists of the
so-called dark matter. The direct evidence for the presence of
the dark matter are flat rotation curves of spiral
galaxies~\cite{rotcurve} (see Fig.~\ref{gal}). To explain these
curves one has to assume the existence of a galactic halo made
of non-shining matter which takes part in the gravitational
interaction. The halo has a size more than twice bigger than
a visible galaxy.
The other manifestation of existence of the dark matter is the
so-called gravitational lensing caused by invisible gravitating
matter in the sky~\cite{lensing}, which leads to the appearance
of circular images of distant stars when the light from them
passes through the dark matter.

There are two possible types of the dark matter: the hot one,
consisting of light relativistic particles and the cold one,
consisting of massive weakly interacting particles
(WIMPs)~\cite{WIMP}. The hot dark matter might consist of
neutrinos, however, this has problems with the galaxy formation.
As for the cold dark matter, it has no candidates within the
SM. At the same time, SUSY provides an excellent candidate
for the cold dark matter, namely, the neutralino, the
lightest superparticle~\cite{abun}. It is neutral, heavy, stable
and takes part in weak interactions, precisely what is needed
for a WIMP.

%------------------------------------------
\subsection{Integrability and superstrings}

Considerable progress in the SUSY field theories in recent
years has shown that they possess some remarkable and
attractive properties. For instance, the $N=4$ maximally
supersymmetric Yang-Mills theory has all the features and seems
to provide the first integrable model in 4 space-time
dimensions. This model, though being unphysical, attracts much
attention nowadays. It has no ultraviolet divergences, keeps
conformal invariance at the quantum level and seems to provide
exact solutions for the amplitudes. Duality of this theory to
the string theory in higher dimensions (AdS/CFT correspondence)
allows to go beyond the perturbation theory thus revealing the
strong coupling regime. This properties distinguish the SYSY
theories by their mathematical nature.

Another motivation for supersymmetry follows from even more
radical changes of the basic ideas related to the ultimate goal
of the construction of the consistent unified theory of
everything. At the moment the only viable conception is the
superstring theory~\cite{string}. In the superstring theory,
the strings are considered as the fundamental objects, closed
or open, and are nonlocal in their nature. The ordinary
particles are considered as string excitation modes. The
interactions of the strings, which are local, generate proper
interactions of the usual particles, including the
gravitational ones.

To be consistent, the string theory should be conformally
invariant in a $D$-dimensional target space and have a stable
vacuum. The first requirement is valid in the classical theory
but may be violated by quantum anomalies. The cancelation of
the quantum anomalies takes place when the space-time dimension
of the target space equals to the critical one which is
$D_c=26$ for the bosonic string and $D_c=10$ for the
fermionic one.

The second requirement is that the massless string excitations
(the particles of the SM) are stable. This assumes the absence
of tachyons, the states with the imaginary mass, which can be
guaranteed only in the supersymmetric string theories!

The low energy limit of string theories is a kind of
supergravity theory which is a local supersymmetric theory.
Besides Einstein gravity it contains new interactions and
particles, among them the superpartner of a graviton --
gravitino, a fermion with spin 3/2. Supergravity itself is not
a consistent quantum field theory and is usually treated as an
effective theory. It is used in supersymmetric models of
particle physics to provide the soft supersymmetry breaking
terms.

%--------------------------
\subsection{Where is SUSY?}

After many years of unsuccessful hunt for supersymmetry in
particle physics experiments the natural question arises: where
is supersymmetry? We try to answer this question describing
searches for SUSY at accelerators, in the deep sky with the help
of telescopes, and with the help of the underground facilities.
It is obvious,  that  only direct detection of superpartners  can  convince
people in discovery of supersymmetry, however combined information
from the sky might give hints to the mass spectra and confirm  the SUSY
interpretation of the data.  

It seems that despite the absence of confirmation supersymmetry stays an unbeatable
candidate for physics beyond the Standard Model. The beauty of SUSY lies in the paradigm of
unification of all forces of Nature, the ultimate theory of everything. Therefore search for supersymmetry 
will continue at LHC and perhaps after it.

%--------------------------------
\section{Basics of supersymmetry}
%\setcounter{equation} 0

%---------------------------
\subsection{Algebra of SUSY}

Combined with the usual Poincar\'e and internal symmetry
algebra the Super-Poincar\'e Lie algebra contains additional
SUSY generators
$Q_{\alpha}^i$ and $\bar Q_{\dot\alpha}^i$~\cite{WessB}
\begin{equation}
\begin{split}
&[P_\mu, P_\nu] = 0, \\
&[P_\mu, M_{\rho\sigma}] = i \,(g_{\mu\rho} P_\sigma - g_{\mu\sigma} P_\rho), \\
&[M_{\mu\nu}, M_{\rho\sigma}] =
    i \,(g_{\nu\rho} M_{\mu\sigma} - g_{\nu\sigma} M_{\mu\rho}
    - g_{\mu\rho} M_{\nu \sigma} + g_{\mu\sigma} M_{\nu\rho}), \\
&[B_r, B_s] = i \, C_{rs}^t B_{t}, \\
&[B_r , P_{\mu}] = [B_r, M_{\mu\sigma}] = 0, \\
&[Q_{\alpha}^i , P_{\mu}] = [\bar Q_{\dot\alpha}^i, P_{\mu}] = 0, \\
&[Q_{\alpha}^i , M_{\mu\nu}] = \frac12 \, (\sigma_{\mu\nu})_{\alpha}^{\beta}Q_{\beta}^i, \ \ \
 [\bar Q_{\dot\alpha}^i, M_{\mu\nu}] = -\frac12 \, \bar Q_{\dot\beta}^i
   (\bar\sigma_{\mu\nu})_{\dot\alpha}^{\dot \beta}, \\
&[Q_{\alpha}^i, B_r] = (b_r)_j^i Q_{\alpha}^j, \ \ \
 [\bar Q_{\dot\alpha}^i, B_r] = - \bar Q_{\dot\alpha}^j (b_r)_j^i, \\
&\{ Q_{\alpha}^i, \bar Q_{\dot\beta}^j \} = 2 \, \delta^{ij} (\sigma^\mu)_{\alpha \dot\beta} P_\mu, \\
&\{ Q_{\alpha}^i, Q_{\beta}^j \} = 2 \, \epsilon_{\alpha \beta}Z^{ij}, \ \ \
    Z_{ij} = a_{ij}^r b_r, \ \ \ Z^{ij} = Z_{ij}^+, \\
&\{ \bar Q_{\dot\alpha}^i, \bar Q_{\dot\beta}^j \} =
    - 2 \, \epsilon_{\dot\alpha \dot\beta} Z^{ij}, \ \ \ [Z_{ij}, anything] = 0, \\
&\alpha, \dot\alpha  =  1,2 \ \ \ \ i,j = 1,2, \ldots , N.
\end{split}
\label{group}
\end{equation}
Here $P_{\mu}$ and $M_{\mu \nu}$ are the four-momentum and
angular momentum operators, respectively, $B_r$ are the
internal symmetry generators, $Q^i$ and $\bar Q^i$ are the
spinorial SUSY generators and $Z_{ij}$ are the so-called
central charges; $\alpha, \dot\alpha, \beta, \dot\beta$ are the
spinorial indices. In the simplest case one has one spinor
generator $Q_\alpha$ (and the conjugated one
$\bar Q_{\dot{\alpha}}$) that corresponds to the ordinary or
$N=1$ supersymmetry. When $N>1$ one has the extended
supersymmetry.

A natural question arises: how many SUSY generators are
possible, i.~e. what is the value of~$N$? To answer this
question, consider massless states. Let us start with the
ground state labeled by the energy and the helicity, i.~e.
the projection of the spin on the direction of momenta, and
let it be annihilated by $Q_i$
\begin{equation*}
{\rm Vacuum} =|E,\lambda \rangle, \ \ \ \ \ \ \
Q_i|E,\lambda \rangle = 0.
\end{equation*}
Then one- and many-particle states can be constructed with the
help of creation operators as
\begin{equation*}
\begin{array}{llc}
\mbox{\underline{State}~~~~~~~~~~~~~~} &
\mbox{\underline{Expression}~~~~~~~~~~~~~~~~~~~~~~~~~~} &
\# \ \mbox{\underline{of states}} \\  \\
\mbox{vacuum} & |E,\lambda \rangle & 1 \\
\mbox{1-particle} & \bar Q_i |E,\lambda \rangle =
|E,\lambda + \frac12 \rangle_i & N \\
\mbox{2-particle} & \bar Q_i \bar Q_j |E,\lambda  =
|E,\lambda + 1 \rangle_{ij} & \frac{N(N-1)}{2} \\
\ldots & \ldots & \ldots \\[1mm]
N\mbox{-particle} & \bar Q_1 \ldots \bar Q_N |E,\lambda \rangle =
|E,\lambda + \frac{N}{2} \rangle & 1
\end{array}
\end{equation*}
The total $\#$ of states is:
$\displaystyle
\sum_{k=0}^{N}\left(
\begin{array}{c}
N \\ k
\end{array}
\right) =2^N=2^{N-1}$
bosons + $2^{N-1}$ fermions.

The energy $E$ is not changed, since according to~(\ref{group})
the operators $\bar Q_i$ commute with the Hamiltonian.

Thus, one has a sequence of bosonic and fermionic states
and the total number of the bosons equals to that of the
fermions. This is a generic property of any supersymmetric
theory. However, in CPT invariant theories the number of states
is doubled, since CPT transformation changes the sign of the
helicity. Hence, in the CPT invariant theories, one has to add
the states with the opposite helicity to the above mentioned
ones.

Let us consider some examples. We take $N=1$ and $\lambda=0$.
Then one has the following set of states:
\begin{equation*}
\begin{array}{lllllc}
N=1 & \lambda=0 & & & & \\
\mbox{helicity} & 0 \ \frac12 & & \mbox{helicity} & 0 \ -\frac 12 \\
& & \stackrel{CPT}{\Longrightarrow} & & \\
\# \ \mbox{of states} & 1 \ 1 & & \# \ \mbox{of states} & 1 \ \ \ \ 1
\end{array}
\end{equation*}
Hence, the complete $N=1$ multiplet is
\begin{equation*}
\begin{array}{llccc} N=1 \ \ &
\mbox{helicity}&-1/2&0&1/2  \\ &
\# \ \mbox{of states}&\ 1& 2&1
\end{array}
\end{equation*}
which contains one complex scalar and one spinor with two
helicity states.

This is an example of the so-called self-conjugated multiplet.
There are also the self-con\-ju\-gated multiplets with $N>1$
corresponding to the extended supersymmetry. Two particular
examples are the $N=4$ super Yang-Mills multiplet and the $N=8$
supergravity multiplet
\begin{equation*}
N=4 \ \ \
\mbox{SUSY YM} \  \ \  \lambda=-1
\end{equation*}
\begin{equation*}
\begin{array}{cccccccccc}
\mbox{helicity}&& &-1&-1/2&0&1/2&1 &&  \\
\# \ \mbox{of states}&&&\ \ 1&\ 4&6 & 4 & 1 &&\\
\end{array}
\end{equation*}
\begin{equation*}
N=8  \ \ \  \mbox{SUGRA} \ \ \   \lambda=-2
\end{equation*}
\begin{equation*}
\begin{array}{ccccccccc}
-2&-3/2&-1&-1/2&0&1/2&1&3/2&2 \\
 1&8&28 & 56 & 70 &56&28&  8& 1
\end{array}
\end{equation*}
One can see that the multiplets of extended supersymmetry
are very rich and contain a vast number of particles.

The constraint on the number of the SUSY generators comes from
the requirement of consistency of the corresponding QFT. The
number of supersymmetries and the maximal spin of the particle
in the multiplet are related by
\begin{equation*}
N \leq 4S,
\end{equation*}
where $S$ is the maximal spin. Since the theories with the spin
greater than 1 are non-renorma\-li\-zable and the theories with the
spin greater than 5/2 have no consistent coupling to gravity,
this imposes a constraint on the number of the SUSY generators
\begin{equation*}
\begin{split}
&N \leq 4 \ \ \ \ \ \text{for renormalizable theories (YM),} \\
&N \leq 8 \ \ \ \ \ \text{for (super)gravity}.
\end{split}
\end{equation*}

In what follows, we shall consider the simple supersymmetry, or
the $N=1$ supersymmetry, contrary to the extended
supersymmetries with $N > 1$. In this case, one has the
following types of the supermultiplets which are used for the
construction of the SUSY generalization of the SM
\begin{equation*}
\begin{array}{ll}
\hspace*{1.2cm}
(\phi,\ \ \psi) & \hspace*{1cm} (\lambda, \ \ A_\mu)\\
Spin=0, \ Spin=1/2 & Spin=1/2, \ Spin=1 \\
\ \ scalar\  \ \ \ \ \ chiral & Majorana\ \ \ \   vector  \\
\hspace*{1.7cm} fermion & fermion
\end{array}
\end{equation*}
each of them contains two physical states, one boson and one
fermion. They are called chiral and vector multiplets,
respectively. To construct the generalization of the SM one
has to put all the particles into these multiplets. For
instance, the quarks should go into the chiral multiplet and
the photon into the vector multiplet.

%------------------------------------------
\subsection{Superspace and supermultiplets}

An elegant formulation of the supersymmetry transformations and
invariants can be achieved in the framework of the superspace
formalism~\cite{sspace}. The superspace differs from the
ordinary Euclidean (Minkowski) space by adding two new
coordinates, $\theta_{\alpha}$ and $\bar\theta_{\dot\alpha}$,
which are Grassmannian, i.~e. anti\-com\-muting, variables
\begin{equation*}
\{ \theta_{\alpha}, \theta_{\beta} \} = 0 , \ \ \{\bar
\theta_{\dot \alpha}, \bar \theta_{\dot \beta} \} = 0, \ \
\theta_{\alpha}^2 = 0,\ \ \bar \theta_{\dot \alpha}^2=0,
\end{equation*}
\begin{equation*}
\alpha,\beta, \dot\alpha, \dot\beta =1,2.
\end{equation*}
Thus, we go from the space to the superspace
\begin{equation*}
\begin{array}{cc}
\text{Space} & \ \Longrightarrow \ \ \text{Superspace} \\
~x_{\mu} & \ \ \ \ \ \ \ \, ~x_{\mu}, \theta_{\alpha} ,
\bar\theta_{\dot \alpha}
\end{array}
\end{equation*}
A SUSY group element can be constructed in the superspace in
the same way as the ordinary translation in the usual space
\begin{equation}
G(x,\theta ,\bar\theta ) = e^{\displaystyle
\, i (-x^{\mu}P_{\mu} + \theta Q + \bar\theta \bar Q)}.
\label{st}
\end{equation}
It leads to a supertranslation in the superspace
\begin{equation}
\begin{split}
x_{\mu} &\rightarrow x_{\mu} + i \, \theta \sigma_{\mu}
\bar\varepsilon - i \, \varepsilon \sigma_{\mu} \bar\theta, \\
\theta &\rightarrow \theta + \varepsilon, \\
\bar\theta &\rightarrow \bar\theta + \bar\varepsilon,
\end{split}
\label{sutr}
\end{equation}
where $\varepsilon$ and $\bar\varepsilon$ are the Grassmannian
transformation parameters. From Eqn.~(\ref{sutr}) one can
easily obtain the representation for the
supercharges~(\ref{group}) acting on the superspace
\begin{equation}
Q_\alpha = \frac{\partial}{\partial\theta_\alpha} -
i \, \sigma^\mu_{\alpha \dot\alpha}\bar{\theta}^{\dot \alpha}
\partial_\mu , \ \ \ \ \
\bar{Q}_{\dot\alpha} =
- \, \frac{\partial}{\partial \bar{\theta}_{\dot\alpha}} +
i \, \theta_\alpha\sigma^\mu_{\alpha \dot\alpha}\partial_\mu .
\label{q}
\end{equation}

To define the fields on the superspace, consider the
representations of the Super-Poincar\'e
group~(\ref{group})~\cite{WessB}. The simplest $N=1$ SUSY
multiplets that we discussed earlier are: the chiral one
$\Phi(y,\theta)$ ($y =x + i\theta \sigma \bar\theta $)
and the vector one $V(x,\theta,\bar\theta)$. Being expanded
in the Taylor series over the Grassmannian variables $\theta$
and $\bar\theta$ they give:
\begin{equation}
\begin{split}
\Phi (y, \theta) &= \, A(y) + \sqrt{2} \, \theta \psi(y) +
\theta \theta F(y) = \\
&= \, A(x) +
i \, \theta \sigma^{\mu} \bar \theta \partial_{\mu}A(x) +
\frac{1}{4} \, \theta \theta \bar\theta \bar\theta \, \Box A(x) \\
& \, \, + \sqrt{2} \, \theta \psi (x) -
\frac{i}{\sqrt{2}} \, \theta \theta \partial_{\mu} \psi(x)
\sigma^{\mu} \bar\theta +
\theta \theta F(x).
\end{split}
\label{eqn:field}
\end{equation}
The coefficients are the ordinary functions of $x$ being the
usual fields. They are called the {\em components} of the
superfield. In Eqn.~(\ref{eqn:field}) one has 2~bosonic (the
complex scalar field $A$) and 2~fermionic (the Weyl spinor
field $\psi$) degrees of freedom. The component fields $A$
and $\psi$ are called the {\em superpartners}. The field
$F$ is an {\em auxiliary} field, it has the "wrong" dimension
and has no physical meaning. It is needed to close the
algebra~(\ref{group}). One can get rid of the auxiliary
fields with the help of equations of motion.

Thus, the superfield contains an equal number of the bosonic
and fermionic degrees of freedom. Under the SUSY
transformation they convert one into another
\begin{equation}
\begin{split}
\delta_\varepsilon A &=
\sqrt2 \, \varepsilon \psi, \\
\delta_\varepsilon \psi &=
i \sqrt2 \, \sigma^\mu \bar\varepsilon \partial_\mu A +
\sqrt2 \, \varepsilon F, \\
\delta_\varepsilon F &=
i \sqrt2 \, \bar\varepsilon \sigma^\mu \partial_\mu \psi.
\end{split}
\label{transf}
\end{equation}
Notice that the variation of the $F$-component is a total
derivative, i.~e. it vanishes when integrated over the
space-time.

The vector superfield is real $V = V^\dagger$. It has the
following Grassmannian expansion:
\begin{align}
V(x, \theta, \bar\theta) &=
C(x) + i\, \theta \chi(x) - i \, \bar\theta \bar\chi(x) +
\frac{i}{2} \, \theta \theta \, \bigl[M(x) + iN(x)\bigr] \notag \\
& \,\, - \frac{i}{2} \, \bar\theta \bar\theta \, \bigl[M(x) - iN(x)\bigr] -
\theta \sigma^{\mu} \bar\theta \, v_{\mu}(x) +
i \, \theta \theta \bar\theta \bigl[\lambda (x) +
\frac{i}{2} \, \bar\sigma^{\mu} \partial_{\mu} \chi(x)\bigr] \\
& \,\, - i \, \bar\theta \bar\theta \theta \bigl[\lambda +
\frac{i}{2} \, \sigma^{\mu} \partial_{\mu} \bar\chi(x)\bigr] +
\frac{1}{2} \, \theta \theta \bar\theta \bar\theta \bigl[D(x) +
\frac{1}{2} \, \Box C(x)]. \notag
\label{p}
\end{align}
The physical degrees of freedom corresponding to the real
vector superfield $V$ are the vector gauge field $v_{\mu}$ and
the Majorana spinor field $\lambda$. All other components are
unphysical and can be eliminated. Indeed, one can choose a
gauge (the Wess-Zumino gauge) where $C = \chi = M = N =0 $,
leaving one with only physical degrees of freedom except for
the auxiliary field $D$. In this gauge
\begin{align}
V \phantom{^2} &= - \, \theta \sigma^{\mu} \bar\theta v_{\mu}(x) +
i \, \theta \theta \bar\theta \bar\lambda (x) -
i \, \bar\theta \bar\theta \theta \lambda (x) +
\frac{1}{2} \, \theta \theta \bar\theta \bar\theta D(x),
\notag \\
V^2 &= - \frac{1}{2} \theta \theta \bar\theta \bar\theta
\, v_{\mu}(x)v^{\mu}(x), \\
V^3 &= 0, \ \ \ etc. \notag
\end{align}
One can define also a field strength tensor (as the analog of
$F_{\mu \nu}$ in the gauge theories)
\begin{equation}
W_{\alpha} = - \frac{1}{4} \, \bar D^2 e^V D_{\alpha} e^{-V},
\ \ \ \
\bar W_{\dot\alpha} = - \frac{1}{4} \, D^2 e^V \bar D_{\alpha}
e^{-V},
\label{str}
\end{equation}
Here $D$ and $\bar D$ are the supercovariant derivatives. The
field strength tensor in the chosen Wess-Zumino gauge is a polynomial
over the component fields:
\begin{equation}
W_\alpha = T^a \Bigl( - i \, \lambda^a_\alpha + \theta_\alpha D^a -
\frac{i}{2} \, (\sigma^\mu \bar{\sigma}^\nu \theta)_\alpha
F^a_{\mu\nu} +
\theta^2 (\sigma^\mu D_\mu \bar{\lambda}^a)_{\alpha} \Bigr),
\end{equation}
where
\begin{equation*}
F^a_{\mu\nu} = \partial_\mu v^a_\nu - \partial_\nu v^a_\mu +
f^{abc} v^b_\mu v^c_\nu, \ \ \
D_\mu \bar{\lambda }^a = \partial\bar{\lambda }^a +
f^{abc}v^b_\mu \bar{\lambda }^c.
\end{equation*}
In the Abelian case Eqs.~(\ref{str}) are simplified and take
the form
\begin{equation*}
W_\alpha = - \frac{1}{4} \bar D^2 D_{\alpha} V, \ \ \
\bar W_{\dot\alpha} = - \frac{1}{4} D^2 \bar D_{\alpha} V.
\end{equation*}

%-----------------------------------------------
\subsubsection{Construction of SUSY Lagrangians}

Let us start with the Lagrangian which has no local gauge
invariance. In the superfield notation the SUSY invariant
Lagrangians are the polynomials of the superfields.  In the
same way, as the ordinary action is the integral over the
space-time of the Lagrangian density, in the supersymmetric
case the action is the integral over the superspace.
The space-time Lagrangian density is~\cite{WessB,sspace}
\begin{equation}
{\cal L}  = \int d^2 \theta \, d^2 \bar\theta \, \Phi_i^+ \Phi_i +
\int d^2\theta \, \bigl[ \lambda_i \Phi_i +
\frac{1}{2} \, m_{ij} \Phi_i \Phi_j +
\frac{1}{3} \, y_{ijk} \Phi_i \Phi_j \Phi_k \bigr] + h.c.
\label{l}
\end{equation}
where the first part is the kinetic term and the second one is
the {\em superpotential}  ${\cal W}$. We use here the
integration over the superspace according to the rules of the
Grassmannian integration~\cite{ber}
\begin{equation*}
\int d\theta_\alpha = 0 , \ \ \ \
\int \theta _\alpha \, d\theta _\beta = \delta_{\alpha\beta}.
\end{equation*}

Performing the explicit integration over the Grassmannian
parameters, we get from Eqn.~(\ref{l})
\begin{equation}
\begin{split}
{\cal L} &=
\,\, i \, \partial_{\mu} \bar\psi_i \, \bar\sigma^{\mu} \psi_i +
A_i^* \Box A_i + F_i^{\ast}F_i \\
& \, + \Bigl[\lambda_i F_i + m_{ij} \Bigl(A_i F_j -
\frac{1}{2} \, \psi_i \psi_j \Bigr) +
y_{ijk} \bigl(A_i A_j F_k - \psi_i \psi_j A_k \bigr) + h.c. \Bigr].
\end{split}
\label{20}
\end{equation}
The last two terms are the interaction ones. To obtain the
familiar form of the Lagrangian, we have to solve the
constraints
\begin{equation}
\begin{split}
\frac{\partial {\cal L}}{\partial F_k^*} &=
F_k + \lambda_k^* + m_{ik}^* A_i^* + y_{ijk}^* A_i^* A_j^* = 0, \\
\frac{\partial {\cal L}}{\partial F_k} &=
F_k^* + \lambda_k + m_{ik} A_i + y_{ijk} A_i A_j = 0.
\end{split}
\end{equation}
Expressing the auxiliary fields $F$ and $F^*$  from these
equations, one finally gets
\begin{equation}
\begin{split}
{\cal L} &=
\,\, i \, \partial_{\mu} \bar\psi_i \bar\sigma^{\mu} \psi_i +
A_i^* \Box A_i - \frac{1}{2} \, m_{ij} \psi_i \psi_j -
\frac{1}{2} \, m_{ij}^* \bar\psi_i \bar\psi_j \\
& \, - y_{ijk} \psi_i \psi_j A_k -
y_{ijk}^* \bar\psi_i \bar\psi_j A_k^* - V(A_i,A_j),
\end{split}
\label{m}
\end{equation}
where the scalar potential $V = F_k^* F_k $. We will return to
the discussion of the form of the scalar potential in the SUSY
theories later.

Consider now the gauge invariant SUSY Lagrangians. They should
contain the gauge invariant interaction of the matter fields
with the gauge ones and the kinetic term and the
self-interaction of the gauge fields.

Let us start with the gauge field kinetic terms. In the
Wess-Zumino gauge one has
\begin{equation}
W^{\alpha} W_{\alpha} \big|_{\theta \theta} =
\, - \, 2\, i \lambda \sigma^{\mu} D_{\mu} \bar\lambda -
\frac{1}{2} F_{\mu\nu} F^{\mu\nu} + \frac{1}{2} D^2 +
\frac{i}{4} F^{\mu\nu} F^{\rho\sigma}
\epsilon_{\mu\nu\rho\sigma},
\end{equation}
where
$D_\mu \bar\lambda = \partial_\mu + ig [v_\mu, \bar\lambda]$
is the usual covariant derivative and the last, the so-called
topological $\theta$-term, is the total derivative. The gauge
invariant Lagrangian now has the familiar form
\begin{equation}
\begin{split}
{\cal L} &=
\, \frac{1}{4} \int d^2\theta \, W^{\alpha} W_{\alpha} +
\frac{1}{4} \int d^2 \bar\theta \,
\bar W^{\dot\alpha} \bar W_{\dot\alpha}  \\
&= \, \frac{1}{2} D^2 - \frac{1}{4} F_{\mu \nu} F^{\mu \nu} -
i \, \lambda \sigma^{\mu} D_{\mu} \bar\lambda.
\end{split}
\label{29}
\end{equation}
To obtain the gauge-invariant interaction with the matter chiral
superfields, one has to modify the kinetic term by inserting the
bridge operator
\begin{equation}
\Phi_i^+  \Phi_i \ \ \Longrightarrow \ \ \Phi_i^+ e^{gV} \Phi_i.
\end{equation}

The complete SUSY and gauge invariant Lagrangian then looks like
\begin{equation}
\begin{split}
{\cal L}_{SUSY \, YM} &=
\, \frac{1}{4} \int d^2 \theta \, {\rm Tr} (W^{\alpha} W_{\alpha}) +
\frac{1}{4}\int d^2 \bar\theta \, {\rm Tr}(\bar{W}^{\alpha} \bar{W}_{\alpha}) \\
& \,\, + \int d^2 \theta \, d^2 \bar\theta \,
\bar\Phi_{ia} (e^{gV})_b^a \Phi_i^b \, + \! \int d^2 \theta \, {\cal W}(\Phi_i) \, +
\! \int d^2 \bar\theta \, \bar{\cal W}(\bar\Phi_i),
\end{split}
\label{nonab}
\end{equation}
where ${\cal W}$ is the superpotential, which should be
invariant under the group of symmetry of the particular model.
In terms of thecomponent fields the above Lagrangian takes the
form
\begin{equation}
\begin{split}
{\cal L}_{SUSY \ YM}  &=
-\frac{1}{4} F^a_{\mu\nu} F^{a\mu\nu} -
 i \, \lambda^a \sigma^\mu D_\mu \bar\lambda^a + \frac{1}{2} D^a D^a  \\
& \,\, + \bigl(\partial_\mu A_i -
 i \, g v^a_\mu T^a A_i\bigr)^\dagger \bigl(\partial_\mu A_i - i \, g v^{a}_\mu T^a A_i\bigr) -
 i \, \bar\psi_i \bar\sigma^\mu \bigl(\partial_\mu \psi_i - i \, g v^{a}_\mu T^a \psi_i\bigr) \\
& \,\, -  D^a A^\dagger_i \, T^a A_i - i \sqrt{2} A^\dagger_i \, T^a \lambda^a \psi_i +
 i \sqrt{2} \, \bar\psi_i T^a A_i \bar\lambda^a + F^\dagger_i F_i \\
& \,\, + \frac{\partial {\cal W}}{\partial A_i} F_i +
 \frac{\partial \bar{\cal W}}{\partial A_i^\dagger} F^\dagger_i -
 \frac{1}{2} \, \frac{\partial^2 {\cal W}}{\partial A_i \partial A_j} \, \psi_i \psi_j -
 \frac{1}{2} \, \frac{\partial^2 \bar{\cal W}}{\partial A_i^\dagger \partial A_j^\dagger} \, \bar\psi_i \bar\psi_j.
\end{split}
\label{sulag}
\end{equation}
Integrating out the auxiliary fields $D^a$ and $F_i$, one
reproduces the usual Lagrangian.

%-----------------------------------
\subsubsection{The scalar potential}

Contrary to the SM, where the scalar potential is arbitrary
and is defined only by the requirement of the gauge
invariance, in the supersymmetric theories it is completely
defined by the superpotential. It consists of the contributions
from the $D$-terms and $F$-terms. The kinetic energy of the
gauge fields (recall Eqn.~(\ref{29}) yields the
$\frac12 D^aD^a$ term, and the matter-gauge interaction (recall
Eqn.~(\ref{sulag}) yields the $g D^a T^a_{ij} A^*_i A_j$ one.
Together they give
\begin{equation}
{\cal L}_D = \frac{1}{2} D^a D^a + g D^a T^a_{ij} A^*_ iA_j.
\label{d}
\end{equation}
The equation of motion reads
\begin{equation}
D^a = - g T^a_{ij} A^*_i A_j.
\label{sol}
\end{equation}
Substituting it back into Eqn.~(\ref{d}) yields the $D$-term
part of the potential
\begin{equation}
{\cal L}_D = - \frac{1}{2} D^a D^a \ \ \
\Longrightarrow \ \ \ V_D = \frac{1}{2} D^a D^a,
\end{equation}
where $D$ is given by Eqn.~(\ref{sol}).

The $F$-term contribution can be derived from the matter field
self-in\-ter\-action~(\ref{20}). For a general type
superpotential ${\cal W}$ one has
\begin{equation}
{\cal L}_F = F^*_i F_i +
\Bigl( \, \frac{\partial W}{\partial A_i} F_i + h.c.\Bigr).
\end{equation}
Using the equations of motion for the auxiliary field $F_i$
\begin{equation}
F^*_i = - \frac{\partial W}{\partial A_i}
\label{solf}
\end{equation}
yields
\begin{equation}
{\cal L}_F = - F^*_i F_i \ \ \
\Longrightarrow \ \ \ V_F = F^*_i F_i ,
\end{equation}
where $F$ is given by Eqn.~(\ref{solf}). The full scalar
potential is the sum of the two contributions
\begin{equation}
V = V_D + V_F.
\end{equation}

Thus, the form of the Lagrangian is practically fixed by the
symmetry requirements. The only freedom is the field content,
the value of the gauge coupling $g$, Yukawa couplings $y_{ijk}$
and the masses. Because of the renormalizability constraint
$V \leq A^4 $ the superpotential should be limited by
${\cal W} \leq \Phi^3 $ as in Eqn.~(\ref{l}). All members of
the supermultiplet have the same masses, i.~e. the bosons and
the fermions are degenerate in masses. This property of the
SUSY theories contradicts to the phenomenology and requires
supersymmetry breaking.

%------------------------------------------------------------
\section{SUSY generalization of the Standard Model. The MSSM}

As has been already mentioned, in the SUSY theories the number
of the bosonic degrees of freedom equals that of fermionic. At
the same time,  in the SM one has 28 bosonic and 90 fermionic
degrees of freedom (with the massless neutrino, otherwise 96).
So the SM is to a great extent non-supersymmetric. Trying to
add some new particles to supersymmetrize the SM, one should
take into account the following observations:
\begin{itemize}
\item There are no fermions with quantum numbers of the
gauge bosons;
\item Higgs fields have nonzero vacuum expectation values;
hence, they cannot be the superpartners of the quarks and
leptons, since this would induce a spontaneous violation of the
baryon and lepton numbers;
\item One needs at least two complex chiral Higgs
multiplets in order to give masses to the up and down quarks.
\end{itemize}

The latter is due to the form of the superpotential and the
chirality of the matter superfields. Indeed, the superpotential
should be invariant under the $SU(3)\times SU(2)\times U(1)$
gauge group. If one looks at the Yukawa interaction in the
Standard Model, one finds that it is indeed $U(1)$ invariant
since the sum of hypercharges in each vertex equals zero. For
the up quarks this is achieved by taking the conjugated Higgs
doublet $\tilde{H}=i\tau_2 H^\dagger$ instead of $H$. However,
in SUSY $H$ is the chiral superfield and hence the
superpotential which is constructed out of the chiral fields,
may contain only $H$ but not $\tilde H$ which is the antichiral
superfield.

Another reason for the second  Higgs doublet is related to
chiral anomalies. It is known that the chiral anomalies spoil
the gauge invariance and, hence, the renormalizability of the
theory. They are canceled in the SM between the quarks and
leptons in each generation~\cite{Peskin}
\begin{equation*}
\begin{split}
{\rm Tr} \, Y^3 =& \ \ 3 \times \left( \frac{1}{27} + \frac{1}{27} -
\frac{64}{27} + \frac{8}{27} \right) \, - \, 1 \, - \, 1 \, + \, 8 \, = \, 0 \\
& \text{color} \ \ \ \, u_L \ \ \ \ d_L \ \ \ \ \!  u_R \ \ \ \ d_R
\ \ \ \ \ \ \ \! \nu_L \ \ \, \, e_L \ \ \, \, e_R
\end{split}
\end{equation*}

However, if one introduces the chiral Higgs superfield, it
contains higgsinos, which are the chiral fermions, and contain
the anomalies. To cancel them one has to add the second Higgs
doublet with the opposite hypercharge. Therefore, the Higgs
sector in the SUSY models is inevitably enlarged, it contains
an even number of the Higgs doublets.

{\em Conclusion}: In the SUSY models the supersymmetry
associates the {\em known} bosons with the {\em new} fermi\-ons
and the {\em known} fermi\-ons with the {\em new} bosons.

%-----------------------------
\subsection{The field content}

\begin{figure}[b]
\begin{center}\vspace{-5mm}
\leavevmode
\includegraphics[width=0.40\textwidth]{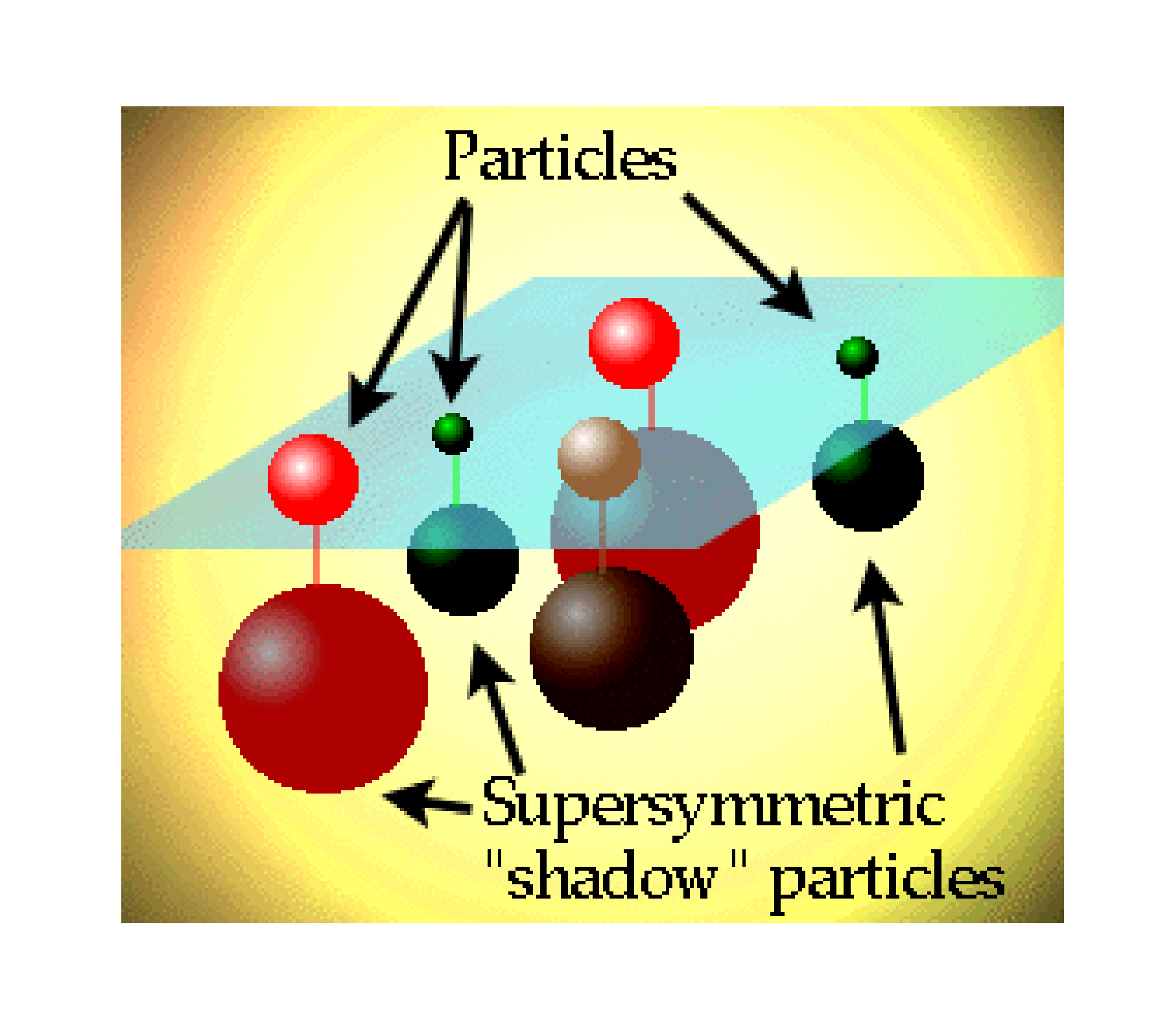}
\end{center}
\vspace{-1.0cm}
\caption{The shadow world of SUSY particles.}
\label{fig:shadow}
\end{figure}
Consider the particle content of the Minimal Supersymmetric
Standard Model~\cite{HT,MSSM,SUSYLHC_GK}. According to the previous
discussion, in the minimal version we double the number of
particles (introducing the superpartner to each particle) and
add another Higgs doublet (with its superpartner).

Thus, the characteristic feature of any supersymmetric
generalization of the SM is the presence of the superpartners
(see Fig.~\ref{fig:shadow})~\cite{shadow}.  If the
supersymmetry is exact, the superpartners of the ordinary
particles should have the same masses and have to be observed.
The absence of them at modern energies is believed to be
explained by the fact that they are very heavy, that means that
the supersymmetry should be broken. Hence, if the energy of
accelerators is high enough, the superpartners will be created.

The particle content of the MSSM then appears as shown in the
Table~\ref{tab:mssm}. Hereafter, a tilde denotes the
superpartner of the ordinary particle.

\begin{table}[t]
\begin{center}
\renewcommand{\tabcolsep}{0.03cm}
\begin{tabular}{|lllccc|}\hline
Superfield & \ \ \ \ \ \ \ Bosons & \ \ \ \ \ \ \ Fermions &
$SU(3)$& $SU(2$ & $U_Y(1)$ \\
\hline \hline
Gauge  &&&&& \\
${\bf G^a}$   & gluon \ \ \ \ \ \ \ \ \ \ \ \ \ \ \  $g^a$ &
gluino$ \ \ \ \ \ \ \ \ \ \ \ \ \tilde{g}^a$ & 8 & 0 & 0 \\
${\bf V^k}$ & Weak \ \ \ $W^k$ \ $(W^\pm, Z)$ & wino, zino \
$\tilde{w}^k$ \ $(\tilde{w}^\pm, \tilde{z})$ & 1 & 3& 0 \\
${\bf V'}$   & Hypercharge  \ \ \ $B\ (\gamma)$ & bino
\ \ \ \ \ \ \ \ \ \ \ $\tilde{b}(\tilde{\gamma})$ & 1 & 1& 0 \\
\hline \hline
Matter &&&&& \\
$\begin{array}{c}
{\bf L_i} \\ {\bf E_i}
\end{array}$ & sleptons
\ $\left\{
\begin{array}{l}
\tilde{L}_i=(\tilde{\nu},\tilde e)_L \\
\tilde{E}_i =\tilde e_R
\end{array} \right. $
& leptons \ $\left\{
\begin{array}{l}
L_i=(\nu,e)_L \\
E_i=e_R^c
\end{array} \right.$
& $\begin{array}{l}
1 \\ 1
\end{array} $
& $\begin{array}{l}
2 \\ 1
\end{array} $ & $
\begin{array}{r} -1 \\ 2
\end{array} $ \\
$\begin{array}{c} {\bf Q_i} \\ {\bf U_i} \\ {\bf D_i}
\end{array}$
& squarks \ $\left\{
\begin{array}{l}
\tilde{Q}_i=(\tilde{u},\tilde d)_L \\ \tilde{U}_i =\tilde u_R \\
\tilde{D}_i =\tilde d_R
\end{array}  \right. $
& quarks \ $\left\{
\begin{array}{l}
Q_i=(u,d)_L \\ U_i=u_R^c \\ D_i=d_R^c
\end{array} \right. $
& $\begin{array}{l}
3 \\ 3^* \\ 3^*
\end{array} $
& $\begin{array}{l}
2 \\ 1 \\ 1
\end{array} $
& $\begin{array}{r}
1/3 \\ -4/3 \\ 2/3
\end{array} $ \\
\hline  \hline
Higgs &&&&& \\
$\begin{array}{c} {\bf H_1} \\ {\bf H_2}
\end{array}$
& Higgses \ $\left\{
\begin{array}{l}
H_1 \\ H_2
\end{array}  \right. $
& higgsinos \ $\left\{
\begin{array}{l}
\tilde{H}_1 \\ \tilde{H}_2
\end{array} \right. $
&
$\begin{array}{l}
1 \\ 1 \end{array} $
& $\begin{array}{l}
2 \\ 2
\end{array} $
& $\begin{array}{r} -1 \\ 1
\end{array} $  \\
\hline \hline
\end{tabular}
\end{center}
\label{tab:mssm}
\caption{Particle content of the MSSM}
\end{table}

The presence of the extra Higgs doublet in the SUSY model is
a novel feature of the theory. In the MSSM one has two doublets
with the quantum numbers (1,2,-1) and (1,2,1), respectively:
\begin{align*}
H_1 &=
\left( \begin{array}{c}
H^0_1 \\[2mm] H_1^-
\end{array}\right) =
\left( \begin{array}{c}
v_1 + \frac{S_1+iP_1}{\sqrt{2}} \\[1mm] H^-_1
\end{array}\right), \\[2mm]
H_2 &= \left( \begin{array}{c}
H^+_2 \\[2mm] H_2^0 \end{array} \right) =
\left( \begin{array}{c}
H^+_2 \\[1mm] v_2 +\frac{S_2+iP_2}{\sqrt{2}}
\end{array} \right),
\end{align*}
where  $v_i$ are the vacuum expectation values of the neutral
components of the Higgs doublets.

Hence, one has $8=4+4=5+3$ degrees of freedom. As in the case
of the SM, 3 degrees of freedom can be gauged away, and one is
left with 5 physical states compared to 1 in the SM. Thus, in
the MSSM, as actually in any two Higgs doublet model, one has
five physical Higgs bosons: two $CP$-even neutral Higgs, one
$CP$-odd neutral Higgs and two charged ones. We consider the
mass eigenstates below.

%----------------------------------
\subsection{Lagrangian of the MSSM}

Now we can construct the Lagrangian of the MSSM. It consists
of two parts; the first part is the SUSY generalization of
the Standard Model, while the second one represents the
SUSY breaking as mentioned above.
\begin{align}
{\cal L}_{MSSM} &= {\cal L}_{SUSY} + {\cal L}_{Breaking},
\intertext{where}
{\cal L}_{SUSY} &= {\cal L}_{Gauge} + {\cal L}_{Yukawa}.
\end{align}

We will not describe the gauge part here, since it is
essentially the gauge invariant kinetic terms, but rather
concentrate on Yukawa terms. They are given by the
superpotential which is nothing else but the usual Yukawa
terms of the SM with the fields replaced by the superfields
as explained above.
\begin{equation}
{\cal L}_{Yukawa} = \epsilon_{ij} \left(
y^U_{ab} Q_a^j U^c_b H_2^i + y^D_{ab} Q_a^j D^c_b H_1^i +
y^L_{ab} L_a^j E^c_b H_1^i + \mu H_1^i H_2^j \right),
\label{R}
\end{equation}
where $i,j=1,2$ are the $SU(2)$ and $a,b=1,2,3$ are the
generation indices; the $SU(3)$ colour indices are omitted.
This part of the Lagrangian almost exactly repeats that of
the SM. The only difference is the last term which describes
the Higgs mixing. It is absent in the SM since there is only
one Higgs field there.

However, one can write down also the different Yukawa terms
\begin{equation}
{\cal L}_{Yukawa} = \epsilon_{ij} \left(
\lambda^L_{abd} L_a^i L_b^j E_d^c +
\lambda^{L\prime}_{abd} L_a^i Q_b^j D_d^c +
\mu'_a L^i_a H_2^j \right) +
\lambda^B_{abd} U_a^c D_b^c D_d^c.
\label{NR}
\end{equation}
These terms are absent in the SM. The reason is very simple: one
can not replace the superfields in Eqn.~(\ref{NR}) by the
ordinary fields like in Eqn.~(\ref{R}) because of the Lorentz
invariance. These terms have also another property, they violate
either the lepton number $L$ (the first 3 terms in
Eqn.~(\ref{NR})) or the baryon number $B$ (the last term). Since
both effects are not observed in Nature, these terms must be
suppressed or excluded. One can avoid such terms introducing a
new special symmetry called $R$-symmetry~\cite{r-symmetry}. The
global $U(1)_R$ invariance
\begin{equation}
U(1)_R: \ \ \theta \to e^{i\alpha} \theta ,\ \  \Phi \to
e^{in\alpha}\Phi,
\label{RS}
\end{equation}
which is reduced to the discrete group $Z_2$, is called
$R$-parity. The $R$-parity quantum number  is
\begin{equation}
R=(-1)^{3(B-L)+2S}
\label{par}
\end{equation}
for the particles with the spin $S$. Thus, all the ordinary
particles have the $R$-parity quantum number equal to $R=+1$,
while all the superpartners have the $R$-parity quantum number
equal to $R=-1$. The first part of the Yukawa Lagrangian is
$R$-symmetric, while the second part is $R$-nonsymmetric.
The $R$-parity obviously forbids the terms~(\ref{NR}). However,
it may well be that these terms are present, though experimental
limits on the couplings are very severe
\begin{equation*}
\lambda^L_{abc}, \ \ \lambda^{L\prime}_{abc} < 10^{-4},
\ \ \ \ \ \lambda^B_{abc} < 10^{-9}.
\end{equation*}

Conservation of the $R$-parity has two important consequences\vspace{-0.3cm}
\begin{itemize}
\item the superpartners are created in pairs;
\item  the lightest superparticle (LSP) is stable.
Usually it is the photino $\tilde \gamma $, the superpartner
of the photon with some admixture of the neutral higgsino. This
is the candidate for the DM particle which should be neutral
and survive since the Big Bang.
\end{itemize}

%--------------------------------------
\subsection{Properties of interactions}

If one assumes that the $R$-parity is preserved, then the
interactions of the superpartners are essentially the same as
in the SM, but two of three particles involved into the
interaction at any vertex are replaced by the superpartners.
The reason for it is the $R$-parity.

\begin{figure}[t]\vspace{-0.5cm}
\begin{center}
\leavevmode
%\hspace*{-5mm}
\includegraphics[width=0.45\textwidth]{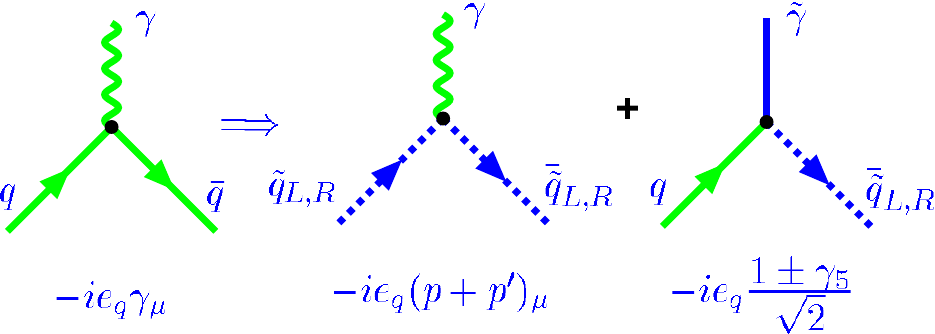}
\hspace*{10mm}
\includegraphics[width=0.45\textwidth]{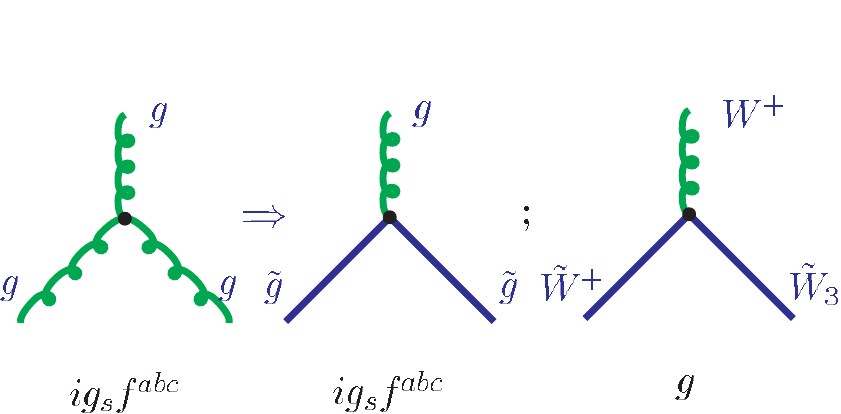}\\
\vspace*{5mm}
\includegraphics[width=0.45\textwidth]{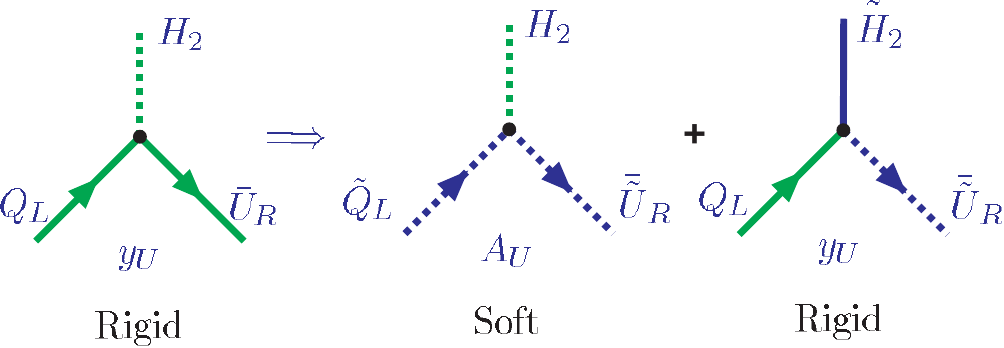}
\end{center}
%\vspace*{-5mm}
\caption{The gauge-matter interaction, the gauge
self-interaction and the Yukawa interaction.}
\label{yukint}
\end{figure}

Typical vertices are shown in Fig.~\ref{yukint}. The tilde
above the letter denotes the corresponding superpartner. Note
that the coupling is the same in all the vertices involving
the superpartners.

\begin{figure}[b]\vspace*{-2mm}
\begin{center}
\leavevmode
\includegraphics[width=0.40\textwidth]{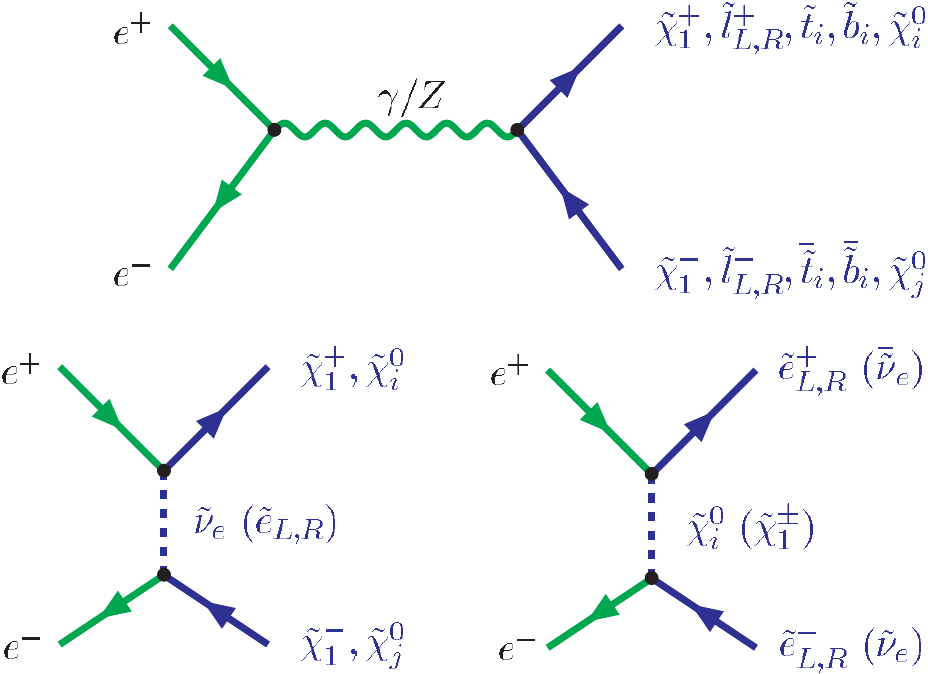}
\end{center}
\vspace*{-3mm}
\caption{Creation of the superpartners at electron-positron colliders.}
\label{creation}
\end{figure}

%-----------------------------------------------
\subsection{Creation and decay of superpartners}

The above-mentioned rule together with the Feynman rules for the
SM enables one to draw diagrams describing creation of the
superpartners. One of the most promising processes is the $e^+e^-$
annihilation (see Fig.~\ref{creation}).
\begin{figure}[htb]
\begin{center}
\leavevmode \hspace*{-2mm}
\includegraphics[width=0.22\textwidth,height=0.1\textwidth]{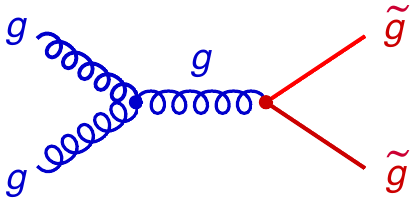}
\includegraphics[width=0.22\textwidth,height=0.1\textwidth]{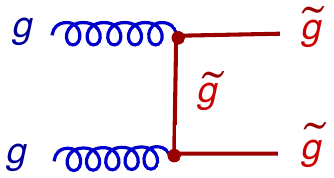}
\includegraphics[width=0.22\textwidth,height=0.1\textwidth]{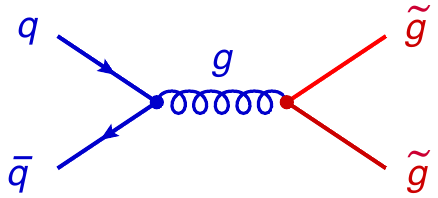}
\includegraphics[width=0.22\textwidth,height=0.1\textwidth]{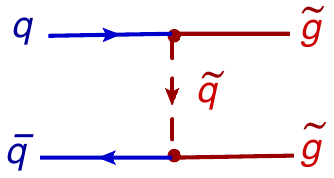}\\[6mm]
\includegraphics[width=0.22\textwidth,height=0.1\textwidth]{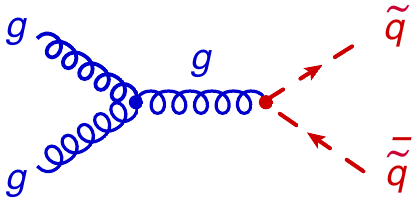}
\includegraphics[width=0.17\textwidth,height=0.1\textwidth]{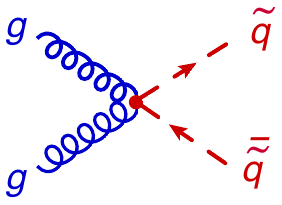}
\includegraphics[width=0.22\textwidth,height=0.1\textwidth]{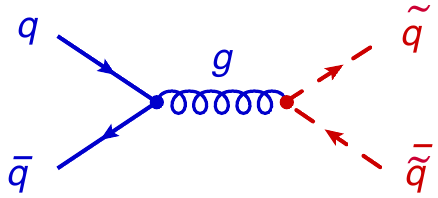}
\includegraphics[width=0.22\textwidth,height=0.1\textwidth]{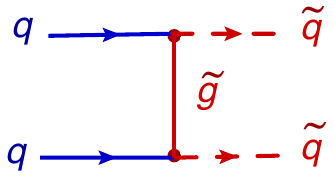}\\[6mm]
\includegraphics[width=0.22\textwidth,height=0.1\textwidth]{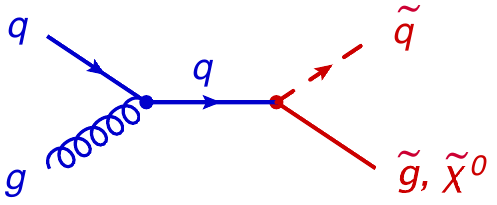}
\includegraphics[width=0.22\textwidth,height=0.1\textwidth]{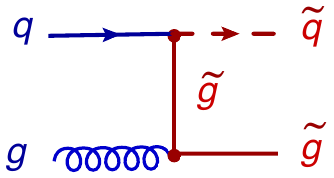}
\includegraphics[width=0.22\textwidth,height=0.1\textwidth]{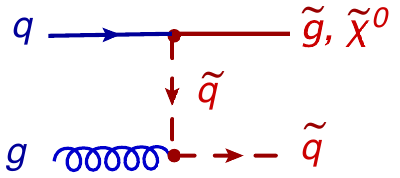}\\[6mm]
\includegraphics[width=0.22\textwidth,height=0.1\textwidth]{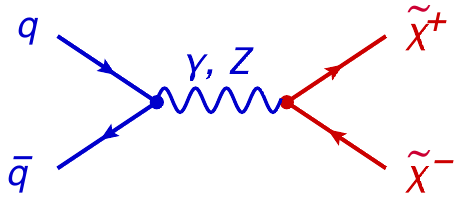}
\includegraphics[width=0.22\textwidth,height=0.1\textwidth]{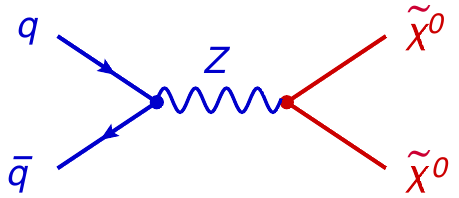}
\includegraphics[width=0.22\textwidth,height=0.1\textwidth]{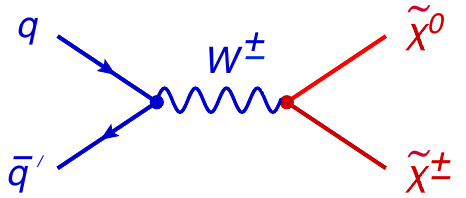}
\includegraphics[width=0.22\textwidth,height=0.1\textwidth]{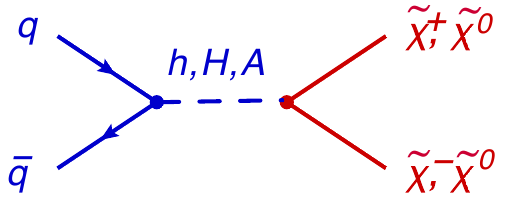}
\end{center}
\caption{Examples of diagrams for the SUSY particle production
via the strong interactions (top rows for $\gl\gl$,
$\sq\overline{\sq}$ and $\gl\sq$, respectively) and the electroweak
interactions (the lowest row).}
\label{f1}
\end{figure}
The usual kinematic restriction is given by the c.m. energy
$m^{max}_{sparticle} \leq \sqrt{s}/2$. Similar
processes take place at hadron colliders with the electrons
and the positrons being replaced by the quarks and the gluons.

Experimental signatures at the hadron colliders are similar to
those at the $e^+e^-$ machines; however, here one has wider
possibilities. Besides the usual annihilation channel, one has
numerous processes of gluon fusion, quark-antiquark and
quark-gluon scattering (see Fig.~\ref{f1}).

Creation of the superpartners can be accompanied by creation
of the ordinary particles as well. We consider various
experimental signatures below. They crucially depend on the
SUSY breaking pattern and on the mass spectrum of the
superpartners.

\begin{figure}[htb]
\begin{center}
$
\begin{array}{llll}
\mbox{squarks} & \tilde q_{L,R}\to q+\tilde \chi^0_i  &&\\
& \tilde q_{L}\to q'+\tilde \chi^\pm_i &&\\
& \tilde q_{L,R}\to q+\tilde g&&\\[4mm]
\mbox{sleptons} & \tilde l\to l+\tilde \chi^0_i &&\\
& \tilde l_{L}\to \nu_l+\tilde \chi^\pm_i &&\\[4mm]
\mbox{chargino}
& \tilde \chi^\pm_i\to e+\nu_e+\tilde \chi^0_i &&\\
& \chi^\pm_i\to q+\bar q'+\tilde \chi^0_i &&\\[4mm]
\mbox{gluino} & \tilde g\to q=\bar q +\tilde \gamma &&\\
& \tilde g\to g+\tilde \gamma &&\\[4mm]
\mbox{neutralino} & \tilde \chi^0_i\to \tilde \chi^0_1+l^+ +l^-
& \hspace{15mm} \mbox{final states} & l^+l^- + \slash \hspace{-0.24cm}E_T\\
& \tilde \chi^0_i\to \tilde \chi^0_1+q +\bar q' & & 2 \mbox{jets} + \slash \hspace{-0.24cm}E_T\\
& \tilde \chi^0_i\to \tilde \chi^\pm_1+l^\pm +\nu_l &  &\gamma + \slash \hspace{-0.24cm}E_T\\
& \tilde \chi^0_i\to \tilde \chi^0_1+\nu_l+\bar \nu_l & & \slash \hspace{-0.24cm}E_T
\end{array}
$
\hspace*{-50mm}
\leavevmode
\raisebox{-8mm}{\includegraphics[width=0.45\textwidth]{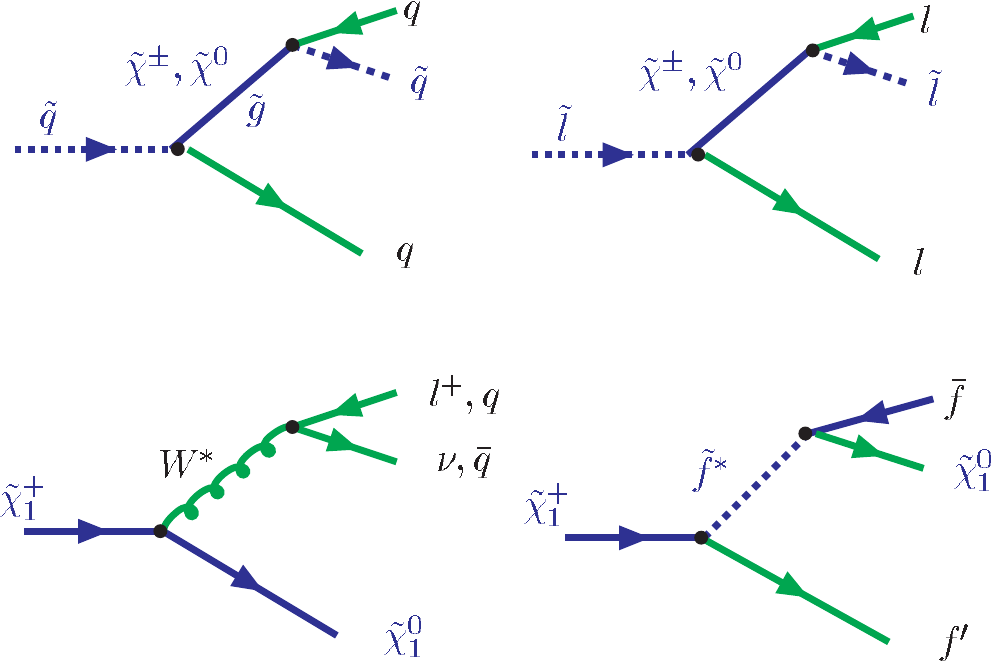}}
\end{center}
\vspace{0cm}
\caption{Decay of superpartners}\label{decay}
\end{figure}

The decay properties of the superpartners also depend on their
masses. For the quark and lepton superpartners the main
processes are shown in Fig.~\ref{decay}.

%-------------------------------------
\section{Breaking of SUSY in the MSSM}

Usually it is assumed that the supersymmetry is broken
spontaneously via the v.e.v.s of some fields.  However, in
the case of supersymmetry one can not use the scalar fields
like the Higgs field, but rather the auxiliary fields present
in any SUSY  multiplet. There are two basic mechanisms of
spontaneous SUSY breaking: the Fayet-Iliopoulos (or $D$-type)
mechanism~\cite{Fayet} based on the  $D$ auxiliary field from
the vector multiplet and the O'Raifeartaigh (or $F$-type)
mechanism~\cite{O'R} based on the $F$ auxiliary field from the
chiral multiplet. Unfortunately, one can not explicitly use
these mechanisms within the MSSM since none of the fields of
the MSSM can develop the non-zero v.e.v. without spoiling the
gauge invariance. Therefore, the spontaneous SUSY breaking
should take place via some other fields.

The most common scenario for producing low-energy supersymmetry
breaking is called the {\em hidden sector
scenario}~\cite{hidden}. According to this scenario, there exist
two sectors: the usual matter belongs to the "visible" one, while
the second, "hidden" sector, contains the fields which lead to
breaking of the supersymmetry. These two sectors interact with
each other by an exchange of some fields called {\em messengers},
which mediate SUSY breaking from the hidden to the visible sector.
There might be various types of the messenger fields: gravity,
gauge, etc. The hidden sector is the weakest part of the MSSM.
It contains a lot of ambiguities and leads to uncertainties of
the MSSM predictions considered below.

So far there are four known main mechanisms to mediate SUSY
breaking from the hidden to the visible sector:
%\vspace{-0.2cm}
\begin{itemize}
\item Gravity mediation (SUGRA)~\cite{gravmed};\\[-0.2cm]
\item Gauge mediation~\cite{gaugemed};\\[-0.2cm]
\item Anomaly mediation~\cite{anommed};\\[-0.2cm]
\item Gaugino mediation~\cite{gauginomed}.
\end{itemize}

All the four mechanisms of soft SUSY breaking are different in
details but are common in results. The predictions for the
sparticle spectrum depend on the mechanism of SUSY breaking.
In what follows, to calculate the mass spectrum of the
superpartners, we need the explicit form of the SUSY breaking
terms. For the MSSM without the $R$-parity violation one has
in general
\begin{align}
&- {\cal L}_{Breaking} = \label{soft} \\
&=  \sum_{i}^{} m^2_{0i} \left|\varphi_i\right|^2 +
\biggl( \frac 12 \sum_{\alpha}^{}
M_\alpha \tilde\lambda_\alpha \tilde\lambda_\alpha +
B H_1 H_2 + A^U_{ab} \tilde Q_a \tilde U^c_b H_2 +
A^D_{ab} \tilde Q_a \tilde D^c_b H_1 +
A^L_{ab} \tilde L_a \tilde E^c_b H_1 \biggr), \notag
\end{align}
where we have suppressed the $SU(2)$ indices. Here $\varphi_i$
are all the scalar fields, $\tilde \lambda_\alpha $ are the
gaugino fields, $\tilde Q, \tilde U, \tilde D$ and
$\tilde L, \tilde E$ are the squark and slepton fields,
respectively, and $H_{1,2}$ are the SU(2) doublet Higgs fields.

Eqn.~(\ref{soft}) contains a vast number of free parameters
which spoils the predictiive power of the model. To reduce
their number, we adopt the so-called {\em universality
hypothesis}, i.~e., we assume the universality or equality of
various soft parameters at the high energy scale, namely, we
put all the spin-0 particle masses to be equal to the universal
value $m_0$, all the spin-1/2 particle (gaugino) masses to be
equal to $m_{1/2}$ and all the cubic and quadratic terms,
proportional to $A$ and $B$, to repeat the structure of the
Yukawa superpotential~(\ref{R}). This is the additional
requirement motivated by the supergravity mechanism of SUSY
breaking. The universality is not the necessary requirement and
one may consider the non-universal soft terms as well. However,
it will not change the qualitative picture presented below; so,
for simplicity, in what follows we consider the universal
boundary conditions. In this case, Eqn.~(\ref{soft}) takes the
form
\begin{align}
&-{\cal L}_{Breaking}= \\
&=m_0^2\sum_{i}^{}|\varphi_i|^2+\left( \frac{ m_{1/2}}{2}
\sum_{\alpha}^{} \tilde \lambda_\alpha\tilde \lambda_\alpha+  B\mu H_1H_2
+A\bigl[y^U_{ab}\tilde Q_a\tilde U^c_bH_2+y^D_{ab}\tilde Q_a
\tilde D^c_bH_1+ y^L_{ab}\tilde L_a\tilde E^c_bH_1\bigr]
\right).\notag
\label{soft2}
\end{align}

Thus, we are left with five free parameters, namely,
$m_0,m_{1/2},A,B$ and $\mu$ versus two parameters of the
Higgs potential in the SM, $m^2$ and $\lambda$. In
the SUSY model the Higgs potential is not arbitrary
but is calculated from the Yukawa and gauge terms as we will
see below.

The soft terms explicitly break the supersymmetry. As will be
shown later, they lead to the mass spectrum of the
superpartners different from that of the ordinary particles.
Remind that the masses of the quarks and leptons remain zero
until the $SU(2)$ symmetry is spontaneously broken.

%------------------------------------------------
\subsection{The soft terms and the mass formulae}

There are two main sources of the mass terms in the
Lagrangian: the $D$-terms and the soft ones. With given values
of $m_0,m_{1/2},\mu,Y_t,Y_b,Y_\tau, A$, and $B$ one can
construct the mass matrices for all the particles. Knowing them
at the GUT scale, one can solve the corresponding RG equations,
thus linking the values at the GUT and electroweak scales.
Substituting these parameters into the mass matrices, one can
predict the mass spectrum of the superpartners~\cite{spectrum,BEK}.

%-------------------------------------------
\subsubsection{Gaugino-higgsino mass terms}

The mass matrix for the gauginos, the superpartners of the gauge
bosons, and for the higgsinos, the superpartners of the Higgs
bosons, is nondiagonal, thus leading to their mixing. The mass
terms look like
\begin{equation}
{\cal L}_{Gaugino-Higgsino}=
-\frac{1}{2}M_3\bar{\lambda}_a\lambda_a
-\frac{1}{2}\bar{\chi}M^{(0)}\chi -(\bar{\psi}M^{(c)}\psi + h.c.),
\end{equation}
where $\lambda_a, a=1,2,\ldots ,8$ are the Majorana gluino
fields and
\begin{equation}
\chi = \left(\begin{array}{c}
\tilde{B}^0 \\
\tilde{W}^3 \\
\tilde{H}^0_1 \\
\tilde{H}^0_2
\end{array}\right), \ \ \
\psi = \left( \begin{array}{c}
\tilde{W}^{+} \\
\tilde{H}^{+}
\end{array}\right)
\end{equation}
are, respectively, the Majorana neutralino and the Dirac
chargino fields.

The neutralino mass matrix is
\begin{equation*}
M^{(0)} =
\begin{pmatrix}
M_1 & 0 & -M_Z\cos\beta \sin\theta_W & \phantom{-}M_Z\sin\beta \sin\theta_W \\
0 & M_2 & \phantom{-}M_Z\cos\beta \cos\theta_W & -M_Z\sin\beta \cos\theta_W \\
-M_Z\cos\beta \sin\theta_W & \phantom{-}M_Z\cos\beta \cos\theta_W & 0 & -\mu \\
\phantom{-}M_Z\sin\beta \sin\theta_W & -M_Z\sin\beta \cos\theta_W & -\mu & 0
\end{pmatrix},
\label{neut}
\end{equation*}
where $\tan\beta = v_2/v_1$ is the ratio of two Higgs v.e.v.s and
$\sin\theta_W$ is the usual sine of the weak mixing
angle. The physical neutralino masses $M_{\tilde{\chi}_i^0}$ are
obtained as eigenvalues of this matrix after diagonalization.

For the chargino mass matrix one has
\begin{equation}
M^{(c)}=
\begin{pmatrix}
M_2 & \sqrt{2} M_W \sin\beta \\
\sqrt{2} M_W \cos\beta & \mu
\end{pmatrix}.
\label{char}
\end{equation}
This matrix has two chargino eigenstates
$\tilde{\chi}_{1,2}^{\pm}$ with mass eigenvalues
\begin{equation*}
M^2_{1,2} =
\frac{1}{2} \left[ M^2_2 + \mu^2 + 2M^2_W \mp
\sqrt{(M^2_2 - \mu^2)^2 + 4 M^4_W \cos^2 2\beta
+4 M^2_W (M^2_2 + \mu^2 + 2 M_2 \mu \sin 2\beta )}\right]\!.
\end{equation*}

%-----------------------------------------
\subsubsection{Squark and slepton masses}

The non-negligible Yukawa couplings cause mixing between the
electroweak eigenstates and the mass eigenstates of the third
generation particles.  The mixing matrices for
$\tilde{m}^{2}_t,\tilde{m}^{2}_b$ and $\tilde{m}^{2}_\tau$ are
\begin{equation*}
\begin{pmatrix}
\tilde m_{tL}^2 & m_t (A_t - \mu\cot\beta) \\
m_t (A_t - \mu\cot\beta) & \tilde m_{tR}^2
\end{pmatrix},
\label{stopmat}
\end{equation*}
\begin{equation*}
\begin{pmatrix}
\tilde  m_{bL}^2 & m_b( A_b - \mu\tan\beta) \\
m_b (A_b - \mu\tan\beta ) & \tilde  m_{bR}^2
\end{pmatrix},
\label{sbotmat}
\end{equation*}
\begin{equation*}
\begin{pmatrix}
\tilde  m_{\tau L}^2 & m_{\tau}(A_{\tau} - \mu\tan\beta) \\
m_{\tau}(A_{\tau} - \mu\tan\beta) & \tilde m_{\tau R}^2
\end{pmatrix}\phantom{,}
\label{staumat}
\end{equation*}
with
\begin{equation*}
\begin{split}
\tilde m_{tL}^2 &= \tilde{m}_Q^2 + m_t^2 + \frac{1}{6} \bigl(4M_W^2 - M_Z^2 \bigr) \cos 2\beta ,\\
\tilde m_{tR}^2&=\tilde{m}_U^2+m_t^2-\frac{2}{3}(M_W^2-M_Z^2)\cos
  2\beta ,\\
\tilde m_{bL}^2&=\tilde{m}_Q^2+m_b^2-\frac{1}{6}(2M_W^2+M_Z^2)\cos
  2\beta ,\\
\tilde m_{bR}^2&=\tilde{m}_D^2+m_b^2+\frac{1}{3}(M_W^2-M_Z^2)\cos
  2\beta ,\\
\tilde m_{\tau L}^2&=\tilde{m}_L^2+m_{\tau}^2-\frac{1}{2}(2M_W^2-M_Z^2)\cos
2\beta ,\\ \tilde m_{\tau
R}^2&=\tilde{m}_E^2+m_{\tau}^2+(M_W^2-M_Z^2)\cos
  2\beta
\end{split}
\end{equation*}
and the mass eigenstates are the eigenvalues of these mass
matrices. For the light generations mixing is negligible.

The first terms here ($\tilde{m}^2$) are the soft ones, which
are calculated using the RG equations starting from their
values at the GUT (Planck) scale. The second ones are the
usual masses of the quarks and leptons and the last ones are
the $D$-terms of the potential.

%-------------------------------
\subsection{The Higgs potential}

As has already been mentioned, the Higgs potential in the MSSM
is totally defined by the superpotential (and the soft terms).
Due to the structure of ${\cal L}_{Yukawa}$ the Higgs
self-interaction is given by the $D$-terms while the $F$-terms
contribute only to the mass matrix. The tree level potential is
\begin{equation}
V_{tree} = m^2_1|H_1|^2 + m^2_2|H_2|^2 - m^2_3(H_1H_2 + h.c.)
+ \frac{g^2+g^{'2}}{8}(|H_1|^2-|H_2|^2)^2 +
\frac{g^2}{2}|H_1^*H_2|^2,
\label{Higpot}
\end{equation}
where $m_1^2 = m^2_{H_1} + \mu^2, m_2^2 = m^2_{H_2} + \mu^2$.
At the GUT scale $m_1^2 = m^2_2 = m_0^2 + \mu^2_0, \ m^2_3 = -B\mu_0$.
Notice that the Higgs self-interaction coupling in
Eqn.~(\ref{Higpot}) is fixed and defined by the gauge interactions
as opposed to the Standard Model.

The Higgs scalar potential in accordance with the
supersymmetry, is positive definite and stable. It has no
nontrivial minimum different from zero.  Indeed, let us write
the minimization condition for  the potential~(\ref{Higpot})
\begin{equation}
\begin{split}
\frac 12\frac{\delta V}{\delta H_1} &= m_1^2v_1 - m^2_3v_2
+ \frac{g^2 + g'^2}4 (v_1^2 - v_2^2)v_1 = 0, \\
\frac 12\frac{\delta V}{\delta H_2} &= m_2^2v_2 - m^2_3v_1
+ \frac{g^2 + g'^2}4 (v_1^2 - v_2^2)v_2 = 0,
\end{split}
\label{mincond}
\end{equation}
where we have introduced the notation
\begin{equation*}
\langle H_1 \rangle \equiv v_1 = v \cos\beta , \ \ \
\langle H_2 \rangle \equiv v_2 = v \sin\beta,
\end{equation*}
\begin{equation*}
v^2 = v_1^2 + v_2^2,\ \ \ \tan\beta \equiv \frac{v_2}{v_1}.
\end{equation*}
Solution to Eqs.~(\ref{mincond}) can be expressed in
terms of $v^2$ and $\sin 2\beta$
\begin{equation}
v^2 = \frac{4(m^2_1 - m^2_2\tan^2\beta)}{(g^2 + g'^2)(\tan^2\beta - 1)},
\ \ \
\sin2\beta = \frac{2m^2_3}{m^2_1 + m^2_2}.
\label{minsol}
\end{equation}
One can easily see from Eqn.~(\ref{minsol}) that if
$m_1^2 = m_2^2 = m_0^2 + \mu_0^2$, $v^2$ happens to be negative, i.~e.
the minimum does not exist. In fact, real positive solutions to
Eqs.~(\ref{mincond}) exist only if the following
conditions are satisfied:
\begin{equation}
m_1^2 + m_2^2 > 2 m_3^2, \ \ \  m_1^2 m_2^2 < m_3^4 ,
\label{cond}
\end{equation}
which is not the case at the GUT scale. This means that
spontaneous breaking of the $SU(2)$  gauge invariance, which is
needed in the SM to give masses for all the particles, does not
take place in the MSSM.

This strong statement is valid, however, only at the GUT scale.
Indeed, going down with the energy, the parameters of the
potential~(\ref{Higpot}) are renormalized.  They become the
``running'' parameters with the energy scale dependence given
by the RG equations.

%---------------------------------------------------
\subsection{Radiative electroweak symmetry breaking}

\begin{figure}[bh]
\begin{center}
\leavevmode
\includegraphics[width=0.45\textwidth]{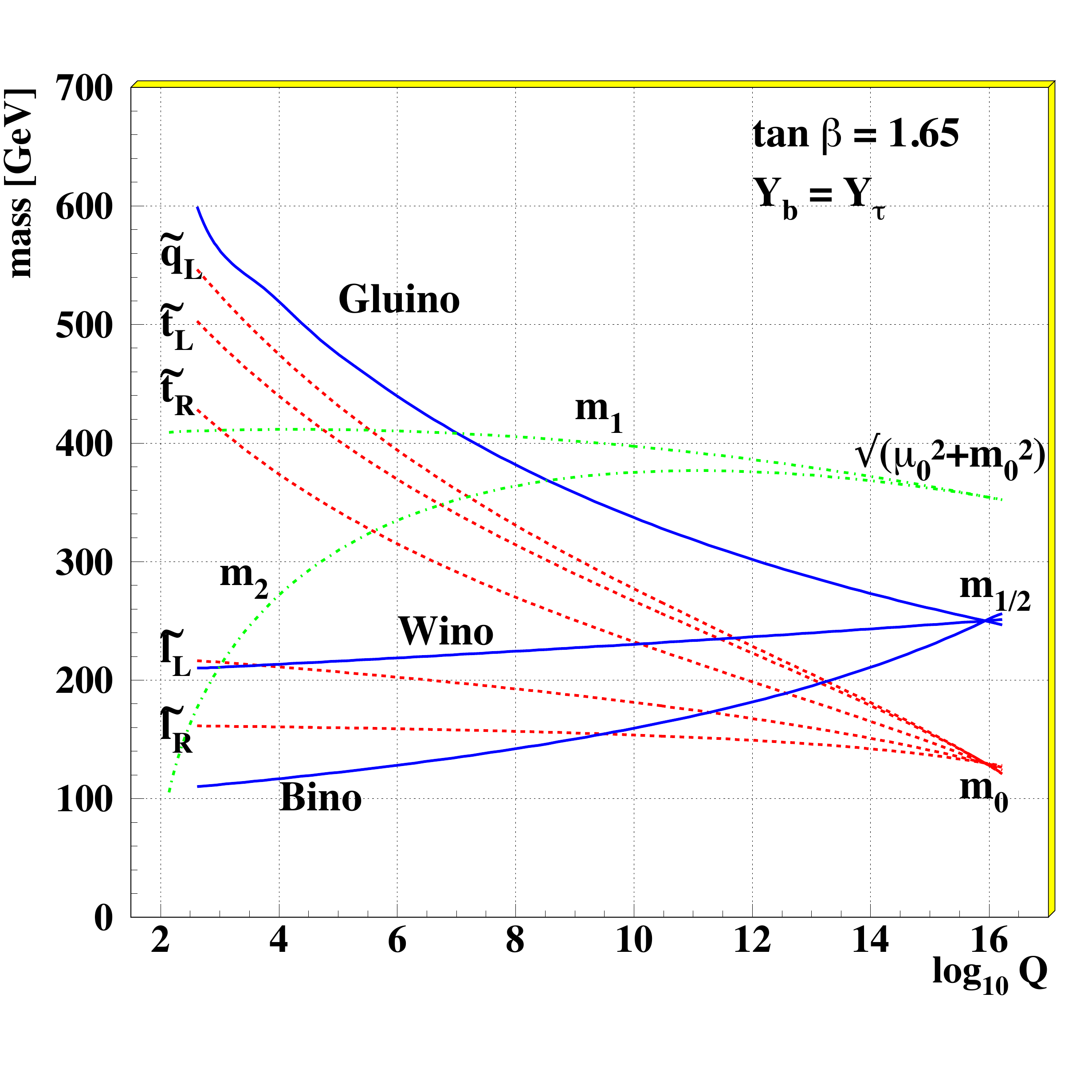}
%\leavevmode
\hspace*{10mm}
\includegraphics[width=0.45\textwidth]{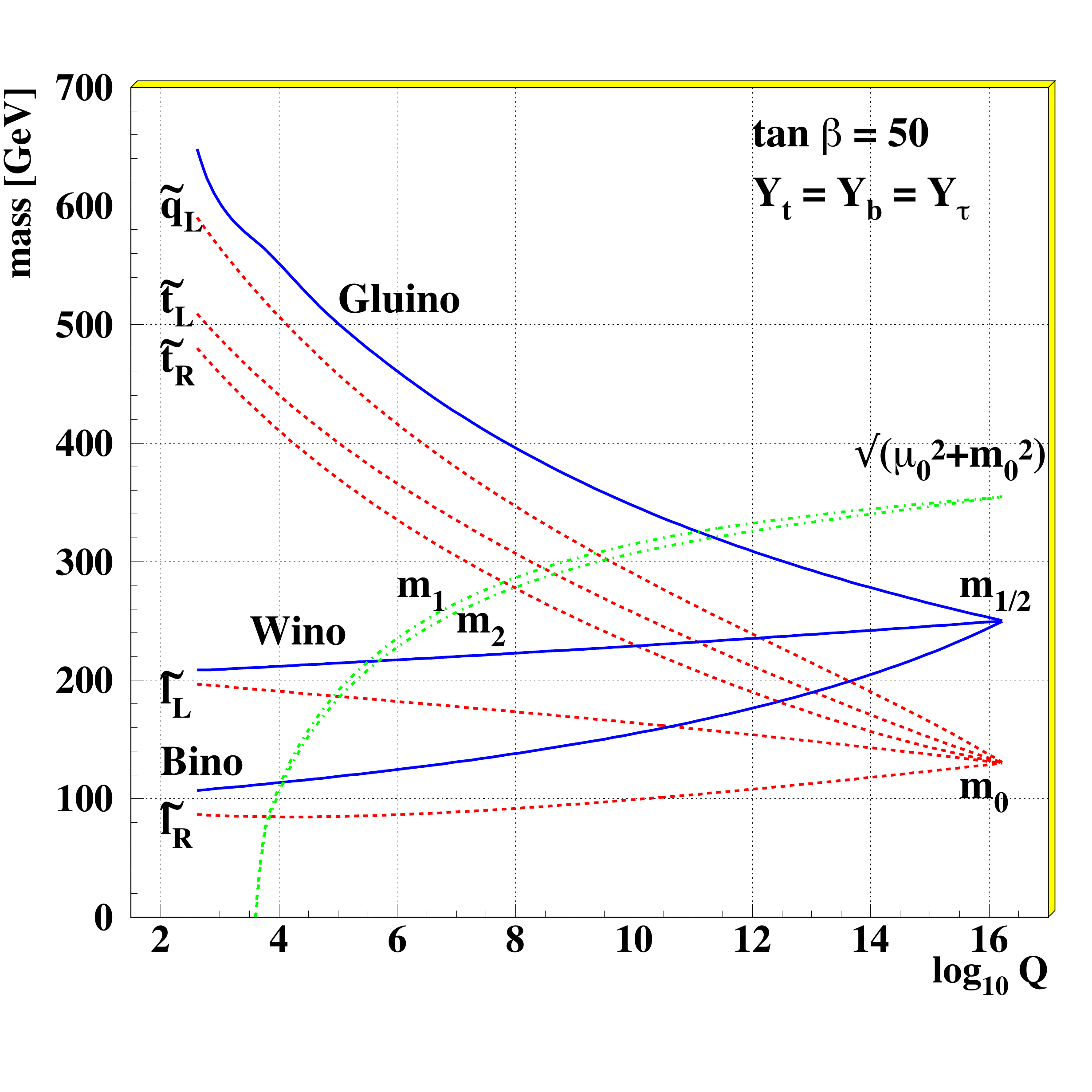}
\end{center}
\vspace*{-12mm}
\caption{An example of evolution of superparticle masses and soft
supersymmetry breaking parameters $m_1^2 = m^2_{H_1} + \mu^2$ and
$m_2^2 = m^2_{H_2} + \mu^2$ for low (left) and high (right) values
of $\tan\beta$.}
\label{16}
\end{figure}

The running of the Higgs masses leads to the remarkable
phenomenon known as {\em radiative electroweak symmetry
breaking}. Indeed, one can see in Fig.~\ref{16} that $m_2^2$
(or both $m_1^2$ and $m_2^2$) decreases when going down from
the GUT scale to the $M_Z$ scale and can even become negative.
As a result, at some value of $Q^2$ the conditions~(\ref{cond})
are satisfied, so that the nontrivial minimum appears. This
triggers spontaneous breaking of the $SU(2)$ gauge invariance.
The vacuum expectations of the Higgs fields acquire nonzero
values and provide masses to the quarks, leptons and $SU(2)$
gauge bosons, and additional contributions to the masses of
their superpartners.

In this way one also obtains the explanation of why the two
scales are so much different. Due to the logarithmic running of
the parameters, one needs a long "running time" to get $m_2^2$
(or both $m_1^2$ and $m_2^2$) to be negative when starting from
a positive value of the order of
$M_{SUSY}\sim 10^2 \div 10^3$~GeV at the GUT scale.
%
%--------------   Evolution of sparticle masses, m1 and m2
%

%-----------------------------
\subsection{The superpartners mass spectrum}

The mass spectrum is defined by the low energy parameters.
To calculate the low energy values of the soft terms, we use the
corresponding RG equations~\cite{Ibanez}. Having all the RG
equations, one can now find the RG flow for the soft terms.
Taking the initial values of the soft masses at the
GUT scale in the interval between $10^2\div 10^3$~GeV consistent
with the SUSY scale suggested by the unification of the gauge
couplings~(\ref{MSUSY}) leads to the RG flow of the soft terms
shown in Fig.~\ref{16}.~\cite{spectrum,BEK}

\begin{figure}[t]\vspace{-0.0cm}
\begin{center}
\leavevmode
\includegraphics[width=0.43\textwidth]{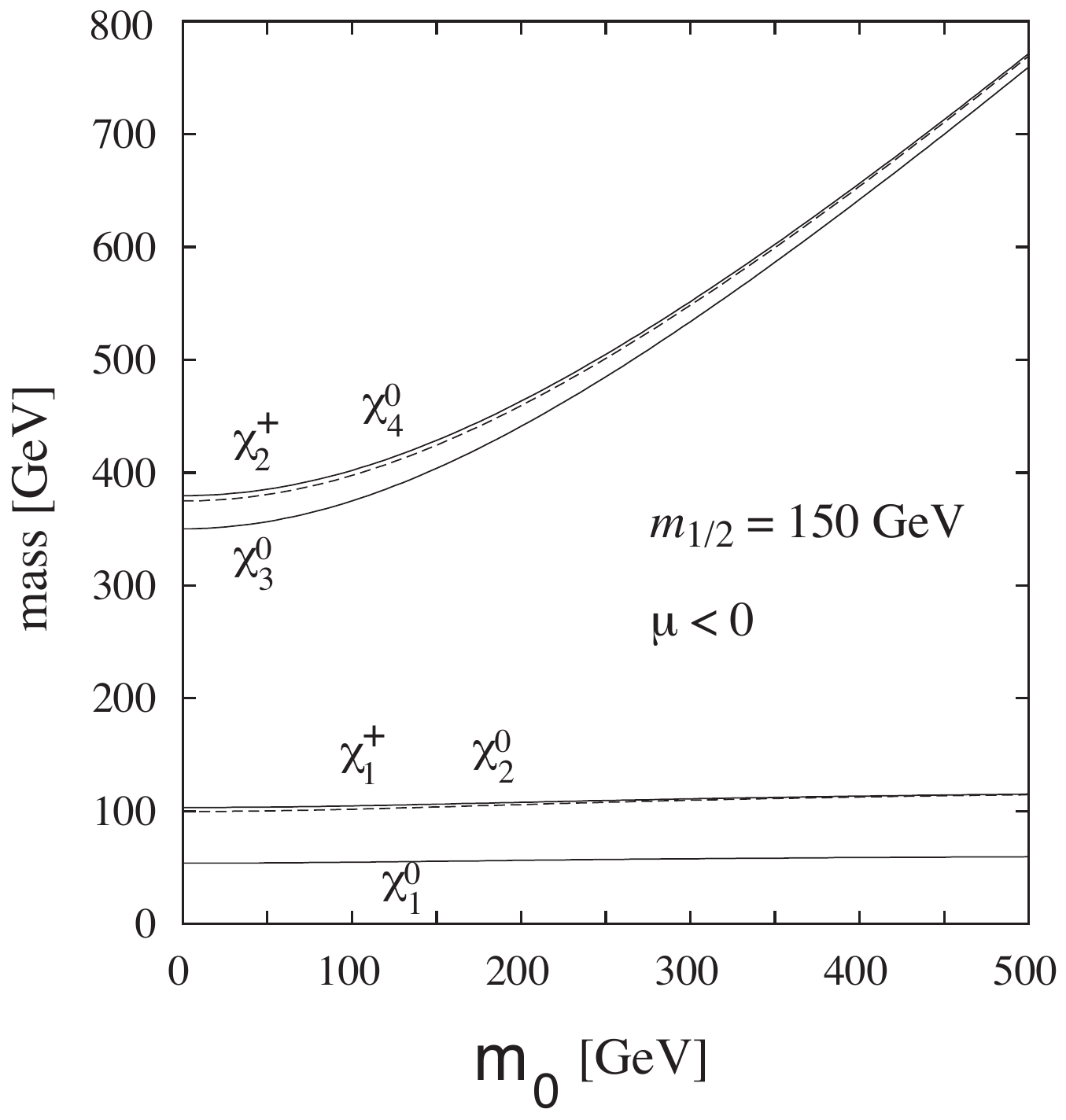}
\hspace*{10mm}
\raisebox{1mm}{\includegraphics[width=0.46\textwidth]{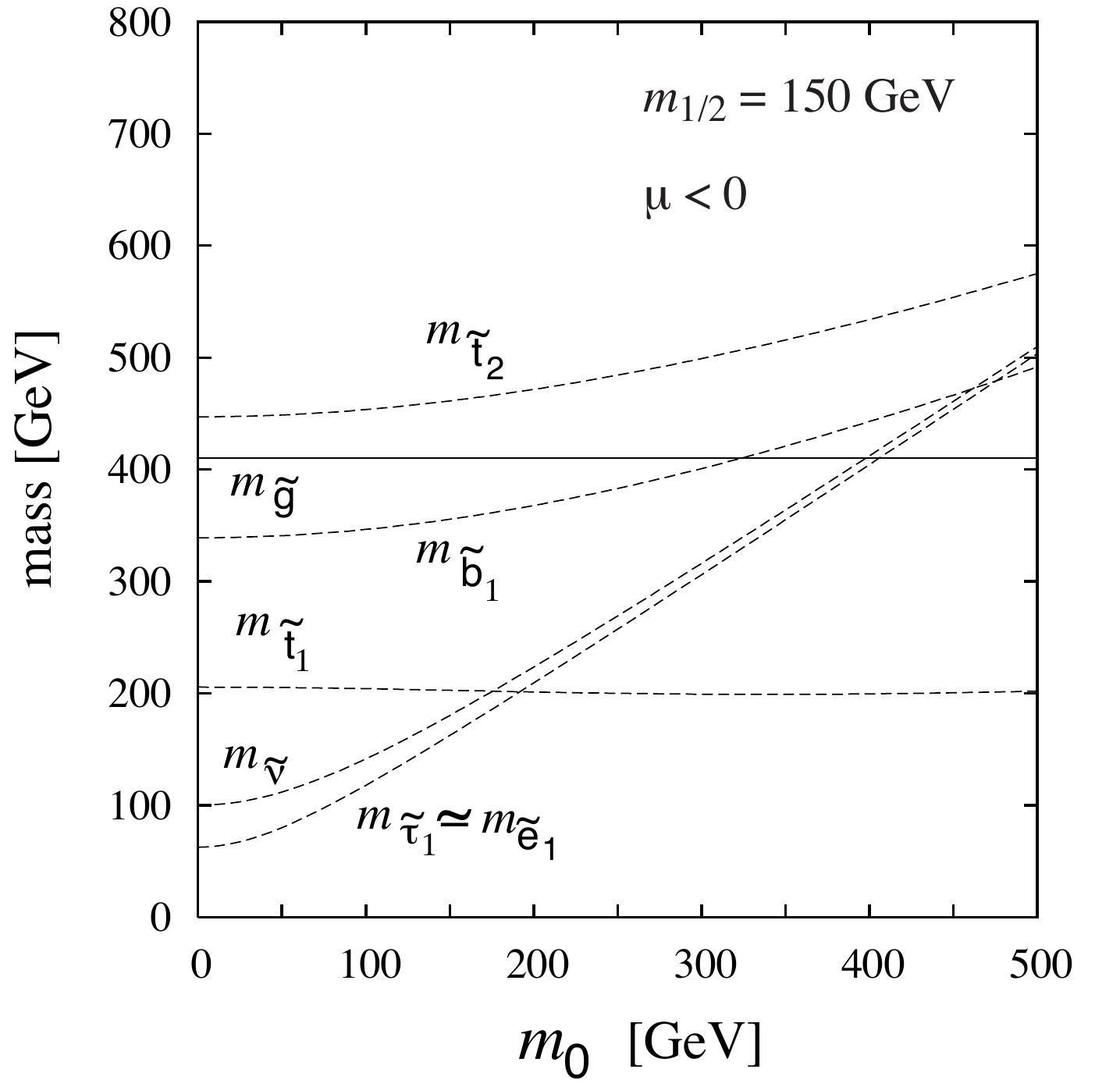}}
\end{center}
\vspace{-0.5cm}
\caption{The masses of sparticles as functions of the initial
value $m_0$.}
\label{fig:barger}
\end{figure}

One should mention the following general features common to any
choice of initial conditions:
\begin{itemize}
\item The gaugino masses follow the running of the gauge couplings
and split at low energies. The gluino mass is running faster
than the other ones and is usually the heaviest due to the
strong interaction.
\item The squark and slepton masses also split at low energies,
the stops (and sbottoms) being the lightest due to the
relatively big Yukawa couplings of the third generation.
\item The Higgs masses (or at least one of them) are running down
very quickly and may even become negative.
\end{itemize}
The typical dependence of the mass spectra on the initial
conditions at the GUT scale ($m_0$) is also shown in
Fig.~\ref{fig:barger}~\cite{Barger,bbog}. For a given value
of $m_{1/2}$ the masses of the lightest particles are
practically independent of $m_0$, while the masses of the
heavier ones increase with it monotonically. One can see that
the lightest neutralinos and charginos as well as the top-squark
may be rather light.

%-----------------------------
\subsection{The Higgs boson masses}

Provided conditions~(\ref{cond}) are satisfied, one can also
calculate the masses of the Higgs bosons taking the second
derivatives of the potential~(\ref{Higpot}) with respect to the
real and imaginary parts of the Higgs fields ($H_i = S_i + iP_i$)
in the minimum. The mass matrices at the tree level are \\[5mm]
$CP$-odd components $P_1$ and $P_2$:
\begin{equation}
{\cal M}^{odd} =
\left.\frac{\partial^2 V}{\partial P_i \partial P_j} \right |_{H_i=v_i} =
\begin{pmatrix}
\tan\beta & 1 \\ 1 & \cot\beta
\end{pmatrix} m_3^2,
\end{equation}
$CP$-even neutral components $S_1$ and $S_2$:
\begin{equation}
{\cal M}^{even} =
\left.\frac{\partial^2 V}{\partial S_i \partial S_j} \right|_{H_i=v_i} =
\begin{pmatrix}
\tan\beta & -1 \\ -1 & \cot\beta
\end{pmatrix} m_3^2 +
\begin{pmatrix}
\cot\beta & -1 \\ -1 & \tan\beta
\end{pmatrix} M_Z^2 \frac{\sin2\beta}{2},
\end{equation}
Charged components $H^-$ and $H^+$:
\begin{equation}
{\cal M}^{ch} =
\left.\frac{\partial^2 V}{\partial H^+_i \partial H^-_j} \right|_{H_i=v_i} =
\begin{pmatrix}
\tan\beta & 1 \\ 1 & \cot\beta
\end{pmatrix} \bigl( m_3^2 + M_W^2 \frac{\sin2\beta}{2} \bigr).
\end{equation}

Diagonalizing the mass matrices, one gets the mass eigenstates:
\begin{equation*}
\hspace{-0.3cm}\begin{array}{l} \left\{\!\!
\begin{array}{l} G^0 \ = -\cos\beta P_1+\sin \beta P_2 ,
\ Goldstone \ boson  \to Z_0, \\ A \ = \sin\beta P_1+\cos \beta P_2 ,  \
Neutral \ CP-odd \ Higgs, \end{array}\right.\\  \\ \left\{\!\!
\begin{array}{l} G^+\! \!= \!-\!\cos\beta (H^-_1)^*\!+\!\sin \beta H^+_2 , \
Goldstone \ boson\!  \to\! W^+, \\ H^+ = \sin\beta (H^-_1)^*+\cos \beta
H^+_2 , \ Charged \ Higgs, \end{array}\right.\\ \\ \left\{\!\!
\begin{array}{l} h \ = -\sin\alpha S_1+\cos\alpha S_2 ,
\ SM \ CP-even \ Higgs, \\ H \ = \cos\alpha S_1+\sin\alpha S_2 ,
\ Extra \ heavy \ Higgs , \end{array}\right.
\end{array}
\end{equation*}
where the mixing angle $\alpha$ is given by
\begin{equation*}
\tan 2\alpha = \tan 2\beta
\left(\frac{m^2_A+M^2_Z}{m^2_A-M^2_Z}\right).
\end{equation*}
The physical Higgs bosons acquire the following masses~\cite{MSSM}:\\[5mm]
$CP$-odd neutral Higgs $A$:
\begin{equation}
m^2_A = m^2_1 + m^2_2,
\end{equation}
Charged Higgses $H^{\pm}$:
\begin{equation}
m^2_{H^{\pm}} = m^2_A + M^2_W,
\end{equation}
$CP$-even neutral Higgses $H, h$:
\begin{equation}
m^2_{H,h} = \frac{1}{2} \left[ m^2_A + M^2_Z \pm
\sqrt{(m^2_A + M_Z^2)^2 - 4m^2_A M_Z^2 \cos^2 2\beta} \right],
\end{equation}
where, as usual,
\begin{equation*}
M^2_W = \frac{g^2}{2} \, v^2, \ \ \
M^2_Z = \frac{g^2+g'^2}{2} \, v^2 .
\end{equation*}
This leads to the once celebrated SUSY mass relations
\begin{gather}
m_{H^{\pm}} \geq M_W, \ \ m_h \leq m_A \leq M_H, \notag \\[1mm]
m_h \leq M_Z |\cos 2\beta| \leq M_Z, \\[1mm]
m_h^2 + m_H^2 = m_A^2 + M_Z^2. \notag
\label{bound}
\end{gather}

Thus, the lightest neutral Higgs boson happens to be lighter than
the $Z$-boson, which clearly distinguishes it from the SM one.
Though we do not know the mass of the Higgs boson in the SM, there
are several indirect constraints leading to the lower boundary of
$m_h^{SM} \geq 135 $ GeV. After including the leading one-loop radiative
corrections,  the mass of the lightest Higgs boson in the MSSM,
$m_h$, reads
\begin{equation}
m_h^2 = M_Z^2 \cos^2 2\beta + \frac{3g^2m_t^4}{16\pi^2M_W^2}
\log\frac{\tilde m^2_{t_1}\tilde m^2_{t_2}}{m_t^4} + \dots
\end{equation}
which leads to about 40~GeV increase~\cite{CW+we}. The second
loop correction is negative but small~\cite{feynhiggs}.

\begin{figure}[t]
\begin{center}
\leavevmode
\includegraphics[width=0.40\textwidth]{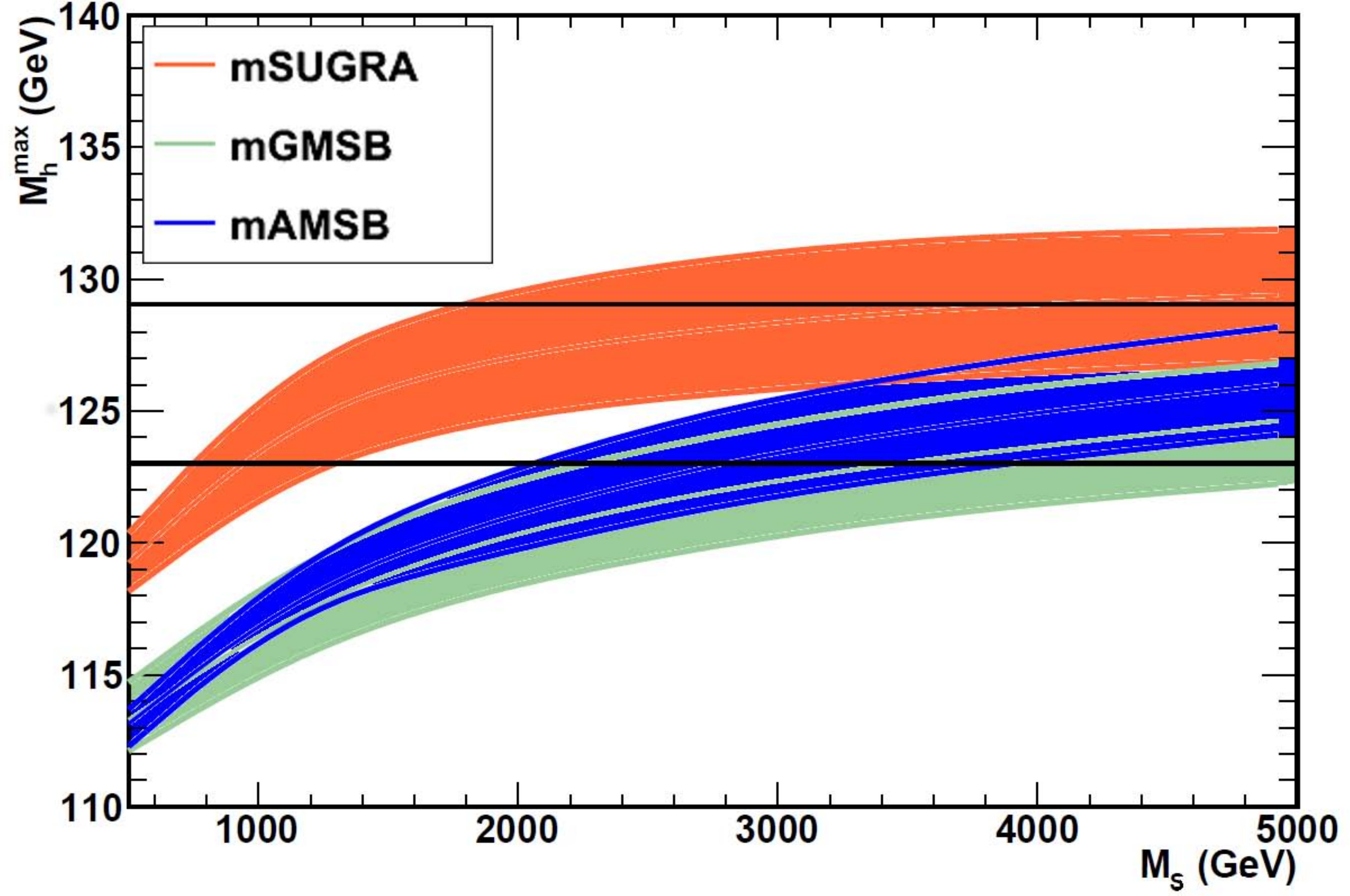}
\raisebox{-11pt}{\includegraphics[width=0.428\textwidth]{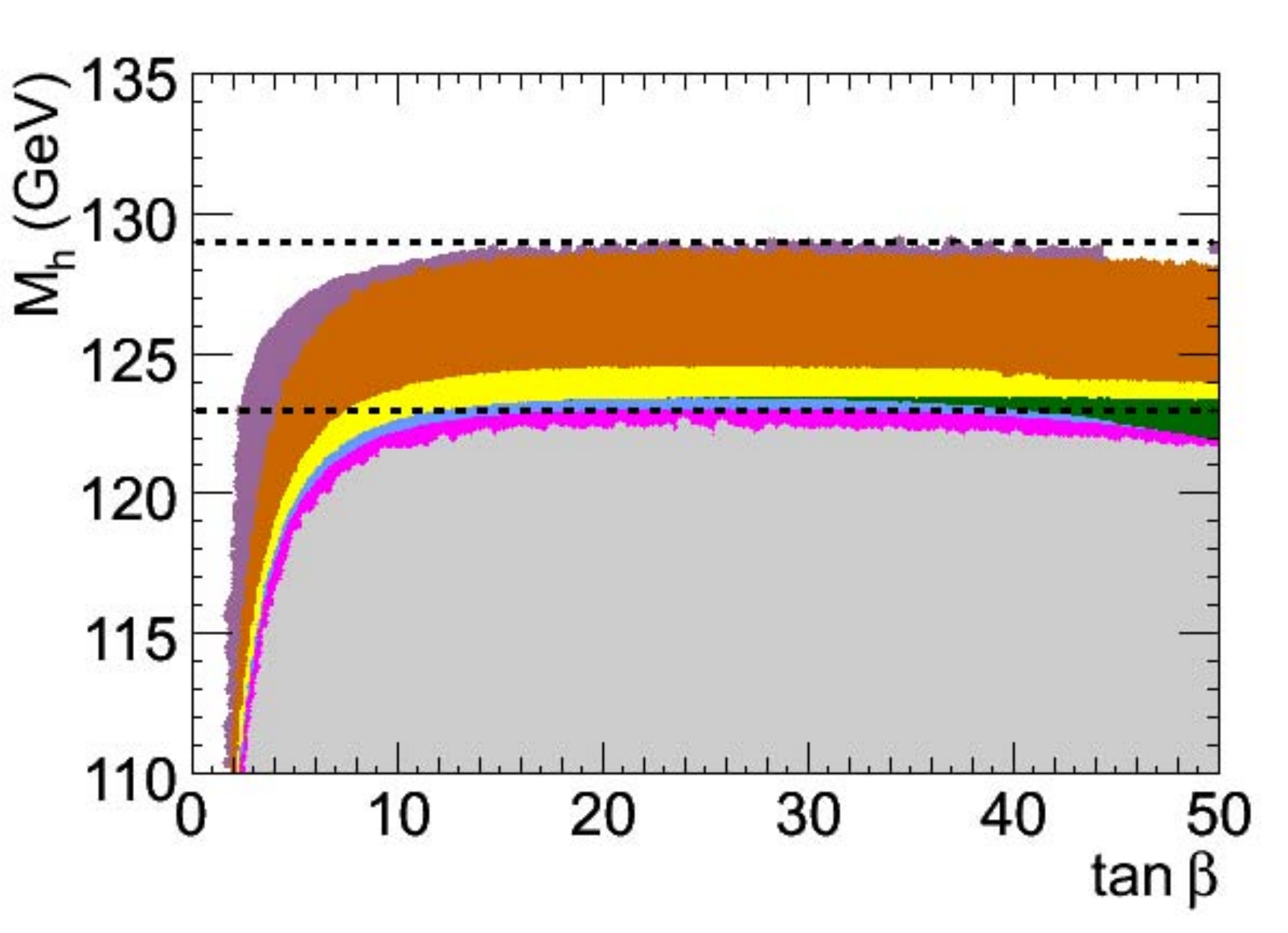}}
\includegraphics[width=0.12\textwidth]{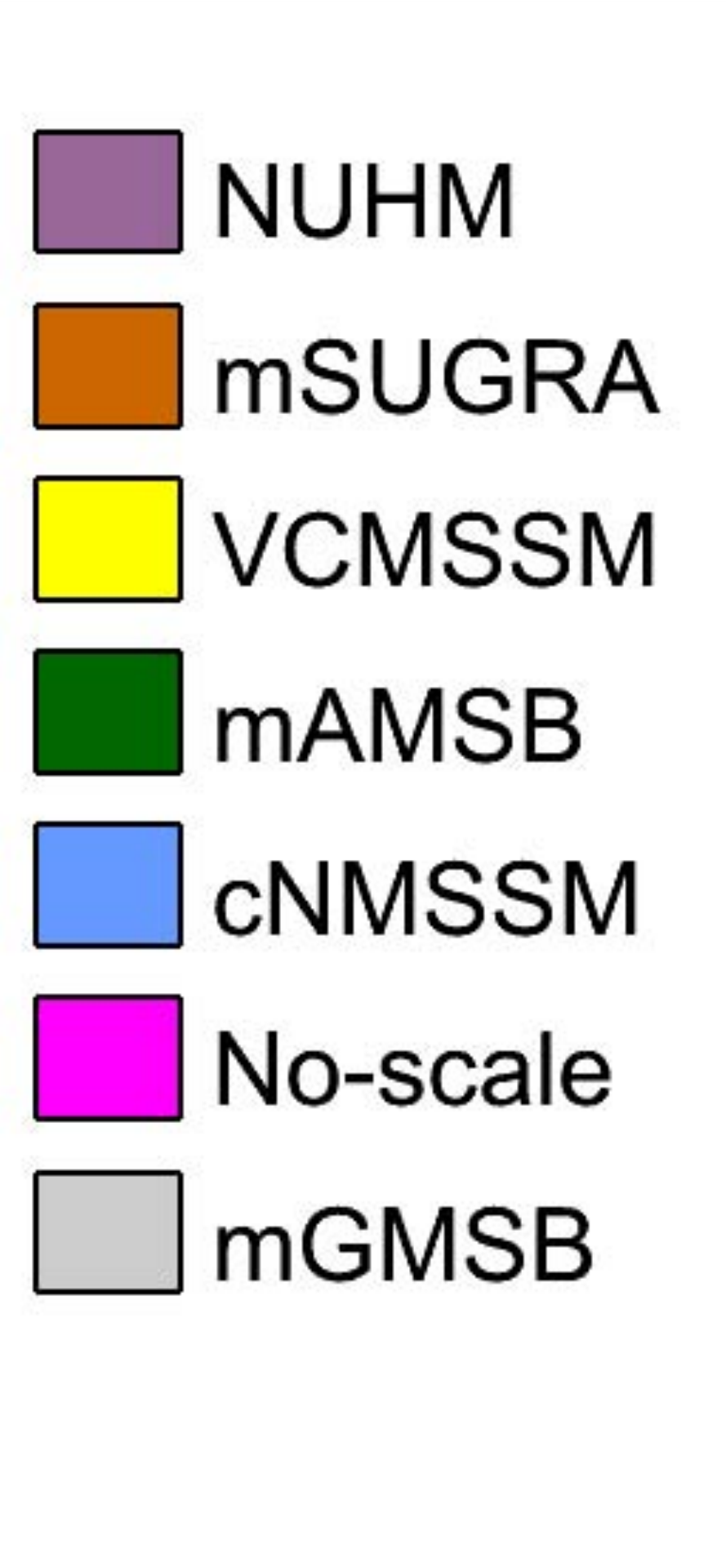}
\end{center}
\caption{The maximal Higgs mass in the constrained MSSM
scenarios mSUGRA, mAMSB and mGMSB, as a function of the scale
$M_S$ when the top quark mass is varied in the range
$m_t=\;$170--176 GeV (left) and as a function of $\tan\beta$
(right).}
\label{fig:higgsmass}
\end{figure}

It is interesting, that the Higgs mass upper bound depends crucially
on some parameters of the model, and is almost independent on the
choice of the other parameters. For example, the 1~GeV change in the mass
of the top quark leads to the $\sim$1~GeV change in the Higgs mass
upper bound. The dependence of the maximal Higgs mass
on the supersymmetry breaking scale $M_S$ is shown in
the left panel of Fig.~\ref{fig:higgsmass}~\cite{ABDM}
for different scenarios of SUSY breaking. The widths of bands corresponds
to the variation of the top mass in the range 170--176~GeV.

The right panel of Fig.~\ref{fig:higgsmass} shows the dependence
of the maximal Higgs mass on $\tan\beta$ for the fixed value of
$m_t=173$~GeV while other parameters of the model vary within
the ranges~\cite{ABDMQ}:
\begin{center}
\begin{tabular}{rccc}
mSUGRA: & 50 GeV $\leq m_0 \leq$ 3 TeV, & 50 GeV $\leq m_{1/2} \leq$ 3 TeV, &
$|A_0|\leq 9 $ TeV; \\[1mm]
GMSB: & 10 TeV $\leq \Lambda\leq$ 1000 TeV, & 1 $\leq M_{\rm mess} / \Lambda
\leq 10^{11}$, & $N_{\rm mess} =$ 1; \\[1mm]
AMSB: & 1 TeV $ \leq m_{3/2}\leq$ 100 TeV, & 50 GeV $\leq m_0\leq$ 3 TeV. &
\end{tabular}
\end{center}

%--------------------------------------
\subsection{The lightest superparticle}

One of the crucial questions is the properties of the lightest
superparticle. Different SUSY breaking scenarios lead to different
experimental signatures and different LSP.
\begin{itemize}
%$\bullet$
\item Gravity mediation

In this case, the LSP is the lightest neutralino
$\tilde{\chi}^0_1$, which is almost 90\% photino for the low
$\tan\beta$ solution and contains more higgsino admixture for
high $\tan\beta$. The usual signature for LSP is the missing
energy; $\tilde{\chi}^0_1$ is stable and is the best candidate
for the cold dark matter particle in the Universe. Typical
processes, where the LSP is created, end up with jets +
$\Big/ \hspace{-0.3cm}E_T$, or leptons +
$\Big/ \hspace{-0.3cm}E_T$, or both jets + leptons +
$/ \hspace{-0.25cm}E_T$.

%$\bullet$
\item Gauge mediation

In this case the LSP is the gravitino $\tilde G$, which also
leads to the missing energy. The actual question here is what
is the NLSP, the next-to-lightest particle, is. There are two
possibilities:

i) $\tilde{\chi}^0_1$ is the NLSP. Then the decay modes are: \
 $\tilde{\chi}^0_1 \to \gamma \tilde G, \ h \tilde G, \ Z \tilde
 G.$\
 As a result, one has two hard photons + $/
 \hspace{-0.25cm}E_T$, or jets + $/ \hspace{-0.25cm}E_T$.

ii) $\tilde l_R$ is the NLSP. Then the decay mode is $\tilde l_R
\to \tau \tilde G$ and the signature is a charged lepton and the
missing energy.

%$\bullet$
\item Anomaly mediation

In this case, one also has two possibilities:

i) $\tilde{\chi}^0_1$ is the LSP and wino-like. It is almost
degenerate with the NLSP.

ii) $\tilde \nu_L$ is the LSP. Then it appears in the decay of
the chargino $\tilde \chi^+ \to \tilde \nu l$ and the signature
is the charged lepton and the missing energy.

%$\bullet$
\item R-parity violation

In this case, the LSP is no longer stable and decays into the SM
particles. It may be charged (or even colored) and may lead to
rare decays like the neutrinoless double $\beta$-decay, etc.
\end{itemize}
Experimental limits on the LSP mass follow from the
non-observation of the corresponding events. The modern lower
limit is around 40~GeV .

%-------------------------
\section{Constrained MSSM}
%\setcounter{equation} 0

%---------------------------------------
\subsection{Parameter space of the MSSM}

The Standard Model has the following set of free parameters:
\begin{itemize}
\item[i)] three gauge couplings $\alpha_i$;
\item[ii)] three (or four if the Dirac neutrino mass term is included)
matrices of the Yukawa couplings $y^i_{ab}$, where
$i = U,D,L (N) $;
\item[iii)] two parameters of the Higgs potential
($\lambda$ and $m^2$).
\end{itemize}
\noindent
The parameters of the Yukawa sector are usually traded for the
masses, mixing angles and phases of the mixing matrices.

In the MSSM one has the same set of parameters except for the
parameters of the Higgs potential which is fixed by
supersymmetry, but in addition one has
\begin{itemize}
\item[iv)] the Higgs fields mixing parameter $\mu$;
\item[v)] the soft supersymmetry breaking terms.
\end{itemize}
The main uncertainty comes from the unknown soft terms. With
the universality hypothesis one is left with the following
set of 5 free parameters defining the mass scales
\begin{equation*}
\mu, \ m_0, \ m_{1/2}, \ A \ \text{and}\
B \leftrightarrow \tan\beta = \frac{v_2}{v_1}.
\end{equation*}
When choosing the set of parameters and making predictions,
one has two possible ways to proceed:
\begin{itemize}
\item[i)] take the low-energy parameters like the superparticle masses
$\tilde{m}_{q1},\tilde{m}_{q2}, m_A$, $\tan\beta$, mixings
$X_{stop},\mu$, etc. as input and calculate the cross-sections
as functions of these parameters. The disadvantage of this
approach is the large number of free parameters.
\item[ii)] take the high-energy parameters like the above mentioned 5
parameters as input, run the RG equations and find the
low-energy values. Now the calculations can be carried out in
terms of the initial parameters. The advantage is that their
number is relatively small. A typical range of these parameters
is
\begin{gather*}
100\ GeV \leq m_0,m_{1/2}, \mu \leq 3 \ TeV, \\
-3m_0 \leq A_0 \leq 3m_0, \ \ \   1\leq \tan\beta \leq 70.
\end{gather*}
The experimental constraints are sufficient to determine these
parameters, albeit with large uncertainties.
\end{itemize}

%-------------------------------------
\subsection{The choice of constraints}

When subjecting constraints on the MSSM, perhaps, the most
remarkable fact is that all of them can be fulfilled
simultaneously. In our analysis we impose the following
constraints on the parameter space of the MSSM:
\begin{itemize}
\item
LEP II experimental lower limits on the SUSY masses;
\item
Limits from the Higgs searches;
\item
Limits from precision measurement of rare decay rates
($B_s\to s\gamma, B_s\to \mu^+\mu^-, B_s\to \tau \nu$);
\item
Relic abundance of the Dark Matter in the Universe;
\item
Direct Dark Matter searches;
\item
Anomalous magnetic moment of the muon;
\item
Radiative electroweak symmetry breaking;
\item
Gauge coupling constant unification;
\item
Neutrality of the LSP;
\item
Tevatron and LHC limits on the superpartner masses.
\end{itemize}

In what follows we use the set of experimental data
shown in the Table~\ref{t1}.
\begin{table}[htb]
\centering
\begin{tabular}{lll}
\hline\noalign{\smallskip}
Constraint & Data & Ref.  \\
\noalign{\smallskip}\hline\noalign{\smallskip}
$\Omega h^2$ & $0.113\pm 0.004$ & \cite{Komatsu:2010fb} \\
$\bsg$ & $(3.55 \pm 0.24)\cdot 10^{-4}$ & \cite{hfag} \\
$\btaunu$ &  $(1.68\pm 0.31)\cdot 10^{-4}$ & \cite{hfag} \\
$\Delta a_\mu$ & $(290~\pm~63(exp)~\pm~61(theo))\cdot 10^{-11}$&
\cite{Bennett:2006fi}\\
$\bsmm$ &  $\bsmm < 4.5\cdot 10^{-9}$ & \cite{Aaij:2012ac}\\
$m_h$  & $ m_h > 114.4$ GeV & \cite{Schael:2006cr}\\
$m_A$ & $m_A > 510$ GeV for $\tb \approx 50$&
\cite{Chatrchyan:2012vp}\\
ATLAS & $ \sigma^{SUSY}_{had} < 0.001-0.03 $ pb &
\cite{ATLAS-CONF-2012-033}\\
CMS & $\sigma^{SUSY}_{had} < 0.003-0.03 $ pb &
\cite{CMS-PAS-SUS-12-005} \\
XENON100 &
$\sigma_{\chi N} < 1.8 \cdot 10^{-45}-6\cdot 10^{-45} cm^2$&
\cite{Aprile:2011hi}\\
\noalign{\smallskip}\hline
\end{tabular}
\caption{List of all constraints used in the fit to determine
the excluded region of the CMSSM parameter space.}
\label{t1}
\end{table}

Having in mind the above mentioned constraints one can find the
most probable region of the parameter space by minimizing the
$\chi^2$ function~\cite{BEK}. Since the parameter space is 5
dimensional one can not plot it explicitly and is bounded to
use various projections. We will accept the following strategy:
we first choose the value of the Higgs mixing parameter $\mu$
from the requirement of radiative EW symmetry breaking and then
take the plane of parameters $m_0 - m_{1/2}$ adjusting the
remained parameters $A_0$ and $\tan\beta$ at each point
minimizing the $\chi^2$. We present the restrictions coming
from various constraints in the $m_0-m_{1/2}$ plane.

The most probable region of the parameter space is determined
by the minimum $\chi^2_{min}$ value. The 95\% C.L. (90\% C.L.)
limit is reached for the values of $\chi^2$ of 5.99 (4.61),
respectively.  The $\chi^2$ function is defined as
\begin{equation}
\begin{split}
\chi^2 &= \frac{\left(\Omega h^2-0.1131 \right)^2}{\sigma^2_{\Omega h^2}}+
\frac{\left(\bsg-3.55\cdot 10^{-4}\right)^2}{\sigma^2_{\bsg}} \\
&+\frac{\left(\btaunu-1.68\cdot 10^{-4}\right)^2}{\sigma^2_\btaunu} +
\frac{\left(\Delta a_\mu-302~\cdot 10^{-11}\right)^2}{
\sigma^2_{\Delta a_\mu}} \\
&+ \chi^2_{\bsmm} +\chi^2_{m_h}+\chi^2_{CMS}+\chi^2_{ATLAS}+
\chi^2_{m_A}+\chi^2_{DDMS}
\end{split}
\label{chi_def}
\end{equation}

In what follows we show the influence of various constraints
and determine the allowed region of the parameter space with
95\% C.L.

%---------------------------------
\section {Electroweak Constraints}

%--------------------------------------------------------------
\subsection{Region excluded by the $B_s\to s\gamma$ decay rate}

\begin{figure}[hbt]
\begin{center}
\leavevmode
\includegraphics[width=0.34\textwidth]{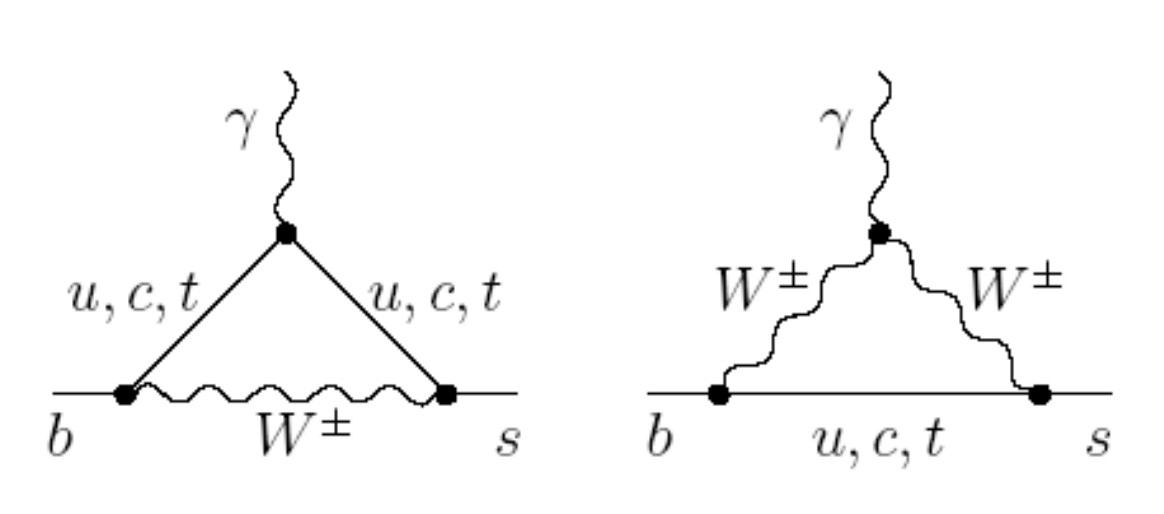}
\hspace{2mm}
\raisebox{2pt}{\includegraphics[width=0.34\textwidth]{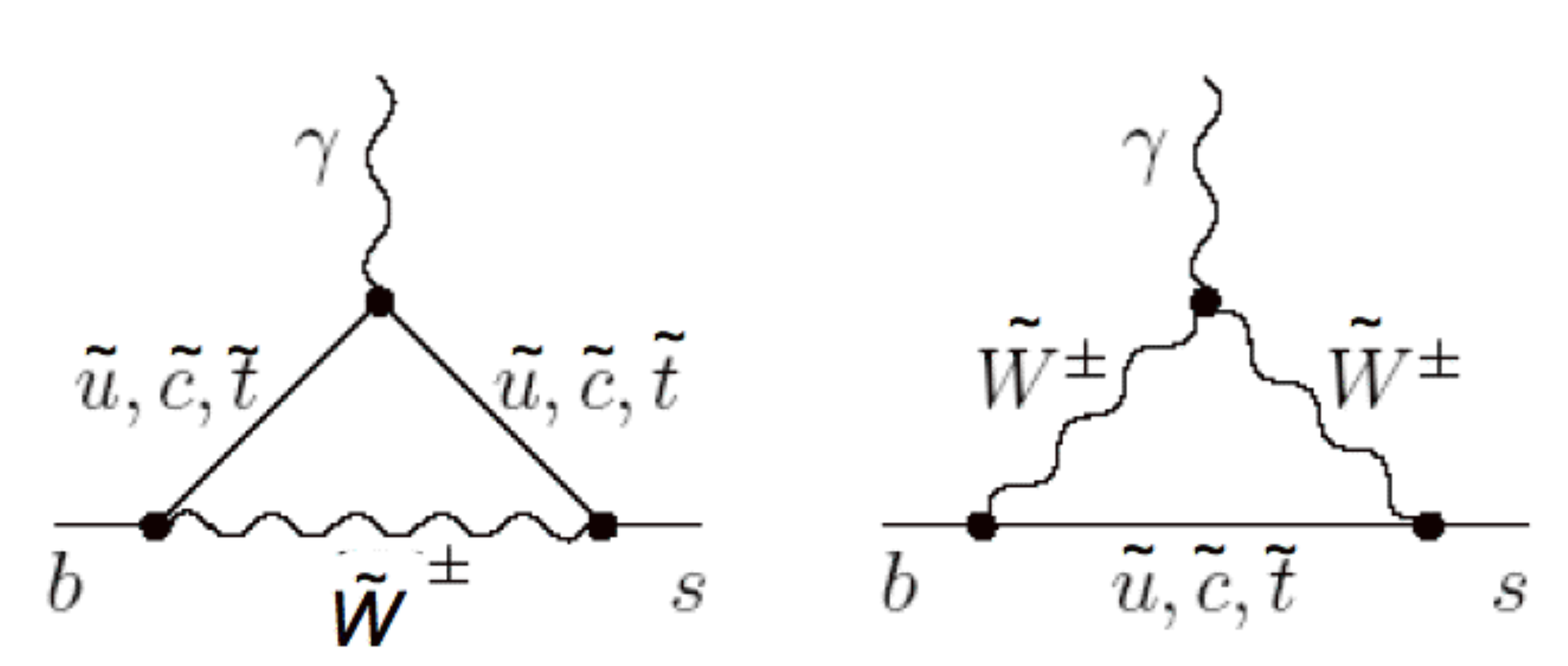}}
\hspace{2mm}
\raisebox{-1pt}{\includegraphics[width=0.17\textwidth]{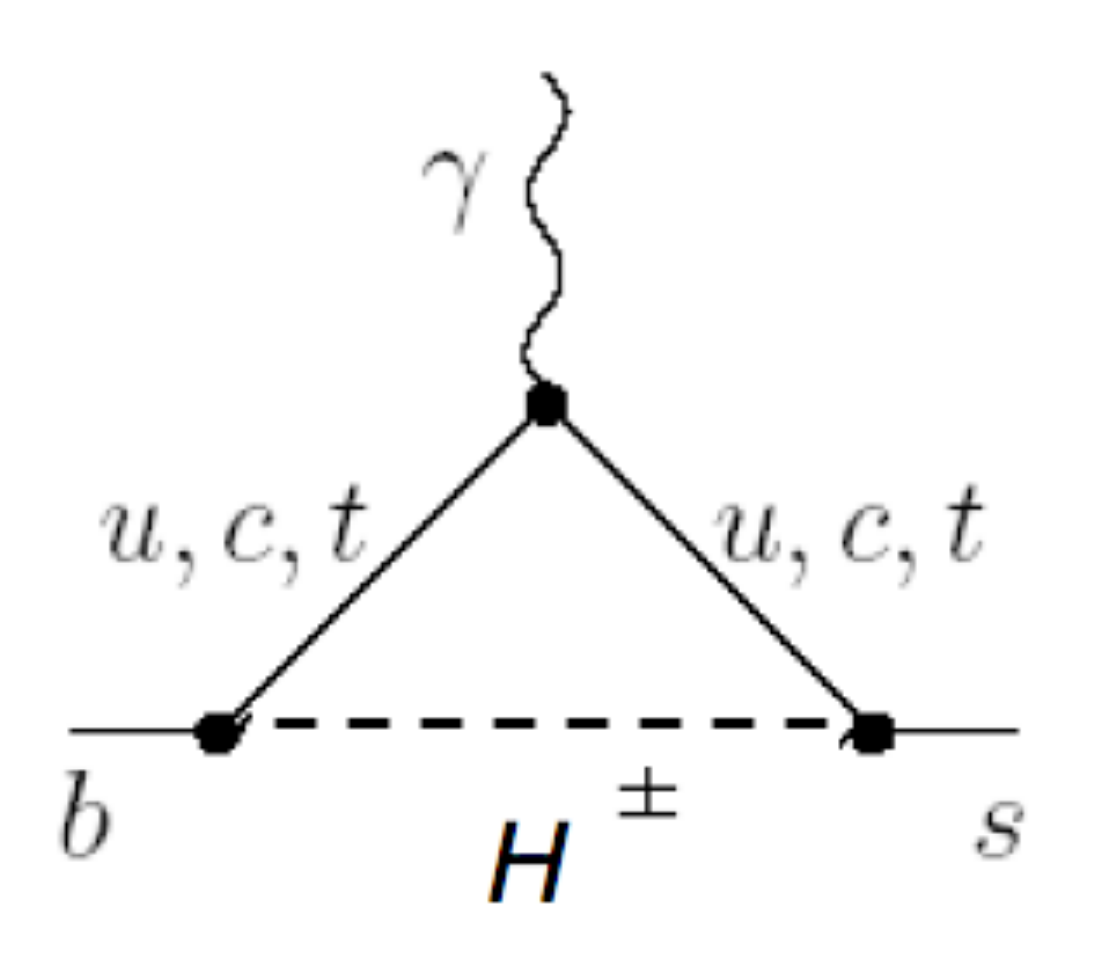}}
\end{center}
\caption{The diagrams contributing to $b\to s\gamma$ decay
in the SM and in the MSSM.}
\label{fig:bsg}
\end{figure}

The next two constraints are related to the rare decays where
SUSY may contribute. The first one is the $b\to s\gamma$ decay
which in the SM given by the first two diagrams shown in
Fig.~\ref{fig:bsg} and leads to~\cite{Misiak}
\begin{equation*}
BR^{SM}(b\to s\gamma)=(3.15\pm0.23) \cdot 10^{-4}
\end{equation*}
while experiment gives ~\cite{hfag}
\begin{equation*}
BR^{exp}(b\to s\gamma)=(3.55\pm0.24) \cdot 10^{-4}.
\end{equation*}
These two values almost coincide but still leave some room for SUSY.

SUSY contribution comes from the last three diagrams shown in
Fig.~\ref{fig:bsg} The $\tan\beta$-enhanced corrections to the
chargino and charged Higgs contributions can be summarized as
follows: the $\tan\beta$-enhanced chargino contributions to
$BR(b\to s\gamma)$  is~\cite{bsgsusy}
\begin{equation}
BR^{SUSY} (b\to s\gamma)\Big|_{\chi^\pm} \propto
\mu A_t  \tan\beta f(\tilde m^2_{t_1},\tilde m^2_{t_1},
m_{\chi^\pm})\frac{m_b}{v(1+\delta m_b)},
\end{equation}
where all dominant higher-order contributions are included
through $\delta m_b$, and $f$ is the integral appearing in the
one-loop diagram. The relevant charged-Higgs contributions
to $BR(b\to s\gamma)$ in the large $\tan\beta$ regime
is~\cite{bsgsusy}
\begin{equation}
BR^{SUSY} (b\to s\gamma )\Big|_{H^\pm} \propto
\frac{m_b(h_t\cos\beta-\delta h_t\sin\beta)}{v\cos\beta
(1+\delta m_b)}g(m_H^\pm,m_t),
\end{equation}
where $g$ is the loop integral appearing in the diagram.

The influence of this constraint is shown below together with
the $B_s\to \mu^+\mu^-$ one.

%-----------------------------------------------------------------
\subsection{Region excluded by the $B_s\to \mu^+\mu^-$ decay rate}

\begin{figure}[ht]
\begin{center}
\leavevmode
\hspace*{-5mm}
\includegraphics[width=0.35\textwidth]{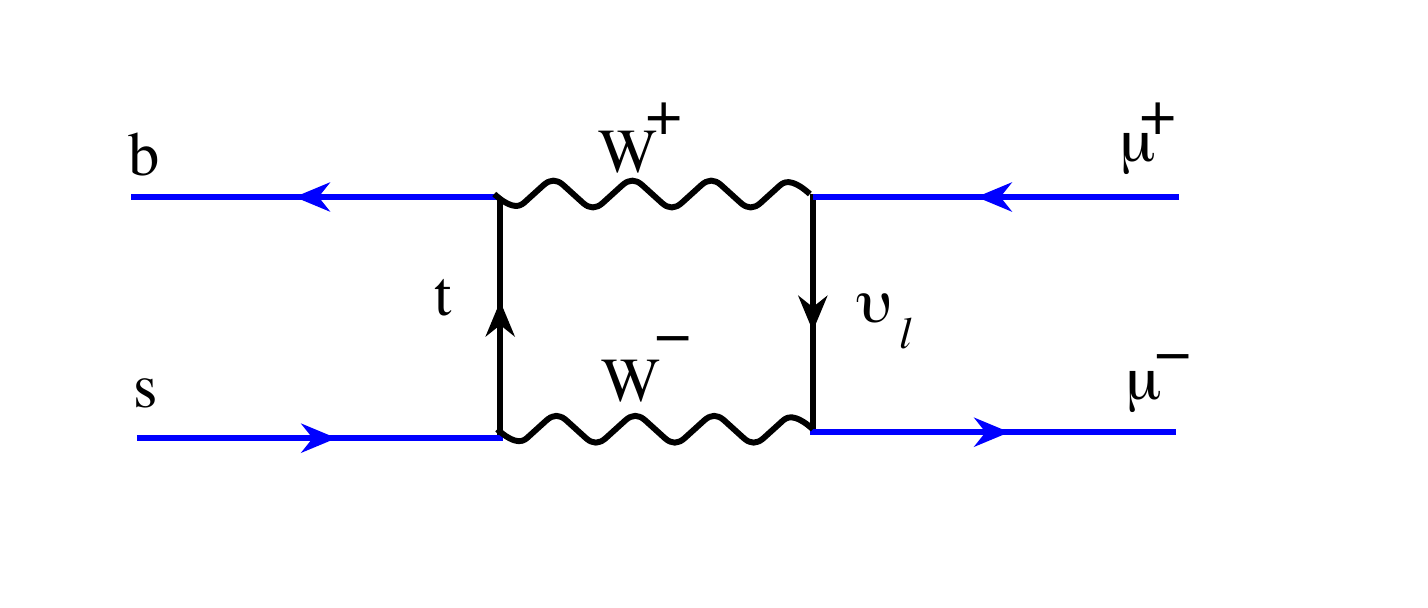}
\hspace*{-5mm}
\raisebox{1pt}{\includegraphics[width=0.36\textwidth]{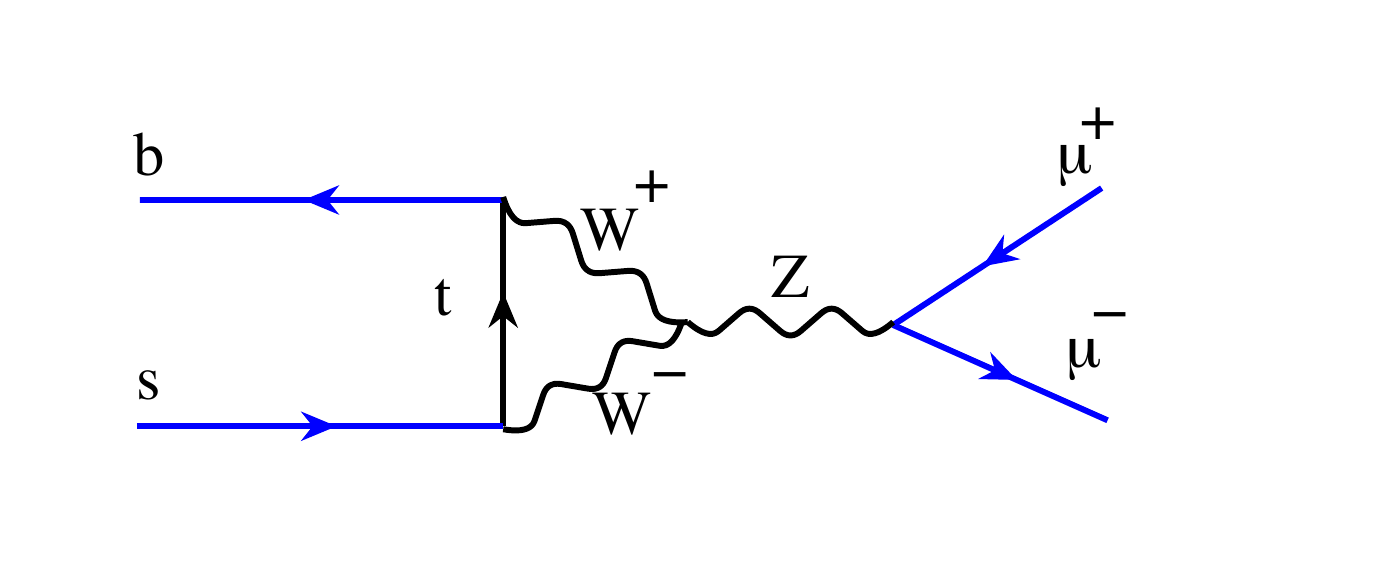}}
\includegraphics[width=0.30\textwidth]{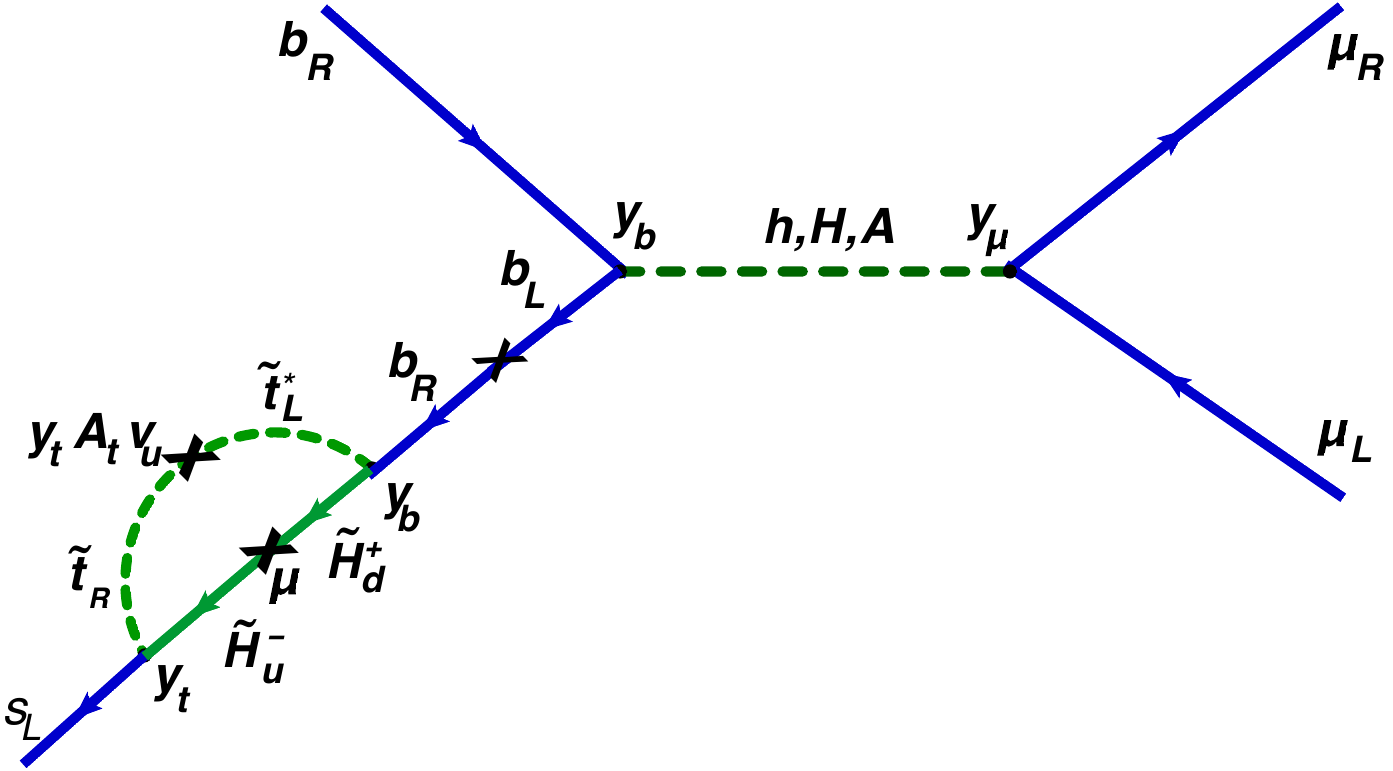}
\end{center}
\caption{The diagrams contributing to $B_s\to \mu^+\mu^-$ decay in
the SM and in the MSSM.}
\label{fig:bmu}
\end{figure}

The second example is the $B_s\to \mu^+\mu^-$ decay. In the SM
it is given by the first two diagrams shown in
Fig.~\ref{fig:bmu}. The branching ratio is~\cite{bsmumu_theor}
\begin{equation*}
BR^{SM}(B_s\to\mu^+\mu^-)=(3.23 \pm 0.27)\cdot  10^{-9},
\end{equation*}
while the recent experiment gives only the lower
bound~\cite{bmu}
\footnote{While the Lectures have been prepared
the first evidence for the decay
$B_s \to \mu^+ \mu^-$ based on 1.1 fb$^{-1}$ of
data recorded in 2012 at $\sqrt{s}=8$~TeV has been
reported~\cite{bsmumu_exp_hcp2012}.
The data show an excess of events with respect
to the background-only prediction with a statistical
significance of 3.5$\sigma$. A fit to the data gives
$BR(B_s \to \mu^+ \mu^-)=(3.2^{+1.5}_{-1.2})\cdot 10^{-9}$
which is in agreement with the SM prediction, thus leaving
less room for SUSY.}
\begin{equation*}
BR^{exp}(B_s\to\mu^+\mu^-)<4.5 \cdot 10^{-9}.
\end{equation*}
In the MSSM one has several additional diagrams but the main
contribution enhanced by $\tan^6\beta$~(!) comes from the one
shown on the right of Fig.~\ref{fig:bmu}.

The branching ratio for $B_s\to \mu^+\mu^-$ is given
in~\cite{Bobeth:2001sq,bmususy} which we write in the form
\begin{align}
BR(B_s\to \mu^+\mu^-) &= \frac{2\tau_BM_B^5}{64\pi} f^2_{B_s}
\sqrt{1-\frac{4m_l^2}{M_B^2}} \\
&\times \left[ \left( 1-{{4m_l^2}\over{M_B^2}} \right)
\Biggl| {{(C_S-C_S')}\over{(m_b+m_s)}} \Biggr|^2+
\Biggl| {{(C_P-C_P')}\over{(m_b+m_s)}}+
2{m_{\mu}\over M_{B_s}^2}(C_A-C_A') \Biggr|^2 \,
\right] \notag
\label{eq:bsmm_br_msugra}
\end{align}
where $f_{B_s}$ is the $B_s$ decay constant,
$M_{B}$ is
the $B$-meson mass, $\tau_B$ is the mean life time and $m_l$ is
the mass of the lepton. $C_S$, $C^{\prime}_S$, $C_P$,
$C^{\prime}_P$ include the SUSY loop contributions due to the
diagrams  involving the particles such as stop, chargino,
sneutrino, Higgs etc.  For large $\tan\beta$, the dominant
contribution to $C_S$ is given approximately by
\begin{equation}
C_S \simeq
{{G_F\alpha}\over {\sqrt 2\pi}}V_{tb}V_{ts}^*
\left( \frac{\tan^3\beta}{4\sin^2\theta_W} \right)
\left(\frac{m_b m_{\mu} m_t \mu}{M_W^2 M_A^2} \right)
\frac{\sin2\theta_{\tilde t}}{2}
\left( \frac{ m_{\tilde t_1}^2
\log\left( m_{\tilde t_1}^2 / \mu^2\right)}
{\mu^2-m_{\tilde t_1}^2} -
\frac{ m_{\tilde t_2}^2
\log\left({m_{\tilde t_2}^2 / \mu^2}\right)}
{\mu^2-m_{\tilde t_2}^2} \right)
\end{equation}
where $m_{\tilde t_{1,2}}$ are the two stop masses, and
$\theta_{\tilde t}$  is the rotation angle to diagonalize the
stop mass matrix. We need to multiply the above expression by
the factor $1/(1+\epsilon_b)^2$ to include the SUSY QCD
corrections. $\epsilon_b$ is proportional to
$\mu\tan\beta$~\cite{Carena:1999py}. Thus, for large
$\tan\beta$ the amplitude grows like $\tan^6\beta$ and might
come in contradiction with experiment. One observes, however,
that the $\tan\beta$ dependence can be compensated by the
strong suppression in the last term if the stop masses become
equal. This means that in order to get not too large branching
ratio the stop masses have to be degenerate.

\begin{figure}[ht]
%\vspace*{10mm}
%\hspace*{0mm}
\leavevmode
\raisebox{-9pt}{\includegraphics[width=0.34\textwidth]{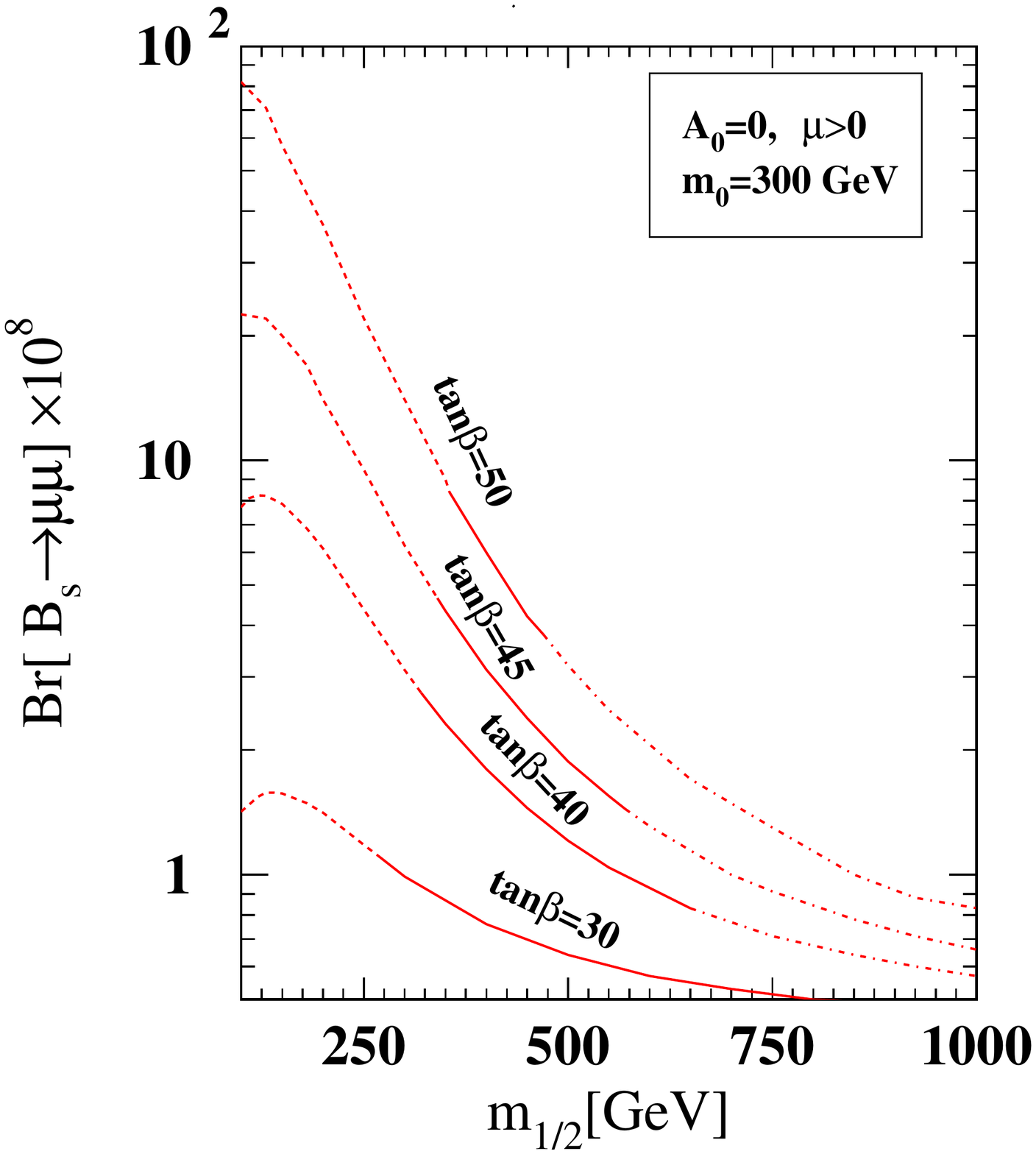}}
%\hspace{1mm}
%\begin{minipage}{5cm}
%\vspace{-6cm}
%\leavevmode
\raisebox{-17pt}{\includegraphics[width=0.64\textwidth]{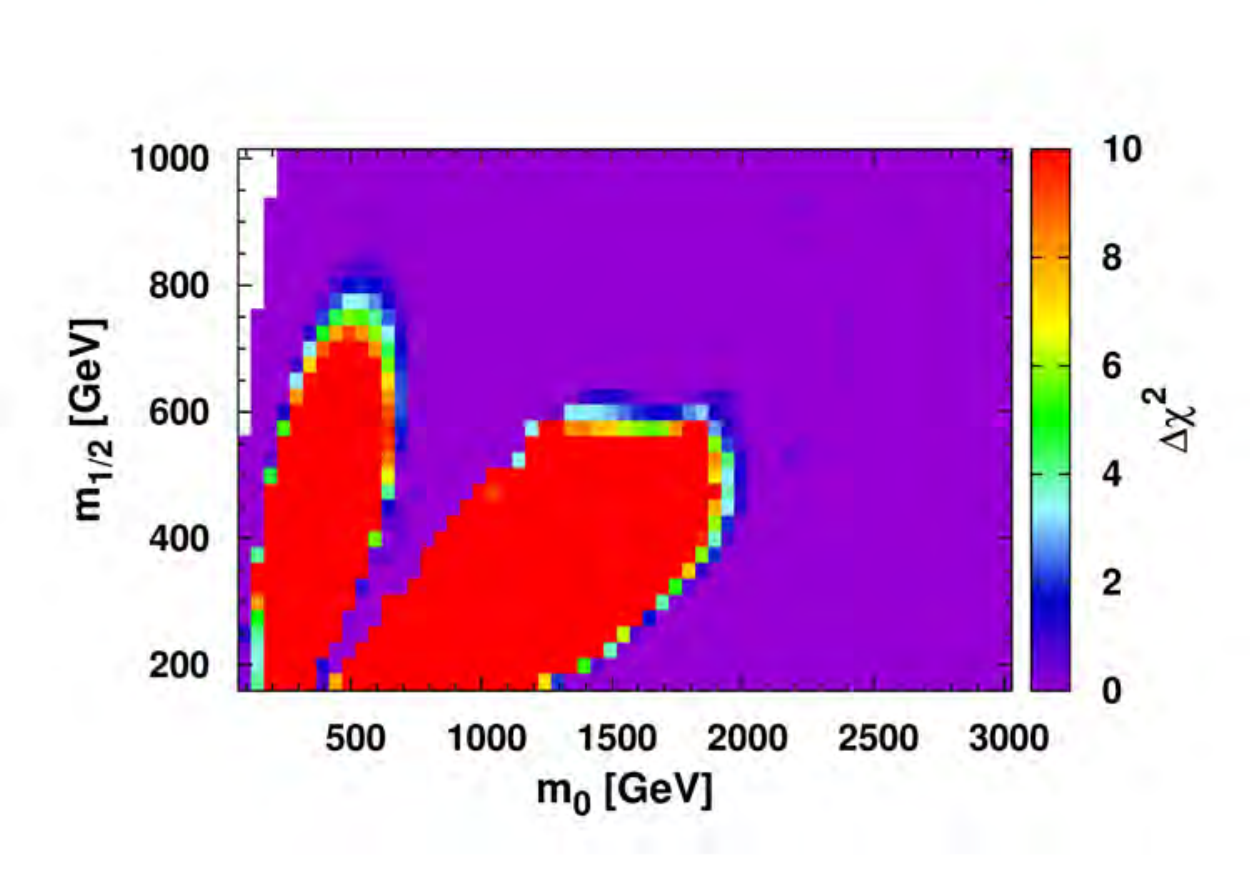}}
%\end{minipage}
\vspace*{-0mm}
\caption{The values of the branching ratio of the
$B_s\to \mu^+\mu^-$ decay in the MSSM (left) and constraints
on parameters space of the MSSM from electroweak observables
(right).}
\label{br}
\end{figure}

The values of the branching ratio for various parameters are
shown on the left part of Fig.~\ref{br}~\cite{bmususy} and the
restrictions on the parameter space -- on the right of
Fig.~\ref{br}.

%--------------------------------------------------------------------
\subsection{Region excluded by the anomalous magnetic moment of muon}

\begin{figure}[ht]\vspace{-0.1cm}
\begin{center}
\leavevmode
\includegraphics[width=0.18\textwidth]{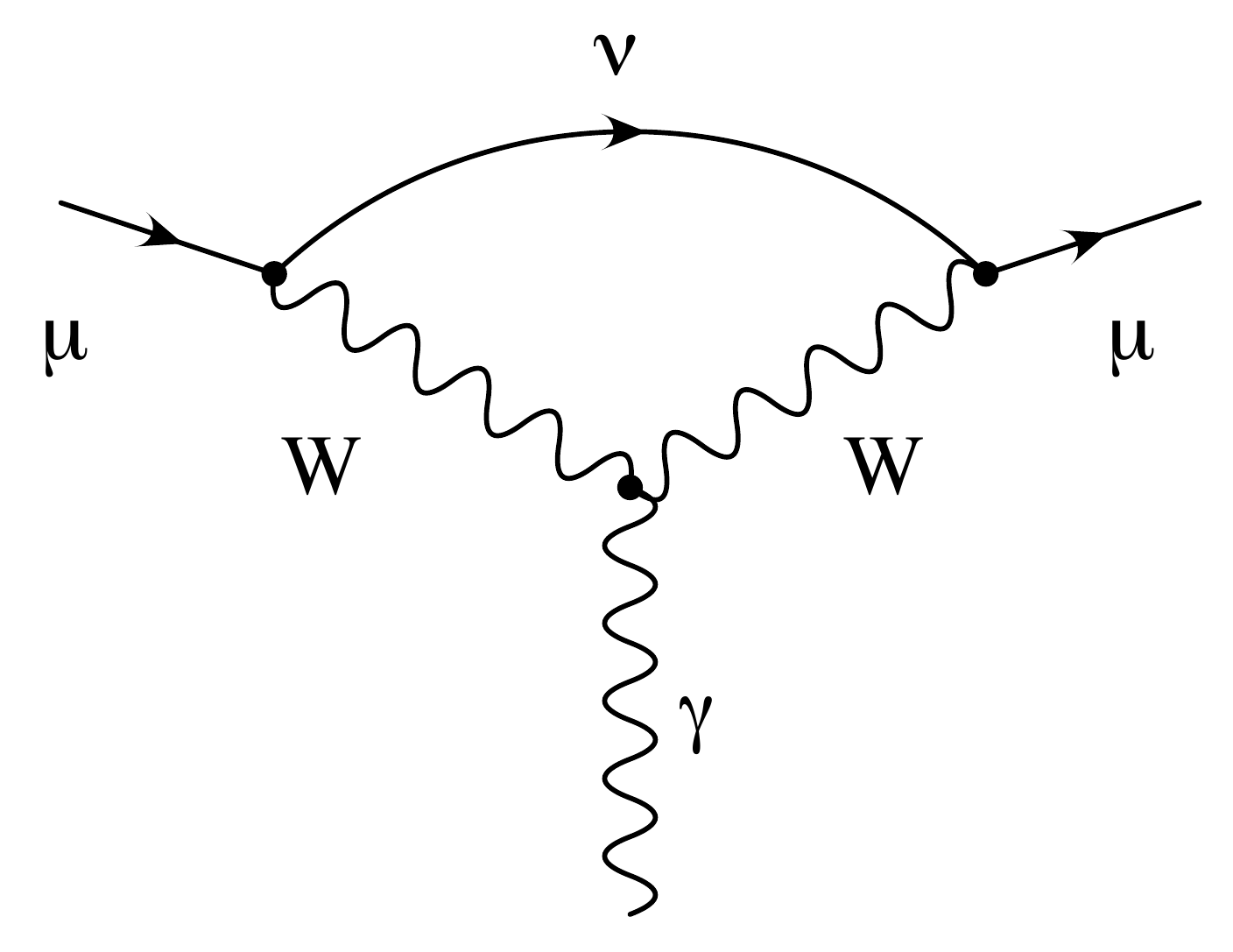}
\includegraphics[width=0.18\textwidth]{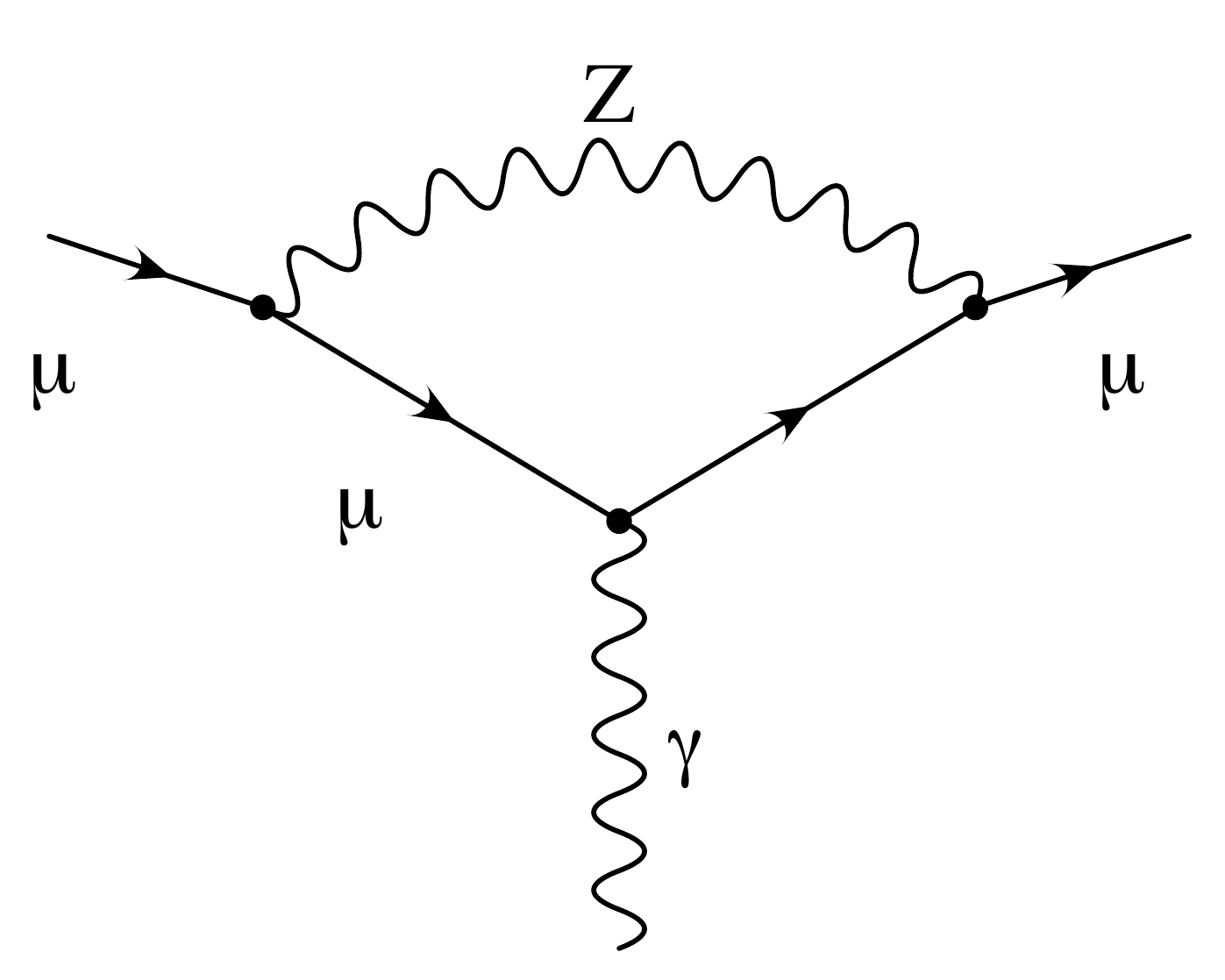}
\includegraphics[width=0.18\textwidth]{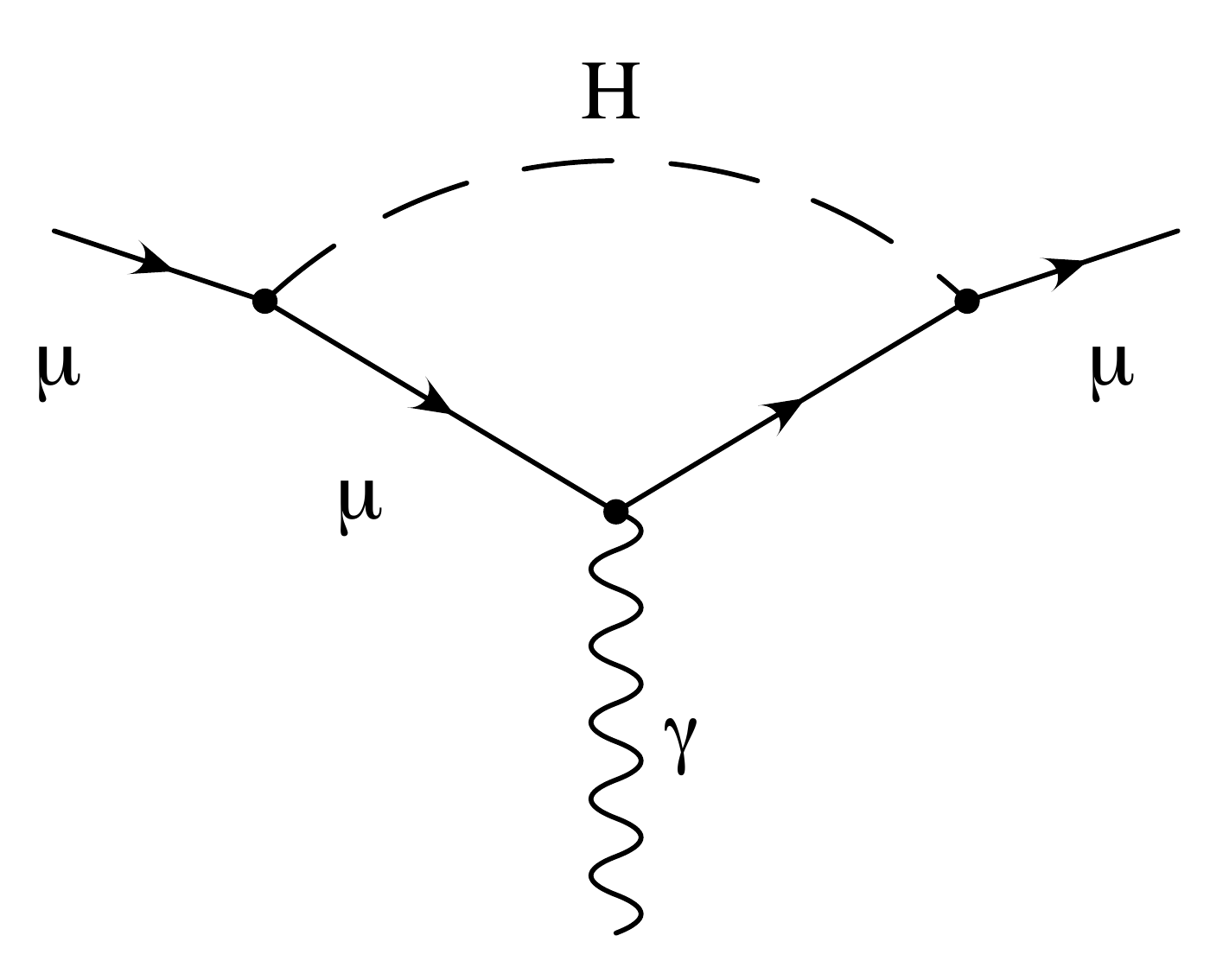}
\includegraphics[width=0.18\textwidth]{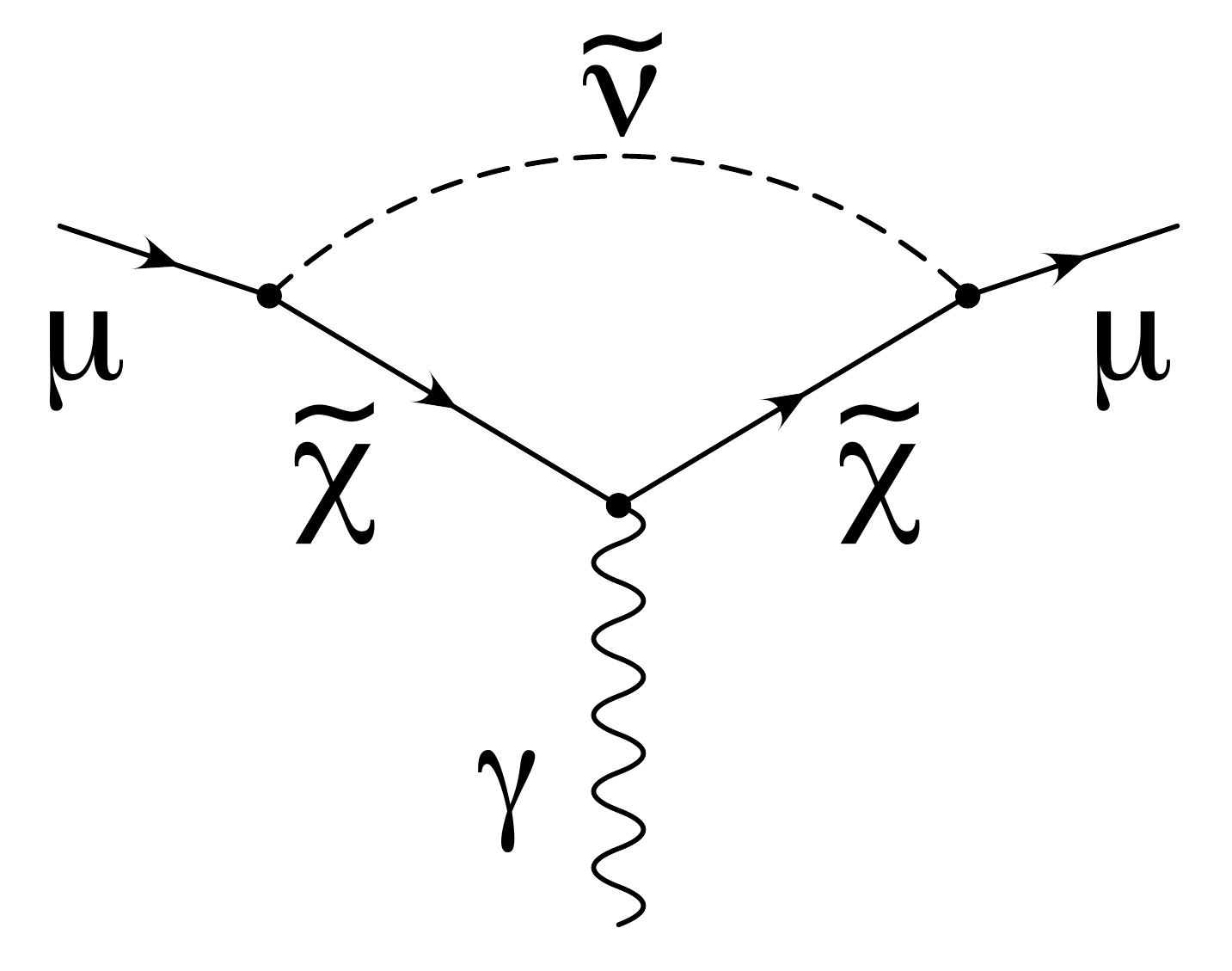}
\includegraphics[width=0.18\textwidth]{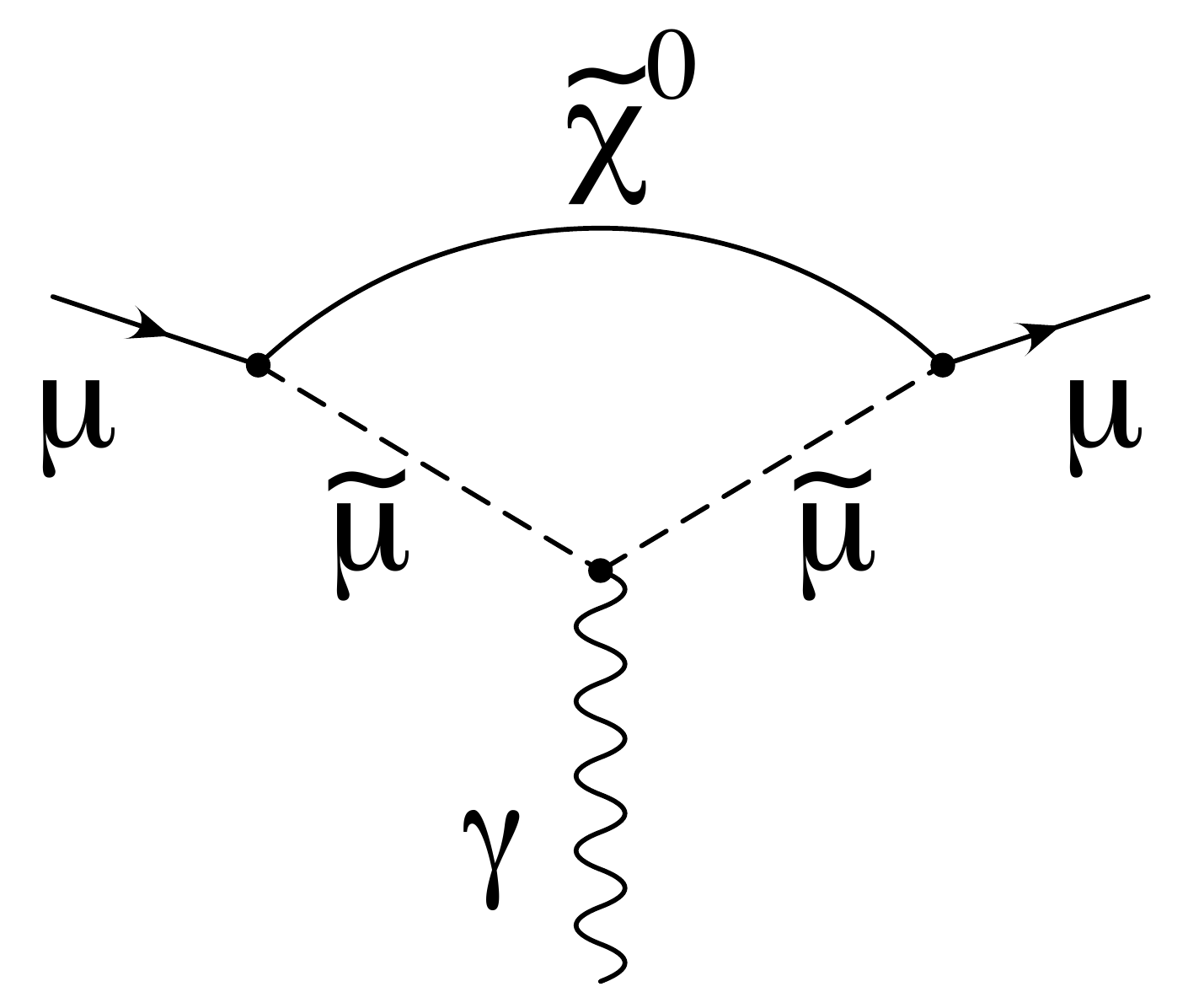}
\end{center}
\vspace{-0.1cm}
\caption{The diagrams contributing to $a_\mu$ in the SM and in
the MSSM.}
\label{anom}
\end{figure}

The theoretical value of $g-2$ has been reviewed in Ref.\cite{JN} which is in agreement with the latest values from \cite{DHM}.
Recent measurement of the anomalous magnetic moment of the muon
indicates small deviation from the SM of the order of
3~$\sigma$~\cite{Bennett:2006fi}:
\begin{align*}
a_\mu^{exp} &= 11\ 659\ 2080(63)\cdot  10^{-11} \\
a_\mu^{SM} &= 11\ 659\ 1790(64) \cdot 10^{-11} \\
\Delta a_\mu &= a_\mu^{exp}-a_\mu^{theor} =
(290\pm90)\cdot 10^{-11},
\intertext{where the SM contribution comes from}
a_\mu^{QED} &= 11\ 658\ 4718.1\ (0.2)\cdot 10^{-11} \\
a_\mu^{weak} &= 153.2\ (1.8) \cdot10^{-11} \\
a_\mu^{hadron} &= 6918.7\ (65) \cdot 10^{-11},
\end{align*}
so that the accuracy of the experiment finally reaches the order
of the weak contribution. The corresponding diagrams are shown
in Fig.~\ref{anom}.

The deficiency may be easily filled with the SUSY contribution
coming from the last two diagrams  of
Fig.~\ref{anom}. They are similar to that of the weak
interactions after replacing the vector bosons by the charginos
and neutralinos.

The total contribution to $a_\mu$ from these diagrams
is~\cite{Lopez}
\begin{equation}
\begin{split}
a^{SUSY}_\mu &= -\frac{g^2_2}{8\pi^2}
\Biggl\{ \sum_{\chi^0_i,\tilde \mu_j} \frac{m_\mu}{m_{\chi^0_i}}
\Bigl[ (-1)^{j+1}\sin 2\theta B_1(\eta_{ij})\tan\theta_W N_{i1}
\bigl[\tan\theta_W N_{i1}+N_{i2}\bigr] \\
&+\left.\left.\frac{m_\mu}{2M_W\cos\beta}
B_1(\eta_{ij})N_{i3}[3\tan\theta_W N_{i1}+N_{i2}]\right.\right.
\\
&+\left.\left.\left(\frac{m_{\mu}}{m_{\chi^0_i}}\right)^{2}
A_1(\eta_{ij})
\left\{\frac{1}{4}[\tan\theta_W N_{i1}+N_{i2}]^{2}+
[\tan\theta_W N_{i1}]^{2}\right\}\right]\right. \\
&-  \left.\sum_{\chi_j^\pm}\left[\frac{m_{\mu}m_{\chi_j^\pm}}
{m_{\tilde \nu}^2}
\frac{m_{\mu}}{\sqrt{2}M_W\cos\beta} B_2(\kappa_j)V_{j1}U_{j2}+
\left(\frac{m_{\mu}}
{m_{\tilde\nu}}\right)^{2}
\frac{A_1(\kappa_j)}{2}V^2_{j1}\right]\right\}.
\end{split}
\label{formula}
\end{equation}
where $N_{ij}$ are elements of the matrix diagonalizing the
neutralino mass matrix, and $U_{ij},V_{ij}$ are the
corresponding ones for the chargino mass matrix, the functions
$A$ and $B$ are the one-loop triangle integrals.

\begin{figure}[htb]
\vspace*{-0mm}
\begin{center}
\leavevmode
\includegraphics[width=0.32\textwidth]{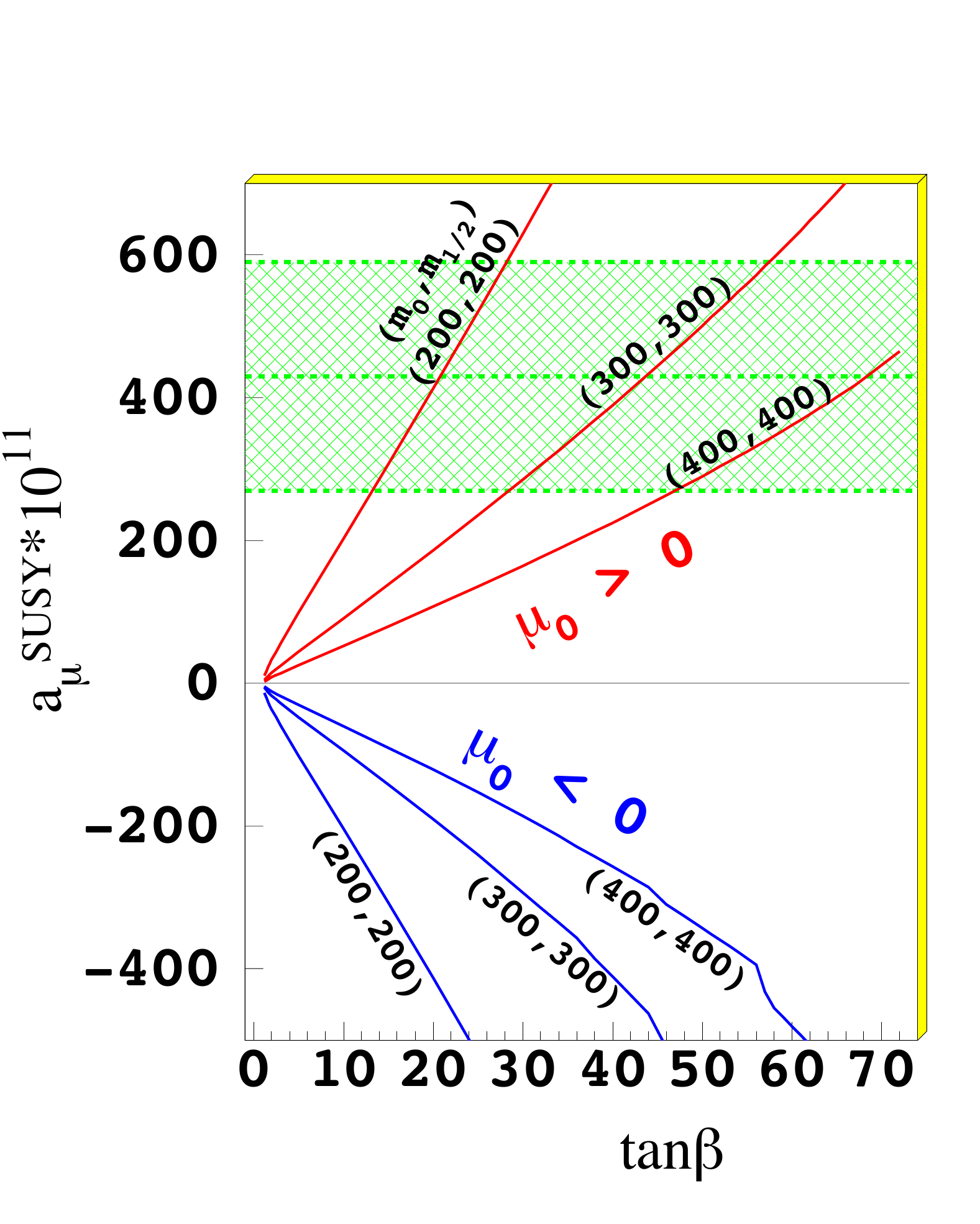}
\raisebox{-16pt}{\includegraphics[width=0.65\textwidth]{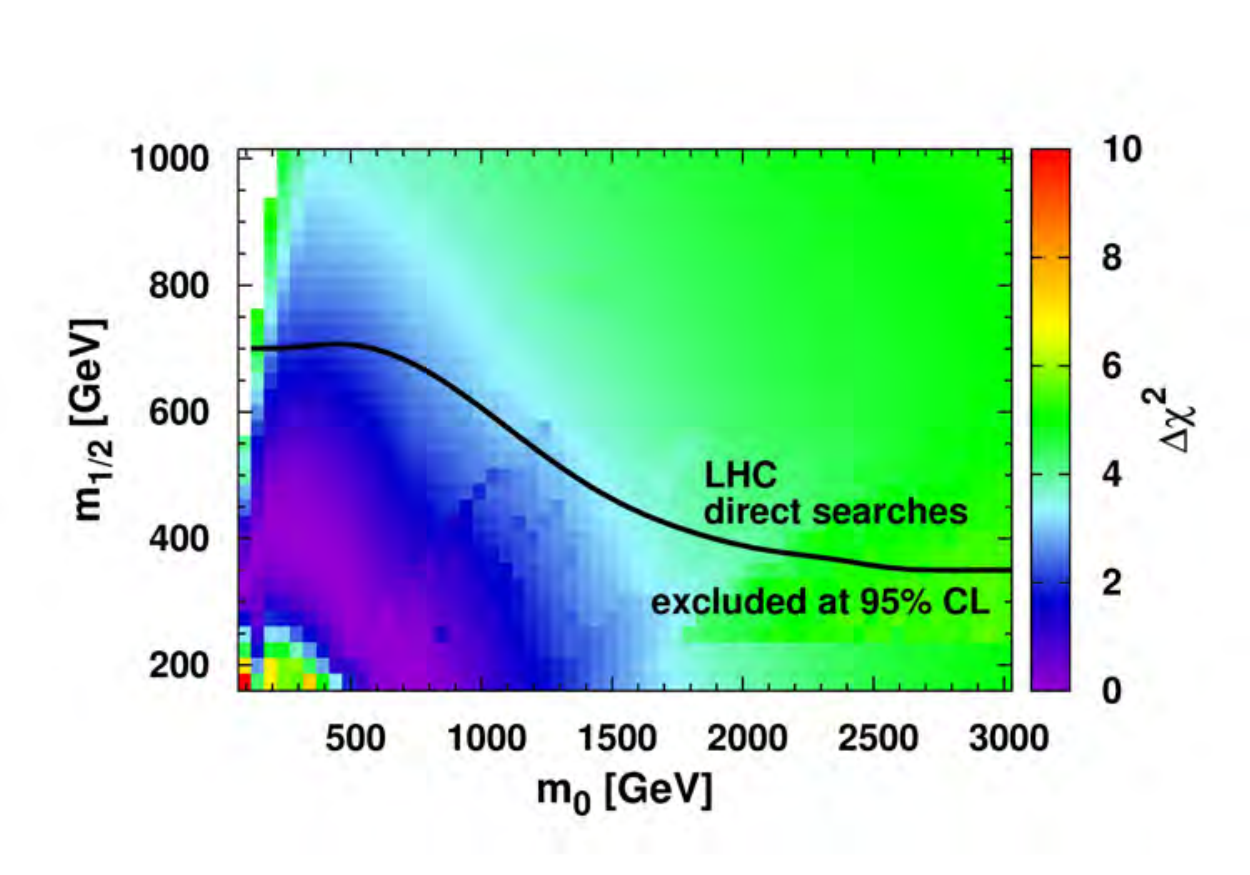}}
\end{center}
\vspace*{-10mm}
\caption{The dependence of $a_\mu^{SUSY}$ versus $\tan\beta$
for various values of the SUSY breaking parameters $m_0$ and
$m_{1/2}$ (left). The horizontal band shows the discrepancy
between the experimental data and the SM estimate. The allowed
regions of the parameters space (right). The black contour shows
the constraint from the LHC searches.}
\label{value}
\end{figure}

For large $\tan\beta$ it can be approximated as~\cite{CM}
\begin{eqnarray*}
&&\hspace*{-0.7cm}
\left| a_\mu^{SUSY} \right| \simeq \frac{\alpha(M_Z)}{8\pi \sin^2\theta_W}
\frac{m_\mu^2\tan\beta}{m_{SUSY}^2}
\left( 1 - \frac{4\alpha}{\pi}
\log\frac{m_{SUSY}}{m_\mu} \right)
 \simeq 14.0 \cdot 10^{-10}
 \left(\frac{100 \ \text{GeV}}{m_{SUSY}}\right)^2 \tan\beta,
\end{eqnarray*}
where $m_\mu$ is the muon mass, $m_{SUSY}$ is an average mass
of the supersymmetric particles in the loop (essentially the
chargino mass). It is proportional to $\mu$ and $\tan\beta$
and requires the positive sign of $\mu$ that kills a half of
the parameter space of the MSSM~\cite{Anom}.

If the SUSY particles are light enough they give the desired
contribution to the anomalous magnetic moment. However, if
they are too light the contribution exceeds the gap between
the experiment and the SM. For too heavy particles the
contribution is too small. The values of $a_\mu^{SUSY}$ versus
$\tan\beta$ for various values of the SUSY breaking parameters
$m_0$ and $m_{1/2}$ are shown on the left of Fig.~\ref{value}
and the restrictions on the parameter space are presented on
the right panel of Fig.~\ref{value}. However, the allowed region
is almost excluded by the direct SUSY searches at the LHC as can
be seen in Fig.~\ref{value} on the right panel. So the observed
deviation from the SM might be caused by the other reasons.

%-----------------------------------------------------------------
\subsection{Region excluded by the pseudo-scalar Higgs mass $m_A$}

\begin{figure}[ht]
\begin{center}
\leavevmode
\raisebox{15.5pt}{\includegraphics[width=0.45\textwidth]{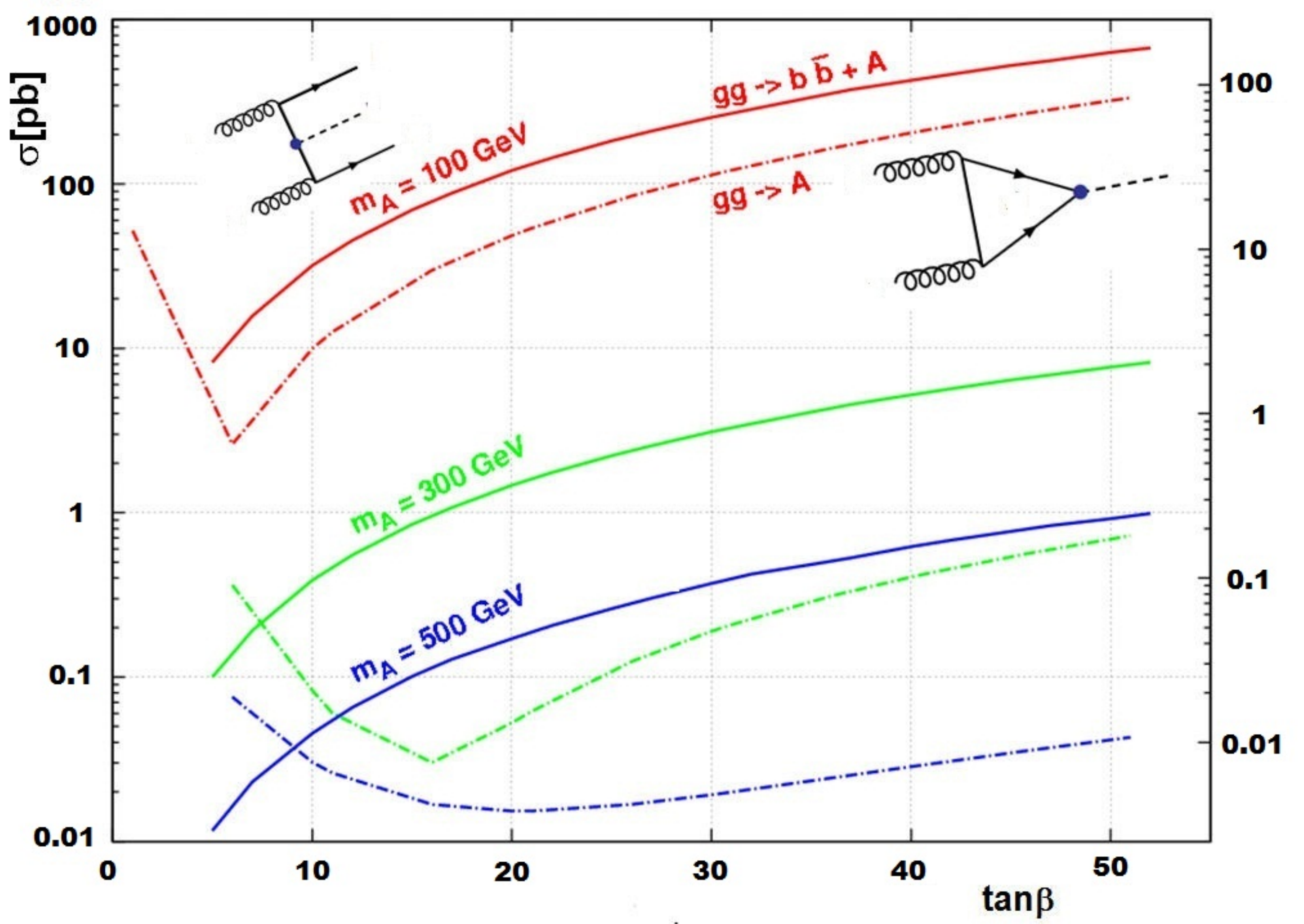}}
\includegraphics[width=0.48\textwidth]{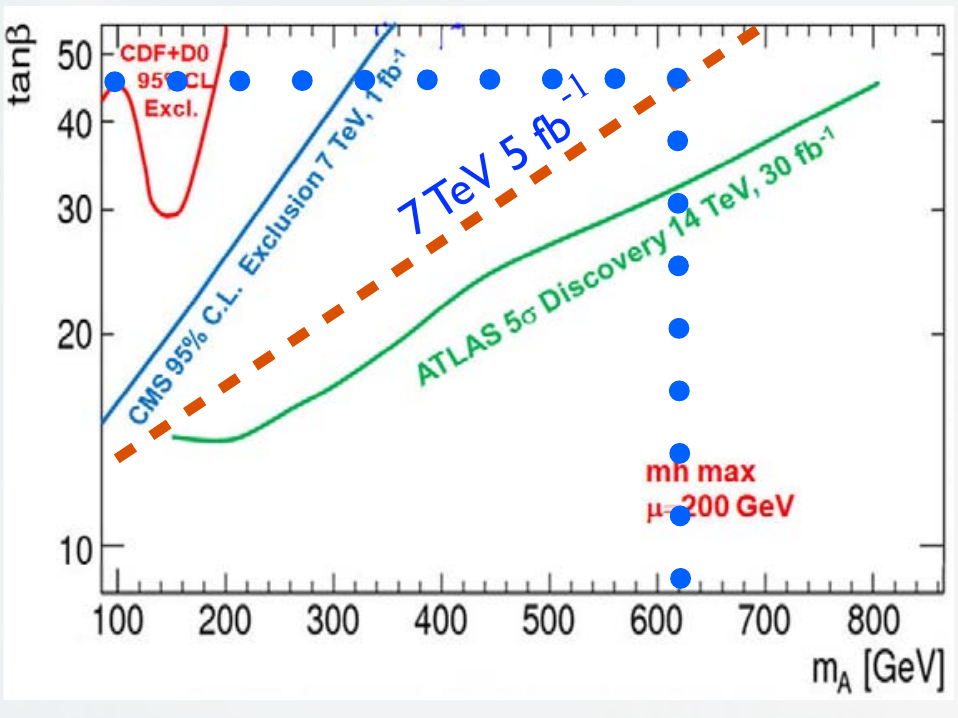}
\end{center}
\vspace*{-5mm}
\caption{Left: The pseudoscalar Higgs production cross section
as function of  $\tan\beta$, both for the gluon fusion diagram
and associated Higgs production with a $b$-quark for the different
Higgs mass values indicated. Right: expected discovery reach for
the ATLAS detector at 14~TeV  and a luminosity of
30~fb$^{-1}$~\cite{atlashiggs}. The region already excluded at
the Tevatron~\cite{tevatronhiggs} and the expected exclusion
reach after the initial 7~TeV run  at the LHC~\cite{cmshiggs}
have been indicated as well (assuming a luminosity of
1~fb$^{-1}$). These sensitivity projections  for future LHC
running of the ATLAS and CMS detectors  are preliminary.}
\label{ff4}
\end{figure}

The pseudo-scalar Higgs boson production is enhanced  by $\tan\beta$.
The main diagrams for for the gluon fusion
and associated Higgs production with a $b$-quark are shown in Fig.~\ref{ff4}
together with the corresponding cross-sections. Since the $b$-quark production is
mostly in the forward direction, the scale on the right-hand
side indicates if at least one $b$-quark is required to be in the
acceptance, defined by $\eta<2.5$, and have a transverse
momentum above 20 GeV/c.

The pseudo-scalar Higgs boson mass is determined by the relic
density constraint, because the dominant neutralino annihilation
contribution comes from the $A$-boson exchange in the region
outside the small co-annihilation regions. One expects
$m_A \propto m_{1/2}$ from the relic density constraint, which
can be fulfilled with $\tan\beta$ values around 50 in the whole
($m_0 - m_{1/2}$) plane~\cite{Beskidt:2010va}. Since the $A$
production cross section at the LHC is proportional to
$\tan^2\beta$ the pseudo-scalar mass limit increases up to
496~GeV for the large values of $\tan\beta$ preferred by the
relic density (see Fig.~\ref{ff4} right panel).
The corresponding
$m_A$-values are displayed in the left panel of Fig.~\ref{f5_100b}
and the $m_A$ values excluded by the LHC searches lead to the
excluded region, shown by the contour line in Fig.~\ref{f5_100b}.

The rather strong limits on the
pseudo-scalar Higgs boson mass from
LHC~\cite{Chatrchyan:2012vp,Chatrchyan:2011nx}, especially at large
values of $\tan\beta$, lead then to constraints on $m_{1/2}$ of
about 400~GeV for intermediate values of $m_0$, as shown in the
left panel of Fig.~\ref{f5_100b}.

%----------------------------------------------------------
\subsection{Effect of a SM Higgs mass $m_h$ around 125 GeV}

\begin{figure}[tb]
\begin{center}
\leavevmode
\hspace*{-5mm}
\includegraphics[width=0.53\textwidth]{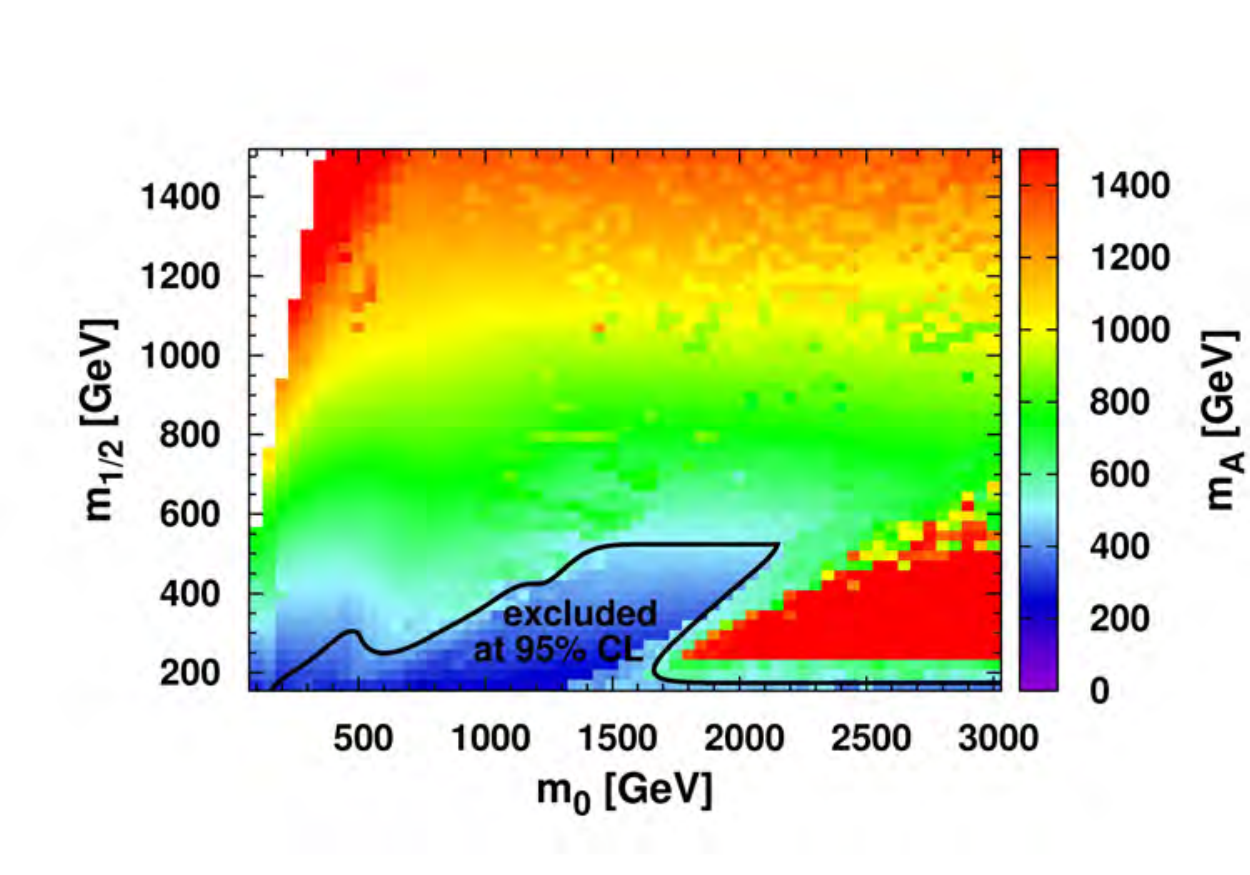}
\raisebox{9pt}{\includegraphics[width=0.47\textwidth]{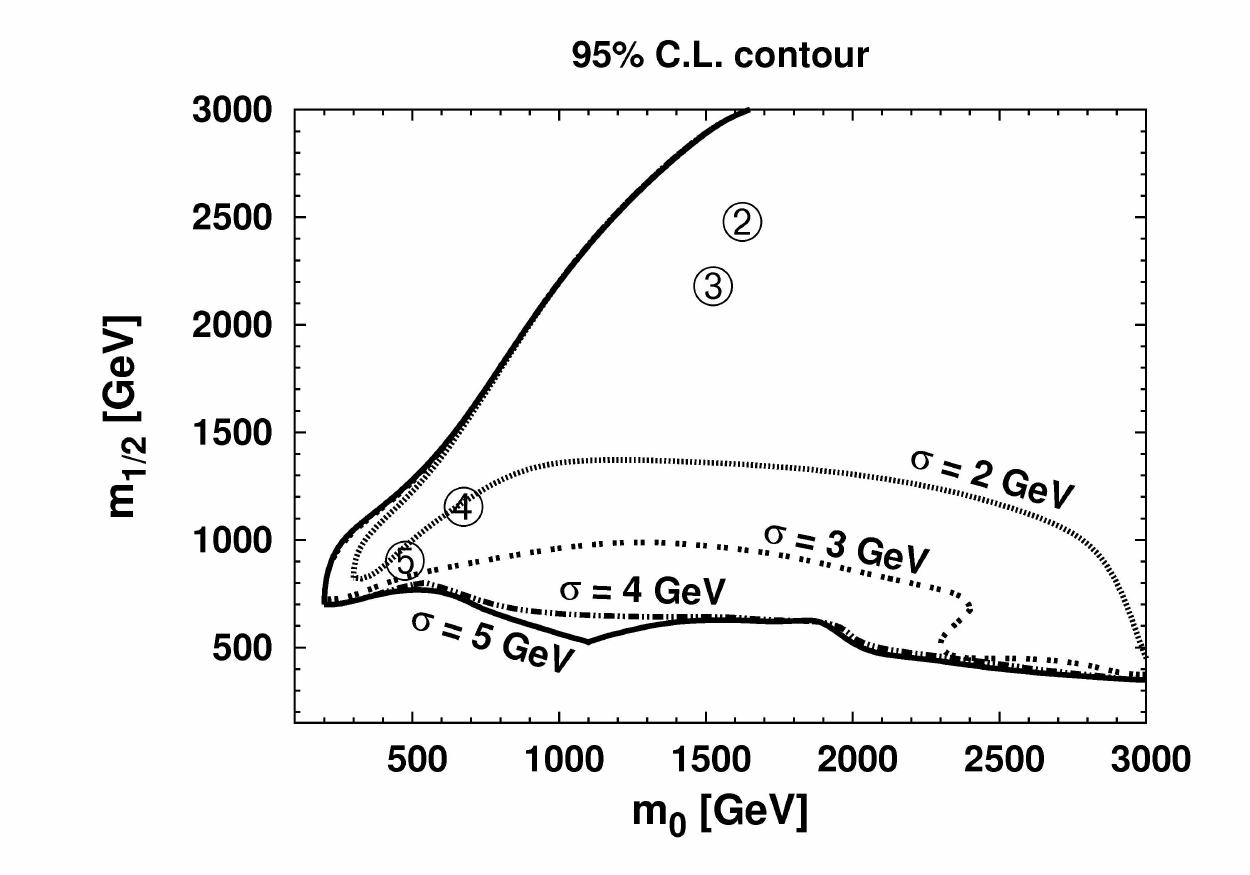}}
\end{center}
\caption{Left: values of $m_A$ in the ($m_0 - m_{1/2}$) plane
after optimizing $\tan\beta$ and $A_0$. The region below the
solid line is excluded at 95\% C.L. Right: The influence of a
Higgs mass of 125~GeV. If it is imposed in the fit, the best-fit
point moves to higher SUSY masses, but the location is strongly
dependent on the assumed error for the calculated Higgs mass.
This error is indicated by the number inside the circle for
the best-fit point: $\Delta\chi^2=5.99 (2\sigma)$ contour.}
\label{f5_100b}
\end{figure}

The 95\%~C.L. LEP limit of 114.4~GeV contributes for the small
and intermediate SUSY masses to the $\chi^2$ function.
In recent publications CMS~\cite{Chatrchyan:2012tx} and
ATLAS~\cite{ATLAS:2012ae} collaborations show evidence for the
Higgs with a mass around 125~GeV. If we assume this to be the
evidence for the SM Higgs boson, which has similar properties
as the lightest SUSY Higgs boson in the decoupling regime, we
can check the consequences for the CMSSM. If the Higgs mass of
125~GeV is included to the fit, the best-fit point moves to
higher SUSY masses, but there is rather strong tension between
the relic density constraint, $B_s \to \mu^+ \mu^-$ and the
Higgs mass, so the best-fit point depends strongly on the error
assigned to the Higgs mass, as shown in Fig.~\ref{f5_100b}
(right panel). The experimental
error on the Higgs mass is about 2~GeV, but the theoretical
error can be easily 3~GeV. Therefore, we have plotted the
best-fit point for Higgs uncertainties between 2~and 5~GeV.
One sees that the best-fit points wanders by several TeV.
Clearly this needs a more detailed investigation in the future.
It should be noted that the fit does not provide the maximum
mixing scenario. If we exclude all other constraints, the
maximum value of the Higgs mass can reach 125~GeV, albeit also
at similarly large values of $m_{1/2}$. A negative sign of the
mixing parameter $\mu$ shows similar results.

%-------------------------------------------------------
\section{The problem of the dark matter in the Universe}

As has been already mentioned the shining matter does not
compose all the matter in the Universe. According to the latest
precise data~\cite{WMAP} the matter content of the Universe is
the following:
\begin{eqnarray*}
&&\Omega_{total}=1.02 \pm 0.02 ,\\
&&\Omega_{vacuum}=0.73 \pm 0.04, \\
&&\Omega_{matter}=0.23 \pm 0.04,\\
&& \Omega_{baryon}=0.044 \pm 0.004,
\end{eqnarray*}
so that the dark matter prevails the usual baryonic matter by
factor of 6.

Besides the rotation curves of spiral galaxies the dark matter
manifests itself in the observation of gravitational lensing
effects~\cite{lensing} and the large structure formation. It is
believed that the dark matter played the crucial role in the
formation of large structures like clusters of galaxies and the
usual matter just fell down in a potential well attracted by
the gravitational interaction afterwards. The dark matter can
not make compact objects like the usual matter since it does not
take part in the strong interaction and can not lose energy by
the photon emission since it is neutral. For this reason the
dark matter can be trapped in much larger scale structures
like galaxies.

In general one may assume two possibilities: either the dark
matter interacts only gravitationally, or it participates also
in the weak interaction. The latter case is preferable since
then one may hope to detect it via the methods of the particle
physics. What makes us to believe that the dark matter is
probably the Weakly Interacting Massive Particle (WIMP)? This is
because the cross-section of the DM annihilation which can be
figured out of the amount of the DM in the Universe is close to
a typical weak interaction cross-section. Indeed, let us assume
that all the DM is made of particles of a single type. Then the
amount of the DM can be calculated from the Boltzman
equation~\cite{Kolb,Rub}
\begin{equation}
\frac{dn_\chi}{dt}+ 3 H n_\chi =
- \langle \sigma v \rangle ( n^2_\chi- n^2_{\chi,eq}),	
\end{equation}
where $H = \dot{R}/ R$ is the Hubble constant and $n_{\chi,eq}$
is the equilibrium concentration. The relic abundance is
expressed in terms of $n_\chi$ as
\begin{equation}
\Omega_\chi h^2 =\frac{m_\chi n_\chi}{\rho_c} \approx
\frac{2\cdot 10^{-27}\ \text{cm}^3\ \text{sec}^{-1}}{\langle \sigma v \rangle}.
\end{equation}
Having in mind that $\Omega_\chi h^2 \approx 0.113\pm 0.009$
and $v\sim 300$ km/sec one gets
\begin{equation}
\sigma\approx 10^{-34}\ \text{cm}^2 = 100\ \text{pb},
\end{equation}
which is a typical electroweak cross-section.

%------------------------------------------------------------
\subsection{Supersymmetric interpretation of the Dark Matter}

Supersymmetry offers several candidates for the role of the
cold dark matter particle. If one looks at the particle content
of the MSSM from the point of view of a heavy neutral particle,
one finds several such particles, namely: the superpartner of
the photon (the photino $\tilde \gamma$), the superpartner of
the $Z$-boson (the particle called zino $\tilde z$), the
superpartner of the neutrino (the sneutrino $\tilde \nu$) and
the superpartners of the Higgs bosons (the higgsinos
$\tilde H$). The DM particle can be the lightest of them, the
LSP. The others decay to the LSP and the SM particles, while
the LSP is stable and can survive since the Big Bang. As a rule
the lightest supersymmetric particle is the neutralino, the
spin 1/2 particle which is the combination of the photino, zino
and two neutral higgsinos and is the eigenstate of the mass
matrix
$$
|\tilde \chi^0_1\rangle =N_1|B_0\rangle
+N_2|W^3_0\rangle +N_3|H_1\rangle +N_4|H_2\rangle.
$$

Thus, supersymmetry actually predicts the existence of the dark
matter. Moreover, one can choose the parameters of soft
supersymmetry breaking in such a way that one gets the right
amount of the DM. This requirement serves as a constraint for
these parameters and is consistent with the requirements coming
from the particle physics.

\begin{figure}[htb]
\begin{center}
\hspace*{3mm}
\raisebox{9pt}{\includegraphics[width=.48\textwidth]{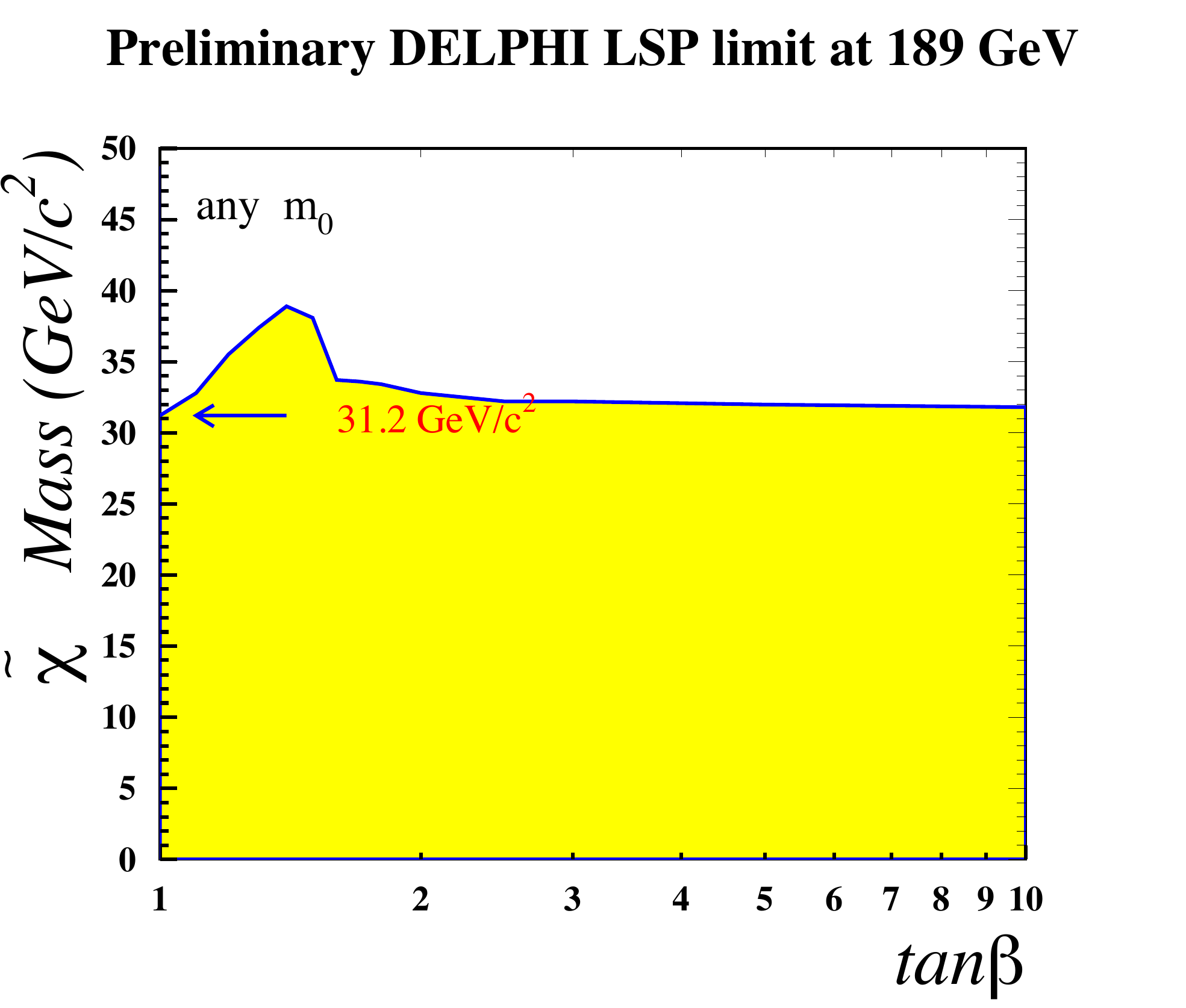}}
\includegraphics[width=.41\textwidth]{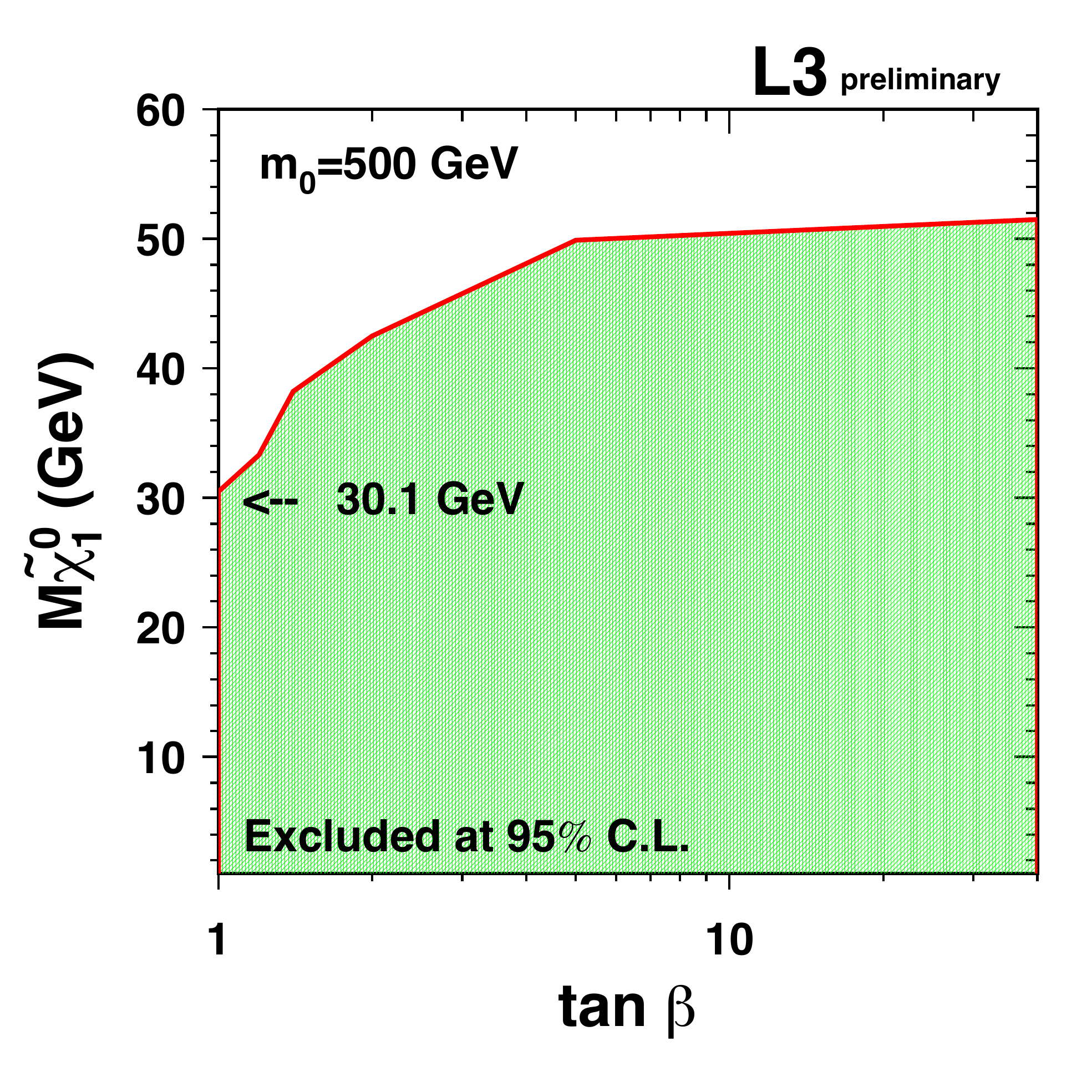}
\vspace{-0.4cm}
\caption{Exclusion limits on the LSP mass from Delphi and L3
Collaboration (LEP)~\cite{LEP}.}
\label{delphi}
\end{center}
\end{figure}

The search for the LSP was one of the tasks of LEP. They were
supposed to be produced as a result of the chargino decays and
be detected via the missing transverse energy and momentum.
Negative results defined the limit on the LSP mass as shown
in Fig.~\ref{delphi}.

\begin{figure}[htb]
\begin{center}
\includegraphics[width=.55\textwidth]{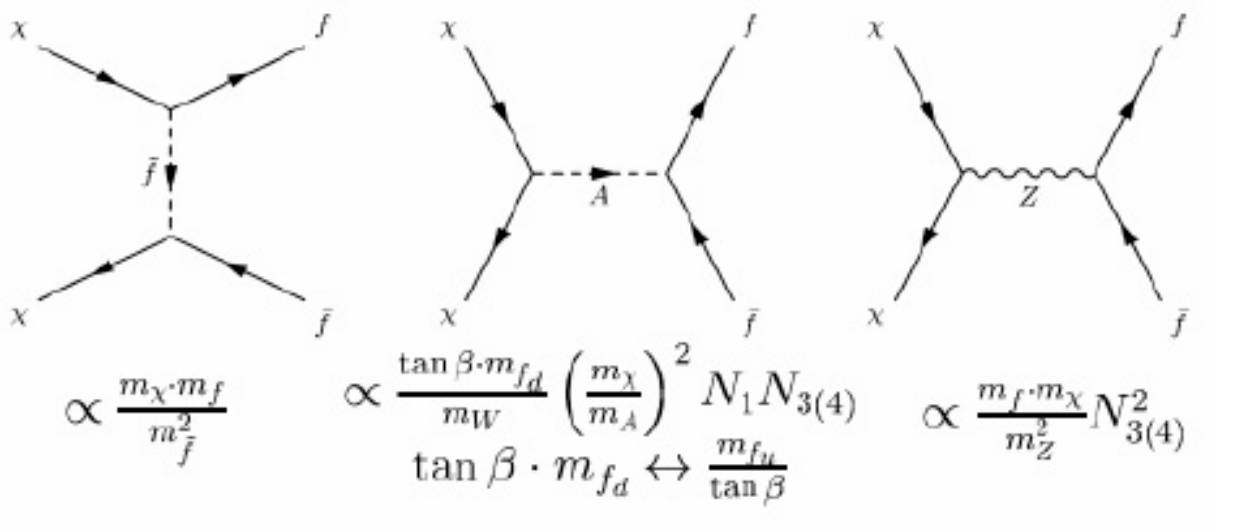}
\includegraphics[width=.40\textwidth]{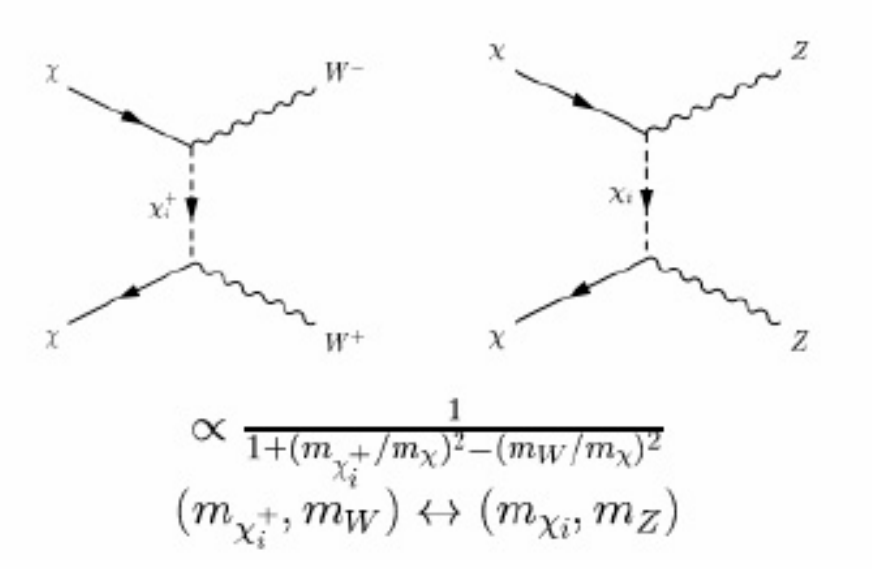}
\end{center}
\caption{The dominant annihilation diagrams for the lightest
neutralino in the MSSM.}
\label{annihilation}
\end{figure}

The DM particles which form the halo of the galaxy annihilate
to produce the ordinary particles in the cosmic rays.
Identifying them with the LSP from a supersymmetric model one
can calculate the annihilation rate and study the secondary
particle spectrum. The dominant annihilation diagrams of the
neutralino LSP are shown in Fig.~\ref{annihilation}. The usual
final states are either the quark-antiquark pairs or the $W$
and $Z$ bosons. Since the cross sections are proportional to
the final state fermion mass, the heavy fermion final states,
i.~e. the third generation quarks and leptons, are expected to
be dominant. The $W$ and $Z$ final states from the $t$-channel
chargino and neutralino exchange have usually a smaller cross
section.

The dominant contribution comes from the $A$-boson exchange:
$\chi+\chi \to A \to b\bar b$. The sum of the diagrams should
yield $\langle \sigma v \rangle=2\cdot  10^{-26}\
\text{cm}^3/\text{sec}$ to get the correct relic density.

The spectral shape of the secondary particles the from DM
annihilation is well known from the fragmentation of the
mono-energetic quarks studied at the electron-positron
colliders, like LEP at CERN, which has been operating up to
the centre-of-mass energy of about 200~GeV, i.~e. it
corresponds to the neutralino mass up to 100~GeV.
%(see Fig.\ref{SMfinal}).
The different quark flavours all yield similar gamma spectra
at high energies. Hence, the specrta of the positrons, photons
and antiprotons is known. The relative amount of
$\gamma, p^-$ and $e^+$ is also known. One expects around
37~photons per collision.

The gamma rays from the DM annihilation can be distinguished
from the background by their completely different spectral
shape: the background originates mainly from cosmic rays hitting
the gas of the disc and producing abundantly $\pi^0$-mesons,
which decay into two photons. The initial cosmic ray spectrum is
a steep power law spectrum, which yields a much softer gamma ray
spectrum than the fragmentation of the hard mono-energetic
quarks from the DM annihilation. The spectral shape of the gamma
rays from the background is well known from fixed target
experiments given the known cosmic ray spectrum.

Unfortunately, modern data on diffuse galactic gamma rays, do
not indicate statistically significant departure from the
background. Local excess observed in some experiments like
EGRET space telescope~\cite{egret} and FERMI~\cite{fermi} is
well inside the uncertainties of the background.

%------------------------------------------------
\subsection{Region excluded by the relic density}
\label{relic}

\begin{figure}[htb]
\begin{center}
\includegraphics[width=0.48\textwidth]{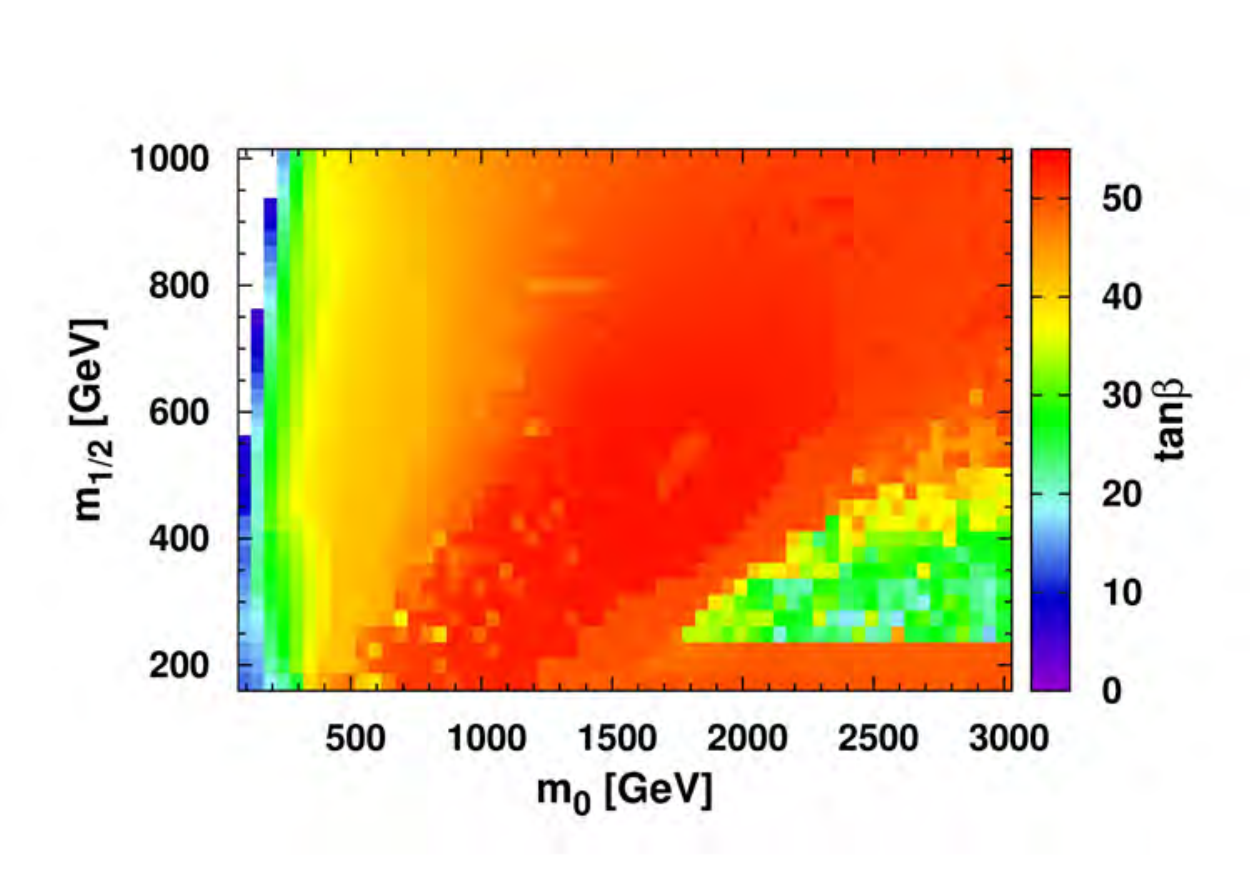}
\includegraphics[width=0.48\textwidth]{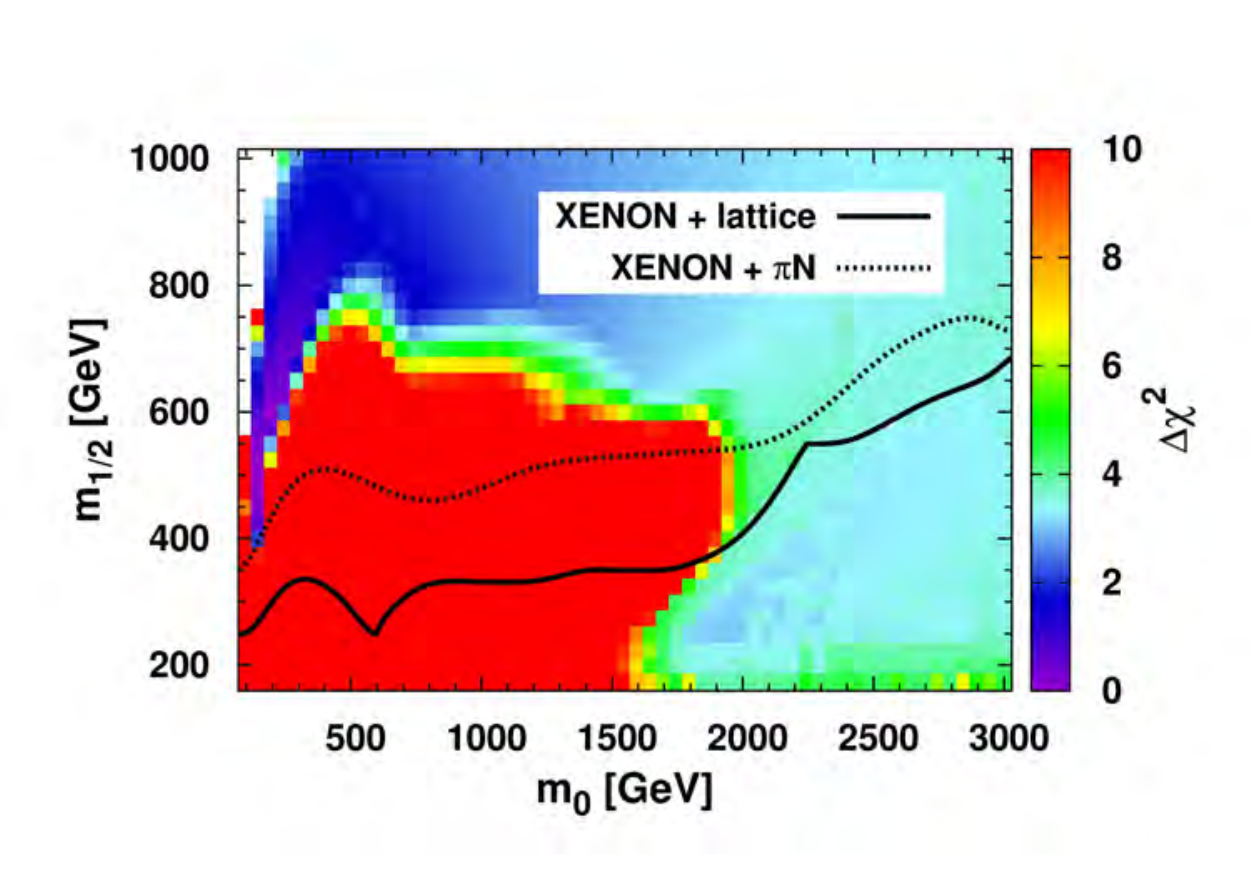}
\end{center}
\caption{Left: Fitted values of $\tan\beta$  in the ($m_0 - m_{1/2}$) plane after optimizing
$A_0$ to fulfil the relic density and EWSB
constraints at every point. The relic density requires
$\tan\beta \approx 50$ in most of the parameter space, where
pseudo-scalar Higgs exchange dominates. In the (non-red) edges
where $\tan\beta$ is lower, the co-annihilation  contributes.
Right: $\Delta\chi^2=\chi^2-\chi^2_{min}$ distribution
in the ($m_0 - m_{1/2}$) plane after imposing the electroweak
constraints in comparison with the XENON100
limits~\cite{Aprile:2011hi} on the direct WIMP-nucleon
cross-section for the two values of the form factors (dotted
line: $\pi N$ scattering, dashed dotted line: lattice gauge
theories).}
\label{f6_100a}
\end{figure}

The observed relic density of the dark matter corresponds to
$\Omega h^2=0.113\pm 0.004$~\cite{Komatsu:2010fb}. This number
is inversely proportional to the annihilation cross section.
The dominant annihilation contribution comes from $A$-boson
exchange in most of the parameter space. The cross section for
$\chi+\chi \to A \to b\bar b$ can be written as:
\begin{equation}
\langle \sigma v \rangle \sim  \frac{M_\chi^4
m_b^2 \tan^2\beta}{\sin^4 2\theta_W
\,M_Z^2}\frac{ \left( N_{31}\sin\beta -N_{41}\cos\beta
\right)^2\left( N_{21}\cos\theta_W - N_{11}\sin\theta_W
\right)^2}{\left( 4M_\chi^2 - M_A^2
\right)^2+M_A^2\Gamma_A^2}.
\end{equation}
As have been mentioned, the correct relic density requires
$\langle \sigma v \rangle = 2 \cdot 10^{-26}$ cm$^3$/s, which
implies that the annihilation cross section $\sigma$ is of the
order of a 100~pb. Such a high cross section can be obtained
only close to the resonance. Actually on the resonance the cross
section is too high, so one needs to be in the tail of the
resonance, i~.e. $m_A \approx 2.2\, m_\chi$ or
$m_A \approx 1.8\, m_\chi$. So one expects $m_A \propto m_{1/2}$
from the relic density constraint. This constraint can be
fulfilled with $\tan\beta$ values around 50 in the whole
($m_0 - m_{1/2}$) plane, except for the narrow
co-annihilation regions~\cite{Beskidt:2010va}. The results can
be extended to larger values of $m_0$, as shown in the left
panel of Fig.~\ref{f6_100a}.

%-------------------------------------------------
\subsection{Region excluded by direct DM searches}
%\label{direct}

There are two methods of the dark matter detection: direct and
indirect. In the direct detection one assumes that the particles
of the dark matter to the Earth and interact with the nuclei of
a target. In the underground experiments one can hope to observe
such events measuring the recoil energy. There are several
experiments of this type: DAMA, Zeplin, CDMS and Edelweiss.
Among them only the DAMA collaboration claims to observe a
positive outcome in the annual modulation of the signal with the
fitted dark matter particle mass around 50 GeV~\cite{DAMA}.
\begin{figure}[hb]
\begin{center}
\includegraphics[width=.39\textwidth]{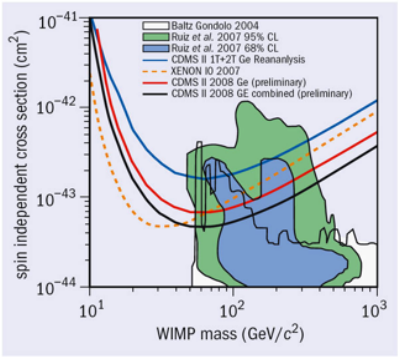}
\hspace{7mm}
\raisebox{2pt}{\includegraphics[width=.4\textwidth]{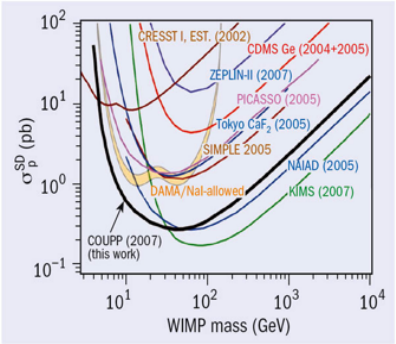}}
\caption{The exclusion plots from the direct dark matter
detection experiments. The spin-independent case (left) from
Chicagoland Observatory for Underground Particle Physics (COUPP)
and the spin-dependent case (right) from Cryogenic Dark Matter
Search (CDMS).}
\label{direct}
\end{center}
\end{figure}\vspace{-0.8cm}

All the other experiments do not see it though CDMS
collaboration recently announced about a few events of a desired
type~\cite{CDMS}. The reason of this disagreement might be in
the different methodology and the targets used since the
cross-section depends on the spin of the target nucleus.
The collected statistics is also essentially different.
DAMA has accumulated by far more data and this is the only
experiment which studies the modulation of the signal that may
be crucial for reducing the background.

The cross section for direct scattering of WIMPS on
nuclei has an experimental upper limit of about 10$^{-8}$ pb,
i.~e. many orders of magnitude below the annihilation cross
section. This cross section is related to the annihilation
cross section by similar Feynman diagrams. The many orders of
magnitude are naturally explained in Supersymmetry by the fact
that both cross sections are dominated by Higgs exchange and
the fact that the Yukawa couplings to the valence quarks in the
proton or neutron are negligible. Most of the scattering cross
section comes from the heavier sea-quarks. However, the
density of these virtual quarks inside the nuclei is small,
which is one of the reasons for the small elastic scattering
cross section. In addition, the momentum transfer in elastic
scattering is small, so the propagator leads to a cross section
inversely proportional to the fourth power of the Higgs mass.

The typical exclusion plots for the spin-independent and
spin-dependent cross-sections are shown in Fig.~\ref{direct}
where one can see DAMA allowed region overlapping with the other
exclusion curves. Still today we have no convincing evidence for
direct dark matter detection or exclusion.
\begin{figure}[t]
\begin{center}
\includegraphics[width=0.5\textwidth]{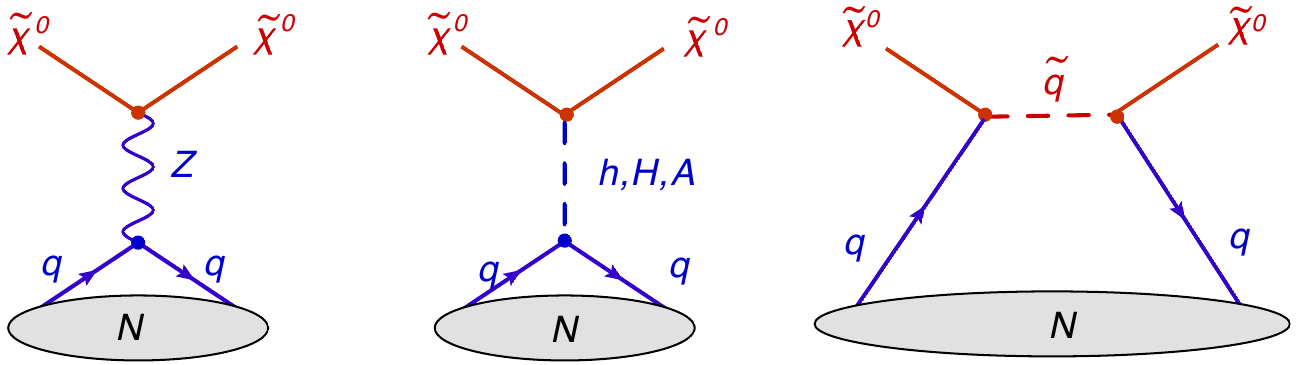}
\end{center}
\caption{Examples of diagrams for elastic neutralino-nucleon
scattering.}
\label{f7}
\end{figure}
Scattering of the LSP on nuclei can only happen via elastic
scattering, provided $R$-parity is
conserved~\cite{abun,Kolb}. The corresponding
diagrams are shown in Fig.~\ref{f7}.

The big blob indicates that one enters a low energy regime, in
which case the protons and neutrons inside the nucleus cannot
be resolved. In this case the spin-independent scattering
becomes coherent on all nuclei and the cross section becomes
proportional to the number of nuclei:
\begin{equation}
\sigma = \frac{4}{\pi} \frac{m_{\rm DM}^2m_N^2}{(m_{\rm DM}+m_N)^2}
\left(Z
f_p+(A-Z)f_n\right)^2
\label{sc}
\end{equation}
where $A$ and $Z$ are the atomic mass and atomic number of the target
nuclei and the form factors are~\cite{Ellis:2008hf}
\begin{equation} \label{fpn}
f_{p,n}=\sum_{q=u,d,s} G_{q}
f^{(p,n)}_{Tq}\frac{m_{p,n}}{m_q}+\frac{2}{27}f^{(p,n)}_{TG}\sum_{q=c,b,t}
G_{q}\frac{m_{p,n}}{m_q},
\end{equation}
where $G_q=\lambda_{\rm DM}\lambda_q / M^2_M$. Here $M$ denotes the
mediator, and $\lambda_{\rm DM}$, $\lambda_{f}$ denote the mediator's
couplings to DM and quark. The parameters $ f^{(p)}_{Tq}$
are defined by
\begin{equation}
m_p  f^{(p)}_{Tq} \equiv \langle p|m_q\bar q q|p \rangle
\label{condensate}
\end{equation}
and similar for $ f^{(n)}_{Tq}$,
whilst $ f^{(p,n)}_{TG}=1-\sum_{q=u,d,s} f^{(p,n)}_{Tq}$.

Since the particle which mediates the scattering is typically
much heavier than the momentum transfer, the scattering can be
written in terms of an effective coupling, which can be
determined phenomenologically from  $\pi N$ scattering or from
lattice QCD calculations.

The default values of the effective couplings in
micrOMEGAs~\cite{Belanger:2008sj} are:
$f^{(p)}_{Tu}=0.033,~f^{(p)}_{Td} =
0.023,~f^{(p)}_{Ts}=0.26,~f^{(n)}_{Tu}=0.042,~f^{(n)}_{Td}=0.018,~f^{(n)}_{Ts}=0.26$.
The lower values from the  lattice calculations~\cite{Cao:2010ph} are:
$f^{(p)}_{Tu}=0.020,~f^{(p)}_{Td} =
0.026,~f^{(p)}_{Ts}=0.02,~f^{(n)}_{Tu}=0.014,~f^{(n)}_{Td}=0.036,~f^{(n)}_{Ts}=0.02$.
Hence the most important coupling to the strange quarks vary from
0.26 to 0.02~\cite{Alarcon:2011zs},  which implies an order of
magnitude  uncertainty in the elastic neutralino-nucleon
scattering cross section.

Another normalization uncertainty in direct dark matter
experiments arises from the uncertainty in the local DM
density, which can take values between 0.3 and 1.3 GeV/cm$^3$, as
determined from the rotation curve of the Milky Way, see
Ref.~\cite{Weber:2009pt,deBoer:2010eh,Salucci:2010qr,Catena:2009mf}.

The excluded region from the XENON100 cross section
limit~\cite{Aprile:2011hi} for two choices of form factors is
shown in Fig.~\ref{f6_100a}. At large values of $m_0$ EWSB
forces the higgsino component of the WIMP to increase and
consequently the exchange via the Higgs, which has an amplitude
proportional to the bino-higgsino mixing, starts to increase.
This leads to an increase in the excluded region at large $m_0$
and has here a similar sensitivity as the LHC. If we would take
the less conservative effective couplings from the default
values of micrOMEGAs the XENON100  limit would be 50\% higher
than the LHC limit.

%-------------------------------------
\section{Search for SUSY at Colliders}

%----------------------------------------------------------
\subsection{Experimental signatures at  $e^+e^-$ colliders}

Experiments are finally beginning to push into a significant
region of supersymmetry parameter space. We know the sparticles
and their couplings, but we do not know their masses and mixings.
Given the mass spectrum one can calculate the cross-sections and
consider the possibilities of observing the new particles at
modern accelerators. Otherwise, one can get restrictions on the
unknown parameters.

We start with the $e^+e^-$ colliders. In the leading order the
processes of creation of the superpartners are given by the
diagrams shown in Fig.~\ref{creation} above. For a given center
of mass energy the cross-sections depend on the masses of the
created particles and vanish at the kinematic boundary.
Experimental signatures are defined by the decay modes which
vary with the mass spectrum. The main ones are summarized below,
see, e.~g.~\cite{SUSYLHC_GK,unexpLHC}

$$
\begin{array}{lll}
\mbox{\underline{Production\phantom{y}\hspace*{-5pt}}}&\mbox{\underline{Decay
Modes}}&~~ \mbox{\underline{Signatures}} \\ && \\ \bullet
~\tilde{l}_{L,R}\tilde{l}_{L,R}~~ &\tilde{l}^\pm_R \to l^\pm
\tilde{\chi}^0_i& \mbox{acompl pair of}
\\ &  \tilde{l}^\pm_L \to l^\pm \tilde{\chi}^0_i  &
\mbox{charged lept} + \Big/ \hspace{-0.3cm E_T} \\ \bullet
~\tilde{\nu}\tilde{\nu}& \tilde{\nu}\to l^\pm \tilde{\chi}^0_1
 & \Big/ \hspace{-0.3cm E_T}\\
\bullet  ~\tilde{\chi}^\pm_1\tilde{\chi}^\pm_1 &\tilde{\chi}^\pm_1 \to
 \tilde{\chi}^0_1 l^\pm \nu &
 \mbox{isol lept} + \mbox{2 jets} + \Big/ \hspace{-0.3cm E_T} \\
&  \tilde{\chi}^\pm_1 \to \tilde{\chi}^0_2 f \bar f' &\mbox{pair
of acompl}\\
 &  \tilde{\chi}^\pm_1 \to l \tilde{\nu}_l &\mbox{leptons} + \Big/ \hspace{-0.3cm E_T}
\\
& ~~~~~~~  \to l\nu_l\tilde{\chi}^0_1 & \\
& \tilde{\chi}^\pm_1 \to \nu_l \tilde{l}&\mbox{4 jets} + \Big/ \hspace{-0.3cm E_T}
\\ & ~~~~~~~ \to \nu_l
l\tilde{\chi}^0_1 & \\[2mm]
 \bullet  ~\tilde{\chi}^0_i\tilde{\chi}^0_j &
\tilde{\chi}^0_i \to \tilde{\chi}^0_1 X& X=\nu_l \bar \nu_l \ \mbox{invisible} \\ &&
~~~= \gamma,2l,\mbox{2 jets} \\ && 2l + \Big/ \hspace{-0.3cm E_T},
l+2j + \Big/ \hspace{-0.3cm E_T} %\\
\end{array}$$
$$\begin{array}{lll}
\bullet ~\tilde{t}_i\tilde{t}_j & \tilde{t}_1 \to c
\tilde{\chi}^0_1 & \mbox{2 jets}+ \Big/ \hspace{-0.3cm E_T} \\ &
\tilde{t}_1 \to b \tilde{\chi}^\pm_1& \mbox{2 $b$-jets} + \mbox{2 lept} + \Big/
\hspace{-0.3cm E_T} \\
& ~~~~~  \to b f\bar
f'\tilde{\chi}^0_1 & \\[1mm]
%&& \mbox{2 $b$-jets}+\mbox{lept} + \Big/ \hspace{-0.3cm E_T}
%\\
\bullet  ~\tilde{b}_i\tilde{b}_j & \tilde{b}_i \to b
\tilde{\chi}^0_1 & \mbox{2 $b$-jets}+ \Big/ \hspace{-0.3cm E_T} \\ &
\tilde{b}_i \to b \tilde{\chi}^0_2
& \mbox{2 $b$-jets} + \mbox{2 lept}\! \!+\!\! \Big/ \hspace{-0.3cm E_T} \\
&~~~~~~  \to b f\bar f'\tilde{\chi}^0_1 &
 \mbox{2 $b$-jets} + \mbox{2 jets} + \Big/ \hspace{-0.3cm E_T}
\end{array}
$$

The characteristic feature of all the possible signatures is
the missing energy and transverse momentum, which is a trade
mark of the new physics.

Numerous attempts to find the superpartners at LEP II gave no
positive result thus imposing the lower bounds on their
masses~\cite{LEPSUSY}. Typical LEP II limits on the
superpartner masses are
\begin{eqnarray*}
\hspace{-0.7cm}&& m_{\chi^0_1} > ~40 \ \text{GeV}, \ m_{\tilde e}>105\ \text{GeV}, \
 m_{\tilde t}> 90\ \text{GeV} \\
\hspace{-0.7cm} &&m_{\chi^\pm_1} > 100 \ \text{GeV}, \ m_{\tilde \mu}>100\ \text{GeV}, \
 m_{\tilde b}> 80\ \text{GeV}, \ m_{\tilde \tau}>80\ \text{GeV}
\end{eqnarray*}

%-------------------------------------------------------
\subsection{Experimental signatures at hadron colliders}

Experimental SUSY signatures at the Tevatron and LHC are
similar. The strategy of the SUSY searches is based on the
assumption that the masses of the superpartners indeed are in
the region of 1~TeV so that they might be created on the mass
shell with the cross-section big enough to distinguish them
from the background of the ordinary particles. Calculation of
the background in the framework of the Standard Model thus
becomes essential since the secondary particles in all the
cases are the same.

There are many possibilities to create the superpartners at
the hadron colliders. Besides the usual annihilation channel
there are numerous processes of the gluon fusion,
quark-antiquark and quark-gluon scattering. The maximal
cross-sections of the order of a few picobarn can be achieved
in the process of gluon fusion.

As a rule all the superpartners are short lived and decay into
the ordinary particles and the lightest superparticle. The main
decay modes of the superpartners which serve as the
manifestation of SUSY are
$$
\begin{array}{lll}
\mbox{\underline{Production\phantom{y}\hspace*{-5pt}}}&\mbox{\underline{Decay
Modes}}&~~ \mbox{\underline{Signatures}} \\ && \\
\bullet
~\tilde{g}\tilde{g}, \tilde{q}\tilde{q},
\tilde{g}\tilde{q}%~~~~~~~~~ ~~
&\begin{array}{l} \tilde{g}
\to q\bar q \tilde{\chi}^0_1   \\
 ~~~~~ q\bar q' \tilde{\chi}^\pm_1  \\
 ~~~~~ g\tilde{\chi}^0_1 \end{array}
  &
\begin{array}{c} \Big/ \hspace{-0.3cm E_T} + \mbox{multijets}\\
 (+\mbox{leptons}) \end{array} \\
& \begin{array}{l}\tilde{q} \to q \tilde{\chi}^0_i \\
    \tilde{q} \to q' \tilde{\chi}^\pm_i \end{array}
  &\\
 \bullet
~\tilde{\chi}^\pm_1\tilde{\chi}^0_2 &\tilde{\chi}^\pm_1 \to
 \tilde{\chi}^0_1 l^\pm \nu &
 \mbox{Trilepton} + \Big/ \hspace{-0.3cm E_T} \\
 & ~~~~\ \tilde{\chi}^0_2 \to
 \tilde{\chi}^0_1 ll & \\
%\end{array}$$
 %$$\begin{array}{lll}
&  \tilde{\chi}^\pm_1 \to \tilde{\chi}^0_1 q \bar q'&\mbox{Dilept}
+\mbox{ jet} + \Big/ \hspace{-0.3cm E_T}\\
 & ~~~~ \tilde{\chi}^0_2 \to
\tilde{\chi}^0_1 ll &\\
%\end{array}$$
% $$\begin{array}{lll}
\bullet  ~\tilde{\chi}^+_1\tilde{\chi}^-_1 &
\tilde{\chi}^+_1 \to l \tilde{\chi}^0_1 l^\pm \nu &
\mbox{Dilepton} + \Big/ \hspace{-0.3cm E_T} \\ \bullet
~\tilde{\chi}^0_i\tilde{\chi}^0_i & \tilde{\chi}^0_i \to
\tilde{\chi}^0_1 X &
\Big/ \hspace{-0.3cm E_T} + \mbox{Dilept+jets}%\\
\end{array}$$
 $$\begin{array}{lll}
\bullet  ~\tilde{t}_1\tilde{t}_1 & \tilde{t}_1 \to c
\tilde{\chi}^0_1 & \mbox{2 acollin jets} + \Big/ \hspace{-0.3cm
E_T} \\ & \tilde{t}_1 \to b \tilde{\chi}^\pm_1 &
 \mbox{sing lept} + \Big/ \hspace{-0.3cm E_T} + b's\\
 & ~~~~  \tilde{\chi}^\pm_1\to \tilde{\chi}^0_1 q\bar q'  & \\
 &\tilde{t}_1 \to b \tilde{\chi}^\pm_1
& \mbox{Dilept} + \Big/ \hspace{-0.3cm E_T} + b's\\
& ~~~ \tilde{\chi}^\pm_1 \to \tilde{\chi}^0_1
l^\pm \nu &\\
\bullet
~\tilde{l}\tilde{l},\tilde{l}\tilde{\nu},\tilde{\nu}\tilde{\nu}
&
 \tilde{l}^\pm \to l\pm \tilde{\chi}^0_i& \mbox{Dilepton}+ \Big/ \hspace{-0.3cm E_T} \\
 & ~~ \tilde{l}^\pm \to \nu_l \tilde{\chi}^\pm_i  &  \mbox{Single lept} +
/ \hspace{-0.25cm E_T} \\
& \tilde{\nu} \to \nu \tilde{\chi}^0_1& / \hspace{-0.25cm E_T}
\end{array}
$$

Note again the typical events with the missing energy and
transverse momentum that is the main difference from the
background processes of the Standard Model. Contrary to the
$e^+e^-$ colliders, at hadron machines the background is
extremely rich and essential. The missing energy is carried
away by the heavy particle with the mass of the order of
100~GeV that is essentially different from the processes with
the neutrino in the final state. In hadron collisions the
superpartners are always created in pairs and then further
quickly decay creating a cascade with the ordinary quarks
(i.~e. hadron jets) or leptons in the final state plus the
missing energy. For the case of the gluon fusion with the
creation of gluino it is presented in Table~\ref{tab:n}
(right panel).

The chargino and neutralino can also be produced in pairs
through the Drell-Yang mechanism
$pp \to \tilde \chi^\pm_1 \tilde \chi^0_2$ and can be detected
via their lepton decays
$\tilde \chi^\pm_1 \tilde \chi^0_2 \to \ell\ell\ell+
\Big/ \hspace{-0.3cm E_T}$. Hence the main signal of their
creation is the isolated leptons and the missing energy,
see Table~\ref{tab:n} (left panel). The main background in
the trilepton channel comes from the creation of the standard
particles $WZ/ZZ,t\bar t, Zb\bar b$ è $b\bar b$. There might be
also the supersymmetric background from the cascade decays of
the squarks and gluinos in multilepton modes.

\begin{table}[p]
\begin{tabular}{|c|p{1.7cm}|} \hline
Process & final  states \\
\hline
\includegraphics[width=50mm,height=35mm]{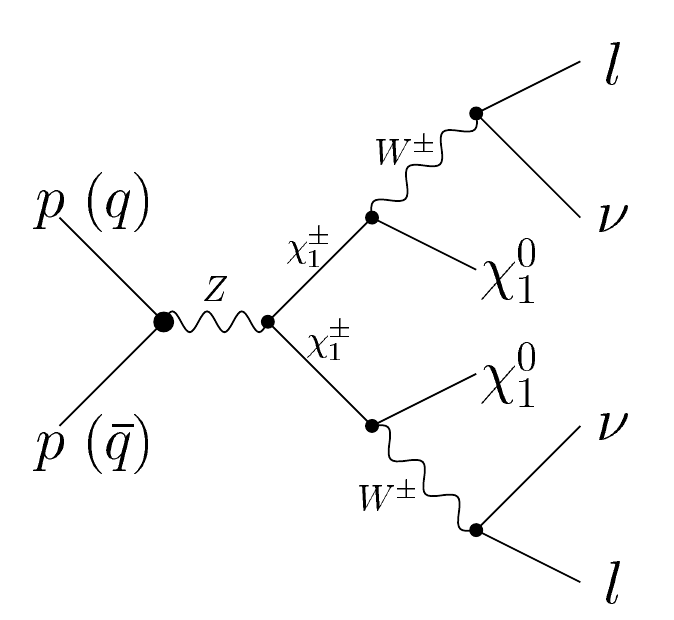}
& \vspace*{-30mm}
\begin{minipage}[t]{1.8cm}
$\begin{array}{c}
2\ell \\ 2\nu \\ \Big/\hspace{-0.3cm E_T}
\end{array}$
\end{minipage} \\ \hline
\includegraphics[width=50mm,height=35mm]{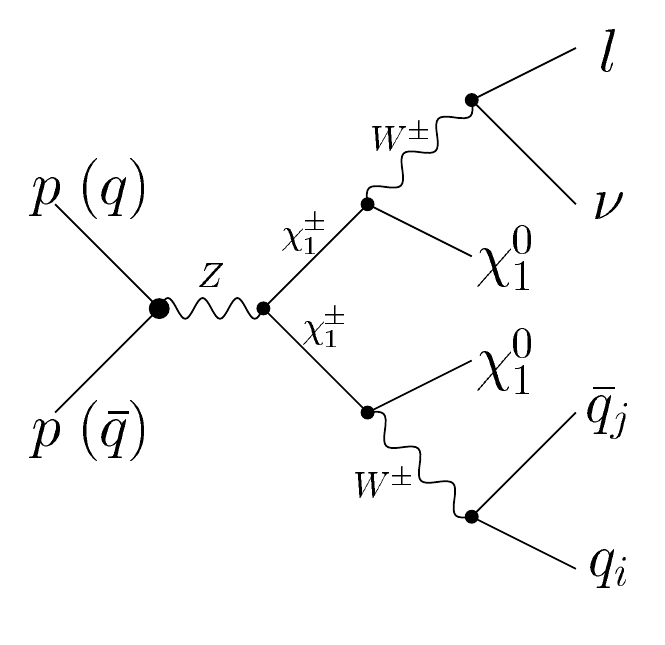}
& \vspace*{-35mm}
\begin{minipage}[t]{1.8cm}
$\begin{array}{c}
\ell \\ \nu \\ 2j \\ \Big/\hspace{-0.3cm E_T}
\end{array}$
\end{minipage} \\ \hline
\includegraphics[width=50mm,height=35mm]{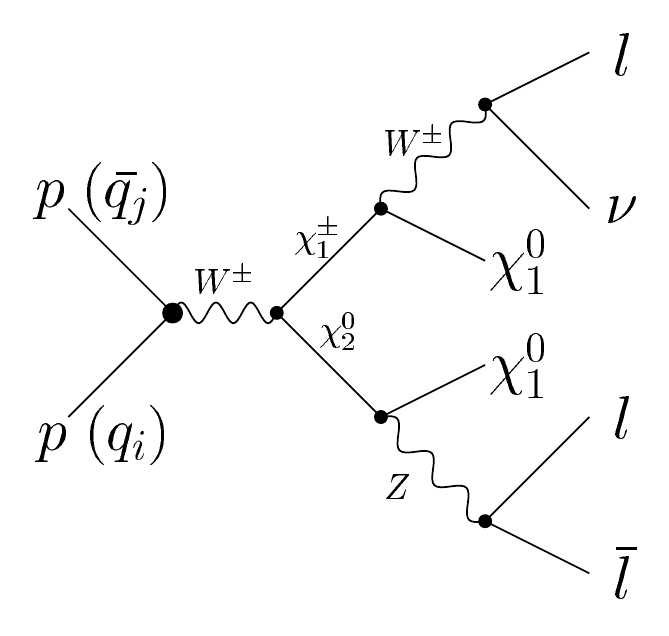}
& \vspace*{-30mm}
\begin{minipage}[t]{1.8cm}
$\begin{array}{c}
3\ell \\ \nu \\ \Big/\hspace{-0.3cm E_T}
\end{array}$
\end{minipage} \\ \hline
\includegraphics[width=50mm,height=35mm]{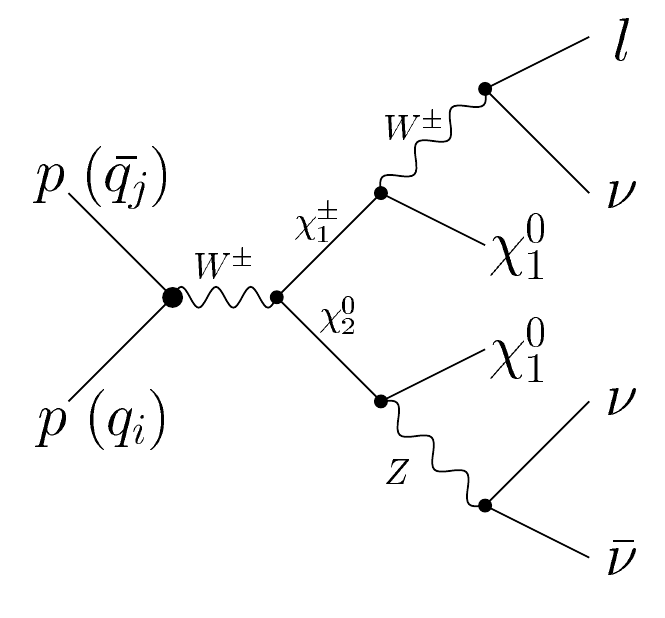}
& \vspace*{-30mm}
\begin{minipage}[t]{1.8cm}
$\begin{array}{c}
\ell \\ 3\nu \\ \Big/\hspace{-0.3cm E_T}
\end{array}$
\end{minipage} \\ \hline
\includegraphics[width=50mm,height=35mm]{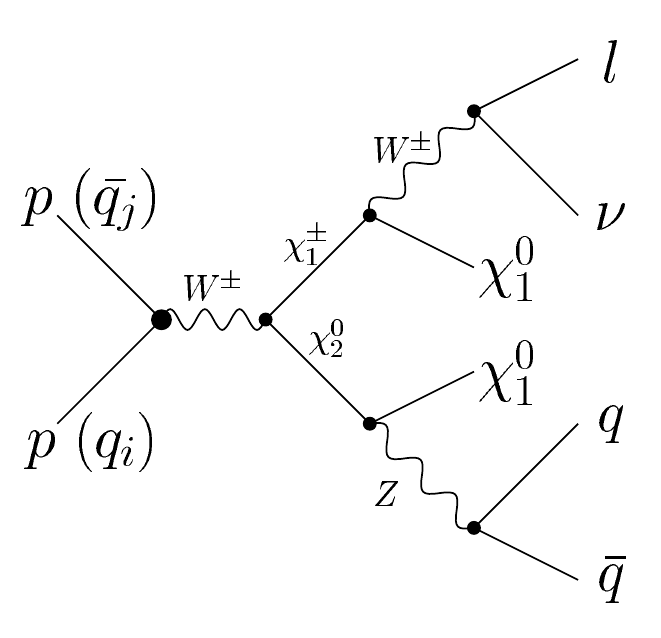}
& \vspace*{-35mm}
\begin{minipage}[t]{1.8cm}
$\begin{array}{c}
\ell \\ \nu \\ 2j \\ \Big/\hspace{-0.3cm E_T}
\end{array}$
\end{minipage} \\ \hline
\end{tabular}
\begin{tabular}{|c|p{1.7cm}|}
\hline Process & final states \\
 \hline
\includegraphics[width=50mm,height=35mm]{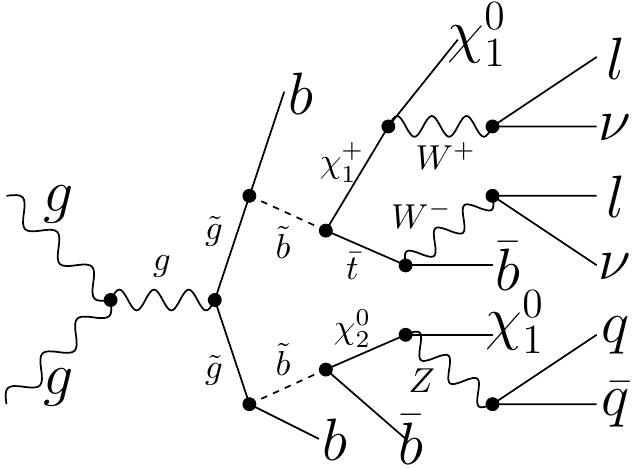}
& \vspace*{-35mm}
\begin{minipage}[t]{1.8cm}
$\begin{array}{c}
2\ell \\ 2\nu \\ 6j \\ \Big/\hspace{-0.3cm E_T}
\end{array}$
\end{minipage} \\ \hline
\includegraphics[width=50mm,height=35mm]{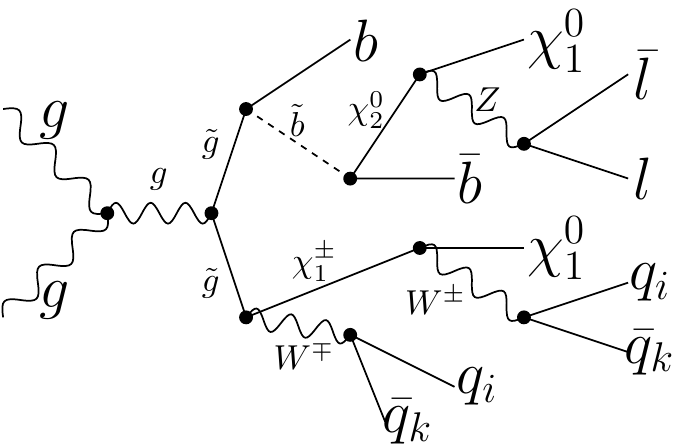}
& \vspace*{-30mm}
\begin{minipage}[t]{1.8cm}
$\begin{array}{c}
2\ell \\ 6j \\ \Big/ \hspace{-0.3cm E_T}
\end{array}$
\end{minipage} \\ \hline
\includegraphics[width=50mm,height=35mm]{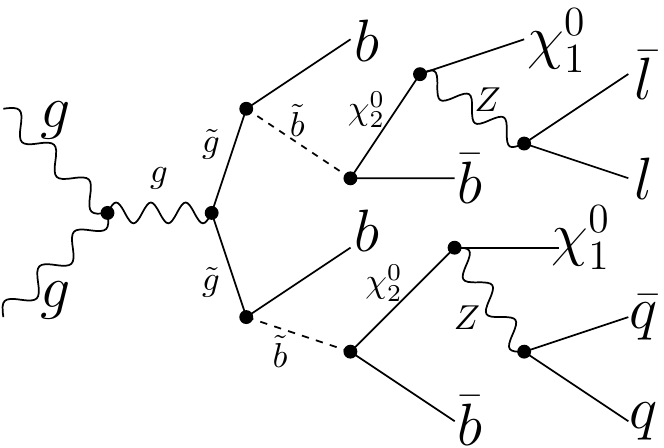}
& \vspace*{-30mm}
\begin{minipage}[t]{1.8cm}
$\begin{array}{c}
2\ell \\ 6j \\ \Big/ \hspace{-0.3cm E_T}
\end{array}$
\end{minipage} \\ \hline
\includegraphics[width=50mm,height=35mm]{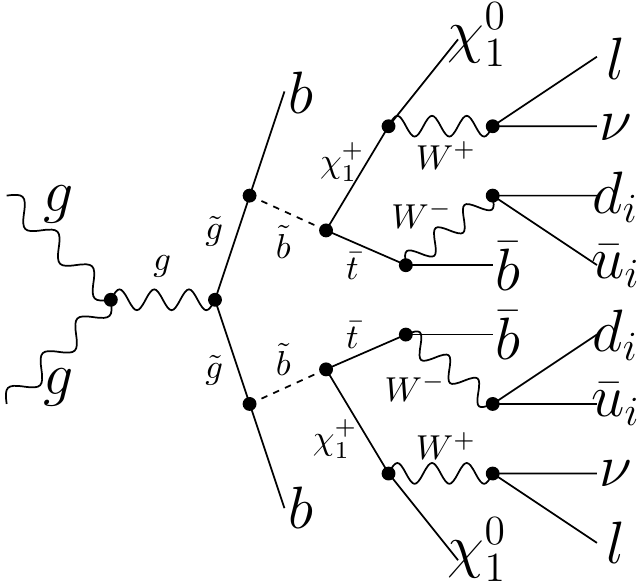}
& \vspace*{-35mm}
\begin{minipage}[t]{1.8cm}
$\begin{array}{c}
2\ell \\ 2\nu \\ 8j \\ \Big/\hspace{-0.3cm E_T}
\end{array}$
\end{minipage} \\ \hline
\includegraphics[width=50mm,height=35mm]{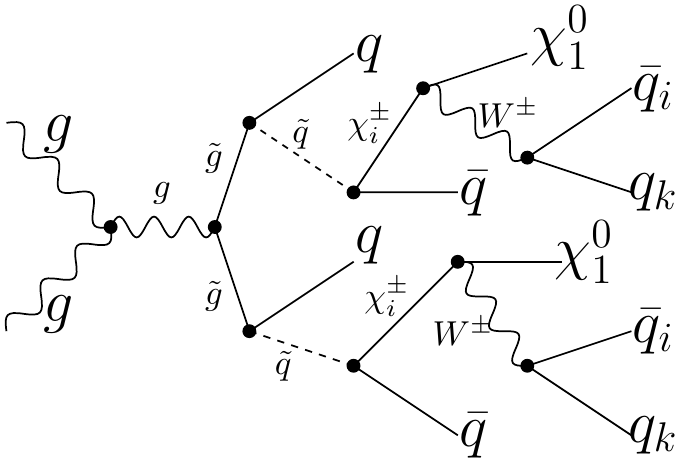}
& \vspace*{-25mm}
\begin{minipage}[t]{1.8cm}
$\begin{array}{c}
8j \\ \Big/ \hspace{-0.3cm E_T}
\end{array}$
\end{minipage} \\ \hline
\end{tabular}
\bigskip
\bigskip
\caption{Creation of the lightest chargino and the second
neutralino with further cascade decay (left). Creation of the
pair of gluinos with further cascade decay (right).}
\label{tab:n}
\end{table}

%-----------------------------------------------------------------
\subsection{Excluded region by direct searches for SUSY at the LHC}
\label{lhc}

The background from the SM processes results in the same final
states although with different kinematics. The missing energy
in this case is taken away by the light neutrinos. The
corresponding processes are shown in Table~\ref{tab:b1}.

\begin{table}[p]
\begin{tabular}{|c|p{1.7cm}|} \hline
Process & final  states \\
\hline
\includegraphics[width=50mm,height=35mm]{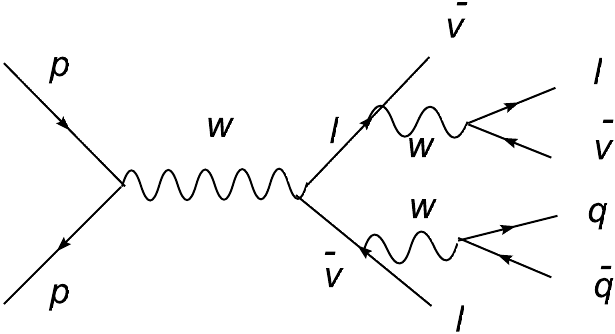}
& \vspace*{-30mm}
\begin{minipage}[t]{1.8cm}
$\begin{array}{c} \\
2\ell \\ 2 j \\ \Big/\hspace{-0.3cm E_T}
\end{array}$
\end{minipage} \\ \hline
\includegraphics[width=50mm,height=35mm]{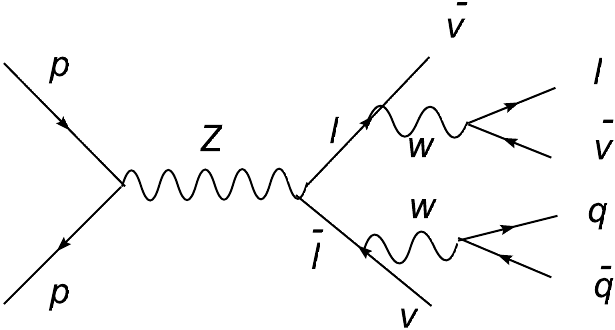}
& \vspace*{-35mm}
\begin{minipage}[t]{1.8cm}
$\begin{array}{c} \\ \\
\ell  \\ 2j \\ \Big/\hspace{-0.3cm E_T}
\end{array}$
\end{minipage} \\ \hline
\end{tabular}\bigskip
\hspace{1cm}
\begin{tabular}{|c|p{1.7cm}|} \hline
Process & final  states \\
\hline
\includegraphics[width=50mm,height=35mm]{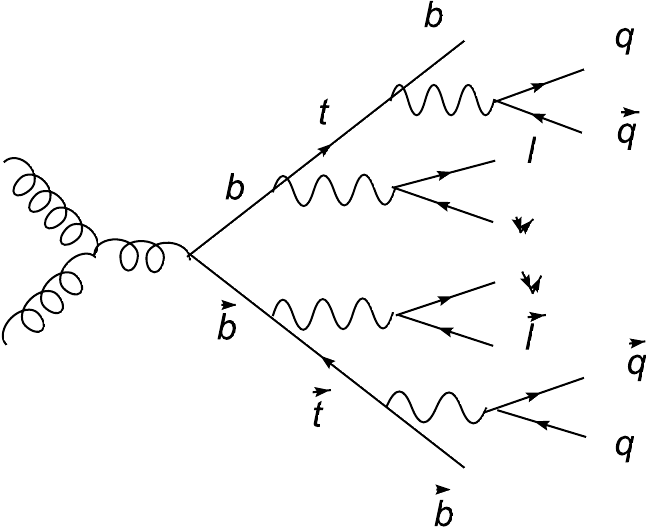}
& \vspace*{-30mm}
\begin{minipage}[t]{1.8cm}
$\begin{array}{c}\\
2\ell \\ 6j \\ \Big/\hspace{-0.3cm E_T}
\end{array}$
\end{minipage} \\ \hline
\includegraphics[width=50mm,height=35mm]{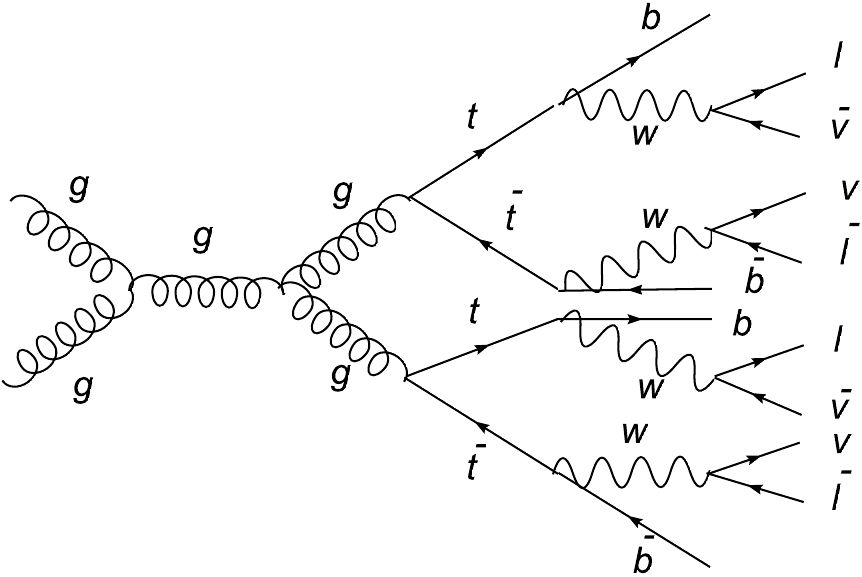}
& \vspace*{-35mm}
\begin{minipage}[t]{1.8cm}
$\begin{array}{c}\\
4\ell \\  4j \\ \Big/\hspace{-0.3cm E_T}
\end{array}$
\end{minipage} \\ \hline
\end{tabular}
\caption{The background at the hadron colliders: the weak
interaction processes (left), and the strong interaction
processes (right).}
\label{tab:b1}
\end{table}

\begin{figure}[p]
\begin{center}
\includegraphics[width=0.45\textwidth]{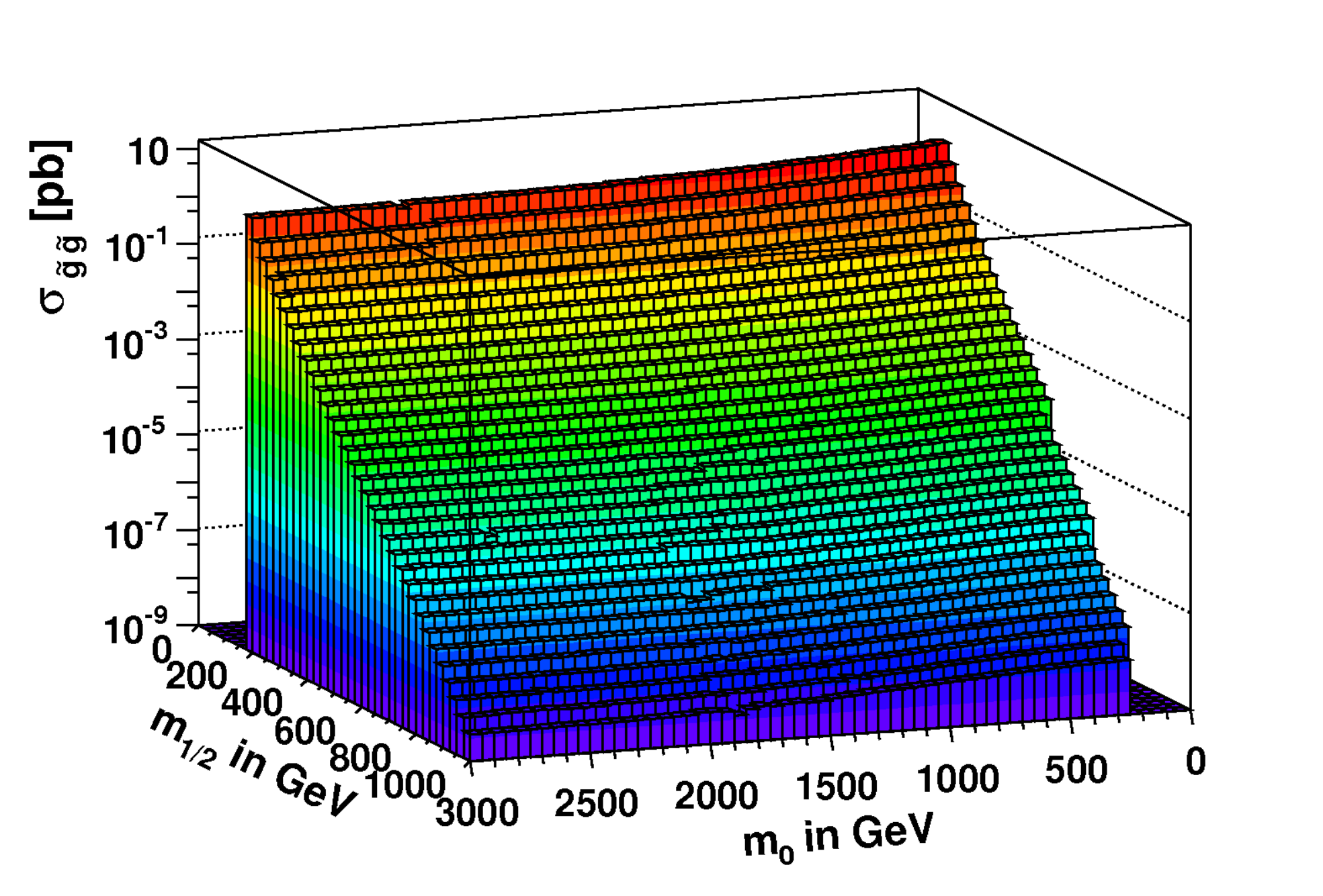}
\includegraphics[width=0.45\textwidth]{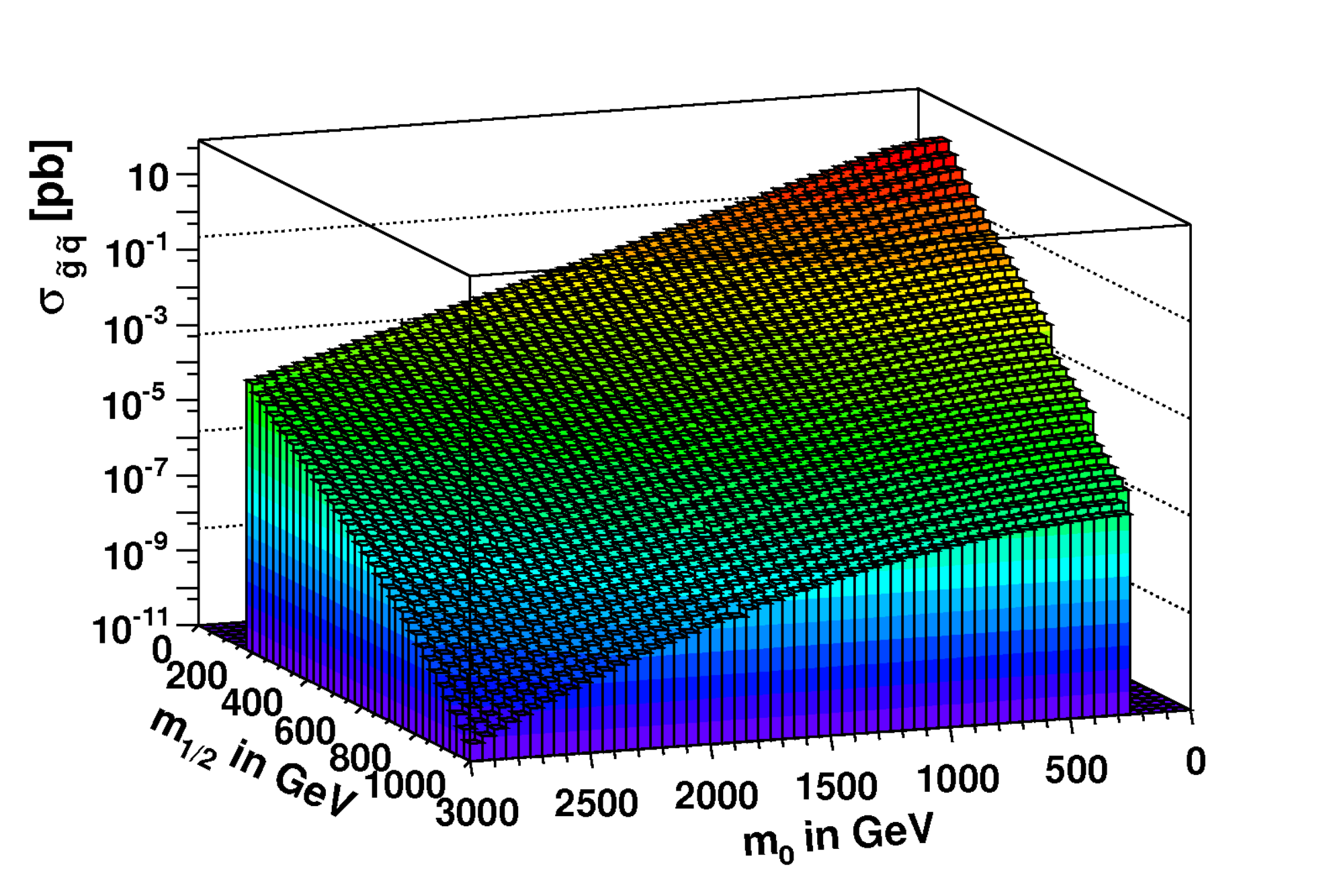}\\
\includegraphics[width=0.45\textwidth]{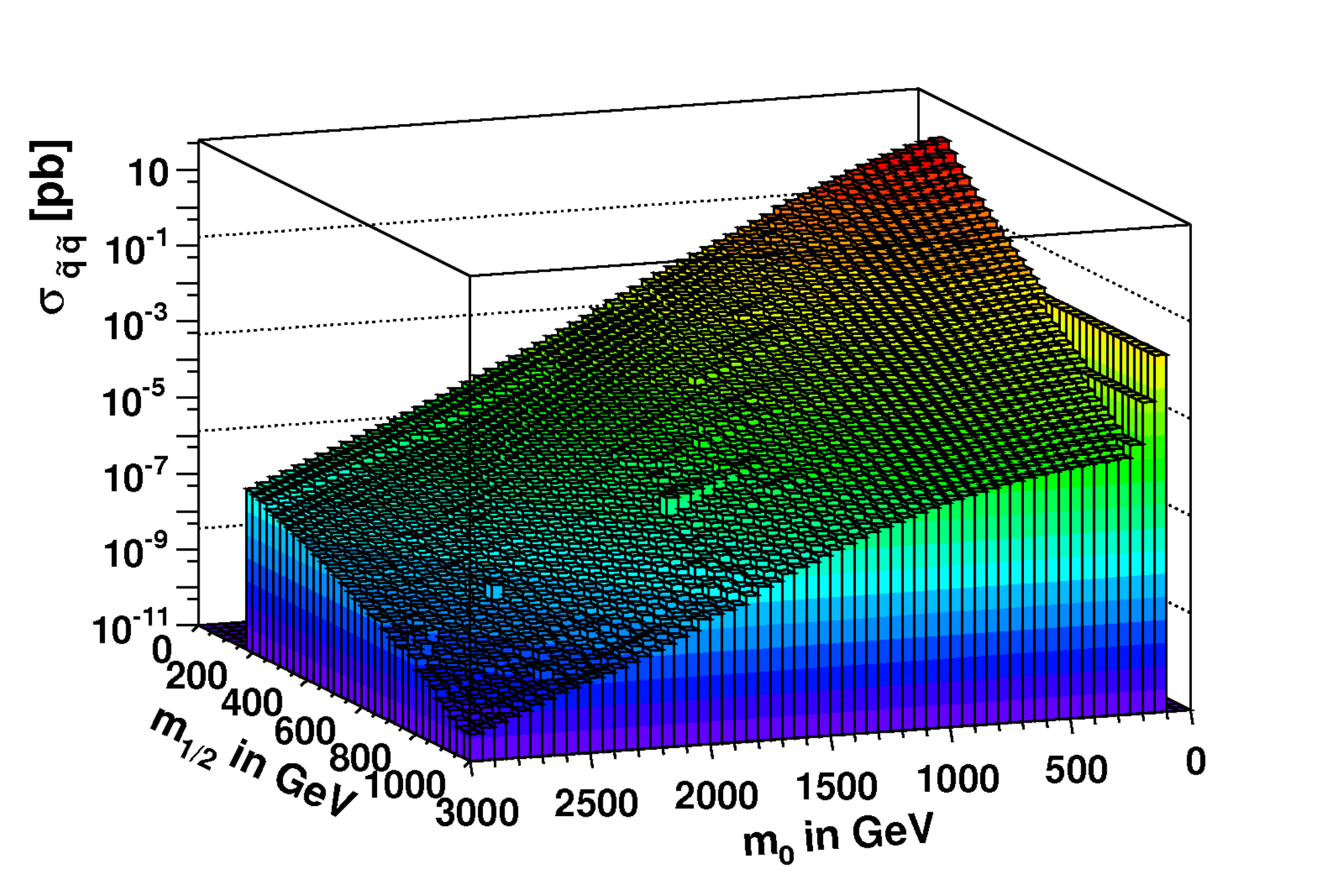}
\includegraphics[width=0.45\textwidth]{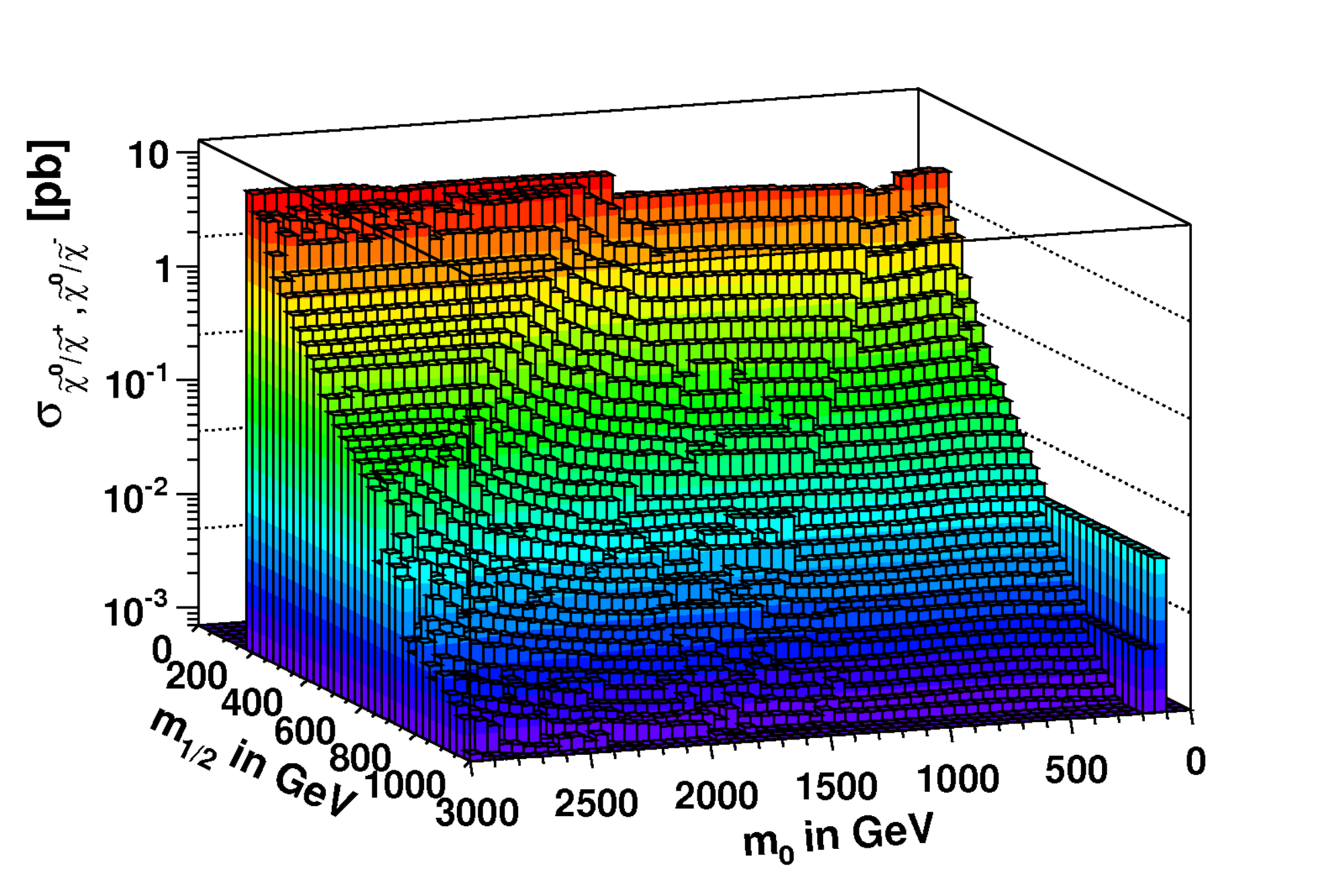}
\end{center}%\vspace{-0.2cm}
\caption{The cross-sections for the SUSY particles production
for the diagrams shown in Fig.~\ref{f1}: clockwise via the
strong interactions ($\gl\gl$, $\gl\sq$ and
$\sq\overline{\sq}$, respectively) and the electroweak
interactions.}
\label{f2}
\end{figure}

Numerous SUSY searches have been already performed at the
Tevatron.  The pair-produced squarks and gluinos have at least
two large-$E_T$ jets associated with the large missing energy.
The final state with the lepton(s) is possible due to the
leptonic decays of the $\tilde \chi^\pm_1$ and/or $\tilde
\chi^0_2$.

In the trilepton channel the Tevatron is sensitive up to
$m_{1/2} \leq 250$~GeV if $m_0\ \leq 200$~GeV and up to
$m_{1/2} \leq 200$~GeV if $m_0\ \geq 500$~GeV. The existing
limits on the squark and gluino masses at the Tevatron
are~\cite{tevatron}:\
$m_{\tilde q} \geq 300$~GeV, $m_{\tilde g} \geq 195$~GeV.

In the proton-proton collisions at the LHC the supersymmetric
particles can be produced according to the main diagrams shown
in the first three rows of Fig.~\ref{f1}, while the main
diagrams for the electroweak production are shown in the last
row. The corresponding cross-sections are shown in Fig.~\ref{f2}
for the centre-of-mass energy of 7~TeV. One observes that the
cross-section for the ``strong'' production of
$\sq\overline{\sq}$ and  $\gl\sq$ are large for the low values
of $m_0$ and $m_{1/2}$, the gluino production $\gl\gl$ is the
strongest at the small values of $m_{1/2}$ and the electroweak
production of gauginos starts to increase at the large values
of $m_0$. The reason for the increase of the electroweak
production at large $m_0$ is the decrease of the Higgs mixing
parameter $\mu$, as determined from the EWSB, which leads to
stronger mixing of the Higgsino component in the gauginos and
so the coupling to the weak gauge bosons and Higgs bosons
increases, thus increasing the amplitudes for the diagrams in
the last row of Fig.~\ref{f1}.

The ``strong'' production cross sections are characterized by
a large number of jets from the long decay chains and the
missing energy from the escaping neutralino. These
characteristics can be used to efficiently suppress the
background. For the electroweak production, both the number
of jets and the missing transverse energy is low, since the
LSP is not boosted so strongly as in the decay of the heavier
strongly interacting particles. Hence, the electroweak gaugino
production needs the leptonic decays to reduce the background,
so these signatures need more luminosity and cannot compete at
present with the sensitivity of the ``strong'' production of the
squarks and gluinos.

\begin{figure}[t]
\begin{center}
\includegraphics[width=0.48\textwidth]{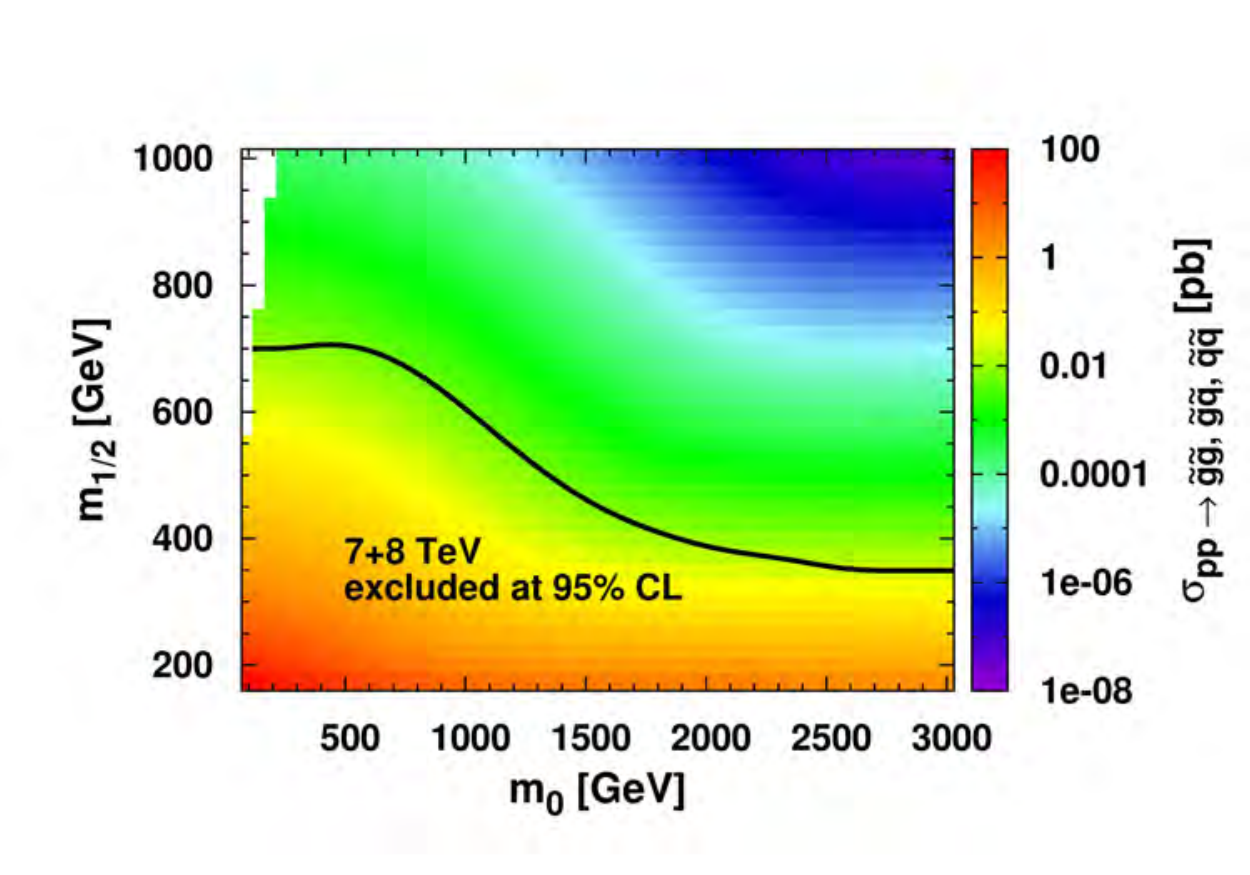}
\includegraphics[width=0.48\textwidth]{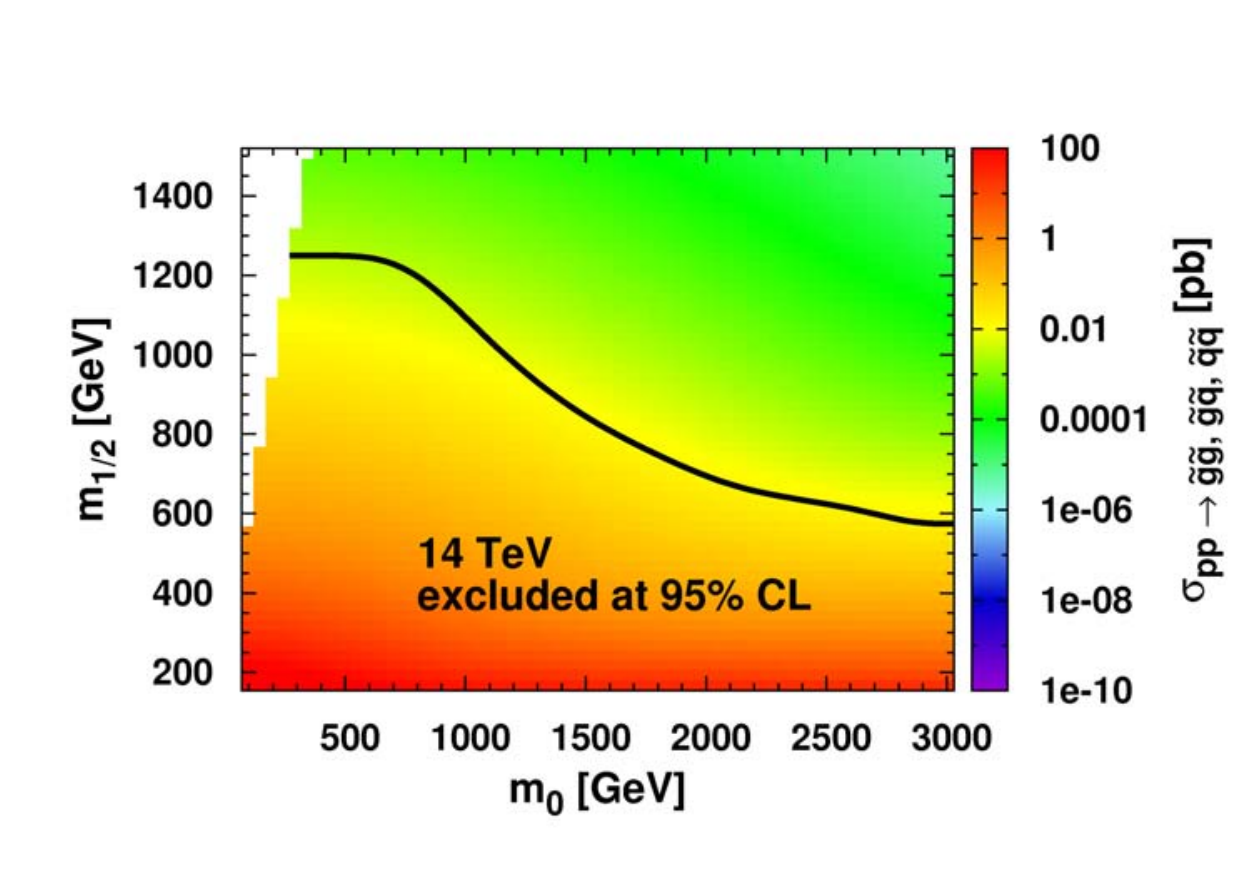}
\vspace{-5mm}
\end{center}
\caption{Left: The total production cross-section of the
strongly interacting particles in comparison with the LHC
excluded limits for 7+8~TeV. Here the data from ATLAS and CMS
were combined and correspond to the integrated luminosity of
1.3 and 1.1~fb$^{-1}$, respectively. One observes that the
cross-section of 0.1 to 0.2~pb is excluded at 95\% CL.
Right: the cross~sections at 14~TeV and expected
exclusion for the same limit on the cross-section as at 7~TeV.}
\label{f4}
\end{figure}
\begin{figure}[t]
\begin{center}
\includegraphics[width=0.46\textwidth]{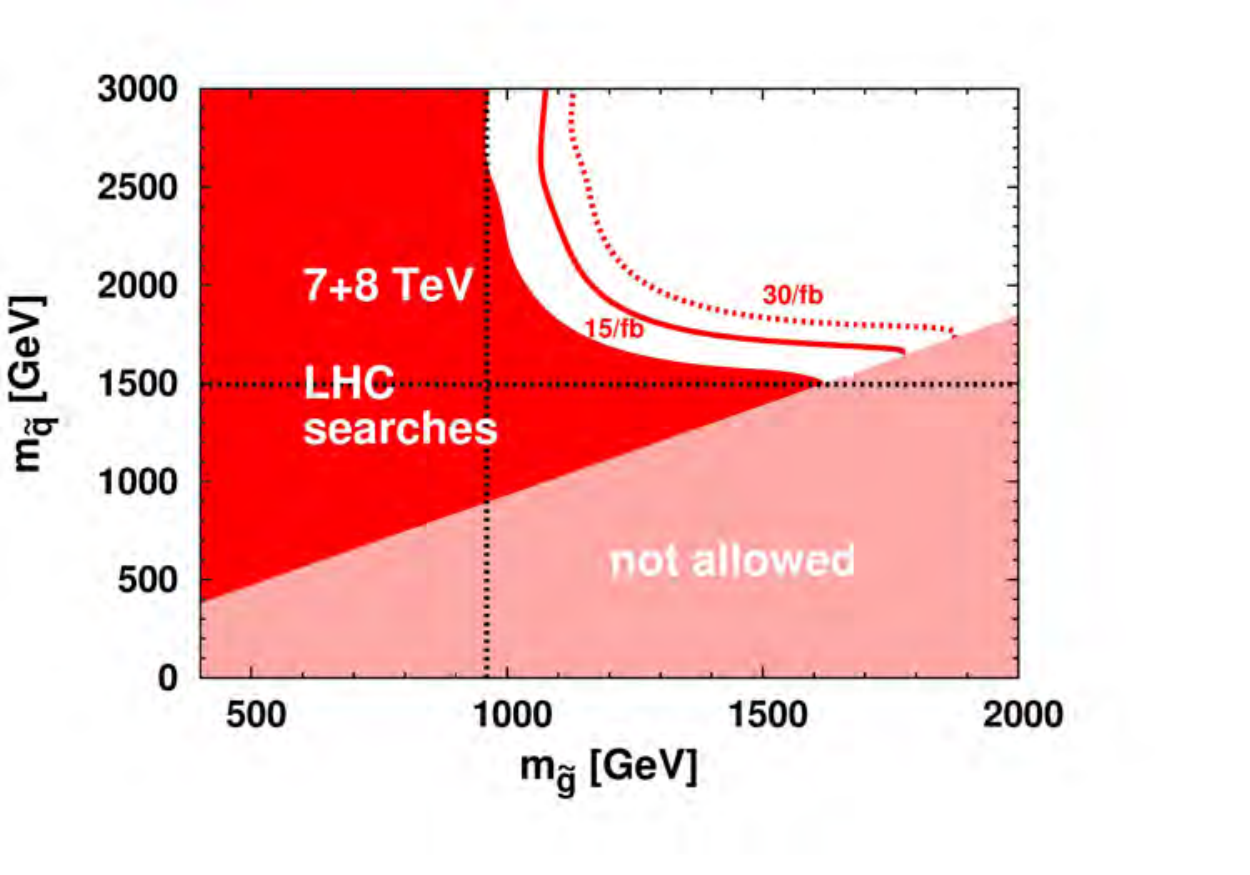}
\raisebox{-5pt}{\includegraphics[width=0.49\textwidth]{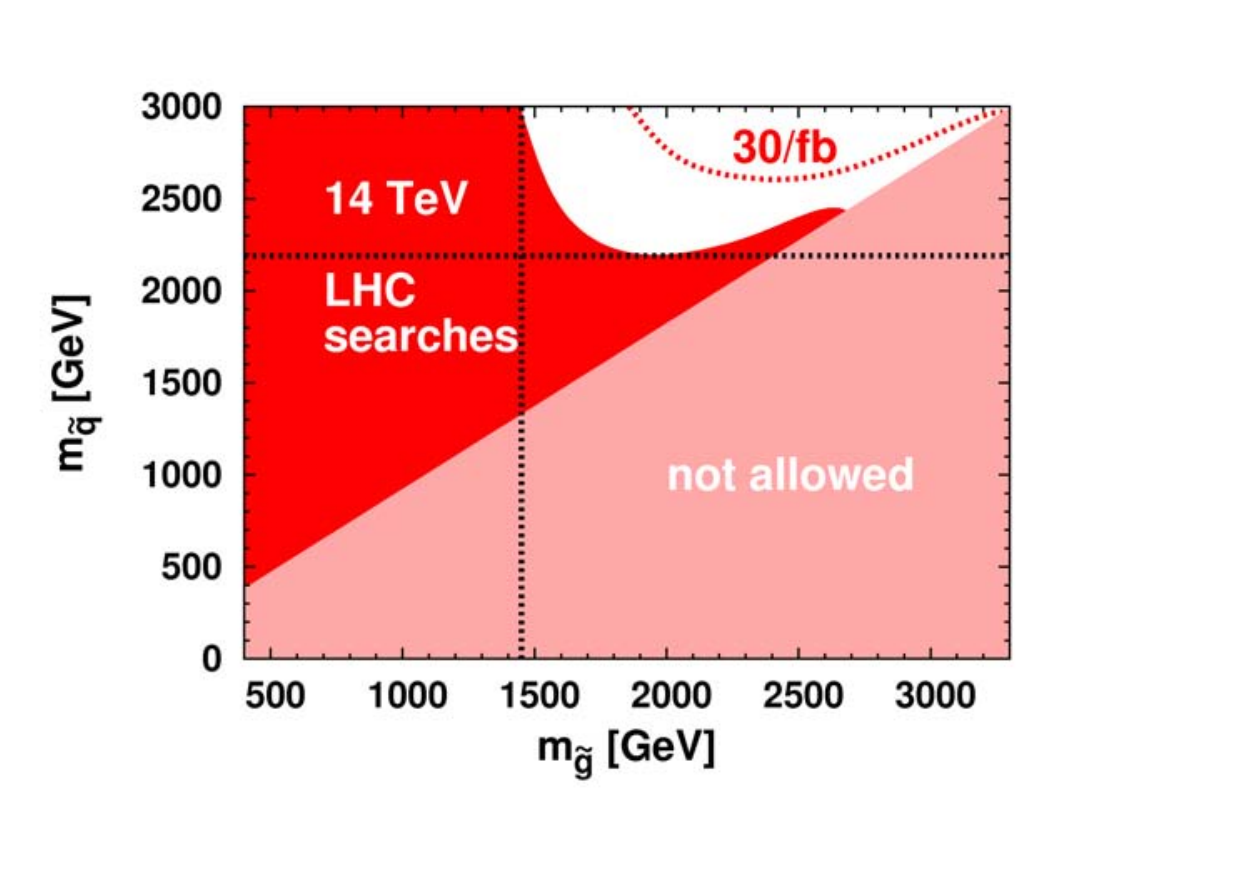}}
\vspace*{-5mm}
\end{center}
\caption{As in Fig.~\ref{f4}, but the excluded region is
translated into the $m_{\tilde q}, m_{\tilde g}$ plane. The
red area corresponds to the excluded regions for the integrated
luminosity slightly above 1~fb$^{-1}$; the expectations for the
higher luminosities have been indicated as well.}
\label{f5}
\end{figure}

The total cross-section for the strongly interacting particles
are shown in Fig.~\ref{f4} together with the excluded region
from the direct searches for SUSY particles at the LHC. One
observes that the excluded region (below the solid line) follows
rather closely the total cross-section, indicated by the colour
shading. From the colour coding one observes that the excluded
region corresponds to the cross-section limit of about
$0.1-0.2$~pb.

The drop of the excluded region at large values of $m_0$ is due
to the fact that in this region the squarks become heavy, which
means that the contributions from the diagrams in the second
and third rows of Fig.~\ref{f1} start to diminish. Here only the
higher energies will help and doubling the LHC energy from 7 to
14~TeV, as planned in the coming years, quickly increases the
cross-section for the gluino production by orders of magnitude,
as shown in the right panel of Fig.~\ref{f4}. The expected
sensitivity at 14~TeV, plotted as the exclusion contour in case
nothing is found, assumes the same efficiency and luminosity
(slightly above one fb$^{-1}$ per experiment) as at 7~TeV.

These limits can be translated to the squark and gluino masses
as follows. The squark masses have a starting value at the GUT
scale equal to $m_0$, but have important contributions from the
gluinos in the colour field, so the squark masses are given by
$m_{\sq}^2 \approx m_0^2+6.6m_{1/2}^2$, as was determined from
the renormalization group equations~\cite{BEK}.
Similarly the gluino mass is given by 2.7$m_{1/2}$. The term
proportional to $m_{1/2}$ in the squark mass corresponds to the
self-energy diagrams, which imply that if the "gluino-cloud" is
heavy, the squarks cannot be light. This leads to the regions
indicated as not allowed ones in Fig.~\ref{f5}. Note that these
regions are forbidden in any model with the coupling between the
squarks and gluinos, so they are not specific to the CMSSM.
The squark masses below 1.1~TeV and the gluino masses below
0.62~TeV are excluded for the LHC data at 7~TeV, as shown in
the left panel of Fig.~\ref{f5}. Expected sensitivities for the
higher integrated luminosities at 7 and 14~TeV have been
indicated as well. One observes that increasing the energy is
much more effective than increasing the luminosity. At 14~TeV
the squarks with masses of 1.7~TeV and gluinos with masses of
1.02~TeV are within reach with 1~fb$^{-1}$ per experiment,
as shown in the right panel of Fig.~\ref{f5}.

%----------------------------------------------------------
\subsection{Excluded region for combination of constraints}

\begin{figure}[htb]
\begin{center}
\includegraphics[width=0.49\textwidth]{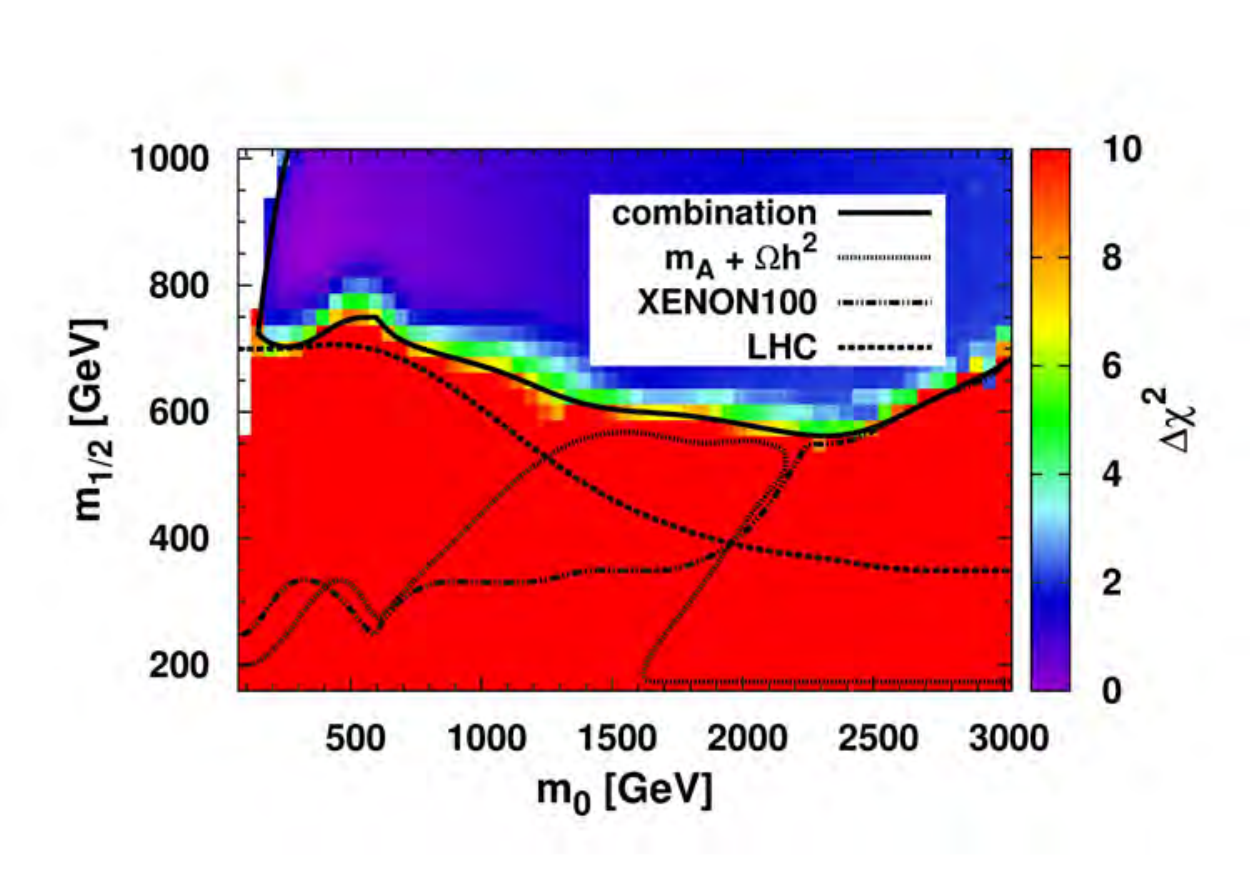}
\includegraphics[width=0.49\textwidth]{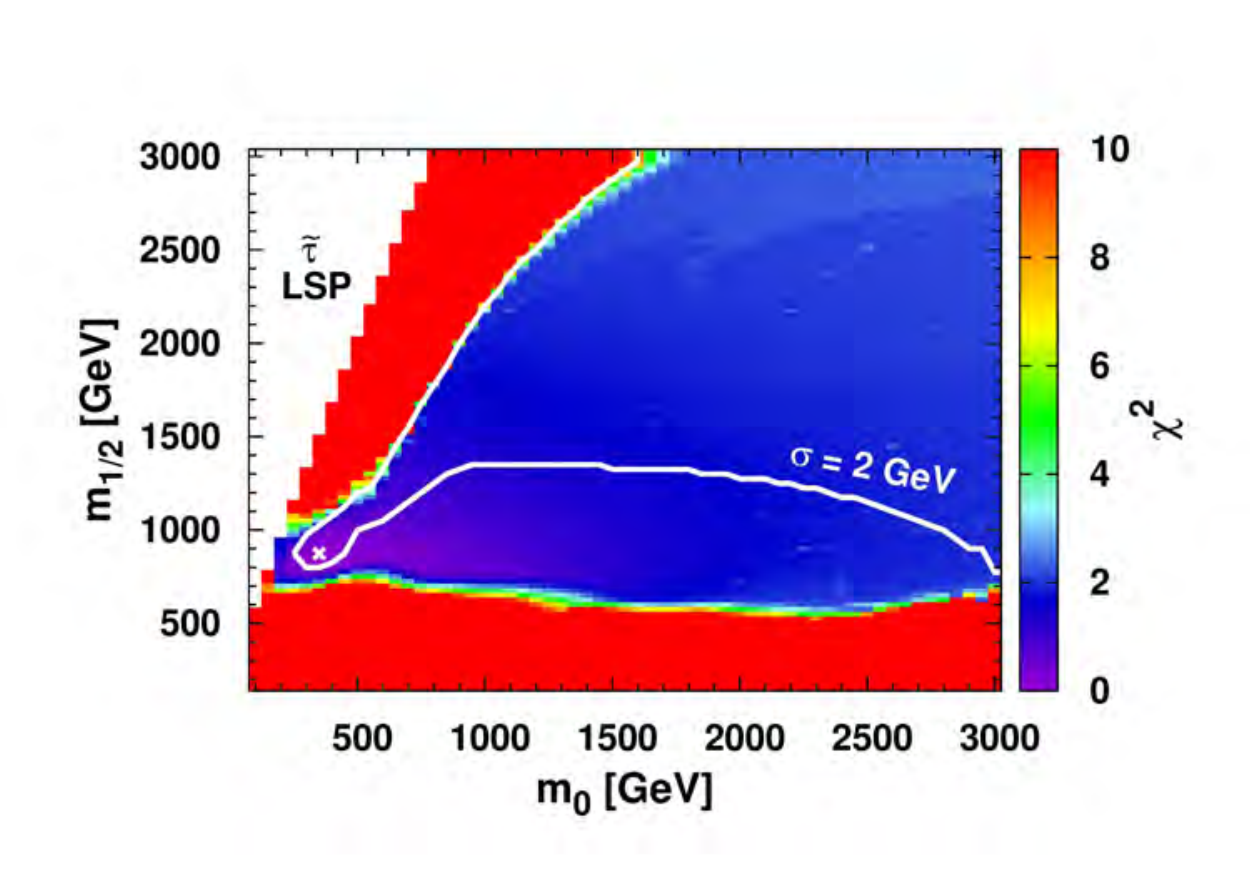}
\end{center}\vspace{-10mm}
\caption{Left: Combined constraints from the LHC searches, the
relic density from WMAP, the direct DM searches from XENON100,
limits on the pseudo-scalar Higgs mass and $g-2$ of muon (without
the 125~GeV Higgs boson mass constraint). Right: The account of
the 125~GeV Higgs boson mass constraint with 2 GeV mass uncertainty.
The region below the white line is excluded at 95\% C.L.}
\label{comb}
\end{figure}

If one combines the excluded regions from the direct searches at
the LHC, the relic density from the WMAP, the already stringent
limits on the pseudo-scalar Higgs mass with the XENON100 limits
one obtains the excluded region shown in the left panel of
Fig.~\ref{comb}. Here the $g-2$ limit is included with the
conservative linear addition of theoretical and experimental
errors. One observes that the combination excludes $m_{1/2}$
below 525 GeV in the CMSSM for  $m_0<1500$ GeV, which implies the
lower limit on the WIMP mass of 230~GeV and a gluino mass of
1370 GeV, respectively.

As discussed earlier, the LHC becomes rather insensitive to the
large $m_0$ region because of the decreasing cross-section for
the production of strongly interaction particles and the large
background for the production of gauginos. However, in this
region one obtains the increased sensitivity above the LHC
limits from the relic density and the direct DM searches.

If a Higgs mass of the lightest Higgs boson of 125~GeV is
imposed, the preferred region is well above this excluded
region, but the size of the preferred region is strongly
dependent on the size of the assumed theoretical uncertainty
as was shown in Fig.~\ref{f5_100b}. Accepting the 2~GeV
uncertainty we get the excluded region shown in
Fig.~\ref{comb} (right panel), which is far above the existing
LHC limits and leads to strongly interacting superpartners above
2~TeV. However, in models with an extended Higgs sector, like
NMSSM~\cite{NMSSM}, a Higgs mass of 125 GeV can be obtained for
lower values of $m_{1/2}$, in which case the regions excluded
in the MSSM become viable.

%-----------------------------
\section{The reach of the LHC}

%--------------------------
\subsection{LHC luminosity}

\begin{figure}[ht]
\begin{center}
\leavevmode
\includegraphics[width=0.49\textwidth]{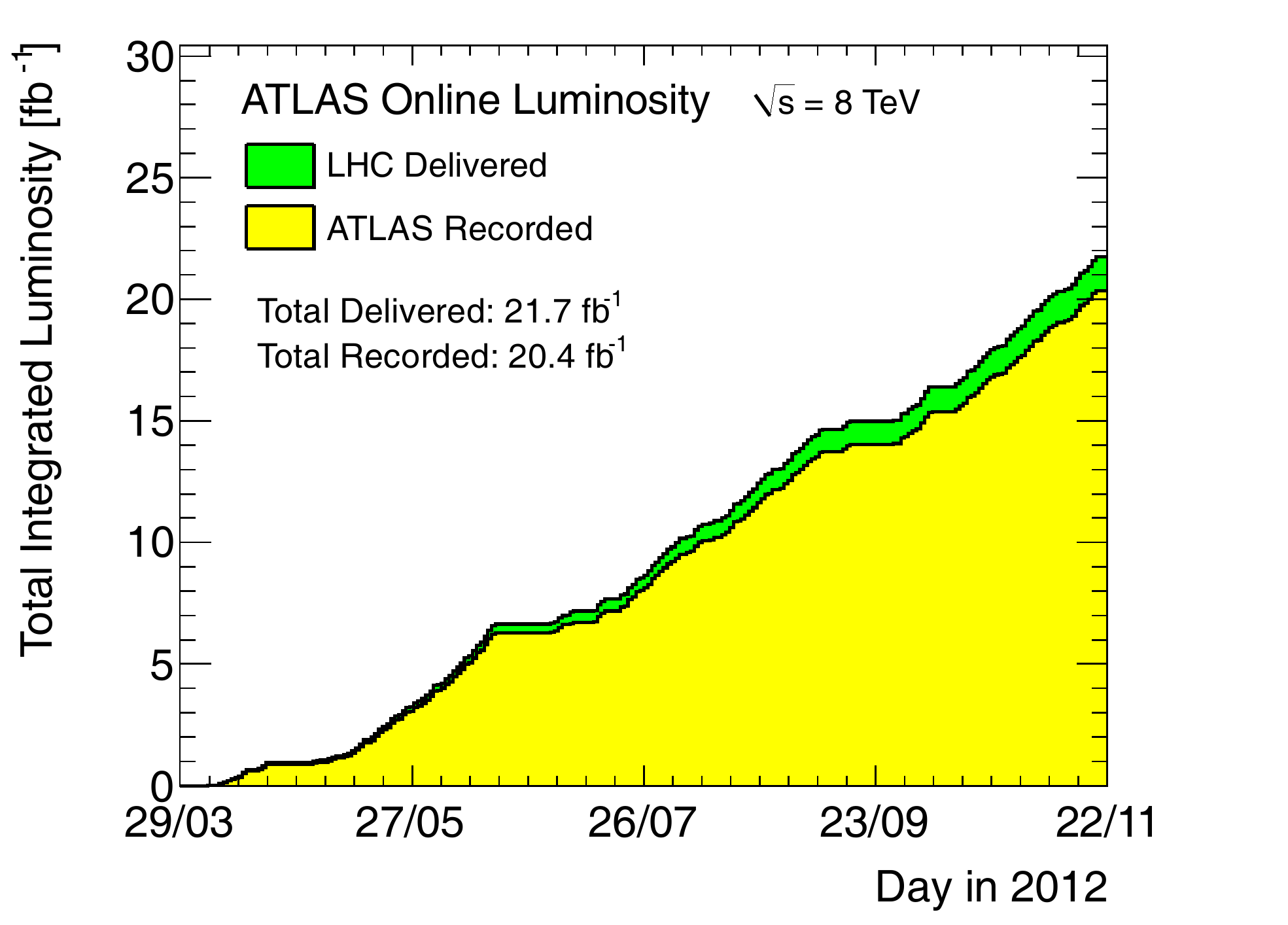}
\includegraphics[width=0.49\textwidth]{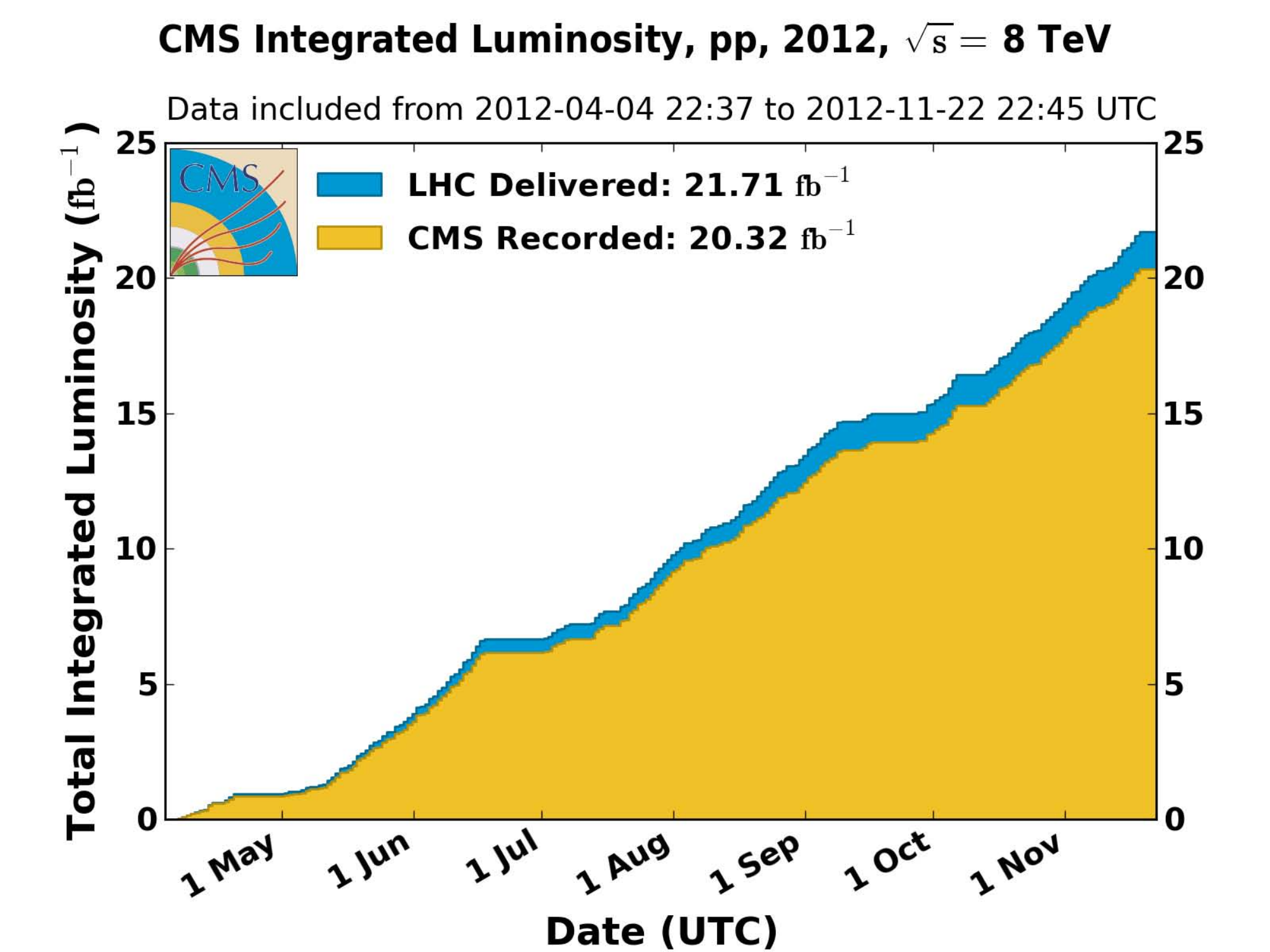}
\end{center}
\caption{Cumulative luminosity versus day delivered by LHC, and
recorded by ATLAS (left) and CMS (right) experiments for $pp$
collisions at $\sqrt s$ = 8~TeV in 2012 from counting rates
measured by the luminosity detectors.}
\label{fig:LHC_lumi2012}
\end{figure}

The Large Hadron Collider is the unique machine for the new
physics searches at the TeV scale. Its c.m. energy is planned to
be 14~TeV with very high luminosity up to a few
hundred~fb$^{-1}$. At the moment the total integrated luminosity
in 2012 is already more than 20~fb$^{-1}$. Fig.~\ref{fig:LHC_lumi2012}
shows the luminosity delivered in 2012 in $pp$ collisions at the
center-of mass energy of 8~TeV and recorded by
ATLAS~\cite{atlas_lumi} and CMS~\cite{cms_lumi} experiments.
\vspace{-1cm}

%---------------------------------------
\subsection{Expected LHC reach for SUSY}

The LHC is supposed to cover a wide range of parameters of the
MSSM (see Figs. below) and discover the superpartners with the
masses below 2~TeV. This will be a crucial test for
the MSSM and the low energy supersymmetry. The LHC potential to
discover supersymmetry is widely discussed in the
literature~\cite{LHCSUSY1,LHCSUSY2}.

\begin{figure}[htb]
\begin{center}
\includegraphics[width=0.50\textwidth]{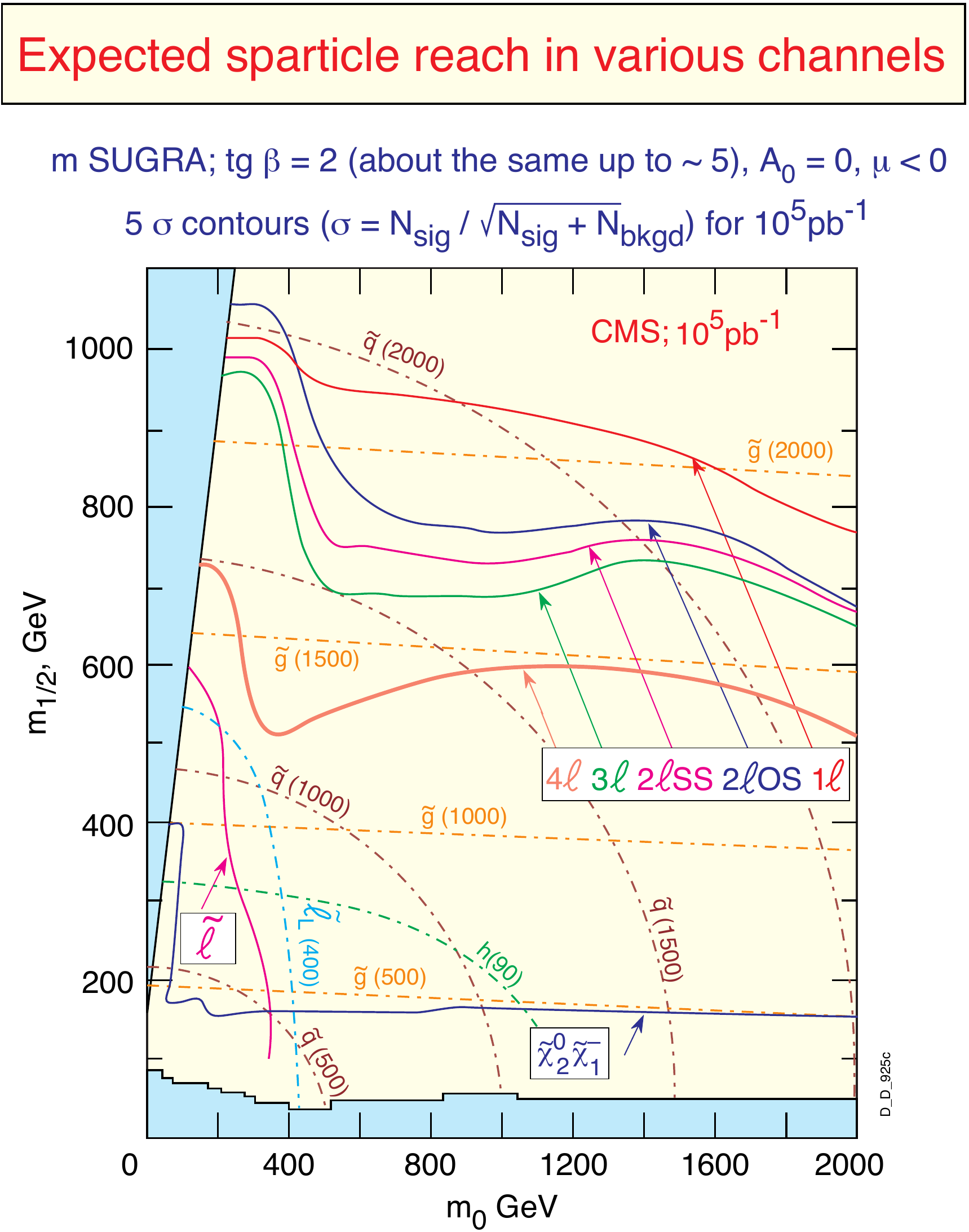}
\hspace{0.04\textwidth}
\includegraphics[width=0.44\textwidth]{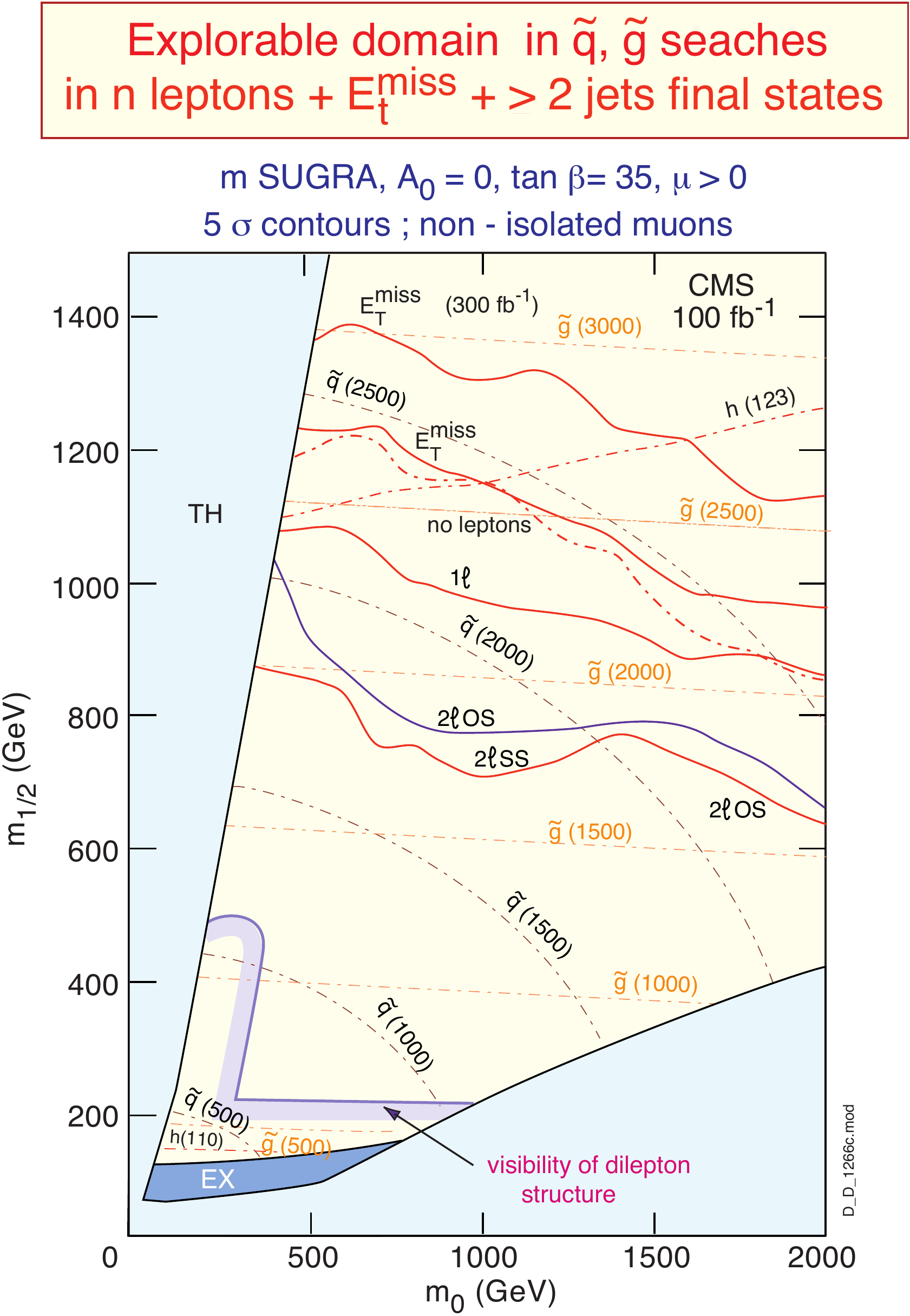}
\end{center}
%\vspace{-1.0cm}
\caption{The expected range of reach for the superpartners in
various channels at the LHC(left) and the expected
domain of searches for the squarks and gluions at the
(right)~\cite{LHC}.}
\label{LHCreach}
\end{figure}

To present the region of reach for the LHC in different
channels of sparticle production it is useful to take the same
plane of soft SUSY breaking parameters $m_0$ and $m_{1/2}$. In
this case one usually assumes certain luminosity which will be
presumably achieved during the accelerator operation. Thus, for
instance, in Fig.~\ref{LHCreach} the regions of reach in
different channels are shown. The lines of the constant squark
mass form the arch curves, and those for the gluino are almost
horizontal. The curved lines show the reach bounds in different
channel of creation of the secondary particles. The theoretical
curves are obtained within the MSSM for a certain choice of the
other soft SUSY breaking parameters. As one can see, for the
fortunate circumstances the wide range of the parameter space up
to the masses of the order of 2~TeV will be examined. The LHC
will be able to discover SUSY with the squark and gluino masses
up to $2\div 2.5$ TeV for the luminosity $L_{tot}=100$~fb$^{-1}$.
The most powerful signature for the squark and gluino detection
are the multijet events; however, the discovery potential
depends on the relation between the LSP, squark, and gluino
masses, and decreases with the increase of the LSP mass.
The same is true for the sleptons. The typical signal used for
the slepton detection is the dilepton pair with the missing
energy without hadron jets. For the luminosity of
$L_{tot}=100$~fb$^{-1}$ the LHC will be able to discover
sleptons with the masses up to 400~GeV~\cite{LHCSUSY1,LHCSUSY2}.

%-------------------------------------------
\subsection{Recent results on SUSY searches}

Direct searches of the superpartners at the LHC in different
channels have pushed the lower limits on their masses, mainly
of the gluinos and the squarks of the light two generations,
upwards to the TeV range. On the other hand, for the third
generation the limits are rather weak and the masses around a
few hundred~GeV are still allowed. The light third generation
squarks are also consistent with the recent observation of the
Higgs-like boson with the mass around 125~GeV.

\begin{figure}[tb]
\begin{center}
\leavevmode
\includegraphics[width=0.49\textwidth]{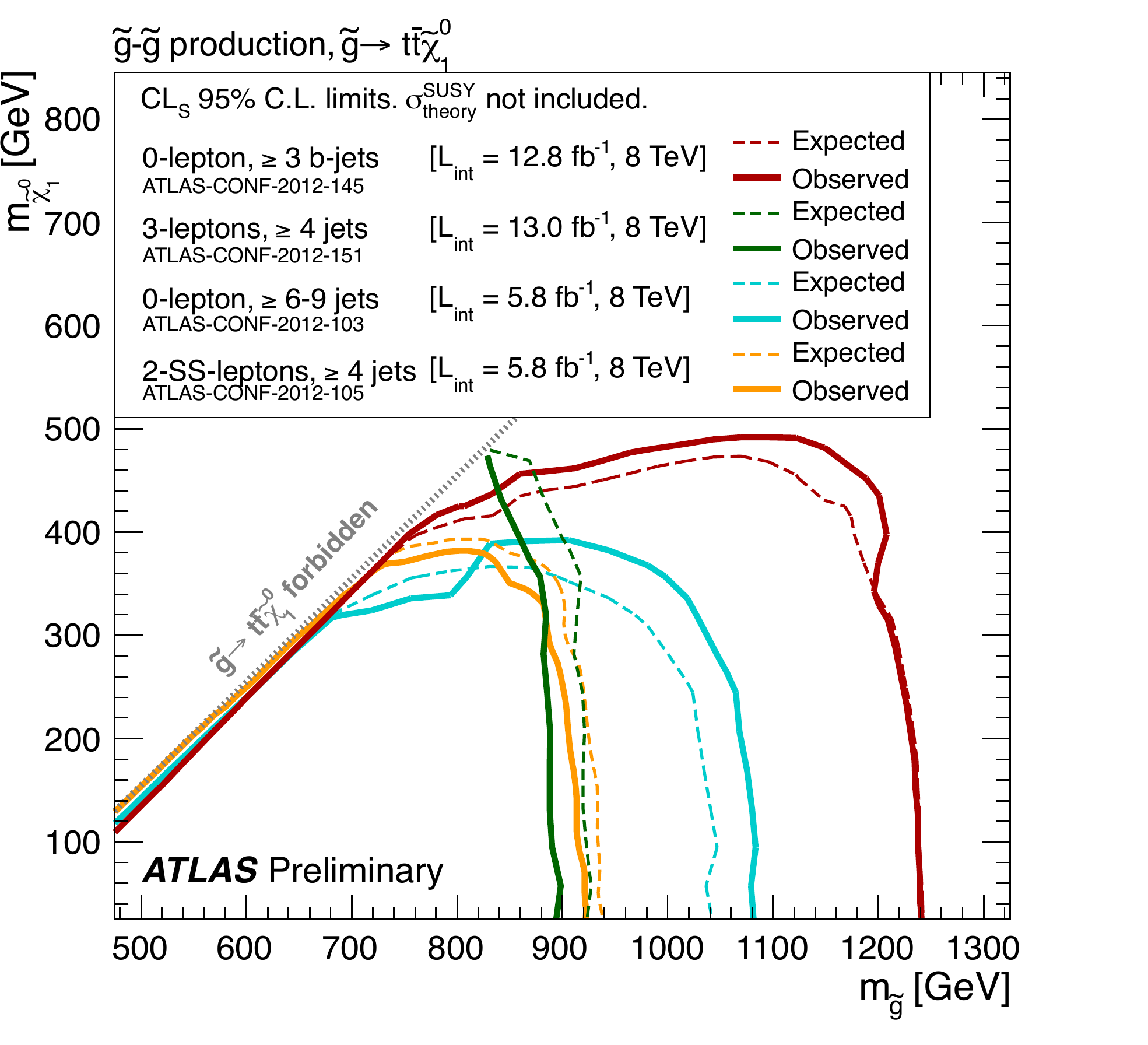}
\includegraphics[width=0.49\textwidth]{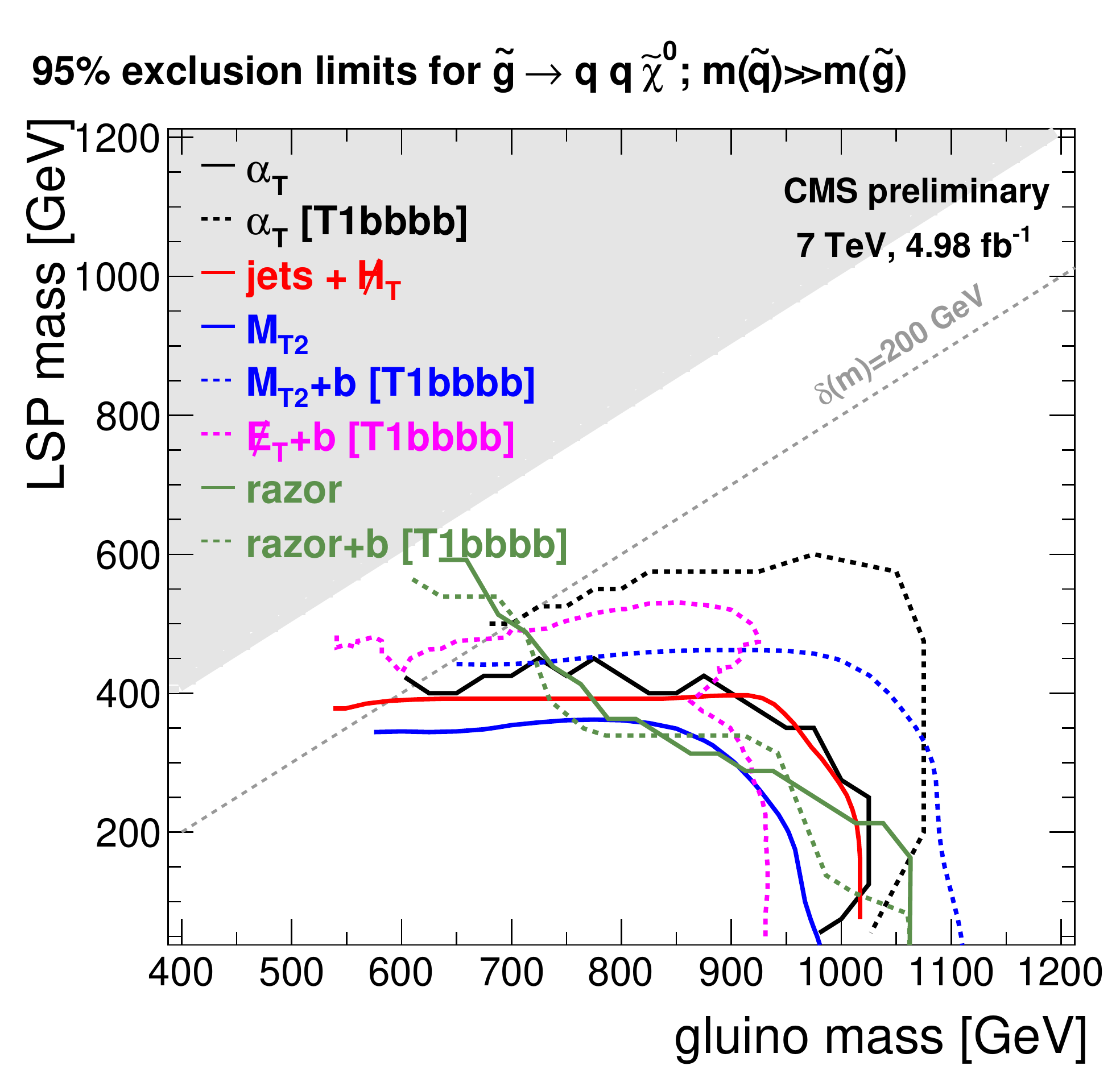}
\end{center}
\caption{Left: ATLAS exclusion limits at 95\%~CL for 8~TeV
analyses in the $m_{\tilde g}-m_{\tilde\chi^0_1}$ plane for the
\emph{Gtt} simplified model where a pair of gluinos decays via
off-shell stop to four top quarks and two neutralinos (LSP).
Right: CMS exclusion limits at 95\%~CL for 7~TeV analyses of
gluino decay $\tilde g \to qq \tilde\chi^0_1$, assuming
$m_{\tilde q} >> m_{\tilde g}$.}
\label{ATLAS_SUSY_Gtt_susy12}
\end{figure}

We present here examples on the superparticle searches in
various scenarios depicted as exclusion plots. Everywhere in
these plots the excluded region is the one below the
corresponding curve (lower masses, lower values of parameters).

The fisrt example is the gluino pair production
$pp \to \tilde g \tilde g$ and
$\tilde g \to t \bar t \tilde\chi^0_1$ decay in the so-called
$Gtt$ simplified model. Four different final states (0 leptons
with $\ge$ 3 $b$-jets~\cite{ATLAS_SUSY_Gtt_1}; 3 leptons with
$\ge$ 4 jets~\cite{ATLAS_SUSY_Gtt_2}; 0 leptons with $\ge$ 6-9
jets~\cite{ATLAS_SUSY_Gtt_3}; and a pair of the same-sign
leptons with more than 4 jets~\cite{ATLAS_SUSY_Gtt_4}) are
considered. The first two analysis performed using
12.8~fb$^{-1}$ and 13.0~fb$^{-1}$ data and the last two ones
using 5.8~fb$^{-1}$ data. The results slightly differ
quantitatively, however, the conclusion is the non-observation
of the gluino lighter than 900~GeV (conservative limit) or even
1200~GeV for the lightest neutralino mass less than around
300~GeV.

\begin{figure}[ht]
\begin{center}
\leavevmode
\includegraphics[width=0.33\textwidth]{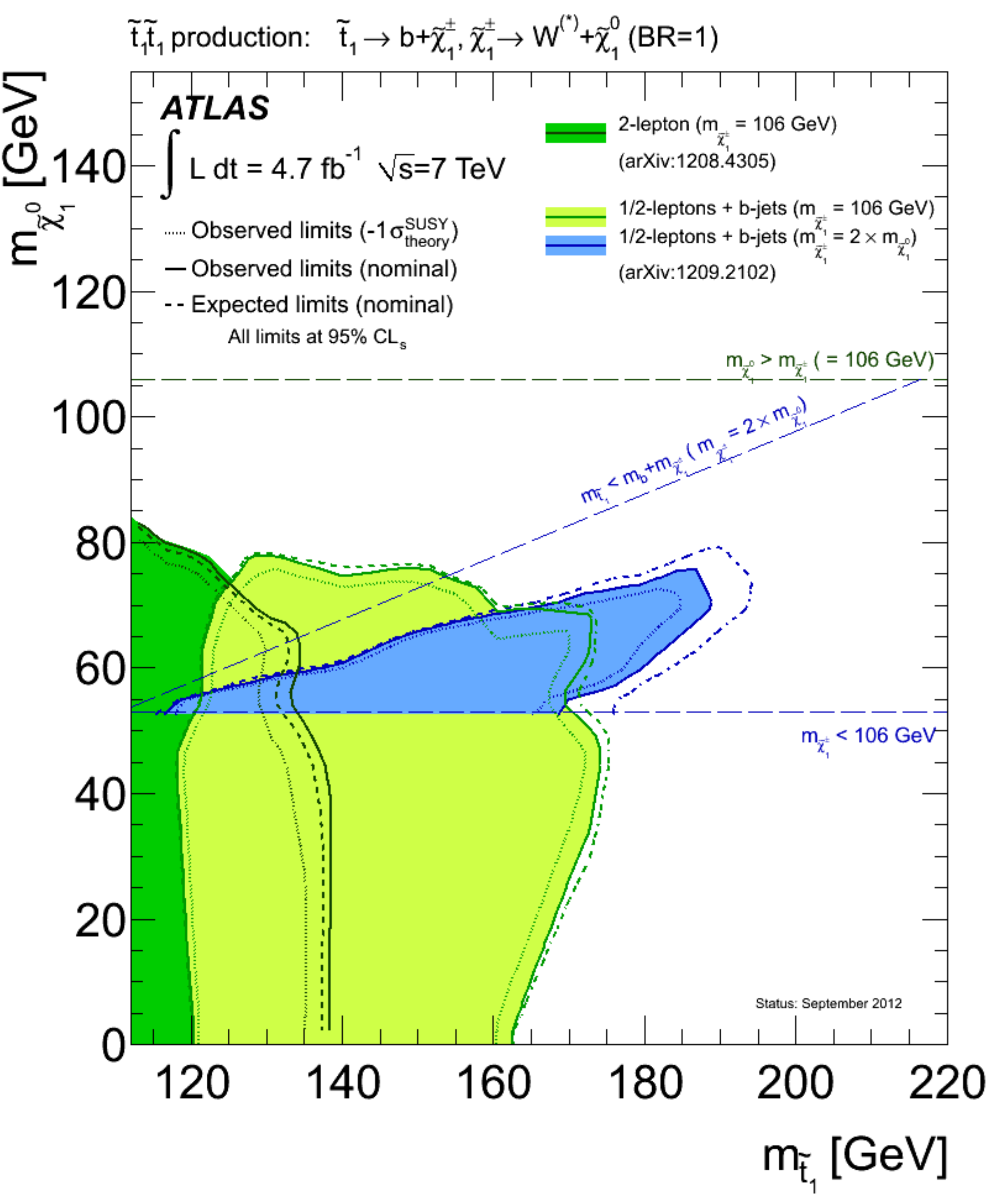}
\includegraphics[width=0.66\textwidth]{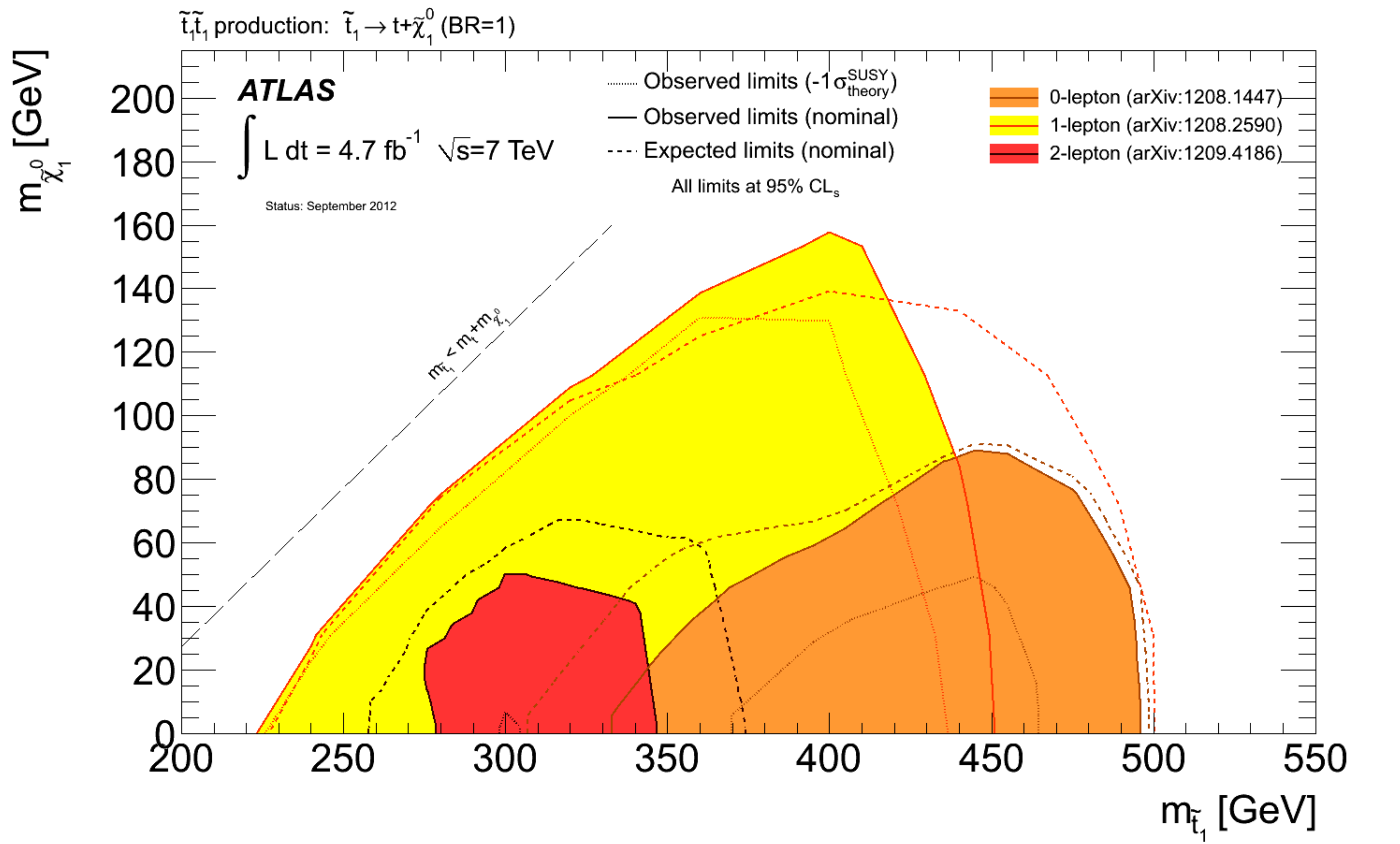}
\end{center}
\caption{Summary of the five dedicated ATLAS searches for the
top-squark pair production based on 4.7~fb$^{-1}$ of the $pp$
collision data taken at $\sqrt{s}$ = 7 TeV.}
\label{fig:ATLAS_directstop}
\end{figure}

Another example is the result of searches of the top-squark pair
production by ATLAS collaboration based on 4.7~$fb^{-1}$ of $pp$
collision data taken at $\sqrt S$ = 7~TeV. The exclusion
limits at 95\%~CL are shown in the $\tilde t_1 - \tilde\chi^0_1$
mass plane. The dashed and solid lines show the expected and
observed limits, respectively, including all uncertainties
except the theoretical signal cross-section uncertainty (PDF and
scale). The dotted lines represent the results obtained when
reducing the nominal signal cross-section by $1\sigma$ of its
theoretical uncertainty. Depending on the stop mass there can be
two different decay channels. For relatively light stops with
masses below 200~GeV, the decay $\tilde t_1 \to b +
\tilde\chi^\pm_1$, $\tilde\chi^\pm_1 \to W^* + \tilde\chi^0_1$
is assumed in all the cases, with two hypotheses on the
$\tilde\chi^\pm_1$, $\tilde\chi^0_1$ mass hierarchy,
$m(\tilde\chi^\pm_1) = 106$~GeV and $m(\tilde\chi^\pm_1) =
2 m(\tilde\chi^0_1)$~\cite{ATLAS_stop1,ATLAS_stop2},
see the left panel of Fig.~\ref{fig:ATLAS_directstop}.
For the heavy stop masses above 200~GeV, the decay $\tilde t_1
\to t + \tilde\chi^0_1$ is assumed to dominate, the excluded
regions are shown in the right panel of
Fig.~\ref{fig:ATLAS_directstop}~\cite{ATLAS_stop1,ATLAS_stop2,ATLAS_stop3}.

\begin{figure}[ht]
\begin{center}
\leavevmode
\includegraphics[width=0.48\textwidth]{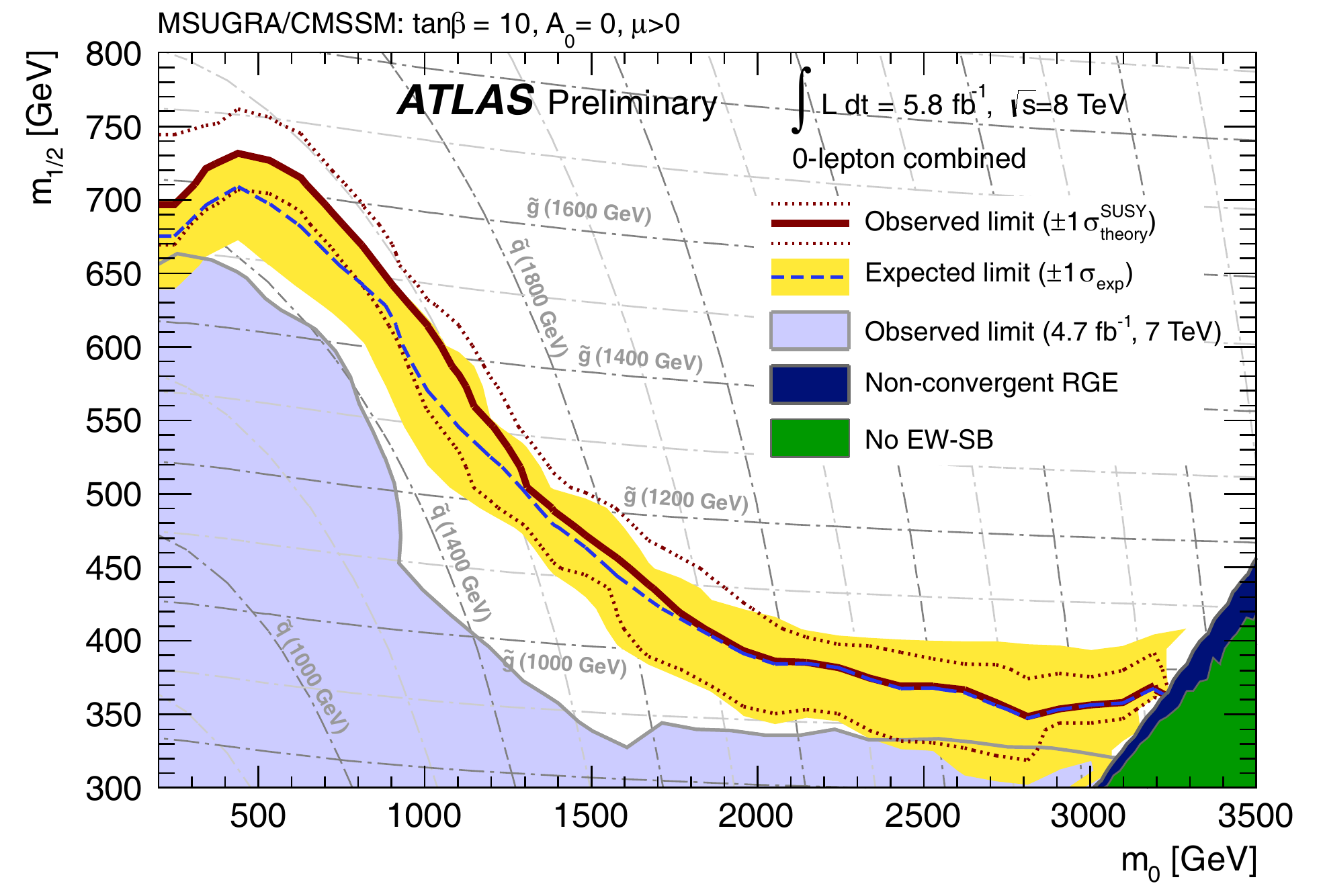}
\includegraphics[width=0.48\textwidth]{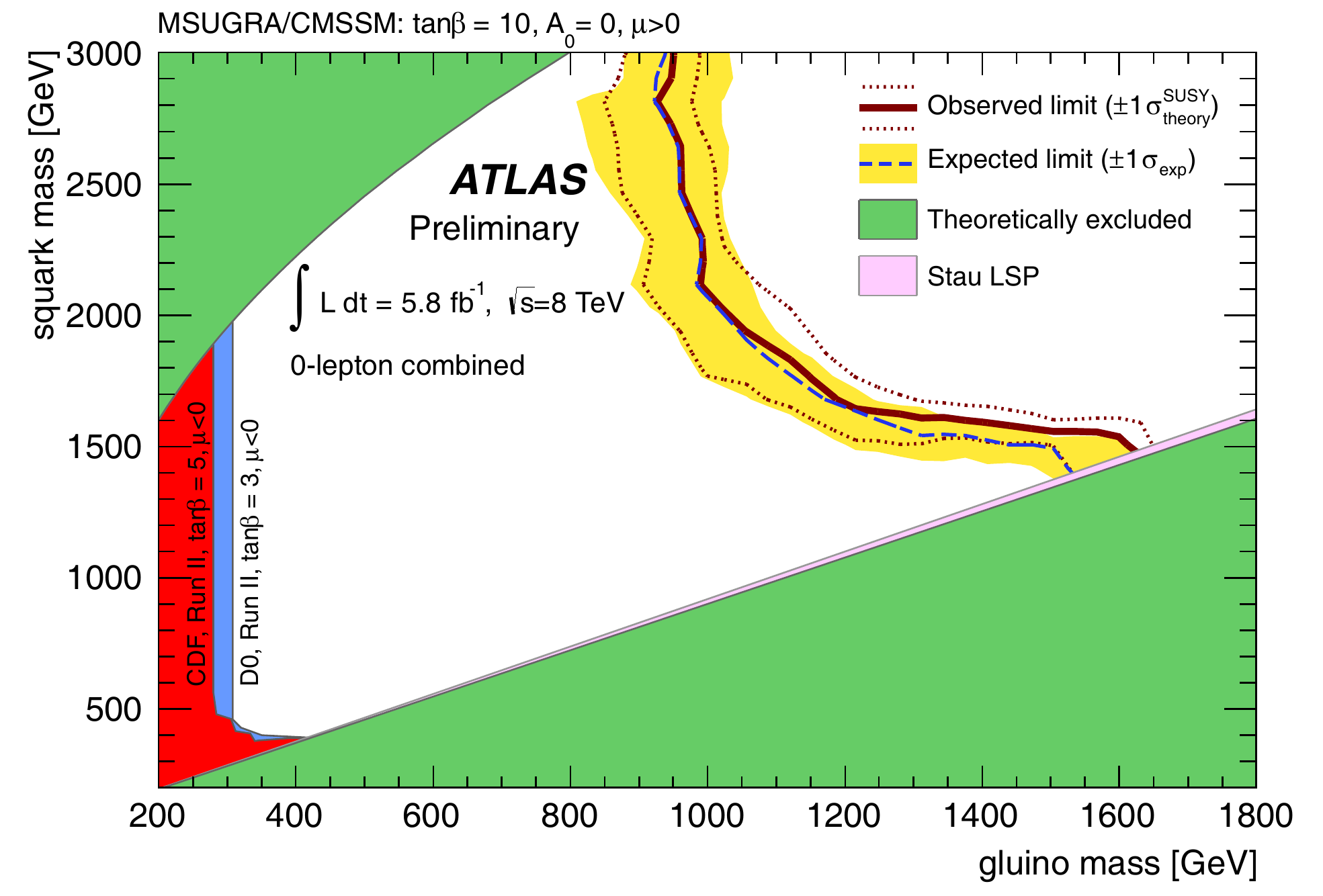}
\end{center}
\caption{95\% CL exclusion limits for MSUGRA/CMSSM models with
$\tan\beta = 10$, $A_0 = 0$ and $mu > 0$ presented in the
$m_0 - m_{1/2}$ plane (left) and in the $m_{gluino}-m_{squark}$
plane (right). The blue dashed lines show the expected limits at
95\% CL, with the light (yellow) bands indicating the $1\sigma$
excursions due to experimental uncertainties. The observed
limits are indicated by medium (maroon) curves. Previous results
from ATLAS~\cite{ATLAS_old} are represented by the shaded (light
blue) area. The theoretically excluded regions (green and blue)
are described in Ref.~\cite{matchev-remington}.}
\label{ATLAS-CONF-2012-109-fig_06}
\end{figure}\vspace{-0.3cm}

\begin{figure}[t]
\begin{center}
\leavevmode
\includegraphics[width=0.98\textwidth]{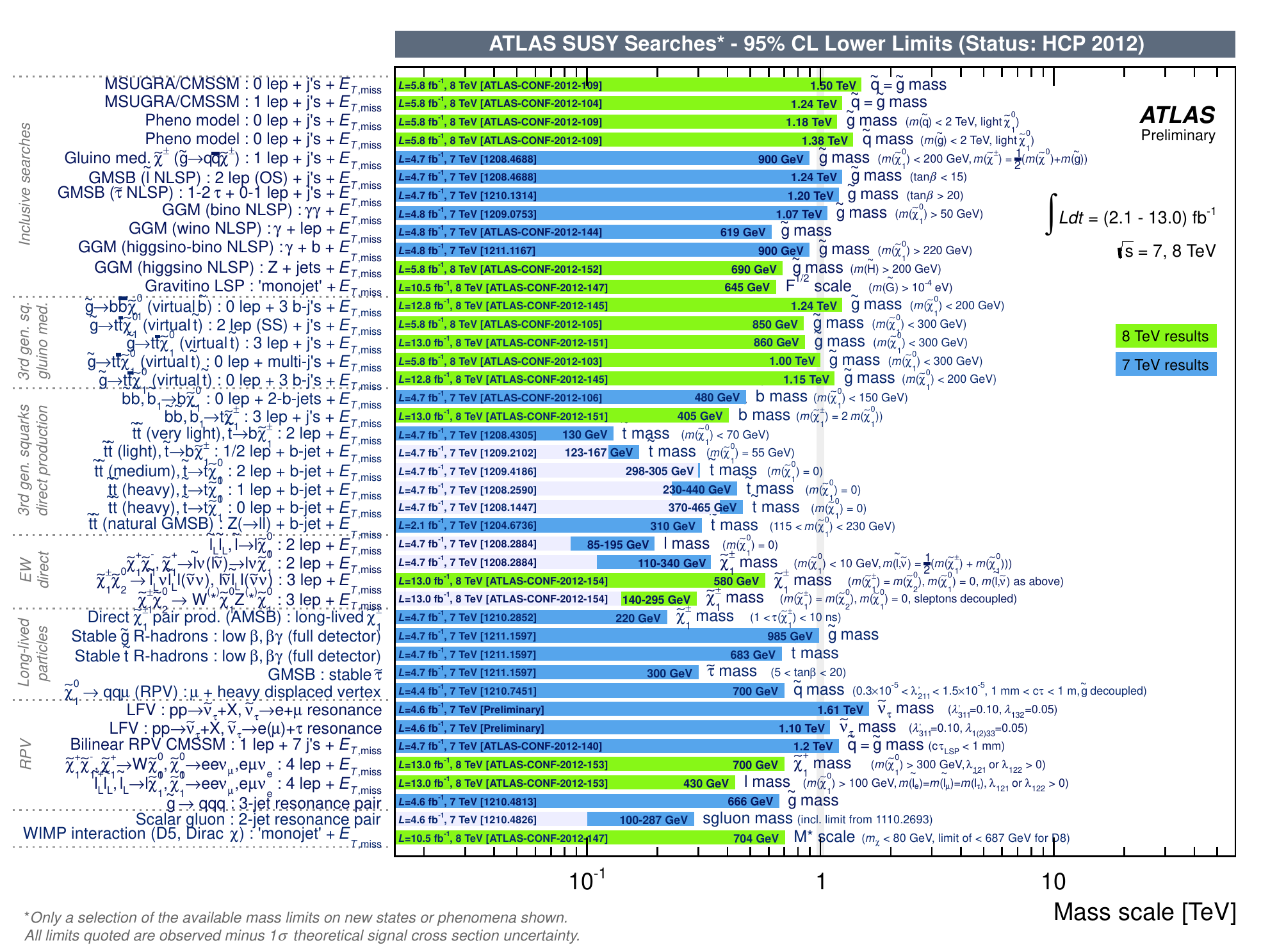}
\vspace*{-5mm}
\end{center}
\caption{Mass reach of ATLAS searches for supersymmetry
(representative selection).}
\label{fig:AtlasSearches_susy12}
\end{figure}

\begin{figure}[t]
\begin{center}
\leavevmode
%\vspace*{10mm}
\raisebox{6pt}{\includegraphics[width=0.4\textwidth]{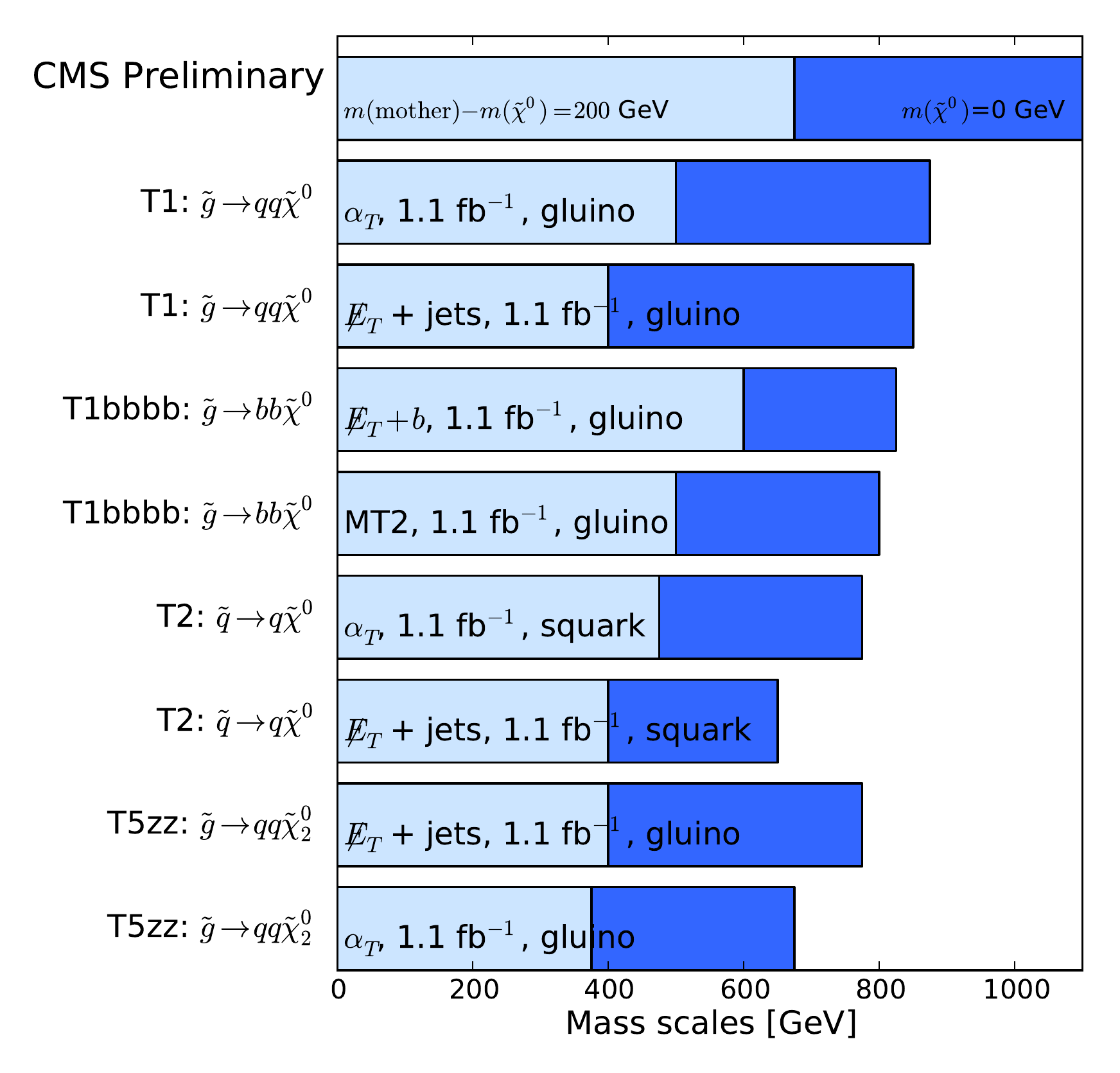}}
\includegraphics[width=0.58\textwidth]{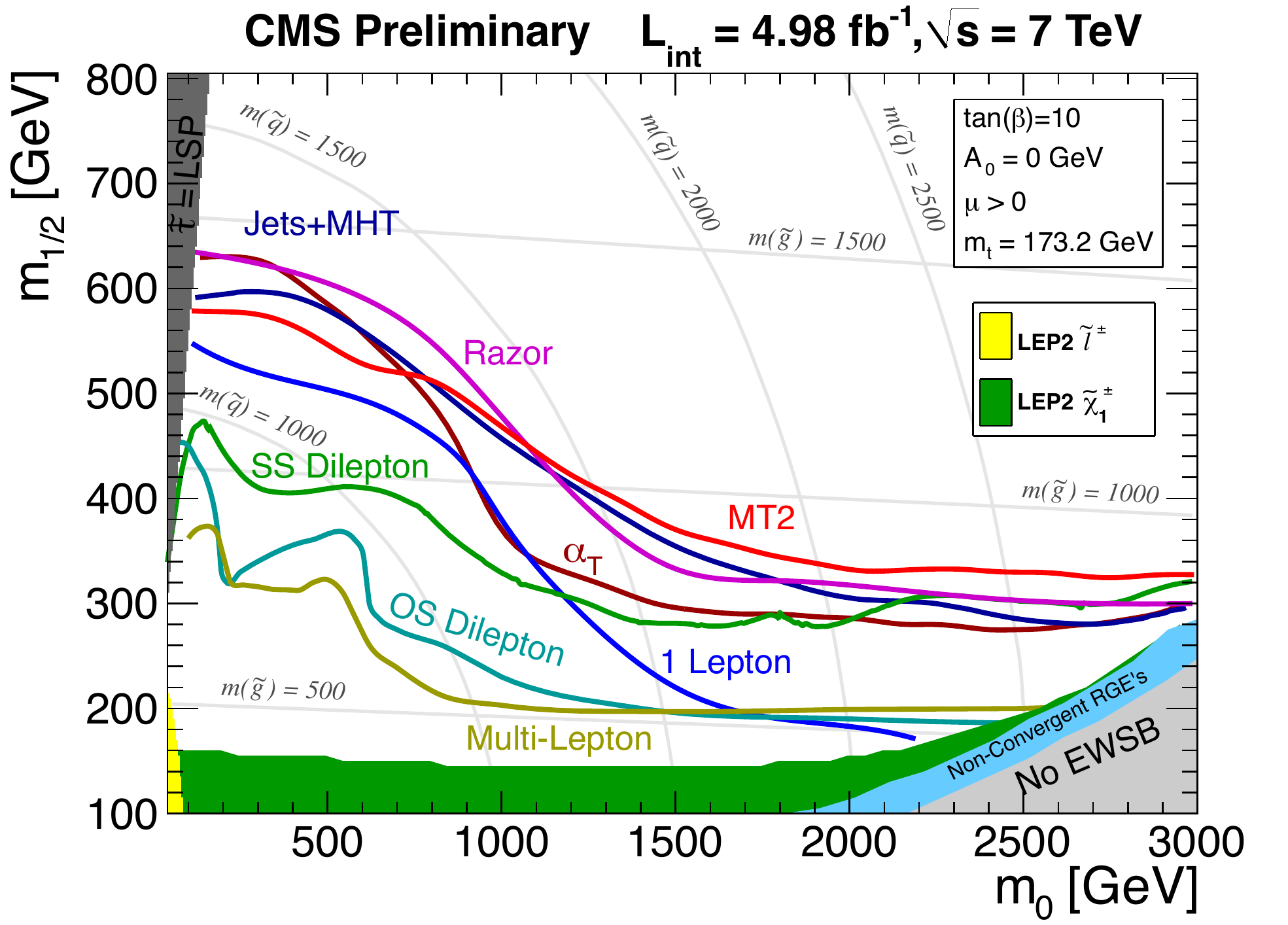}
\vspace*{-3mm}
\end{center}
\caption{Left: Best exclusion limits for the gluino and squark
masses, for $m_{\chi^0}$ = 0~GeV (dark blue) and
$m(mother) - m_{\chi^0}$ = 200~GeV (light blue), for each
topology, for the hadronic results. Right: Observed limits from
several 2011 CMS SUSY searches plotted in the CMSSM
$(m_0-m_{1/2})$ plane.}
\label{fig:CMSSearches_susy12}
\end{figure}

All the exclusion plots discussed above can give direct limits
on the masses of supersymmetric particles under certain
assumptions (mass relations, dominant decay channels, modified
or/and simplified models, etc.). The latest mass limits for the
different models and final state channels obtained by ATLAS are
shown in Fig.~\ref{fig:AtlasSearches_susy12}~\cite{ATLAS_SUSY}.
Fig.~\ref{fig:CMSSearches_susy12}~\cite{CMS_SUSY_pub,CMS_SUSY_web}
shows the best exclusion limits of the CMS collaboration for
4.98~fb$^{-1}$ data and $\sqrt{s}= 7$~TeV as well as observed
limits plotted in the CMSSM $(m_0 - m_{1/2})$ plane.

\clearpage
%-------------------
\section{Conclusion}

Supersymmetry remains the most popular extension of the Standard
Model. Comparison of the MSSM with precision experimental data
 is as good as  for the SM. At the same time, supersymmetry stabilizes the SM
due to the cancellation of quadratic divergences to the Higgs boson mass. 
The prediction of the Higgs boson mass in the MSSM in the region indicated by experimental data
 can be also considered as an argument if favour of supersymmetry. Besides,
 the relic density of the DM is not described in the SM but is naturally explained in the MSSM. 
 What is remarkable, the cross section of neutralino annihilation happens to be precisely equal to what is needed for a correct relic density.
 
 Constrained MSSM with a few free parameters seems to satisfy all experimental and theoretical requirements, though recently some tension with the light Higgs  boson mass has appeared. The natural  way out
would be either to release some constraints thus introducing more free parameters or to extend the minimal model, for instance, enlarging the Higgs sector like in the NMSSM.  Since it is not clear which model might be correct, all possibilities are open. Unfortunately, there is no "model independent" way
of describing SUSY searches, as well as a "smoking gun" process for SUSY except for the discovery of superpartners in the events with missing transverse energy.

 Today after 40 years since the invention of supersymmetry
we have no single convincing evidence that supersymmetry is
realized in particle physics. Still it remains very popular in quantum
field theory and in string theory due to its exceptional
properties but needs experimental justification.

Let us remind the main \emph{pros} and \emph{contras} for
supersymmetry in particle physics

\emph{Pro}:

$\bullet$  Provides natural framework for unification with gravity

$\bullet$  Leads to gauge coupling unification (GUT)

$\bullet$  Solves the hierarchy problem

$\bullet$  Is a solid quantum field theory

$\bullet$  Provides natural candidate for the WIMP cold DM

$\bullet$  Predicts new particles and thus generates new job positions

\emph{Contra}:

$\bullet$ Does not shed new light on the problem of

\ \ \ $*$  Quark and lepton mass spectrum

\ \ \ $*$ Quark and lepton mixing angles

\ \ \ $*$ the origin of CP violation

\ \ \ $*$ Number of flavours

\ \ \ $*$  Baryon asymmetry of the Universe

$\bullet$ Doubles the number of particles

Low energy supersymmetry promises us that new physics is
round the corner at the TeV scale to be exploited at colliders
and astroparticle experiments of this decade. If our
expectations are correct, very soon we will face new
discoveries, the whole world of supersymmetric particles
will show up and the table of fundamental particles will be
enlarged in increasing rate.  This would be a great step in
understanding the microworld. 

Coming back to the question in the title of these lectures, whether SUSY is alive or not,
we can say that so far the parameter space of SUSY models is large enough to incorporate all data.
Slight tension that appears in particular models can be removed by extension of a model. However, there exist some broad prediction of low energy SUSY that is falsifiable. This is the presence of superpartners at 
TeV scale.  At least some of them should be light enough to be discovered at the LHC at full energy run at 14 TeV. Otherwise, if the scale of SUSY exceeds several TeV, we loose the main arguments in favour of
low energy supersymetry, namely, the unification of the gauge couplings and the solution of the hierarchy problem.  Then the need for a low energy supersymmetry becomes questionable and the possibilities to test it become hardly feasible. The future will show whether we
are right in our expectations or not.

\vspace{0.3cm} {\large \bf Acknowledgements} \vspace{0.1cm}

The authors would like to express their gratitude to the
organizers of the School for their effort in creating a pleasant
atmosphere and support. This work was partly supported by
RFBR grant \# 11-02-01177 and Russian Ministry of Education and
Science grant \# 3802.2012.2.


\begin{thebibliography}{100}
\addcontentsline{toc}{section}{~~~~References}

\bibitem{super}%--- 1
Y. A. Golfand and E. P. Likhtman,
{\em JETP Letters} {\bf 13} (1971) 452;\\
D. V. Volkov and V. P. Akulov,
{\em JETP Letters} {\bf 16} (1972) 621;\\
J. Wess and B. Zumino,
{\em Phys. Lett.} {\bf B49} (1974) 52.

\bibitem{Rev}%--- 2
P. Fayet and S. Ferrara,
{\em Phys. Rep.} {\bf 32} (1977) 249;\\
M. F. Sohnius,
{\em Phys. Rep.} {\bf 128} (1985) 41;\\
H. P. Nilles,
{\em Phys. Rep.} {\bf 110} (1984) 1;\\
H. E. Haber and G. L. Kane,
{\em Phys. Rep.} {\bf 117} (1985) 75;\\
A. B. Lahanas and D. V. Nanopoulos,
{\em Phys. Rep.} {\bf 145} (1987) 1.

\bibitem{WessB}%--- 3
J. Wess and J. Bagger,
{\em Supersymmetry and Supergravity}, Princeton Univ. Press, 1983.

\bibitem{sspace}%--- 4
A. Salam, J. Strathdee, {\em Nucl. Phys.}  {\bf B76} (1974) 477;\\
S. Ferrara, J. Wess, B. Zumino, {\em Phys. Lett.} {\bf BS1} (1974) 239.

\bibitem{Books}%--- 5
S. J. Gates, M. Grisaru, M. Ro\v{c}ek and W. Siegel,
{\em Superspace or One Thousand and One Lessons in
Supersymmetry}, Benjamin \& Cummings, 1983;\\
P. West, {\em Introduction to supersymmetry and supergravity},
World Scientific, 1990;\\
S. Weinberg,{\em  The quantum theory of fields},  Vol. 3, Cambridge,
UK: Univ. Press, 2000.

\bibitem{theorem}%--- 6
S. Coleman and J .Mandula,
{\em Phys. Rev.} {\bf 159} (1967) 1251.

\bibitem{GUT}%--- 7
G. G. Ross,
{\em Grand Unified Theories}, Benjamin \& Cummings, 1985.

\bibitem{SM}%--- 8
C. Amsler et al. (Particle Data Group),
{\em Phys. Lett.} {\bf B667} (2008) 1.

\bibitem{ABF}%--- 9
U. Amaldi, W. de Boer and H. F\"urstenau,
{\em Phys. Lett.} {\bf B260} (1991) 447.

\bibitem{rotcurve}%--- 10
Y. Sofue, V. Rubin,
{\em Ann. Rev. Astron. Astrophys.} {\bf 39} (2001) 137,
%\texttt{astro-ph/0010594}, 
and refs therein.

\bibitem{lensing}
C.S. Kochanek,
{\em Astrophys. J.} {\bf 453} (1995) 545;\\
N.Kaiser, G.Squires,
{\em Astrophys. J.} {\bf 404} (1993) 441.

\bibitem{WIMP}%--- 12
V.A. Ryabov, V.A. Tsarev and A.M. Tskhov\-rebov,
{\em Phys. Usp.} {\bf 51} (2008) 1091, and refs therein.

\bibitem{abun}%--- 13
G. Jungman, M. Kamionkowski and K. Griest,
{\em Phys. Rep.} {\bf 267} (1996) 195;\\
H. Goldberg, {\em Phys. Rev. Lett.} {\bf  50} (1983) 1419;\\
%J.R. Ellis, J.S. Hagelin, D.V. Nanopoulos, K.A. Olive, M. Srednicki,
J.R. Ellis, et al.,
{\em Nucl. Phys.} {\bf  B238} (1984) 453.

\bibitem{string}%--- 14
M.B.~Green, J.H.~Schwarz and E.~Witten,
{\em Superstring Theory}, Cambridge, UK: Univ. Press, 1987.
%{\it Cambridge Monographs On Mathematical Physics}.

\bibitem{ber}%--- 15
F.A. Berezin,
{\em The Method of Second Quantization}, Moscow, Nauka, 1965.

\bibitem{Peskin}%--- 16
M. Peskin and D. Schr\"{o}der,
%{\em "An Introduction to Quantum Field Theory"}, Addison-Wesley Pub. Company, 1995
{\em An Introduction to Quantum Field Theory}, Addison-Wesley, 1995.

\bibitem{HT}%--- 17
H. Baer and X. Tata,
{\em Weak Scale Supersymmetry}, Cambridge University Press, 2006.

\bibitem{MSSM}%--- 18
H.E. Haber, {\em Introductory Low-Energy Supersymmetry},
Lectures given at TASI 1992, (SCIPP 92/33, 1993),
\texttt{hep-ph/9306207};\\
D.I. Kazakov,
{\em Beyond the Standard Model (In search of supersymmetry)},
%Lectures at the European school on high energy physics 2000, CERN-2001-003,
Lectures at the ESHEP 2000, CERN-2001-003,
\texttt{hep-ph/0012288};\\
D. I. Kazakov,
\textit{Beyond the Standard Model}, 
%Lectures at the European school on high energy physics 2004, 
Lectures at the ESHEP 2004, 
\texttt{hep-ph/0411064}.

\bibitem{SUSYLHC_GK}
A.V. Gladyshev, D.I. Kazakov,
\emph{Supersymmetry and LHC}, Phys. Atom. Nucl. \textbf{70} (2007) 1553,
\texttt{hep-ph/0606288}.

\bibitem{shadow}%--- 19
\texttt{http://atlasinfo.cern.ch/Atlas/documentation/EDUC/physics14.html}

\bibitem{r-symmetry}%--- 20
P. Fayet, {\em Nucl. Phys.} {\bf B90} (1975) 104;\\
A. Salam and J. Srathdee, {\em Nucl. Phys.} {\bf B87} (1975) 85.

\bibitem{Fayet}%--- 21
P. Fayet and J. Illiopoulos,
{\em Phys. Lett.} {\bf B51} (1974) 461.

\bibitem{O'R}%--- 22
L. O'Raifeartaigh,
{\em Nucl.Phys.} {\bf B96} (1975) 331

\bibitem{hidden}%--- 23
L. Hall, J. Lykken and S. Weinberg,
{\em Phys. Rev.} {\bf D27} (1983) 2359;\\
S.K. Soni and H.A. Weldon,
{\em Phys. Lett.} {\bf B126} (1983) 215;\\
I. Affleck, M. Dine and N. Seiberg,
{\em Nucl. Phys.} {\bf B256} (1985) 557.

\bibitem{gravmed}%--- 24
H. P. Nilles,
{\em Phys. Lett.} {\bf B115} (1982) 193;\\
A.H. Chamseddine, R. Arnowitt and P. Nath,
{\em Phys. Rev. Lett.} {\bf 49} (1982) 970;\\
A.H. Chamseddine, R. Arnowitt and P. Nath,
{\em Nucl. Phys.} {\bf B227} (1983) 121;\\
R. Barbieri, S. Ferrara and C. A. Savoy,
{\em Phys. Lett.} {\bf B119} (1982) 343.

\bibitem{gaugemed}%--- 25
M. Dine and A.E. Nelson,
{\em Phys. Rev.} {\bf D48} (1993) 1277;\\
M. Dine, A.E. Nelson and Y. Shirman,
{\em Phys. Rev.} {\bf D51} (1995) 1362.

\bibitem{anommed}%--- 26
L. Randall and R. Sundrum,
{\em Nucl. Phys}. {\bf B557} (1999) 79;\\
G.F.~Giudice, M.A.~Luty, H.~Murayama and R.~Rattazzi,
{\em JHEP}, {\bf 9812} (1998) 027.

\bibitem{gauginomed}%--- 27
D.E.~Kaplan, G.D.~Kribs and M.~Schmaltz,
{\em Phys. Rev.}  {\bf D62} (2000) 035010;\\
Z.~Chacko, M.A.~Luty, A.E.~Nelson and E.~Ponton,
{\em JHEP}, {\bf 0001} (2000) 003.

\bibitem{spectrum}%--- 28
G.G. Ross and R.G. Roberts,
{\em Nucl. Phys.} {\bf B377} (1992) 571;\\
V. Barger, M.S. Berger and P. Ohmann,
{\em Phys. Rev.} {\bf D47} (1993) 1093.

\bibitem{BEK}%--- 29
W. de Boer, R. Ehret and D. Kazakov,
{\em Z. Phys.} {\bf C67} (1995) 647;\\
W. de Boer et al., {\em Z. Phys.} {\bf C71} (1996) 415.

\bibitem{Ibanez}%--- 30
L.E. Ib\'a\~nez, C. Lop\'ez and C. Mu\~noz,
{\em Nucl. Phys.} {\bf B256} (1985) 218.

\bibitem{Barger}%--- 31
V. Barger, M.S. Berger and P. Ohman,
{\em Phys. Rev.} {\bf D49} (1994) 4908.

\bibitem{bbog}%--- 32
V. Barger, M.S. Berger, P. Ohmann and R. Phillips,
{\em Phys.~Lett.} {\bf B314} (1993) 351;\\
P.~Langacker and N.~Polonsky,
{\em Phys. Rev.} {\bf D49} (1994) 1454;\\
S.~Kelley, J.L. Lopez and D.V. Nanopoulos,
{\em Phys.~Lett.} {\bf B274} (1992) 387.

\bibitem{CW+we}%--- 33
M. Carena, M. Quiros and C.E.M. Wagner,
{\em Nucl. Phys.} {\bf B461} (1996) 407;\\
A.V. Gladyshev, D.I. Kazakov, W. de Boer, G. Burkart, R. Ehret,
%A.V. Gladyshev, et al.,
{\em Nucl. Phys.} {\bf  B498} (1997) 3;\\
A.V. Gladyshev, D.I. Kazakov,
{\em Mod. Phys. Lett.} {\bf A10} (1995) 3129.

\bibitem{feynhiggs}%--- 34
S. Heinemeyer, W. Hollik and G. Weiglein,
{\em Phys. Lett.} {\bf B455} (1999) 179;\\
S. Heinemeyer, W. Hollik and G. Weiglein,
{\em Eur. Phys. J.} {\bf C9} (1999) 343.

\bibitem{ABDM}%--- 35
A. Arbey, M. Battaglia, A. Djouadi, F. Mahmoudi,
\emph{JHEP} \textbf{1209} (2012) 107.

\bibitem{ABDMQ}%--- 36
A. Arbey, M. Battaglia, A. Djouadi, F. Mahmoudi, J. Quevillon,
%A. Arbey, et al.,
\emph{Phys. Lett.} \textbf{B708} (2012) 162.

\bibitem{Komatsu:2010fb}%--- 37
E. Komatsu et~al.,
\emph{Astrophys. J. Suppl.} \textbf{192} (2011) 18.

\bibitem{hfag}%--- 38
\texttt{http://www.slac.stanford.edu/xorg/hfag/rare/ichep10/radll/OUTPUT/TABLES/}
\texttt{radll.pdf}

\bibitem{Bennett:2006fi}%--- 39
Muon G-2 Collaboration,
\textit{Phys. Rev.} \textbf{D73} (2006) 072003.

\bibitem{Aaij:2012ac}%--- 40
LHCb collaboration,
\textit{Phys. Rev. Lett.} \textbf{108} (2012) 231801

\bibitem{Schael:2006cr}%--- 41
ALEPH Collaboration, DELPHI Collaboration, L3 Collaboration,
OPAL Collaborations, LEP WG for Higgs Boson Searches
Collaboration, \textit{Eur. Phys. J.} \textbf{ C47} (2006) 547.

\bibitem{Chatrchyan:2012vp}%--- 42
CMS Collaboration, \textit{Phys. Lett.} \textbf{ B713} (2012) 68.

%\bibitem{Aad:2011rv}%--- 43
ATLAS Collaboration, \emph{Search for neutral MSSM Higgs
bosons decaying to tau tau pairs in proton-proton collisions at
7~TeV with the ATLAS detector}, \texttt{arXiv:1107.5003}.

\bibitem{ATLAS-CONF-2012-033}%--- 44
ATLAS Collaboration, \emph{Search for squarks and gluinos 
using final states with jets and missing transverse momentum 
with the ATLAS detector in $\sqrt{s}$ = 7~TeV proton-proton 
collisions}, Technical Report ATLAS-CONF-2012-033, CERN, 
Geneva, Mar, 2012.

\bibitem{CMS-PAS-SUS-12-005}%---45
ATLAS Collaboration, \emph{Search for supersymmetry with the 
razor variables at CMS}, CMS-PAS-SUS-12-005

\bibitem{Aprile:2011hi}%---46
E. Aprile, et~al.,
\textit{Phys. Rev. Lett.} \textbf{107} (2011) 131302.

\bibitem{Misiak}%--- 47
M. Misiak and M. Steinhauser, {\em Nucl. Phys.} {\bf  B 764}
(2007)  62, [hep-ph/0609241]; {\em Nucl. Phys.} {\bf B 840} (2010) 271,
[1005.1173].

\bibitem{bsgsusy}%--- 48
W. de Boer, H.J. Grimm, A. Gladyshev, D. Kazakov,
{\em Phys. Lett.} {\bf B438} (1998) 281;\\
W. de Boer, M. Huber, A. Gladyshev, D. Kazakov,
{\em Eur. Phys. J.} {\bf C20} (2001) 689;\\
W. de Boer, M. Huber, A. Gladyshev, D. Kazakov,
\emph{The $b \to X(s) \gamma$ decay rate in NLO, Higgs boson 
limits, and LSP masses in the Constrained Minimal 
Supersymmetric Model}, \texttt{hep-ph/0007078},  and 
refs therein

\bibitem{bsmumu_theor}%--- 50
A. J. Buras, J. Girrbach, D. Guadagnoli, and G. Isidori,
\emph{Eur. Phys. J.} \textbf{C72} (2012) 2172.

\bibitem{bmu}%--- 51
F. Abe, et al. [CDF Collaboration],
{\em Phys. Rev.} {\bf  D57} (1998) 3811.

\bibitem{bsmumu_exp_hcp2012}%--- 52
R. Aaij, et al. [The LHCb Collaboration] , \emph{First 
evidence for the decay $B^0_s \to \mu^+ \mu^-$}, 
CERN-PH-EP-2012-335, LHCb-PAPER-2012-043, 
\texttt{arXiv:1211.2674 [hep-ex]}.

\bibitem{Bobeth:2001sq}%--- 53
C. Bobeth, T. Ewerth, F. Kruger and J. Urban, {\em Phys. Rev}. {\bf D 64} (2001) 074014.

\bibitem{bmususy}%--- 54
R.L. Arnowitt, B. Dutta, T. Kamon, and M. Tanaka,
{\em Phys. Lett.} {\bf B538} (2002) 121.

\bibitem{Carena:1999py}%--- 55
M.S. Carena, D. Garcia, U. Nierste et al., 
\emph{Nucl. Phys.} \textbf{B577} (2000) 88.

\bibitem{JN}%--- 56
F.Jegerlehner, A.Nyffeler, {\em Phys.Rep.} {\bf 477} (2009) 1-110, arXiv: 0902.3360.

 \bibitem{DHM} 
 M.Davier, A.Hecker, B.Malaescu, Z.Zhang, {\em Eut.Phys.J.} {\bf C71} (2011) 1515, arXiv: 1010.4180.

\bibitem{Lopez}%--- 57
J.L. Lopez, D. V. Nanopoulos, Xu Wang, {\em Phys.Rev.} {\bf D49} (1994) 366, hep- ph/9308336

\bibitem{CM}%--- 58
A. Czarnecki and W. Marciano,
{\em Phys. Rev.} {\bf D64} (2001) 013014.

\bibitem{Anom}%--- 59
W. de Boer, M. Huber, C. Sander, D.I. Kazakov,
{\em Phys. Lett.} {\bf B515} (2001) 283.

\bibitem{Beskidt:2010va}%--- 60
C.~Beskidt, et~al.,
\textit{Phys. Lett.} \textbf{B695} (2011) 143.

\bibitem{atlashiggs}%--- 61
ATLAS Collaboration, Note ATL-PHYS-PUB-2010-011,\\ 
\texttt{http://cdsweb.cern.ch/record/1279115/files/ATL-PHYS-PUB-2010-011.pdf}.

\bibitem{tevatronhiggs}%--- 62 
D. Benjamin, et al. [Tevatron New Phenomena \& Higgs Working Group],\\ 
\texttt{arXiv:1003.3363 [hep-ex]}.

\bibitem{cmshiggs}%--- 63 
G.L. Bayatian, et al. [CMS Collaboration], 
\emph{J. Phys.} \textbf{G34} (2007) 995.

\bibitem{Chatrchyan:2011nx}%--- 64
CMS Collaboration, 
\textit{Phys. Rev. Lett.} \textbf{106} (2011) 231801.

\bibitem{Chatrchyan:2012tx}%--- 65
CMS Collaboration,
\textit{Phys. Lett.} \textbf{B710} (2012) 26.

\bibitem{ATLAS:2012ae}%--- 66
ATLAS Collaboration, 
\emph{Combined search for the Standard Model Higgs boson in $pp$ 
collisions at $\sqrt{s}$ = 7~TeV with the ATLAS detector}, 
\texttt{arXiv:1207.0319}.

\bibitem{WMAP}%--- 67
C.L. Bennett, et al.,
{\em Astrophys. J. Suppl.} \textbf{148} (2003) 1;\\
D.N. Spergel et al.,
{\em Astrophys. J. Suppl.} \textbf{148} (2003) 175.

\bibitem{Kolb}%--- 68
E. Kolb and M.S. Turner,
{\em The Early Universe}, Frontiers in Physics, Addison Wesley, 1990.

\bibitem{Rub}%--- 69
D.S. Gorbunov, V.A. Rubakov,
{\em Introduction to the theory of the early Universe},
Moscow, URSS, 2008 (in Russian).

\bibitem{LEP}
Delphi Collab., 
{\em Eur. Phys. J.} {\bf C1} (1998) 1;\\
L3 Collab., 
{\em Phys. Lett.} {\bf B472} (2000) 420.

\bibitem{egret}
\texttt{http://heasarc.gsfc.nasa.gov/docs/cgro/egret/}

\bibitem{fermi}
\texttt{http://fermi.gsfc.nasa.gov/}

\bibitem{DAMA}
R. Bernabei, et al. [DAMA Collaboration],
{\em Eur. Phys. J.} {\bf C56} (2008) 333.

\bibitem{CDMS}
Z. Ahmed, et al. [CDMS Collaboration],
{\em Science} {\bf 327} (2010) 1619;\\
Z. Ahmed, et al. [CDMS Collaboration],
{\em Phys. Rev.} {\bf D81} (2010) 042002.

\bibitem{Ellis:2008hf}
J.~R. Ellis, K.~A. Olive, and C.~Savage, 
\textit{Phys. Rev.} \textbf{D77} (2008) 065026.

\bibitem{Belanger:2008sj}
G.~Belanger, F.~Boudjema, A.~Pukhov, et~al., 
\textit{Comput. Phys. Commun.} \textbf{180} (2009) 747.

\bibitem{Cao:2010ph}
J.~Cao, K.-i. Hikasa, W.~Wang et~al., 
\textit{Phys. Rev.} \textbf{D82} (2010) 051701.

\bibitem{Alarcon:2011zs}
J.~M.~Alarcon, J.~Martin Camalich and J.~A.~Oller,
\emph{Phys. Rev.} {\bf D85} (2012) 051503.

\bibitem{Weber:2009pt}
M.~Weber and W.~de~Boer,
\textit{Astron. Astrophys.} \textbf{509} (2010) A25.

\bibitem{deBoer:2010eh}
W.~de~Boer and M.~Weber,
\textit{JCAP} \textbf{1104} (2011) 002.

\bibitem{Salucci:2010qr}
P.~Salucci, F.~Nesti, G.~Gentile and C.~F.~Martins,
\emph{Astron. Astrophys}. {\bf 523} (2010) A83.

\bibitem{Catena:2009mf}
R.~Catena and P.~Ullio,
\emph{JCAP} {\bf 1008} (2010) 004.

\bibitem{unexpLHC}
D.Yu. Bogachev, A.V. Gladyshev, D.I. Kazakov, A.S. Nechaev,
%D.Yu. Bogachev, et al.,
{\em Int. J. Mod. Phys.} {\bf A21} (2006) 5221.

\bibitem{LEPSUSY}
ALEPH Collaboration,
{\em Phys. Lett.} {\bf B499} (2001) 67.

\bibitem{tevatron}
D. Acosta, et al. [CDF Collaboration],
{\em Phys. Rev. Lett.} {\bf 90} (2003) 251801;\\
T. Affolder et al. [CDF Collaboration],
{\em Phys. Rev. Lett.} {\bf 87} (2003) 251803;\\
T.Kamon, Proc. of IX Int. Conf. "SUSY-01", WS 2001, p.196,
\texttt{hep-ex/0301019}.

\bibitem{NMSSM}
U. Ellwanger, C. Hugonie, A.M. Teixeira, 
\emph{Phys. Rept.} \textbf{496} (2010) 1, and refs therein;\\
M. Maniatis, 
\emph{Int. J. Mod. Phys.} \textbf{A25} (2010) 3505, and refs therein. 

\bibitem{atlas_lumi}
\texttt{https://twiki.cern.ch/twiki/bin/view/AtlasPublic/LuminosityPublicResults}

\bibitem{cms_lumi}
\texttt{https://twiki.cern.ch/twiki/bin/view/CMSPublic/LumiPublicResults}


\bibitem{LHCSUSY1}
F.E. Paige,
\emph{SUSY Signatures in ATLAS at LHC}, \texttt{hep-ph/0307342};\\
D.P. Roy,
{\em Acta Phys. Polon.} {\bf B34} (2003) 3417;\\
D.R. Tovey,
{\em Phys. Lett} {\bf B498} (2001) 1;\\
H. Baer, C.Balaz, A. Belyaev, T. Krupov\-nickas and X. Tata,
{\em JHEP} {\bf 0306} (2003) 054;\\
G. Belanger, F. Boudjema, F. Donato, R. Godbole and S. Rosier-Lees,
%G. Belanger, et al.,
{\em Nucl. Phys.} {\bf B581} (2000) 3.

\bibitem{LHCSUSY2}
O. Buchmueller  et al,  {\em Eur.Phys.J.} {\bf C72} (2012) 1878, arXiv:1110.3568 [hep-ph];\\
 S. Sekmen et al,  {\em JHEP} {\bf 1202} (2012) 075, arXiv:1109.5119 [hep-ph];\\ 
 S.S. AbdusSalam et al, {\em Eur.Phys.J.} {\bf C71} (2011) 1835, arXiv:1109.3859 [hep-ph];\\
 H.Baer, V.Barger, P.Huang, A.Mustafayev, {\em Phys.Rev.} {\bf D84} (2011) 091701, \\
 arXiv:1109.3197 [hep-ph];\\
  S. Heinemeyer,  {\em SUSY Predictions for and from the LHC}, arXiv:1103.0952 [hep-ph]; \\
  B.Altunkaynak, M.Holmes, P.Nath, B.D.Nelson, G. Peim, {\em Phys.Rev.} {\bf D82} (2010) 115001, 
 arXiv:1008.3423 [hep-ph];\\
 N.V. Krasnikov, V.A. Matveev, {\em Phys. Atom. Nucl.} {\bf 73} (2010) 191;\\
  M. Spiropulu, {\em Eur.Phys.J.} {\bf C59} (2009) 445;\\
   D.P. Roy, {\em Acta Phys.Polon.} {\bf B34} (2003) 3417,  hep-ph/0303106 ;\\
F. E. Paige,{\em  Czech.J.Phys.} {\bf 55} (2005) B185, hep-ph/0211017;\\ 
M.Dittmar, {\em SUSY discovery strategies at the LHC}, hep-ex/9901004;\\ 
MSSM Working Group, A. Djouadi et al., hep-ph/9901246.

\bibitem{LHC} http://CMSinfo.cern.ch/Welcome.html- /CMSdocuments/CMSplots

\bibitem{ATLAS_SUSY_Gtt_1}
ATLAS collaboration, 
\emph{Search for gluino pair production in events with missing 
transverse momentum and at least three b-jets using 13.0 fb$^-1$
of pp Collisions at $\sqrt{s} = 8$~TeV with the ATLAS Detector}, 
ATLAS-CONF-2012-145.

\bibitem{ATLAS_SUSY_Gtt_2}
ATLAS collaboration, 
\emph{Search for supersymmetry in events with three leptons, 
multiple jets, and missing transverse momentum in 13.0 fb$^-1$ 
of pp collisions with the ATLAS detector at $\sqrt{s} = 8$~TeV}, 
ATLAS-CONF-2012-151.

\bibitem{ATLAS_SUSY_Gtt_3}
ATLAS collaboration, 
\emph{Search for new phenomena using large jet multiplicities 
and missing transverse momentum with ATLAS in 5.8 fb$^{-1}$ of 
$\sqrt{s}$=8~TeV proton-proton collisions}, 
ATLAS-CONF-2012-103.

\bibitem{ATLAS_SUSY_Gtt_4}
ATLAS collaboration, 
\emph{Search for Supersymmetry in final states with two
same-sign leptons, jets and missing transverse momentum with
the ATLAS detector in pp collisions at $\sqrt{s}$=8~TeV}, 
ATLAS-CONF-2012-105.

\bibitem{ATLAS_stop1}
ATLAS collaboration, 
\emph{Search for light scalar top quark pair production in 
final states with two leptons with the ATLAS detector in 
$\sqrt{s}$ = 7~TeV proton-proton collisions}, 
\texttt{arxiv:1208.4305}.

\bibitem{ATLAS_stop2}
ATLAS collaboration, 
\emph{Search for light top squark pair production in final 
states with leptons and b-jets with the ATLAS detector in 
$\sqrt{s}$ = 7~TeV proton-proton collisions}, 
\texttt{arxiv:1209.2102}

\bibitem{ATLAS_stop3}
ATLAS collaboration,
\emph{Search for a supersymmetric partner to the top quark 
in final states with jets and missing transverse momentum at 
$\sqrt{s}$ = 7~TeV with the ATLAS detector},
\texttt{arXiv:1208.1447} (Accepted by PRL).

\bibitem{ATLAS_stop4}
ATLAS collaboration,
\emph{Search for direct top squark pair production in final 
states with one isolated lepton, jets, and missing transverse 
momentum in $\sqrt{s}$ = 7~TeV pp collisions using 4.7~fb$^{-1}$ 
of ATLAS data}, \texttt{arXiv:1208.2590} (Accepted by PRL).

\bibitem{ATLAS_stop5}
ATLAS collaboration,
\emph{Search for direct top squark pair production in final 
states with one isolated lepton, jets, and missing transverse 
momentum in $\sqrt{s}$ = 7~TeV pp collisions using 4.7~fb$^{-1}$ 
of ATLAS data}, \texttt{arXiv:1209.4186}.

\bibitem{ATLAS_old}
ATLAS Collaboration, 
\emph{Search for squarks and gluinos with the ATLAS detector in 
final states with jets and missing transverse momentum using 4
.7~fb$^{-1}$ of $\sqrt{s}=7$~TeV proton-proton collision data},
\texttt{arXiv:1208.0949}.

\bibitem{matchev-remington}
K. Matchev and R. Remington, 
\emph{Updated templates for the interpretation of LHC results on 
supersymmetry in the context of mSUGRA}, 
\texttt{arXiv:1202.6580}.

\bibitem{ATLAS_SUSY}
\texttt{https://twiki.cern.ch/twiki/bin/view/AtlasPublic/CombinedSummaryplots}

\bibitem{CMS_SUSY_pub}
CMS Collaboration, 
\emph{Interpretation of Searches for Supersymmetry with 
Simplified Models}, CMS-PAS-SUS-11-016.

\bibitem{CMS_SUSY_web}
\texttt{https://twiki.cern.ch/twiki/bin/view/CMSPublic/SUSYSMSSummaryplots}
%
%\bibitem{BSK}
%W. de Boer and C. Sander,
%{\em Phys. Lett.} {\bf B585} (2004) 276.
\end{thebibliography}
\end{document}